\documentclass{article}
\usepackage{graphicx} 
\usepackage{multirow}
\usepackage{textcomp}
\usepackage{hhline}
\usepackage{tabularx}
\usepackage[flushleft]{threeparttable}
\usepackage[table]{xcolor}
\usepackage{caption}
\usepackage{subcaption}
\usepackage{comment}
\usepackage{siunitx}
\usepackage{minted}
\usepackage{float}
\usepackage[margin=2.7cm]{geometry}
\usepackage{amsmath}
\usepackage{amsfonts}
\usepackage{bm} 
\usepackage{booktabs,graphicx}

\usepackage{xcolor}
\definecolor{light-gray}{gray}{0.8}
\definecolor{apsblue}{rgb}{0.176, 0.152, 0.57}
\usepackage[colorlinks=true, linkcolor=black, citecolor=apsblue, filecolor=black, urlcolor=apsblue]{hyperref}
\usepackage{tikz}
\usetikzlibrary{calc, arrows.meta, angles, quotes}

\begin{document}

\begin{centering}
{\Large {\bf HALHF interim progress report (Pre-CDR)}}\\
\end{centering}
\vspace{20pt}
\begin{centering}
{\Large {\bf September 2026}}\\

\end{centering}

\begin{figure}[h]
	\centering\includegraphics[width=0.2\textwidth]{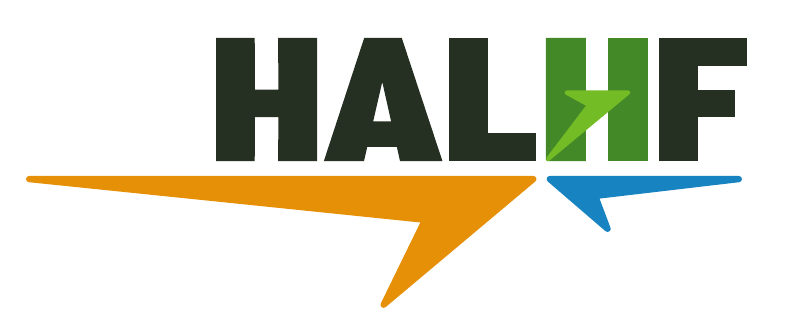}
\end{figure}

\noindent
The HALHF Collaboration: Erik Adli$^1$, Joshua Appleby$^2$, Timothy L. Barklow$^3$, Maria Enrica Biagini$^4$, Jonas Björklund Svensson$^5$, Mikael Berggren$^6$,  Lucy Bishop$^{7,6}$, Simon Bohlen-Blömers$^6$, 
Stewart Boogert$^8$, Philip Burrows$^2$, Allen Caldwell$^9$, Conrad Caliari$^{10}$, Jian Bin Ben Chen$^1$, Vera Cilento$^{2,10}$, 
Richard D’Arcy$^2$, Steffen Doebert$^{10}$, Juan Pablo Díaz$^6$, Wang Dou$^{11}$, Keegan Downham$^3$, Pierre Drobniak$^1$, Calvin Dyson$^{12}$, Sinead Farrington$^{13}$, John Farmer$^{9}$, Angeles Faus-Golfe$^{14}$, Marvin Flores$^{15}$, Manuel Formela$^6$, Arianna Formenti$^{16}$, Louis Forrester$^{12}$, Brian Foster$^{2,6}$, Jie Gao$^{11}$, Spencer Gessner$^3$, Niclas Hamann$^6$, Alexander Harrison$^2$, Eir Eline H{\o}rlyk$^1$, Maryam Huck$^6$, Daniel Kalvik$^1$, Lewis Kennedy$^{2,10}$, Alexander Knetsch$^6$, Antoine Laudrain$^6$, Rafael Legaspi$^{15}$, Remi Lehe$^{16}$, Xiaoping Li$^{11}$, Carl A. Lindstr{\o}m$^1$, Benno List$^6$, Jenny List$^6$, Mike Litos$^{17}$, Xueying Lu$^{18}$, Edward Mactavish$^{10}$, Vasyl Maslov$^6$, Ryan McGuigan$^{19}$, Lukas Müller$^6$, Ibrahim Najmuddin$^2$, Emilio Nanni$^3$, John Osborne$^{10}$, Jens Osterhoff$^{16}$, Fern Pannell$^{2}$, Trupen Parikh$^6$, Felipe Pe\~na$^{20,1}$,
Gudrid Moortgat Pick$^{21,6}$, Shriyansh Ranjan$^{22,6}$, J\"{u}rgen Reuter$^6$, 
Ted Schätzlein Liang$^6$, 
Dmitrii Samoilenko$^6$, Nela Sedlackova$^{12}$, Andrei Seryi$^{23}$, Kyrre Sjobak$^1$, Steinar Stapnes$^{1}$, Lodewyk Steyn$^{24}$, Rogelio Tom\'as Garc\'ia$^{10}$, Maxim Titov$^{25}$, Malte Trautwein$^{6}$, 
Maxence Th\'evenet$^6$, Nicholas J. Walker$^6$, Marc Wenskat$^6$, Matthew Wing$^{7,6}$, Jonathan Wood$^6$.\\
\\
$^1$ Department of Physics, University of Oslo, 0316 Oslo, Norway\\
$^2$ John Adams Institute for Accelerator Science at University of Oxford, Denys Wilkinson Building, Keble Road, Oxford, OX1 3RH, UK\\
$^3$ SLAC National Accelerator Laboratory, 2575 Sand Hill Road, CA 94025, Menlo Park, USA\\
$^4$ Laboratori Nazionali di Frascati, INFN, Via Enrico Fermi, 54, 00044 Frascati RM, Italy\\
$^5$ Department of Physics, Lund University, Box 118, 221 00 Lund, Sweden\\
$^6$ DESY, Notkestrasse 85, 22607, Hamburg, Germany\\
$^{7}$ Department of Physics \& Astronomy, University College London, Gower St, London WC1E 6BT, UK\\
$^8$ Cockcroft Institute, Daresbury Laboratory, STFC, Keckwick Lane, Daresbury, WA4 4AD, Warrington, UK\\
$^{9}$ Max Planck Institut für Physik, Boltzmannstrasse 8, 85748 Garching/Munich, Germany\\
$^{10}$ CERN, CH-1211 Geneva 23, Switzerland\\
$^{11}$ Institute of High Energy Physics, Chinese Academy of Sciences, 9 Yuquan Rd, Shi Jing Shan Qu, Bei Jing Shi, 100039, China.\\ 
$^{12}$ John Adams Institute for Accelerator Science at Department of Physics, Imperial College London, Prince Consort Road, South Kensington, London SW7 2BW, UK\\
$^{13}$ Rutherford Appleton Laboratory, STFC, Harwell Campus, OX11 0QX, Didcot, UK\\
$^{14}$ Laboratoire de Physique des 2 Infinis Irène Joliot-Curie, IJCLab, Orsay, Bât. 100 et 200, 15 rue Georges Clémenceau, F-91405 Orsay, France\\
$^{15}$ National Institute of Physics, University of the Philippines Diliman, Quezon City, Philippines\\
$^{16}$ Lawrence Berkeley National Laboratory, 1 Cyclotron Rd, Berkeley, CA 94720, USA\\
$^{17}$ University of Colorado, Boulder, CO 80309-0390, USA\\
$^{18}$ Argonne National Laboratory, 9700 S Cass Avenue, IL60439, Lemont, USA\\
$^{19}$ University of Lancaster, Bailrigg, Lancaster LA1 4YW, UK\\
$^{20}$ Ludwig-Maximilians-Universität München, Geschwister-Scholl-Platz 1, 80539 München, Germany \\
$^{21}$ II. Institute of Theoretical Physics, University of Hamburg, Luruper Chaussee 149, 22761 Hamburg, Germany \\
$^{22}$ University of Hamburg, Mittelweg 177, 20148 Hamburg, Germany \\
$^{23}$ Old Dominion University, Norfolk, VA 23529 USA\\
$^{24}$ Université Paris-Saclay, 3 Rue Joliot Curie, 91190 Gif-sur-Yvette, France\\
$^{25}$ CEA-IRFU, Bât. 141 CEA - Saclay, 91191 Gif-sur-Yvette, France\\
\vspace{0.5cm}
\\
{\Large{\textbf{Executive Summary}}}
\vspace{0.5cm}
\\
This document reports on the current status of the HALHF project, as an intermediate step towards a full CDR. After defining the structure of the project and the current baseline, the document details progress in several subsystems. It begins with physics, where substantial progress has been made both in understanding the beam--beam interaction for asymmetric collisions and its effects on the design of the interaction region. Representative physics processes have been considered; HALHF performs almost as well as a symmetric collider and no show-stoppers have been found to date. Progress on the plasma linac is reported, including a fully designed achromatic staging optics, simulations of several HALHF cells staged together and a simplified simulation of the full 48-cell accelerator. Studies of the transverse instability and tolerances show that while they impose demanding constraints, solutions are possible. Polarisation can be preserved during the acceleration process. Colliders require high beam power; the design of a suitable cell and cooling is underway; some of the demands  and some solutions are outlined. The radiation environment and the design of beam dumps is reported on; the radiation environment represents a major challenge. A scheme to distribute drivers to the PWFA cells has been designed and simulated; good performance was found. The positron source is a well-known challenge for linear colliders and is reported on in detail; a solution for a polarised electron source is also outlined. Finally, new ideas on the beam-delivery system are shown, together with further possibilities for improving and shortening the system. The topics discussed do not include some subsystems required to build a full collider. This is because of a shortage of both experts and funds; the final section outlines the strategy for future work and estimates the resources required to produce a full CDR. 

\tableofcontents

\section{Introduction}
This document reports on progress on the HALHF design since the Erice workshop in March 2025 and the HALHF submission to the European Particle Physics Strategy Update (EPPSU) a few months later. It represents important steps on the road to a Conceptual Design Report for HALHF but substantial work remains before a self-consistent working point for all parts of the project can be set down. Indeed, aspects of the basic parameter set are still under review, as detailed in this document. 

The general scarcity of effort on the project means that not all components of HALHF are as yet fully coherent. For example, design evolution in the Beam Delivery System (BDS) implies a requirement for reduced emittance parameters from the plasma-acceleration arm that have yet to be implemented in that design. Nevertheless, we consider it important to detail the current status both to emphasise the progress that has been made as well as to indicate the necessity for more resources and collaborators to move to the next stage. This document has been submitted to the Laboratory Directors Group charged by CERN with implementing the European Strategy with regard to accelerator R\&D.
 
\section{Baseline collider design: HALHF 2.0}
The schematic of the current baseline design of HALHF (HALHF2.0) is shown in Fig. 1, while the
baseline parameter set is given in Table~\ref{tab:1}. Further details on the evolution of the design from the original HALHF concept\cite{HALHF} to the current one are given elsewhere~\cite{Foster2025}, as are possible upgrades~\cite{HALHF_upgrades}.

Note that, although work since the Erice workshop has indicated several areas where changes in design and/or parameters would be desirable, for the purposes of the current document, the design and parameter set defined in Erice has been retained. Some minor updates to the ``Luminosity'' and ``Luminosity fraction in the top 1\%'' rows of Table~\ref{tab:1} have been made to reflect more exact higher-statistics results from the beam-beam simulation. The $\Upsilon$-parameter row has been removed, since giving a single number for this is somewhat over-simplified. Possible design changes are indicated at places in this document. In due course, a new design and parameter set will be issued.
\\
\begin{minipage}[hbt!]{\textwidth}
        \centering
\vspace{0.2cm}

    \includegraphics[width=\textwidth]{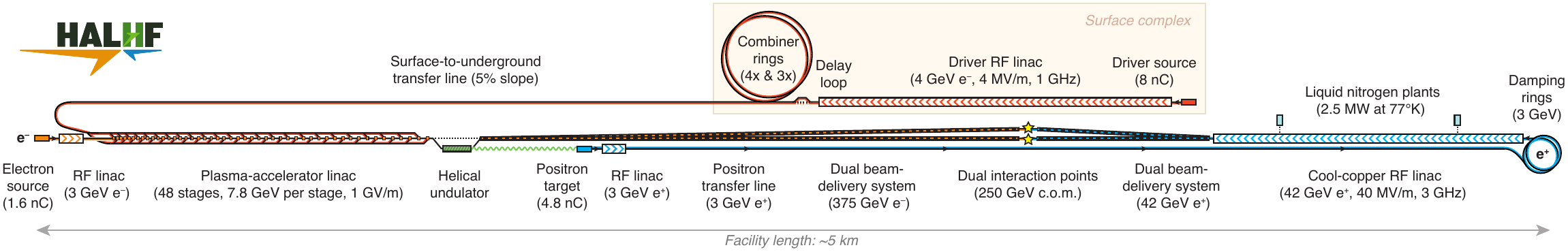}
\captionof{figure}{Schematic view of the HALHF baseline at 250 GeV CoM. The red sections relate to electrons, blue to positrons and green to photons. Other components are as labelled on the figure.}
    \label{Fig-1}

    \end{minipage}
\begin{table*}
    \tiny
    \centering

    \begin{tabular}{p{0.24\linewidth}>{\centering}p{0.05\linewidth}>{\centering}p{0.07\linewidth}>{\centering\arraybackslash}p{0.07\linewidth}>{\centering\arraybackslash}p{0.07\linewidth}>{\centering\arraybackslash}p{0.07\linewidth}>{\centering\arraybackslash}>{\centering\arraybackslash}p{0.07\linewidth}>{\centering\arraybackslash}p{0.07\linewidth}}
    
    \textit{Machine parameters} & \textit{Unit} & 
   \multicolumn{2}{c}{\textit{Value (250 GeV)}}& \multicolumn{2}{c}{\textit{Value (380 GeV)}} & \multicolumn{2}{c}{\textit{Value (550 GeV)}}\\
    \hline
    Centre-of-mass energy & GeV & \multicolumn{2}{c}{250} & \multicolumn{2}{c}{380} &\multicolumn{2}{c}{550}\\
    Centre-of-mass boost & ~ & \multicolumn{2}{c}{1.67} & \multicolumn{2}{c}{1.67} & \multicolumn{2}{c}{1.67}\\
    Bunches per train & & \multicolumn{2}{c}{160} & \multicolumn{2}{c}{160}  & \multicolumn{2}{c}{160}  \\
    Train repetition rate & Hz & \multicolumn{2}{c}{100} & \multicolumn{2}{c}{100} & \multicolumn{2}{c}{100}\\
    Average collision rate & kHz & \multicolumn{2}{c}{16} & \multicolumn{2}{c}{16} & \multicolumn{2}{c}{16}\\
    Luminosity & cm\textsuperscript{--2}~s\textsuperscript{--1} & \multicolumn{2}{c}{1.14$\times$10\textsuperscript{34}} & \multicolumn{2}{c}{1.75$\times$10\textsuperscript{34}} & \multicolumn{2}{c}{2.55$\times$10\textsuperscript{34}} \\
    Luminosity fraction in top 1\% & & \multicolumn{2}{c}{66\%} & \multicolumn{2}{c}{57\%} &\multicolumn{2}{c}{50\%} \\
   \\
    Estimated total power usage & MW & \multicolumn{2}{c}{106} & \multicolumn{2}{c}{154} & \multicolumn{2}{c}{218} \\
        Total site length & km & \multicolumn{2}{c}{4.9} & \multicolumn{2}{c}{6.5} & \multicolumn{2}{c}{8.4} \\
    \hline
     & & & \\[-6pt]
    \textit{Colliding-beam parameters} & & $\rm{e}^-$ & $\rm{e}^+$ & $\rm{e}^-$ & $\rm{e}^+$ & $\rm{e}^-$ & $\rm{e}^+$ \\ 
    \hline
    Beam energy & GeV & 375 & 41.7 & 570 & 63.3 & 825 & 91.7 \\
    Bunch population & 10\textsuperscript{10} & 1 & 3 & 1 & 3 & 1 & 3 \\
    Bunch length in linacs (rms) & {\textmu}m & 40 & 150 & 40 & 150 & 40 & 150 \\ 
    Bunch length at IP (rms) & {\textmu}m & \multicolumn{2}{c}{150} & \multicolumn{2}{c}{150} & \multicolumn{2}{c}{150}  \\
    Energy spread (rms) & \% & \multicolumn{2}{c}{0.15} & \multicolumn{2}{c}{0.15} & \multicolumn{2}{c}{0.15} \\
    Horizontal emittance (norm.) & mm mrad & 90 & 10 & 90 & 10 & 90 & 10 \\
    Vertical emittance (norm.) & mm mrad & 0.32 & 0.035 & 0.32 & 0.035 & 0.32 & 0.035\\
    IP horizontal beta function & mm & \multicolumn{2}{c}{3.3} & \multicolumn{2}{c}{3.3} & \multicolumn{2}{c}{3.3}\\
    IP vertical beta function & mm & \multicolumn{2}{c}{0.1} & \multicolumn{2}{c}{0.1} & \multicolumn{2}{c}{0.1} \\
    IP horizontal beam size (rms) & nm & \multicolumn{2}{c}{636} & \multicolumn{2}{c}{519} & \multicolumn{2}{c}{429} \\
    IP vertical beam size (rms) & nm & \multicolumn{2}{c}{6.6} & \multicolumn{2}{c}{5.2} & \multicolumn{2}{c}{4.4} \\
    Average beam power delivered & MW & 9.6 & 3.2 & 14.6 & 4.9 & 21.1 & 7.0 \\
    Bunch separation & ns & \multicolumn{2}{c}{16} &\multicolumn{2}{c}{16} &\multicolumn{2}{c}{16} \\
    Average beam current & {\textmu}A & 26 & 77 & 26 & 77 & 26 & 77  \\[2pt]
    \hline \\[-6pt]
    \multicolumn{2}{l}{\textit{Positron cool-copper RF linac parameters (S-band)}}  & & & & \\
     \hline
     Average cavity gradient & MV/m & \multicolumn{2}{c}{40}& \multicolumn{2}{c}{40} & \multicolumn{2}{c}{40}\\
        Average gradient & MV/m & \multicolumn{2}{c}{36} & \multicolumn{2}{c}{36} & \multicolumn{2}{c}{36}\\
    Wall-plug-to-beam efficiency & \% & \multicolumn{2}{c}{11} & \multicolumn{2}{c}{11} & \multicolumn{2}{c}{11}\\
    RF power & MW & \multicolumn{2}{c}{11.7} &\multicolumn{2}{c}{17.8} &\multicolumn{2}{c}{25.8} \\
    Cooling power  & MW & \multicolumn{2}{c}{17.9} &  \multicolumn{2}{c}{27.3} & \multicolumn{2}{c}{39.5}\\
     Total power & MW & \multicolumn{2}{c}{29.6} & \multicolumn{2}{c}{45.1} & \multicolumn{2}{c}{65.3} \\
   Klystron peak power & MW & \multicolumn{2}{c}{67} & \multicolumn{2}{c}{67} & \multicolumn{2}{c}{67} \\
     Number of klystrons & ~ & \multicolumn{2}{c}{321} & \multicolumn{2}{c}{452} & \multicolumn{2}{c}{678} \\
    RF frequency & GHz & \multicolumn{2}{c}{3} & \multicolumn{2}{c}{3} & \multicolumn{2}{c}{3} \\
 Operating Temperature & K & \multicolumn{2}{c}{77} & \multicolumn{2}{c}{77} & \multicolumn{2}{c}{77} \\
         Length (after damping ring, starting at 3 GeV) & km & \multicolumn{2}{c}{1.1} & \multicolumn{2}{c}{1.7} & \multicolumn{2}{c}{2.5} \\
    \hline \\[-6pt]
    \multicolumn{2}{l}{\textit{Driver linac RF parameters (L-band)}} & & & & \\
    \hline \\[-6pt]
        Average cavity gradient & MV/m & \multicolumn{2}{c}{4} & \multicolumn{2}{c}{4} & \multicolumn{2}{c}{4}\\
    Average gradient & MV/m & \multicolumn{2}{c}{3} & \multicolumn{2}{c}{3} & \multicolumn{2}{c}{3} \\
    Wall-plug-to-beam efficiency & \% & \multicolumn{2}{c}{55} & \multicolumn{2}{c}{55} & \multicolumn{2}{c}{55} \\
    RF power usage & MW & \multicolumn{2}{c}{42.9} & \multicolumn{2}{c}{66.0} & \multicolumn{2}{c}{96.4}\\
   Klystron peak power & MW & \multicolumn{2}{c}{21} &  \multicolumn{2}{c}{21} & \multicolumn{2}{c}{21}\\
     Number of klystrons & ~ & \multicolumn{2}{c}{409} & \multicolumn{2}{c}{630} & \multicolumn{2}{c}{919} \\
    RF frequency & GHz & \multicolumn{2}{c}{1} & \multicolumn{2}{c}{1} & \multicolumn{2}{c}{1} \\
        Length & km & \multicolumn{2}{c}{1.3} & \multicolumn{2}{c}{1.9} & \multicolumn{2}{c}{2.8} \\
\hline \\[-6pt]
\multicolumn{2}{l}{\textit{Combiner ring parameters}} 
    & & \\
    \hline \\[-3pt]
    ~ & ~ & ~ & ~ & ~ & ~ \\[-8pt]
    Delay loop length & m & \multicolumn{2}{c}{1.5} &\multicolumn{2}{c}{1.5} &\multicolumn{2}{c}{1.5} \\
    CR1 diameter & m & \multicolumn{2}{c}{244} &\multicolumn{2}{c}{244} &\multicolumn{2}{c}{244} \\
    CR2 diameter & m & \multicolumn{2}{c}{244} &\multicolumn{2}{c}{244} &\multicolumn{2}{c}{244} \\
    \hline \\[-6pt]
\multicolumn{2}{l}{\textit{PWFA linac and drive-beam parameters}} 
    & & & & \\
    \hline \\[-6pt]
    Number of stages &  & \multicolumn{2}{c}{48} & \multicolumn{2}{c}{48} & \multicolumn{2}{c}{48} \\
    Plasma density & cm\textsuperscript{--3} & \multicolumn{2}{c}{6$\times$10\textsuperscript{14}} & \multicolumn{2}{c}{6$\times$10\textsuperscript{14}} & \multicolumn{2}{c}{6$\times$10\textsuperscript{14}}\\ 
    In-plasma accel. gradient & GV/m & \multicolumn{2}{c}{1}& \multicolumn{2}{c}{1} & \multicolumn{2}{c}{1} \\
    Av. gradient (incl. optics) & GV/m & \multicolumn{2}{c}{0.33}& \multicolumn{2}{c}{0.38} & \multicolumn{2}{c}{0.43} \\
    Transformer ratio & ~ & \multicolumn{2}{c}{2} & \multicolumn{2}{c}{2} & \multicolumn{2}{c}{2} \\
    Length per stage & m & \multicolumn{2}{c}{7.8} &\multicolumn{2}{c}{11.8} & \multicolumn{2}{c}{17.1} \\
    Energy gain per stage\tnote{a} & GeV & \multicolumn{2}{c}{7.8} & \multicolumn{2}{c}{11.8} & \multicolumn{2}{c}{17.1} \\
    Initial injection energy & GeV & \multicolumn{2}{c}{3} & \multicolumn{2}{c}{3} & \multicolumn{2}{c}{3}\\
    Driver energy & GeV & \multicolumn{2}{c}{4}  & \multicolumn{2}{c}{5.9} & \multicolumn{2}{c}{8.6}\\
    Driver bunch population & 10\textsuperscript{10} & \multicolumn{2}{c}{5.0} & \multicolumn{2}{c}{5.0} & \multicolumn{2}{c}{5.0} \\
    Driver bunch length (rms)  & {\textmu}m &  \multicolumn{2}{c}{253} &  \multicolumn{2}{c}{253} & \multicolumn{2}{c}{253} \\
    Driver average beam power & MW & \multicolumn{2}{c}{23.8} & \multicolumn{2}{c}{36.2} & \multicolumn{2}{c}{52.6} \\
    Driver bunch separation & ns & \multicolumn{2}{c}{4} & \multicolumn{2}{c}{4} & \multicolumn{2}{c}{4} \\
    Driver-to-wake efficiency & \% & \multicolumn{2}{c}{80} & \multicolumn{2}{c}{80} & \multicolumn{2}{c}{80} \\
    Wake-to-beam efficiency & \% & \multicolumn{2}{c}{50} & \multicolumn{2}{c}{50} & \multicolumn{2}{c}{50} \\
    Driver-to-beam efficiency & \% & \multicolumn{2}{c}{40} & \multicolumn{2}{c}{40} & \multicolumn{2}{c}{40} \\
    Wallplug-to-beam efficiency & \% & \multicolumn{2}{c}{22} & \multicolumn{2}{c}{22} & \multicolumn{2}{c}{22} \\
    Cooling req.~per stage length & kW/m & \multicolumn{2}{c}{38.4} & \multicolumn{2}{c}{38.4} & \multicolumn{2}{c}{38.4} \\
    Length & km & \multicolumn{2}{c}{1.1} & \multicolumn{2}{c}{1.5} & \multicolumn{2}{c}{1.9} \\
    \hline
    \end{tabular}
   \caption{HALHF parameters for the baseline designs at 250~GeV, 380~GeV and 550~GeV CoM energies.}
   \label{tab:1}
\end{table*}

\section{Physics in the asymmetric experimental environment at HALHF}
The physics case of Linear Colliders has been described extensively in the literature, see e.g.~\cite{LinearColliderVision:2025hlt, ILCInternationalDevelopmentTeam:2022izu, Fujii:2017vwa, Moortgat-Pick:2015lbx, ILC:2013jhg} for reviews. Key measurements comprise e.g.\ the measurement of the total $e^+e^- \to ZH$ cross-section via the recoil technique, best done close to the production threshold and cross-section maximum, at $\sqrt{s}=250$\,GeV, which strongly depends on the momentum resolution for high-momentum muons  from $Z\to\mu^+\mu^-$~\cite{Yan:2016xyx, Li:2009ybm}. In addition, precision measurements of the various decay modes of the Higgs boson, in particular those which are challenging at the LHC, like $H \to c \bar{c}$, which requires an exquisite impact-parameter resolution~\cite{ILD:IDR_HiggsBRNote}, as well as precision measurements of $W$ and $Z$ processes play important roles. At higher energies, the top quark, in particular the precision mass measurement via a scan over the $t\bar{t}$ production threshold and its electroweak couplings, as well as the top-quark Yukawa coupling and the trilinear self-coupling of the Higgs boson are among the key targets. These typically require the full exploitation of all information provided by the detector, combined in complex algorithms. Thus momentum resolution, impact-parameter resolution, jet-energy resolution, particle identification, hermeticity etc. collectively determine the overall physics performance of the detector.

Like at any $e^+e^-$ collider, polarised beams are a strong asset to the physics case~\cite{LinearColliderVision:2025hlt, Fujii:2018mli, Moortgat-Pick:2015lbx}. Therefore it is of key importance to verify that HALHF can deliver polarised beams to the interaction point, which will be discussed in Sec.~\ref{sec:sources}. Beyond this, there is no specific polarization-related change due to the asymmetric experimental environment at the IP, therefore polarization aspects will not be discussed further here.   

In the context of HALHF, the first goal is to understand the impact of the asymmetric collider setup on the experimental environment, the detector design and the resulting capabilities to perform the well-known linear-collider program. Simple signatures such as the Higgs recoil measurement are focussed on here. After introducing the potential implications of the asymmetric environment, this chapter presents the current understanding of the beam dynamics at the HALHF interaction point and derives the expected luminosity spectra as well as the corresponding beam-induced backgrounds. Taking into account the latter, a variant of the ILD detector concept~\cite{ILDConceptGroup:2020sfq, Behnke:2013lya} adapted to HALHF will be introduced, along with some very first physics projections, demonstrating that the impact of the asymmetric setup on the physics measurements are at a relatively minor level.

\subsection{General implications of asymmetric experimental environment}
The asymmetric setup of HALHF has several implications on the experimental program: 
\begin{itemize}
    \item First and foremost, the laboratory frame is boosted with respect to the centre-of-mass system of the collisions. With the current parameters for HALHF, the corresponding Lorentz $\gamma$ equals 1.67, c.f.\ Table~\ref{tab:1}. Compared to a symmetric collider, this leads to more particles being produced in the direction of the electron beam, referred to as the ``forward'' region, where the momentum resolution and the impact-parameter resolution become worse. On the positive side, the boost is expected to have a beneficial effect on the identification of heavy-flavour jets, since the decay length in the lab frame increases by the boost, as exploited at SuperKEK-B / Belle-II.
    \item In addition, the asymmetric energies and beam charges impact the beam-beam interaction. In particular, the electrons are subject to a more intense field from the higher charge of the positron bunch, tightly compressed in the $z$ direction in the rest frame of the electrons. Therefore, a higher rate of $e^+e^-$ pairs from beamstrahlung is expected in the direction of the electron beam. Of these pairs, the positrons are expected to be deflected to higher angles due to the repulsion by the oncoming positron bunch, thus constraining the instrumentation in the forward direction. 
    \item Finally, also the luminosity (or beam energy) spectra will be influenced, although the expected effect is less clear.
\end{itemize}
In order to assess the impact of these effects on the physics capabilities of HALHF, the beam-beam interaction has to be simulated, providing both luminosity spectra and pair-background predictions, as will be described in Secs.~\ref{sec:physics:beamdynamics} through~\ref{sec:physics:background}. Based on these, the detector design can be adapted to be compatible with the background conditions and re-optimised with increased emphasis on the performance in the forward region. The first steps in this re-optimisation, the resulting physics-performance implications, as well as further questions to be clarified in the future will be discussed in Sec.~\ref{sec:physics:detector}.

\subsection{Beam dynamics at the interaction point}
\label{sec:physics:beamdynamics}
The current beam parameters for HALHF, based on the most recent optimisation of the accelerator layout as discussed in the later sections of this report, are summarised in Table~\ref{tab:1}. Compared to the previous design~\cite{HALHF}, in particular the collider's asymmetry is rebalanced. By shifting the beam energies, the Lorentz boost to the centre-of-mass frame drops from $2.13$ to $1.67$. The positron bunch population is scaled down from $4\times10^{10}$ to $3\times10^{10}$, which further mitigates the asymmetry at the interaction point. Additionally, the bunch length is doubled, boosting the final luminosity. 

In the following, the simulation tools used to study the beam-beam interactions are briefly introduced, before  the resulting luminosity spectra and pair background distributions are presented. 

\subsubsection{Tools for simulation of Beam-Beam Interactions}

The primary simulation tool used to simulate the beam-beam interactions is {\sc GuineaPig}~\cite{Schulte:1997nga}. {\sc GuineaPig} (Generator of Unwanted Interactions for Numerical Experiment Analysis - Program Interfaced to Geant) is one of the canonical programs for modelling beam-beam interactions in high-energy lepton colliders. {\sc GuineaPig} is a 2D PIC (Particle In Cell) code which uses a macro-particle approach and tracks them on a spatial grid to simulate the deflection of the beam particles in the field of the oncoming bunch, as well as the production of beamstrahlung of photons, coherent and incoherent $e^+e^-$ pairs, Compton scattering etc. For this study, {\sc GuineaPig} v1.4.4~\cite{guinea_pig_legacy} is used to obtain the luminosity spectra, pair backgrounds, etc.\ at the different HALHF centre-of-mass energies.

\begin{figure}[htb]
\centering
\begin{subfigure}{.5\textwidth}
    \centering
    \includegraphics[width=0.95\textwidth]{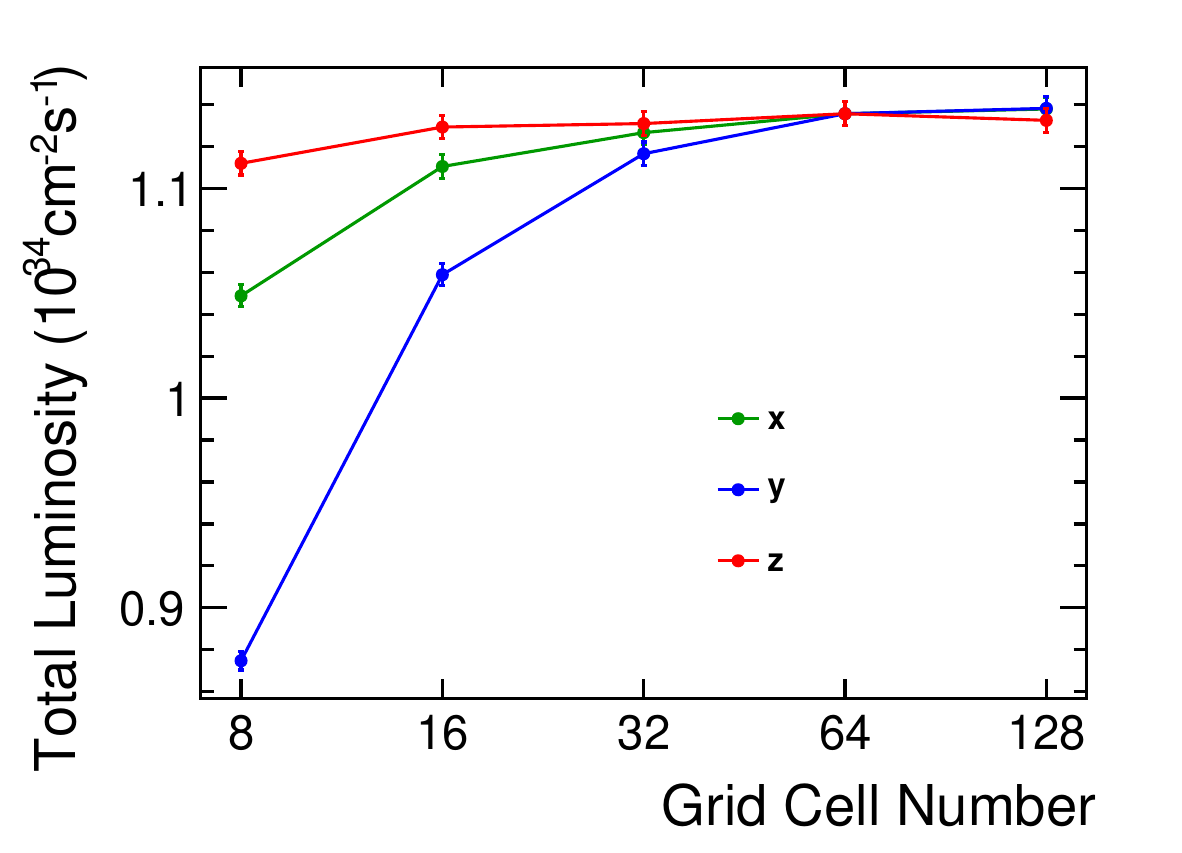}
    \caption{}
    \label{fig:GuineaPig_convscans:lumi}
\end{subfigure}\hfill
\begin{subfigure}{.5\textwidth}
    \centering
    \includegraphics[width=0.95\textwidth]{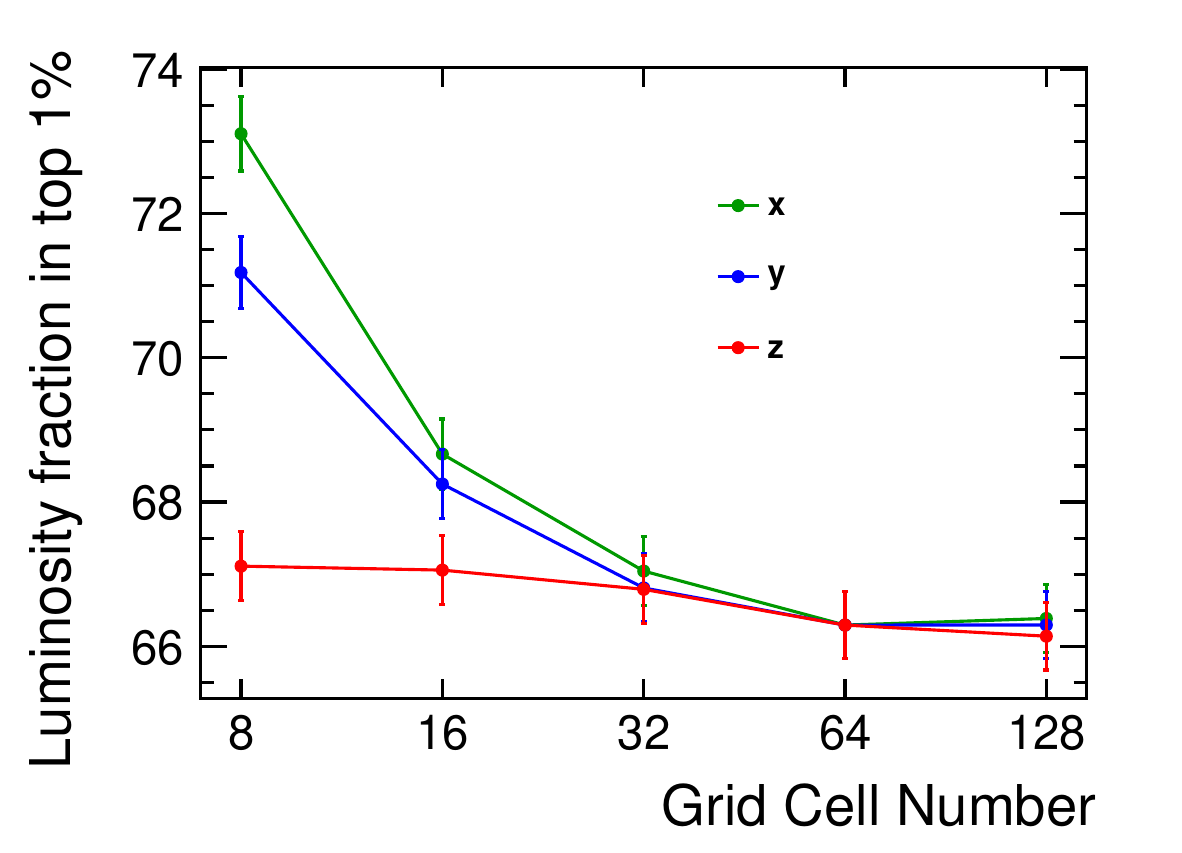}
    \caption{}
    \label{fig:GuineaPig_convscans:l1percent}
\end{subfigure}
\vspace{0.1cm}
\begin{subfigure}{.5\textwidth}
    \centering
    \includegraphics[width=0.95\textwidth]{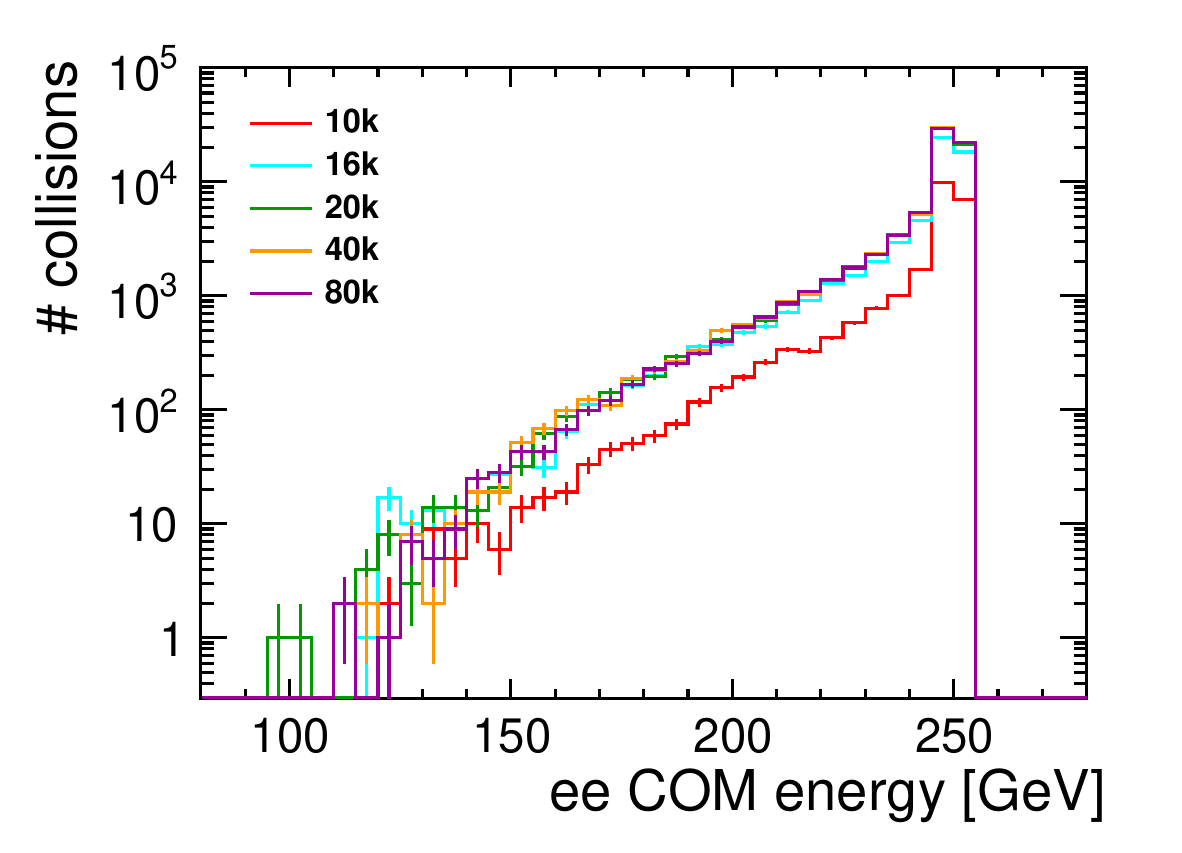}
    \caption{}
    \label{fig:GuineaPig_convscans:macropart}
\end{subfigure}\hfill
\caption{{\sc GuineaPig} convergence scans showing (a) Total luminosity values and (b) L1\% values for different grid values in $x$, $y$ and $z$. For each scan, the other coordinates' grid sizes were held at 128. (c) The effect of the number of macroparticles on the number of $e^+e^-$ collisions during one bunch crossing.}
\label{fig:GuineaPig_convscans}
\end{figure}

Figure~\ref{fig:GuineaPig_convscans} shows convergence scans for the grid cell numbers in $x$, $y$ and $z$ directions and the number of macroparticles used while simulating the collisions at 250\,GeV centre-of-mass energy. The simulation results converge around 128 grid points in each direction and 40000 macroparticles per beam. The final simulation parameters used to obtain the luminosities and pair-background data are stated in the example {\sc GuineaPig} steering file provided in Appendix~\ref{gp_steeringfile}.

\begin{figure}[htb]
\centering
\begin{subfigure}{.53\textwidth}
    \centering
    \includegraphics[width=0.95\textwidth]{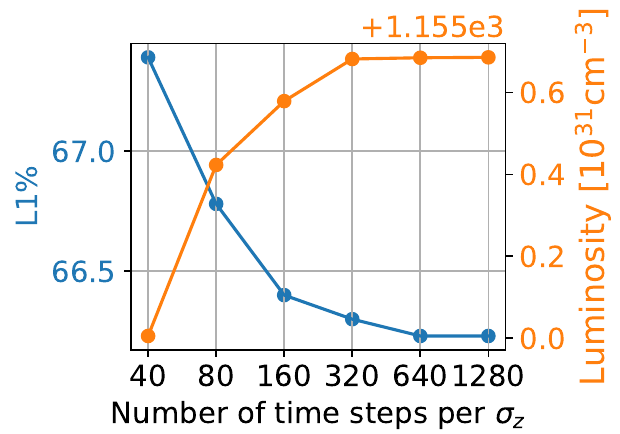}
    \caption{}
    \label{fig:WarpX_convscans:dt}
\end{subfigure}\hfill
\begin{subfigure}{.47\textwidth}
    \centering
    \includegraphics[width=0.95\textwidth]{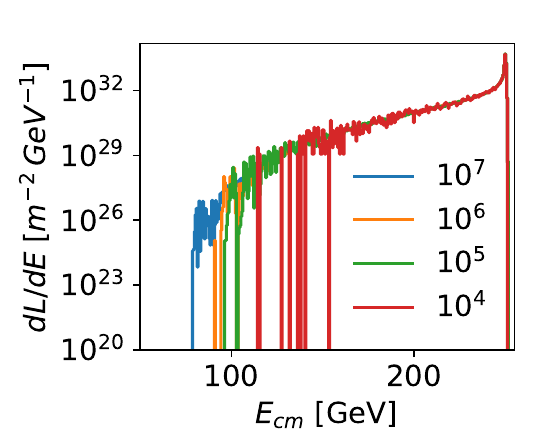}
    \caption{}
    \label{fig:WarpX_convscans:macropart}
\end{subfigure}
\vspace{0.1cm}
\begin{subfigure}{.53\textwidth}
    \centering
    \includegraphics[width=0.95\textwidth]{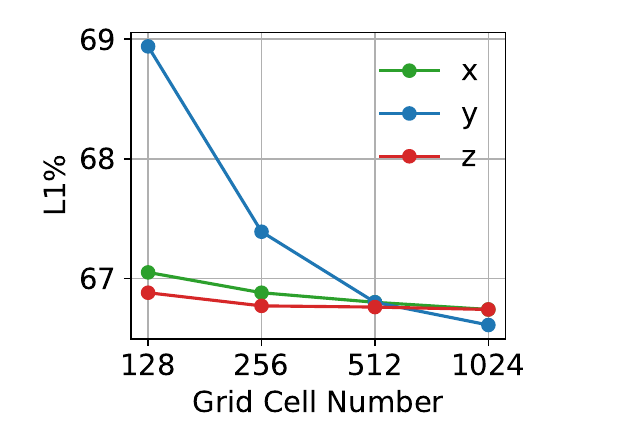}
    \caption{}
    \label{fig:WarpX_convscans:gridL1pct}
\end{subfigure}\hfill
\begin{subfigure}{.47\textwidth}
    \centering
    \includegraphics[width=0.95\textwidth]{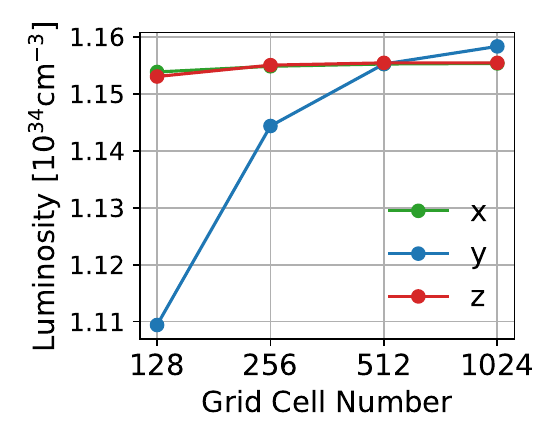}
    \caption{}
    \label{fig:WarpX_convscans:gridL}
\end{subfigure}\hfill
\caption{{\sc WarpX} convergence scans showing (a) L1\% and total luminosity values for different time-step sizes, in units of the number of steps its takes for the beams to move by a distance of $\sigma_{z}$.  (b) The effect of the number of macroparticles on luminosity spectrum shape. A ratio of 1:3 electron to positron macroparticles was used in each case. (c) L1\% and (d) total luminosity values for different grid values in $x$, $y$ and $z$. For each scan, the other coordinates' grid sizes were held at 512.}
\label{fig:WarpX_convscans}
\end{figure}
\clearpage
{\sc WarpX}~\cite{WarpX_gitrepo} is an alternative advanced PIC code commonly used for plasma simulations. It is also capable of simulating beam-beam effects by utilising the PICSAR-QED library~\cite{Fedeli_2022_NJP}. The most relevant module for simulating the luminosity spectrum shape is the Quantum Synchrotron QED module which models beamstrahlung photon emission. The beam macroparticles are each randomly assigned an optical depth which is updated each time step according to total photon emission rate. When a macroparticle’s optical depth reaches 0, a photon macroparticle is created with an energy assigned by sampling the appropriate cumulative probability distribution. The photon emission rate and cumulative probability distribution are not calculated at each step but instead saved in lookup tables that must be generated before the simulation. Similar tables also need to be generated for the Breit-Wheeler QED module which models coherent-pair generation. 
 
The parameters used for these lookup tables can be found in the example input script for modelling HALHF250 given in Appendix~\ref{warpx_steeringfile}. {\sc WarpX}’s in-built Gaussian beam generator does not currently offer the ability to add an independent waist shift in the $y$ direction. Therefore, the beams were generated externally using the python script in Appendix~\ref{warpx_beamgenscript}.

From an initial rough simulation, it was found that luminosity begins to accumulate when the bunches are 4$\sigma_{z}$ before the interaction point and stops accumulating 4$\sigma_{z}$ after the interaction point. From this, it was decided to initialise the beams at 5$\sigma_{z}$ before the IP, and run the simulation for 10 full $\sigma_{z}$. Further convergence scans were performed for timestep size, number of macroparticles, and grid size, shown in Figure~\ref{fig:WarpX_convscans}. The timestep is defined as the number of steps for a bunch to move by 1$\sigma_{z}$ and converges at a value of 640 steps per $\sigma_{z}$. Figure~\ref{fig:WarpX_convscans:macropart} shows how increasing macroparticle number leads to greater resolution of the low-energy tail of the beam-beam luminosity spectrum. In the final simulations, $1\times10^{7}$ and $3\times10^{7}$ macroparticles were used for the electron and positron beams respectively.

Three separate scans were performed for grid size in the $x$, $y$ and $z$ direction by varying only one at a time and holding the others constant at 512. Due to memory constraints, the highest grid size possible in any one direction was 1024. Figures~\ref{fig:WarpX_convscans:gridL1pct} and \ref{fig:WarpX_convscans:gridL} show that the number of grid cells in $y$ has the greatest impact on luminosity values, so it was decided to run with a grid size of 512 in $x$ and $z$ and 1024 in $y$. The luminosity and L1$\%$ approach converged values but do not reach them at this resolution. For each centre-of-mass energy, 5 simulations were run and the average luminosity spectra taken.

\begin{figure}[htp]
\vspace{-1cm}
\centering
\includegraphics[width=0.8\textwidth]{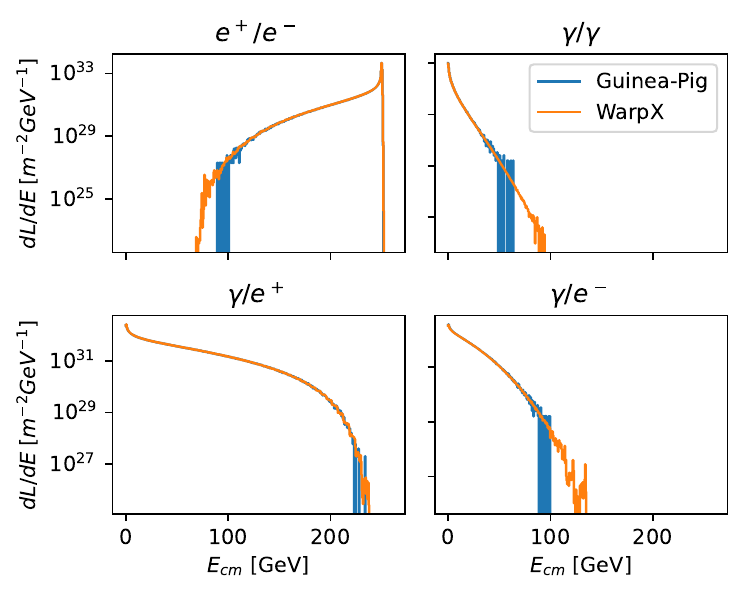}
\caption{Comparison of luminosity spectra for $e^+e^-$, $\gamma\gamma$, $e^+\gamma$ and $e^-\gamma$ collisions at HALHF250 obtained from {\sc WarpX} and {\sc GuineaPig}.}
\label{fig:warpxvsGP_lumi250}
\end{figure}

\begin{figure}[H]
\centering
\includegraphics[width=0.8\textwidth]{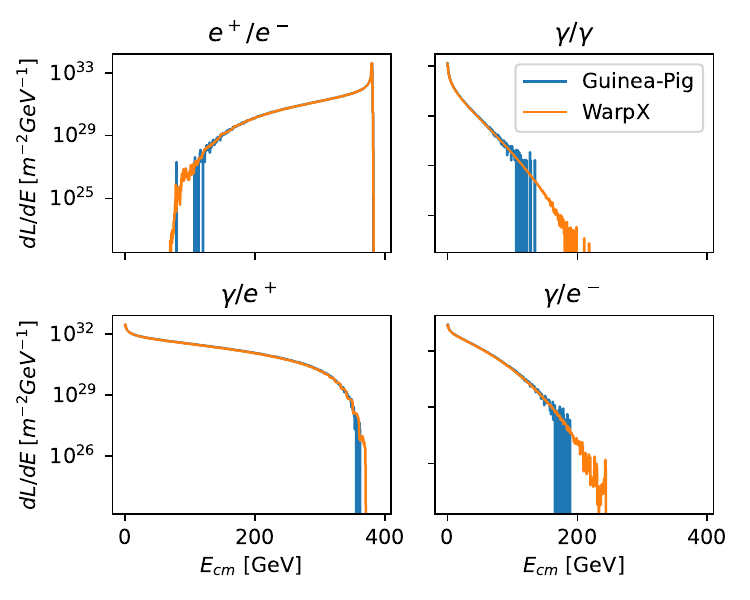}
\caption{Comparison of luminosity spectra for $e^+e^-$, $\gamma\gamma$, $e^+\gamma$ and $e^-\gamma$ collisions at HALHF380 obtained from {\sc WarpX} and {\sc GuineaPig}.}
\label{fig:warpxvsGP_lumi380}
\end{figure}

\begin{figure}[H]
\centering
\includegraphics[width=0.8\textwidth]{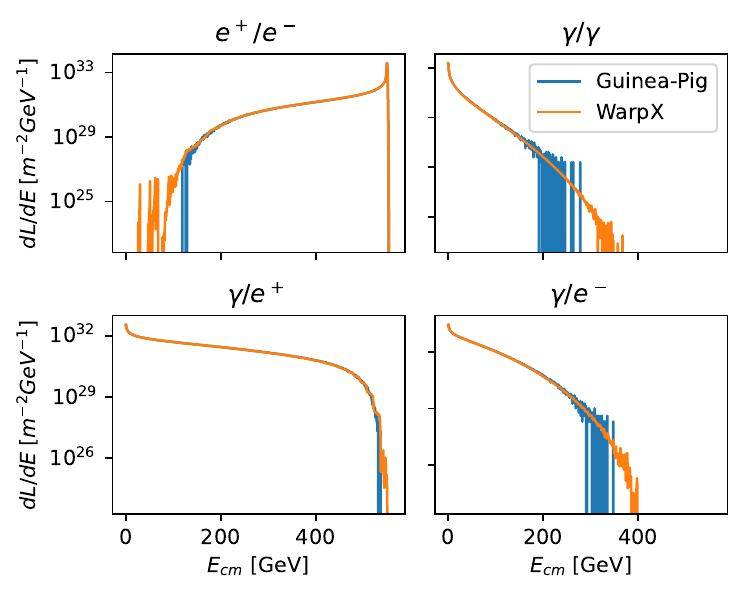}
\caption{Comparison of luminosity spectra for $e^+e^-$, $\gamma\gamma$, $e^+\gamma$ and $e^-\gamma$ collisions at HALHF550 obtained from {\sc WarpX} and {\sc GuineaPig}.}
\label{fig:warpxvsGP_lumi550}
\end{figure}

The luminosity spectra for $e^+e^-$, $e^+\gamma$, $\gamma e^-$ and $\gamma\gamma$ collisions obtained from  {\sc GuineaPig} and {\sc WarpX} with these setups agree very well at all centre-of-mass energies, as shown in Figs.~\ref{fig:warpxvsGP_lumi250} to~\ref{fig:warpxvsGP_lumi550}.

\subsubsection{Asymmetric beam-beam interaction}

As discussed at the beginning of this chapter, the asymmetric energies and charges of the beams impact the beam-beam interaction, and the electrons will see a more intense field from the oncoming positron bunch than vice-versa. Therefore, defining one single beamstrahlung parameter with the well-known formula is no longer sufficient. Instead, the average and maximum beamstrahlung parameters have to be determined for the two beams individually.  

\begin{figure}[htb]
\centering
\begin{subfigure}{.5\textwidth}
    \centering
    \includegraphics[width=0.95\textwidth]{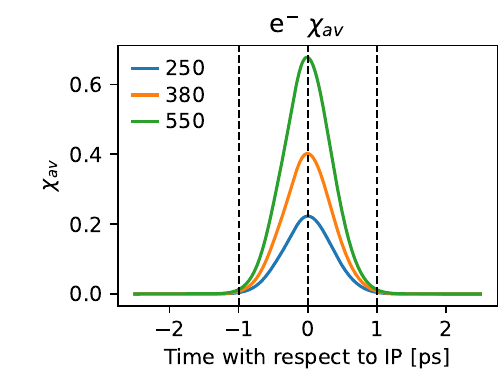}
    \caption{}
    \label{fig:chi:ave_elec}
\end{subfigure}\hfill
\begin{subfigure}{.5\textwidth}
    \centering
    \includegraphics[width=0.95\textwidth]{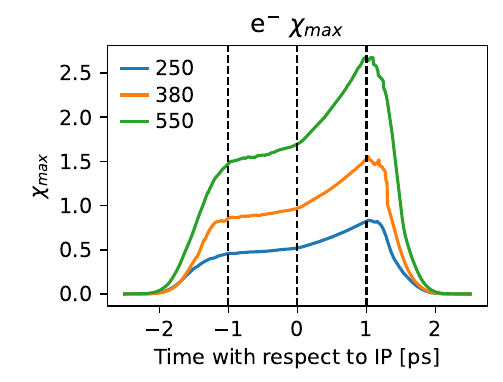}
    \caption{}
    \label{fig:chi:max_elec}
\end{subfigure}
\vspace{0.1cm}
\begin{subfigure}{.5\textwidth}
    \centering
    \includegraphics[width=0.95\textwidth]{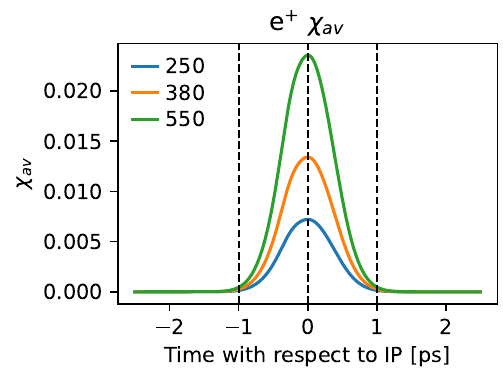}
    \caption{}
    \label{fig:chi:ave_posi}
\end{subfigure}\hfill
\begin{subfigure}{.5\textwidth}
    \centering
    \includegraphics[width=0.95\textwidth]{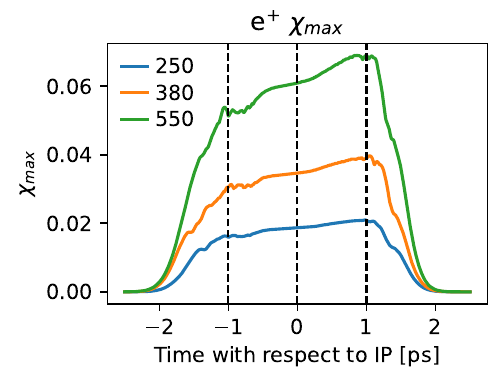}
    \caption{}
    \label{fig:chi:max_posi}
\end{subfigure}
\caption{Evolution of the beamstrahlung parameter $\chi$ (or $\Upsilon$ ) during the collision as obtained from {\sc WarpX} for all three centre-of-mass energies. Vertical dashed lines indicate the nominal IP location $\pm 2\sigma_z/c$ of the bunches. (a) Average $\chi$ seen by the electrons, (b) maximum $\chi$ seen by the electrons, (c) average $\chi$ seen by the positrons, and (d) maximum $\chi$ seen by the positrons.}
\label{fig:chi}
\end{figure}

Figure~\ref{fig:chi} shows the evolution of the beamstrahlung parameter $\Upsilon$ (in the strong-field QED community also referred to as $\chi$) during the collision as obtained from {\sc WarpX} for all three centre-of-mass energies. It clearly shows that due to the lower charge of the  electron bunch, the positrons see relatively weak electromagnetic fields from the electron bunch (c.f.\ Figs.~\ref{fig:chi:ave_posi} and~\ref{fig:chi:max_posi}), far below the Schwinger limit ($\chi$ =1). Thus the production of $e^+e^-$ pairs from the positron beam will be moderate and fully dominated by incoherent, linear processes. The electrons, however, see quite significant electromagnetic fields from the three-times greater charge in the positron beam, reaching up to $\chi$ = 0.8 towards the end of the collision (c.f.\ Fig.~\ref{fig:chi:max_elec}) at 250\,GeV, and even up to $\chi$ = 2.7 at 550\,GeV. Thus, coherent, non-linear pair production starts to play a significant role in the direction of propagation of the electron beam, and significantly higher numbers of pairs with a harder energy spectrum and larger angles are expected. The impact on the detector design will be discussed in Sec.~\ref{sec:physics:background}.

\subsubsection{Luminosity spectra for 
\texorpdfstring{$e^+e^-$, $\gamma\gamma$, $e^-\gamma$, $e^+\gamma$}{e+e-, gamma-gamma, e-gamma, e+gamma} and comparison with ILC}
\label{sec:physics:lumispec}

\begin{figure}[H]
\vspace{-1.3cm}
\centering
\begin{subfigure}{.5\textwidth}
    \centering
    \includegraphics[width=0.95\textwidth]{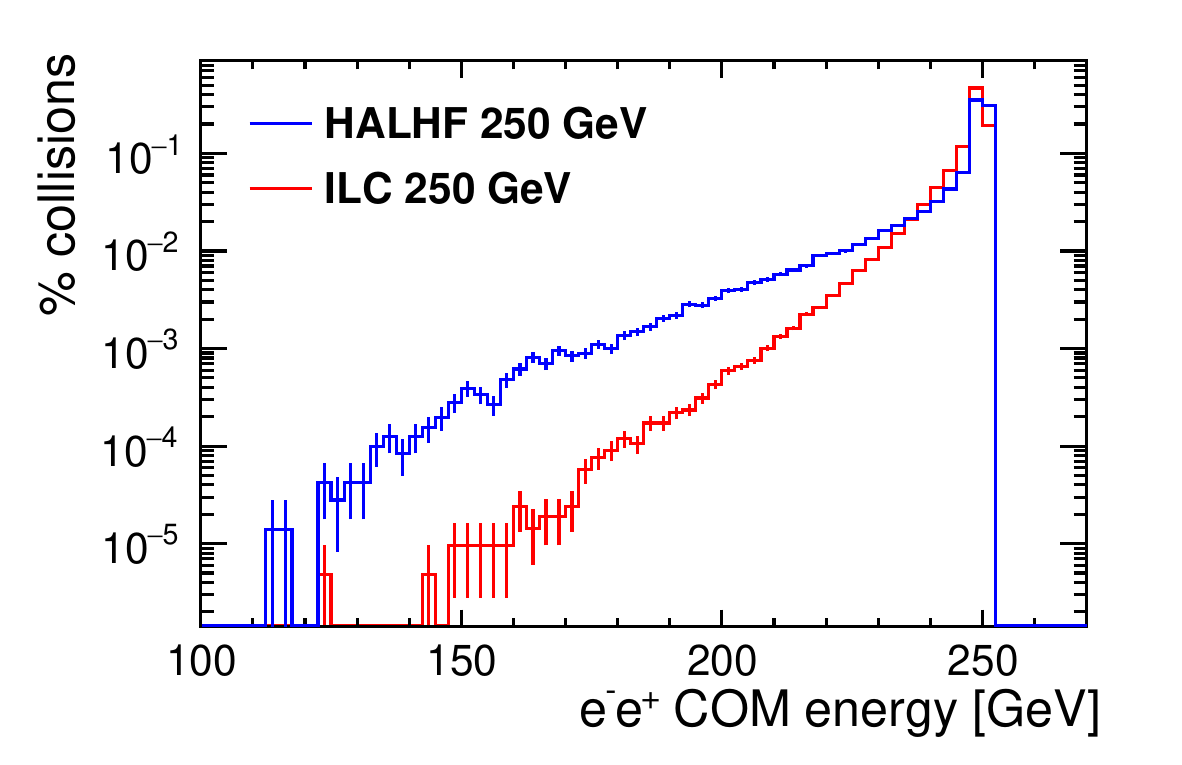}
    \caption{}
    \label{fig:lumi250:ee1}
\end{subfigure}\hfill
\begin{subfigure}{.5\textwidth}
    \centering
    \includegraphics[width=0.95\textwidth]{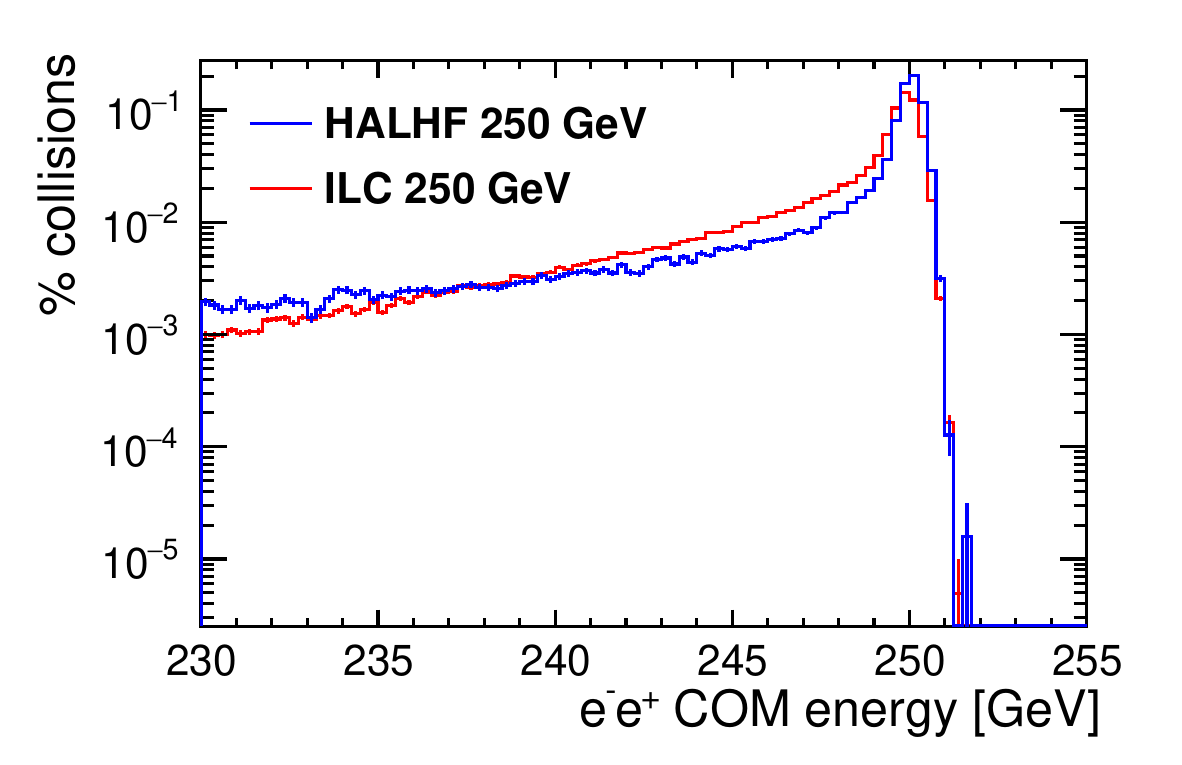}
    \caption{}
    \label{fig:lumi250:ee2}
\end{subfigure}
\vspace{0.1cm}
\begin{subfigure}{.5\textwidth}
    \centering
    \includegraphics[width=0.95\textwidth]{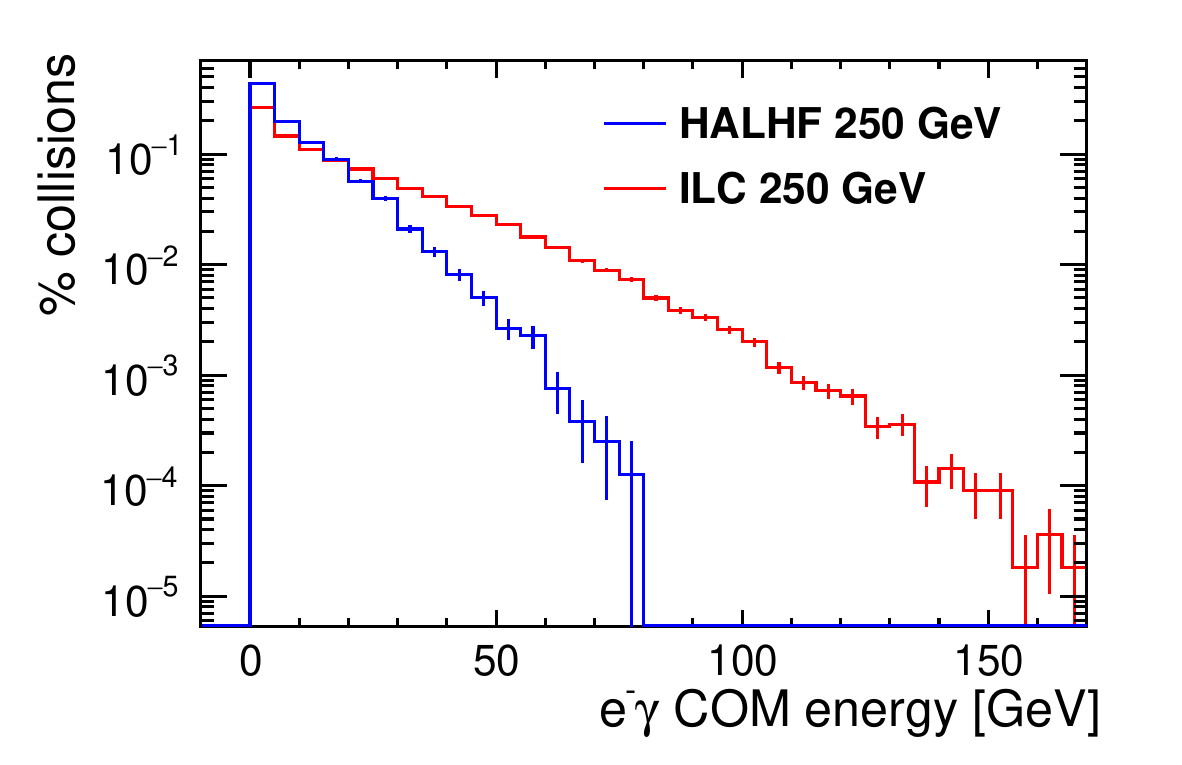}
    \caption{}
    \label{fig:lumi250:eg1}
\end{subfigure}\hfill
\begin{subfigure}{.5\textwidth}
    \centering
    \includegraphics[width=0.95\textwidth]{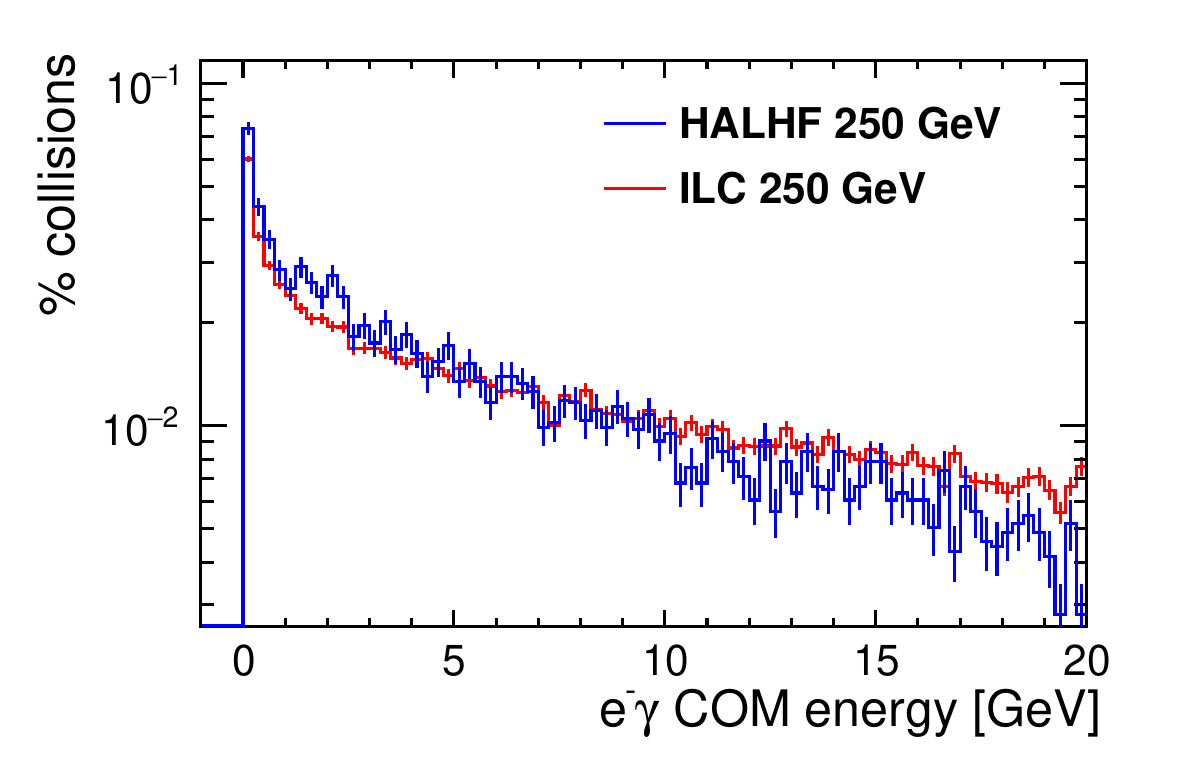}
    \caption{}
    \label{fig:lumi250:eg2}
\end{subfigure}
\vspace{0.1cm}
\begin{subfigure}{.5\textwidth}
    \centering
    \includegraphics[width=0.95\textwidth]{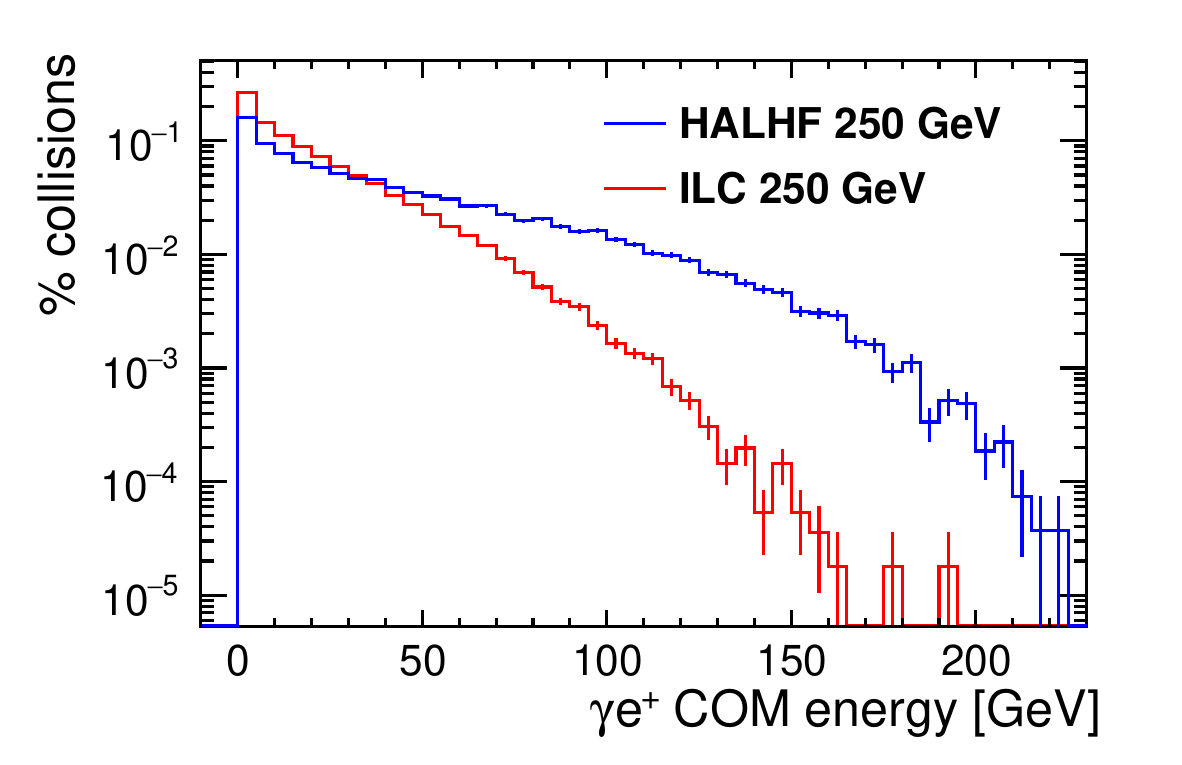}
    \caption{}
    \label{fig:lumi250:ge1}
\end{subfigure}\hfill
\begin{subfigure}{.5\textwidth}
    \centering
    \includegraphics[width=0.95\textwidth]{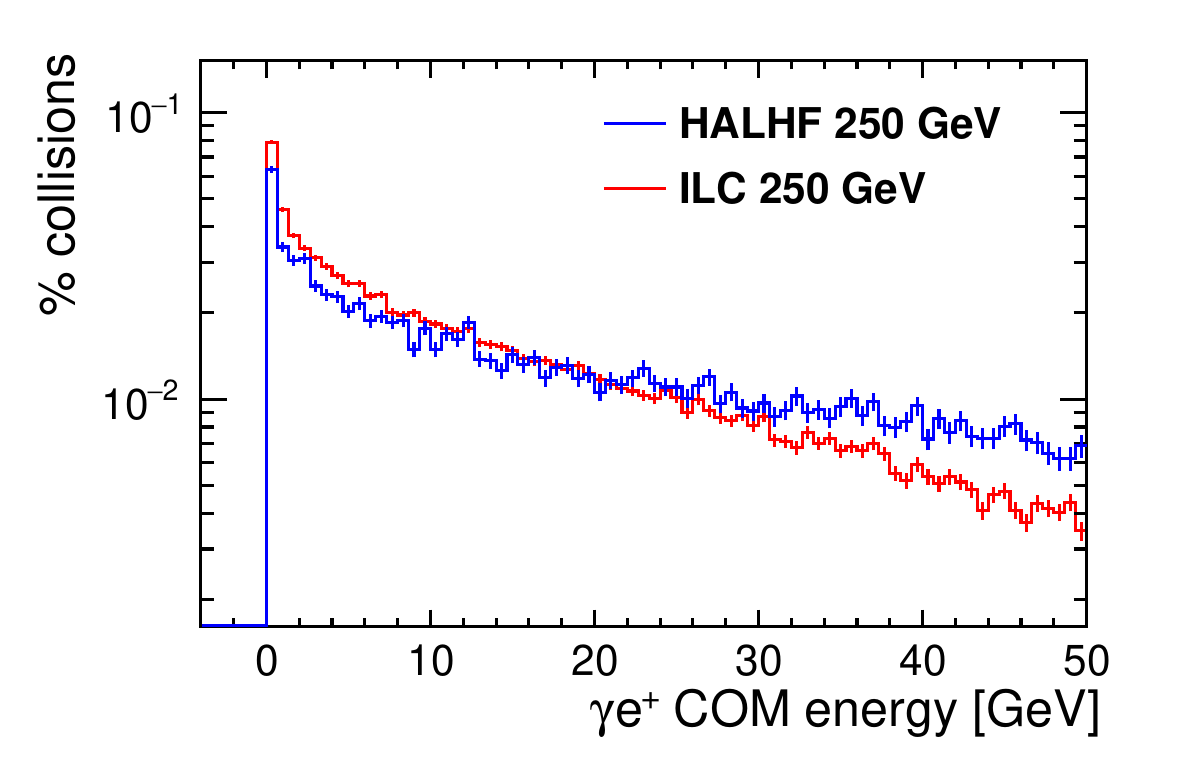}
    \caption{}
    \label{fig:lumi250:ge2}
\end{subfigure}
\vspace{0.1cm}
\begin{subfigure}{.5\textwidth}
    \centering
    \includegraphics[width=0.95\textwidth]{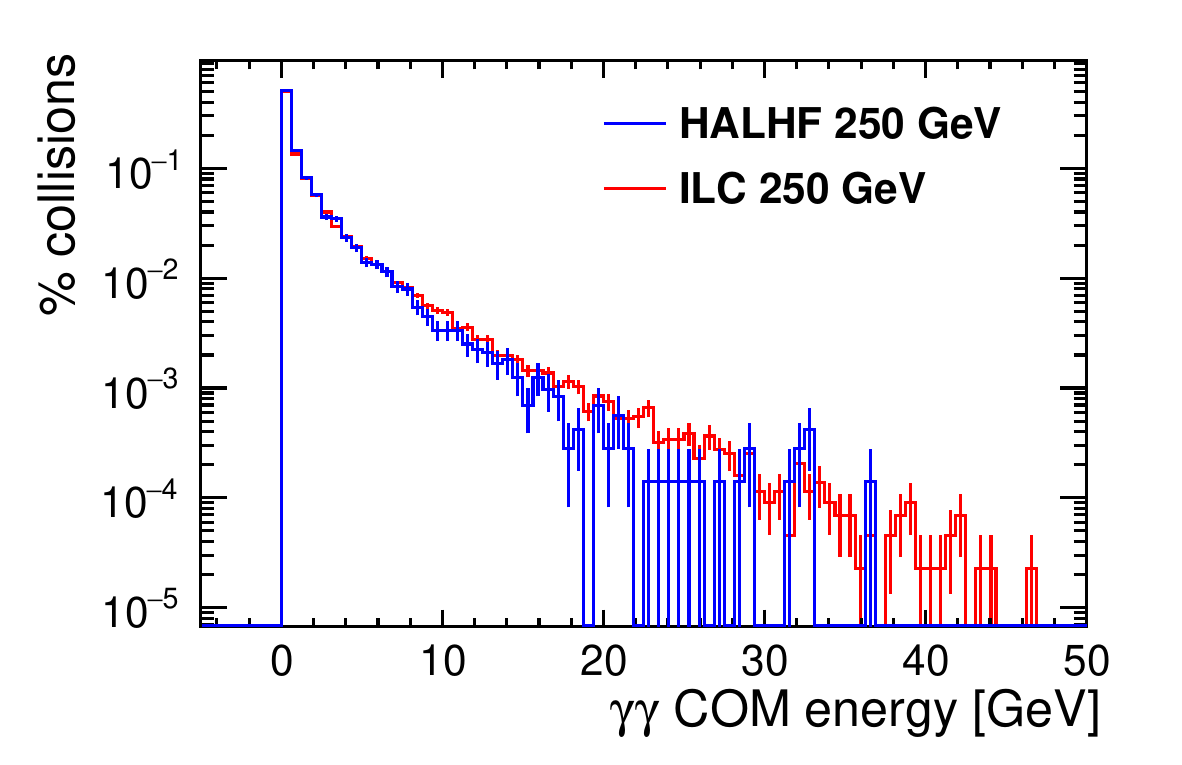}
    \caption{}
    \label{fig:lumi250:gg1}
\end{subfigure}\hfill
\begin{subfigure}{.5\textwidth}
    \centering
    \includegraphics[width=0.95\textwidth]{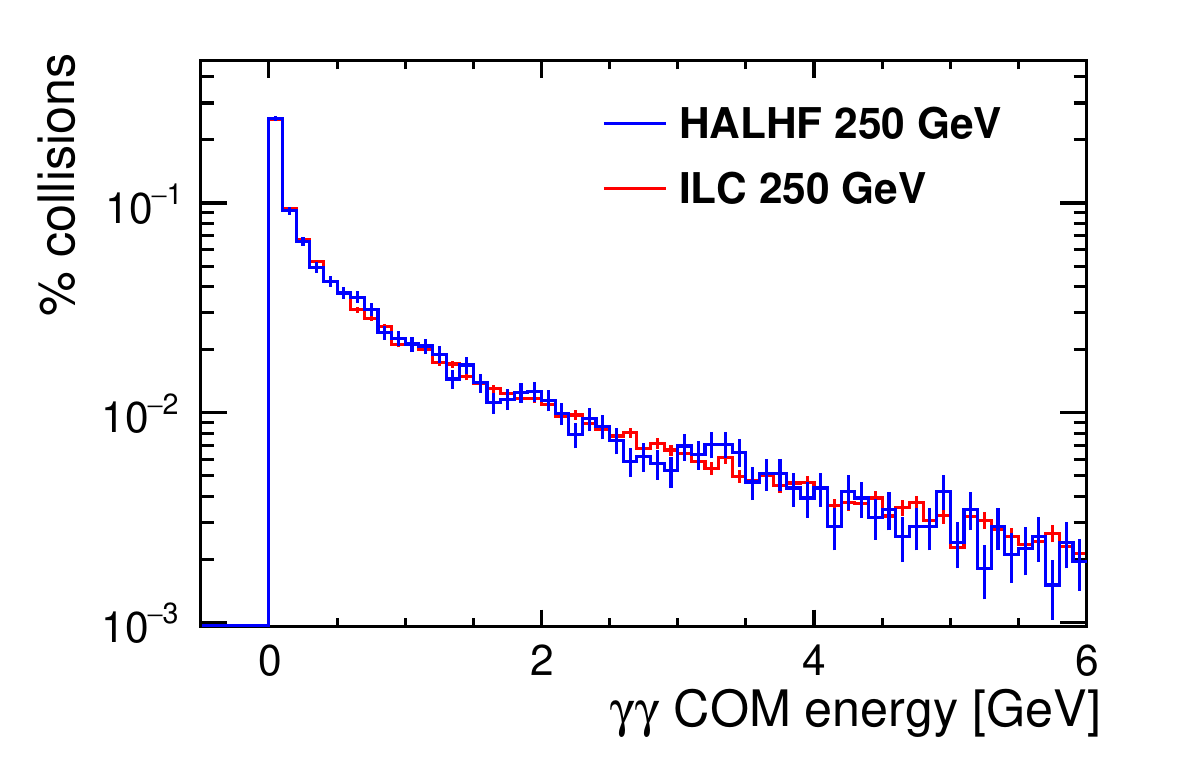}
    \caption{}
    \label{fig:lumi250:gg2}
\end{subfigure}
\caption{Luminosity spectra for (a) $e^+e^-$, (c) $e^-\gamma$, (e) $\gamma e^+$, and (g) $\gamma\gamma$ collisions along with the zoomed-in distributions around the peaks for (b) $e^+e^-$, (d) $e^-\gamma$, (f) $\gamma e^+$, and (h) $\gamma\gamma$ collisions at HALHF250 obtained from {\sc GuineaPig}, compared to the respective distributions for ILC250.} 
\label{fig:lumi250}
\end{figure}

Using the {\sc GuineaPig} setup described above, 100 bunch crossings have been simulated for the HALHF parameters given in Table~\ref{tab:1}. Figure~\ref{fig:lumi250} shows the luminosity spectra for $e^+e^-$, $e^-\gamma$, $\gamma e^+$ and  $\gamma\gamma$ collisions at HALHF250, compared the the corresponding distributions at the symmetric ILC250. The left column shows the full energy range, while the right column shows zoomed version of the peak regions. 

The comparison between the symmetric and the asymmetric collisions allows a number of interesting observations. The $e^+e^-$ luminosity spectrum of HALHF has a much longer tail to low centre-of-mass energies than the ILC, but still the peak at the nominal 250\,GeV is sharper and higher than at ILC, which is an advantage for any measurement relying on the centre-of-mass energy being known, like e.g.\ the Higgs recoil measurement. 

This at a first glance surprising result can be understood in view of Fig.~\ref{fig:chi}: The positron beam sees only rather weak electromagnetic fields, thus emits much less beamstrahlung, and retains a rather sharply peaked energy spectrum. The electron beam, in contrast, is even at 250\,GeV subject to fields approaching the Schwinger limit; it thus suffers much stronger beamstrahlung than at ILC and causes the longer tail at low centre-of-mass energies.

\begin{figure}[htb]
\centering
\begin{subfigure}{.5\textwidth}
    \centering
    \includegraphics[width=0.95\textwidth]{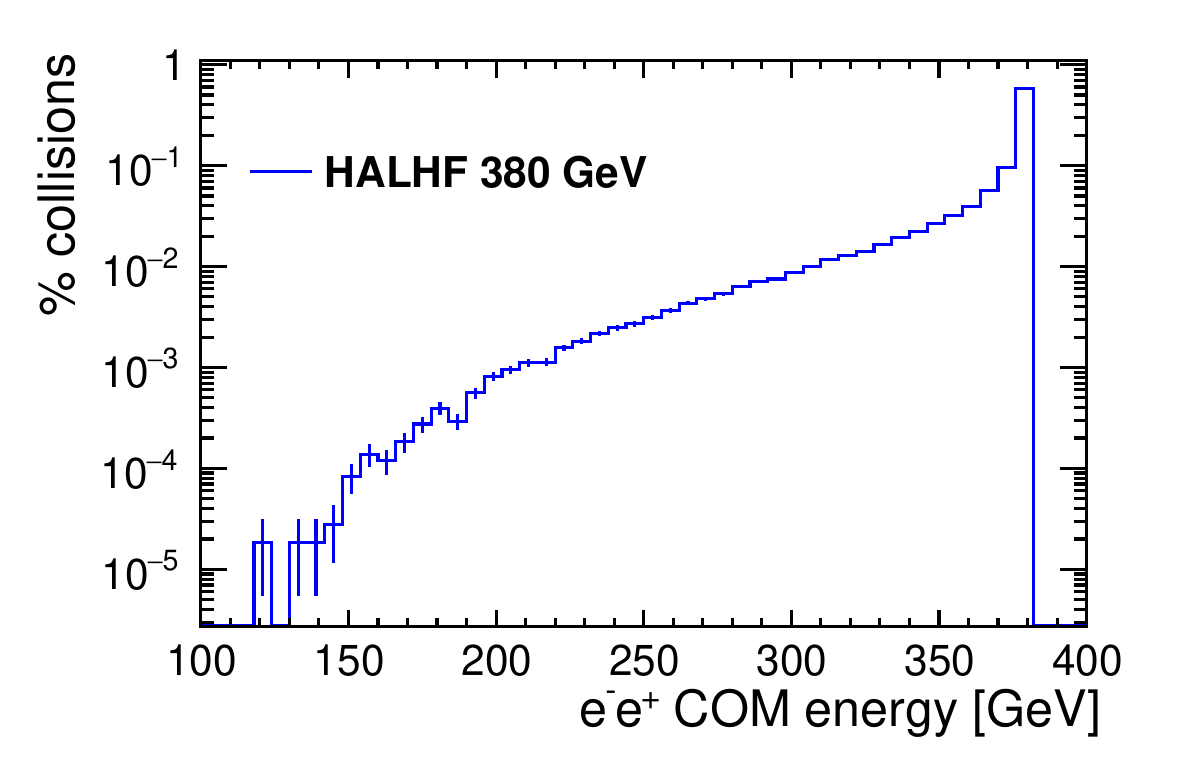}
    \caption{}
    \label{fig:lumi380:ee}
\end{subfigure}\hfill
\begin{subfigure}{.5\textwidth}
    \centering
    \includegraphics[width=0.95\textwidth]{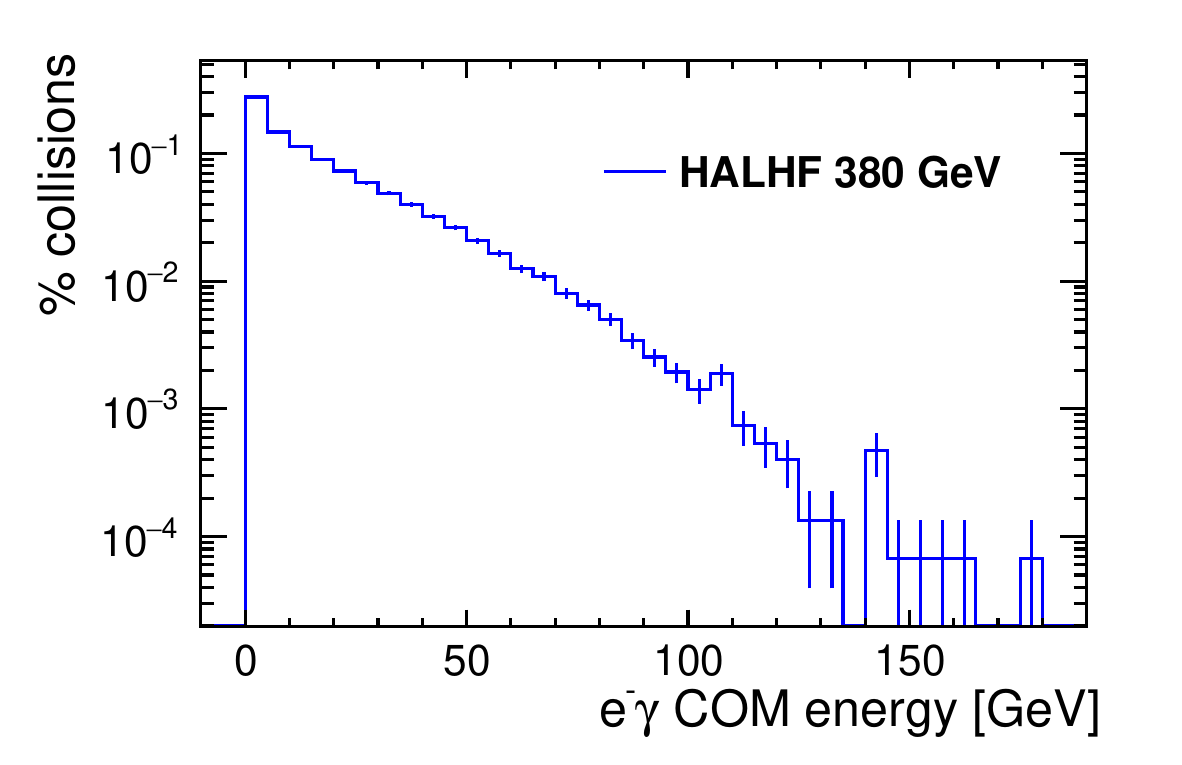}
    \caption{}
    \label{fig:lumi380:eg}
\end{subfigure}
\vspace{0.1cm}
\begin{subfigure}{.5\textwidth}
    \centering
    \includegraphics[width=0.95\textwidth]{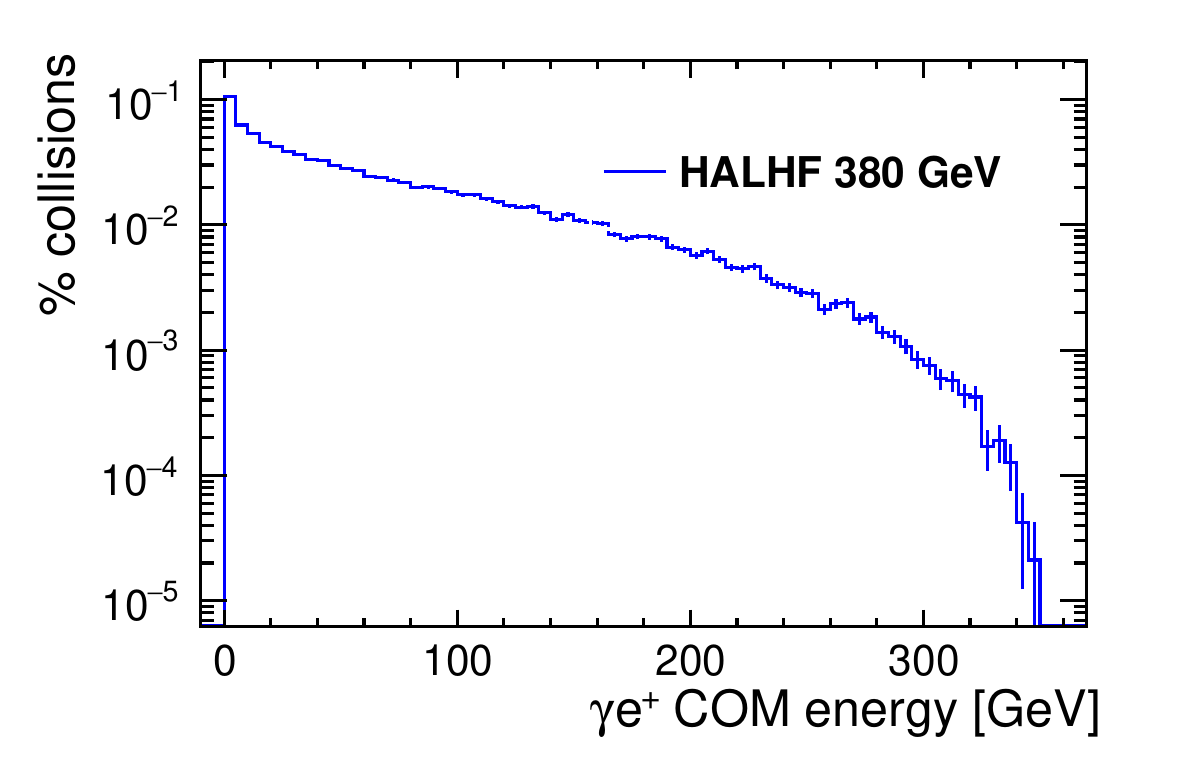}
    \caption{}
    \label{fig:lumi380:ge}
\end{subfigure}\hfill
\begin{subfigure}{.5\textwidth}
    \centering
    \includegraphics[width=0.95\textwidth]{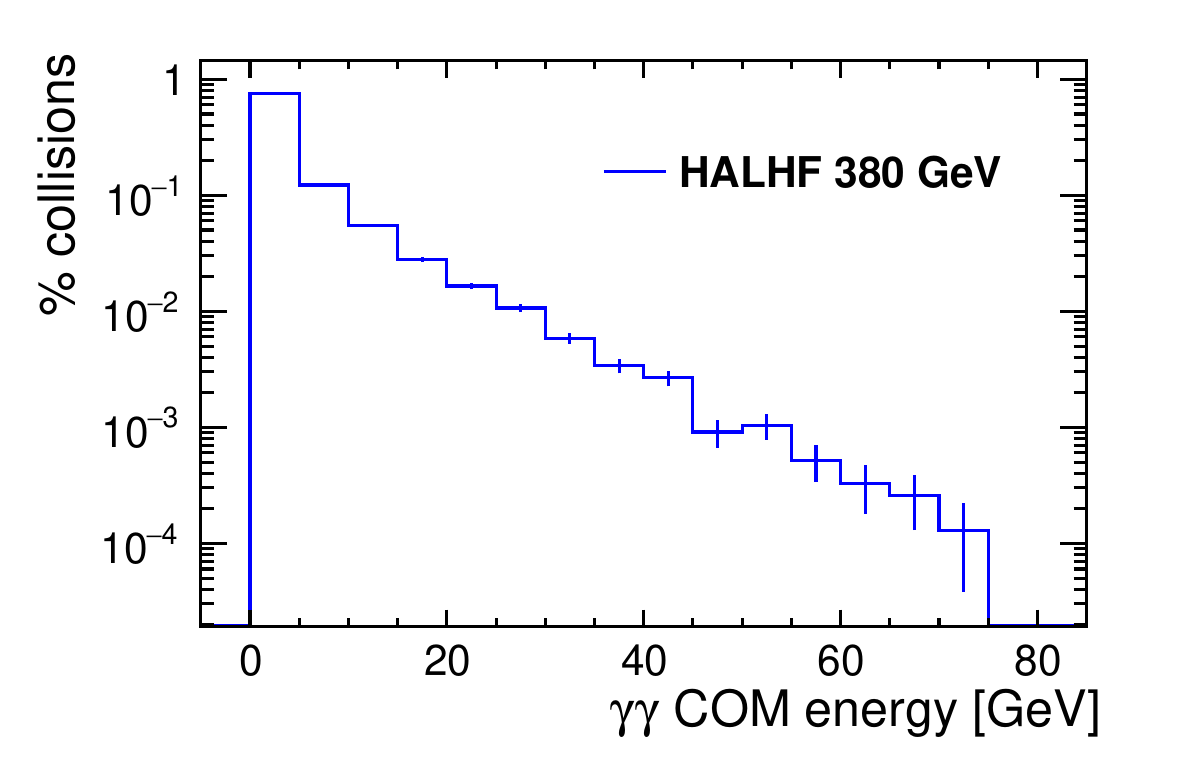}
    \caption{}
    \label{fig:lumi380:gg}
\end{subfigure}
\caption{Luminosity spectra for (a) $e^+e^-$, (b) $e^-\gamma$, (c) $\gamma e^+$, and (d) $\gamma\gamma$ collisions at HALHF380 obtained from {\sc GuineaPig}.} 
\label{fig:lumi380}
\end{figure}

\begin{figure}[htb]
\centering
\begin{subfigure}{.5\textwidth}
    \centering
    \includegraphics[width=0.95\textwidth]{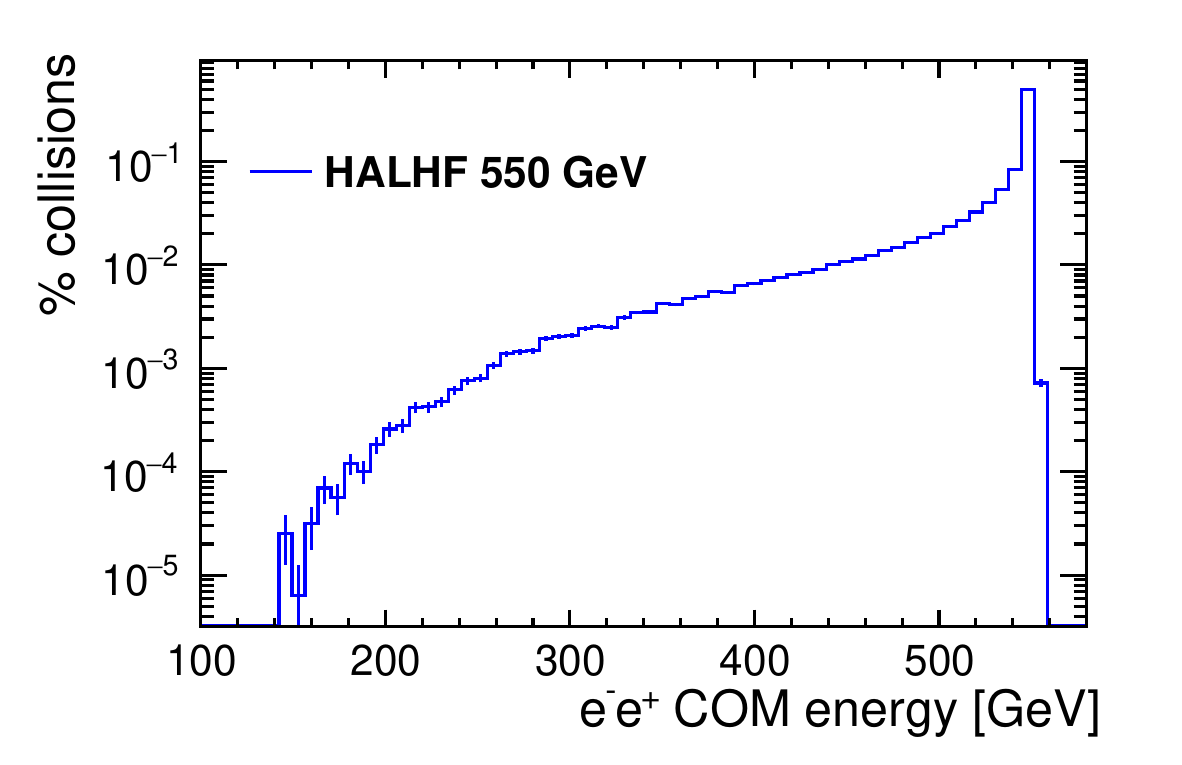}
    \caption{}
    \label{fig:lumi550:ee}
\end{subfigure}\hfill
\begin{subfigure}{.5\textwidth}
    \centering
    \includegraphics[width=0.95\textwidth]{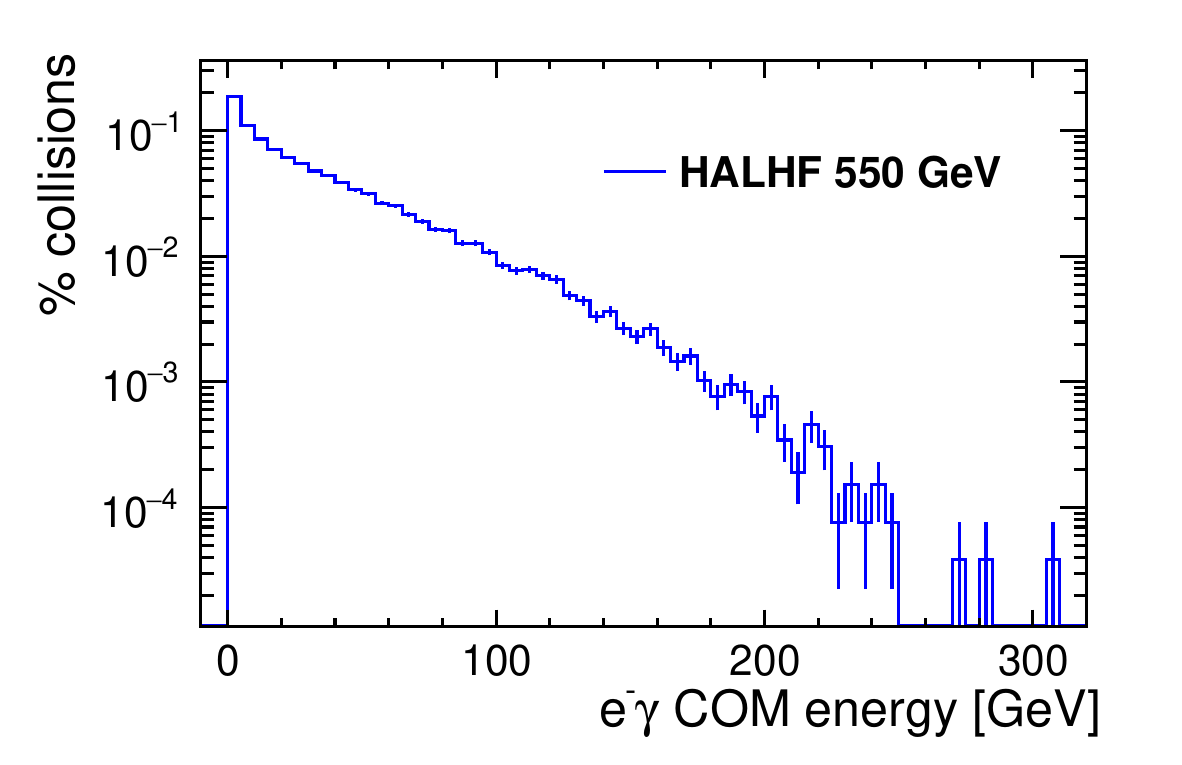}
    \caption{}
    \label{fig:lumi550:eg}
\end{subfigure}
\vspace{0.1cm}
\begin{subfigure}{.5\textwidth}
    \centering
    \includegraphics[width=0.95\textwidth]{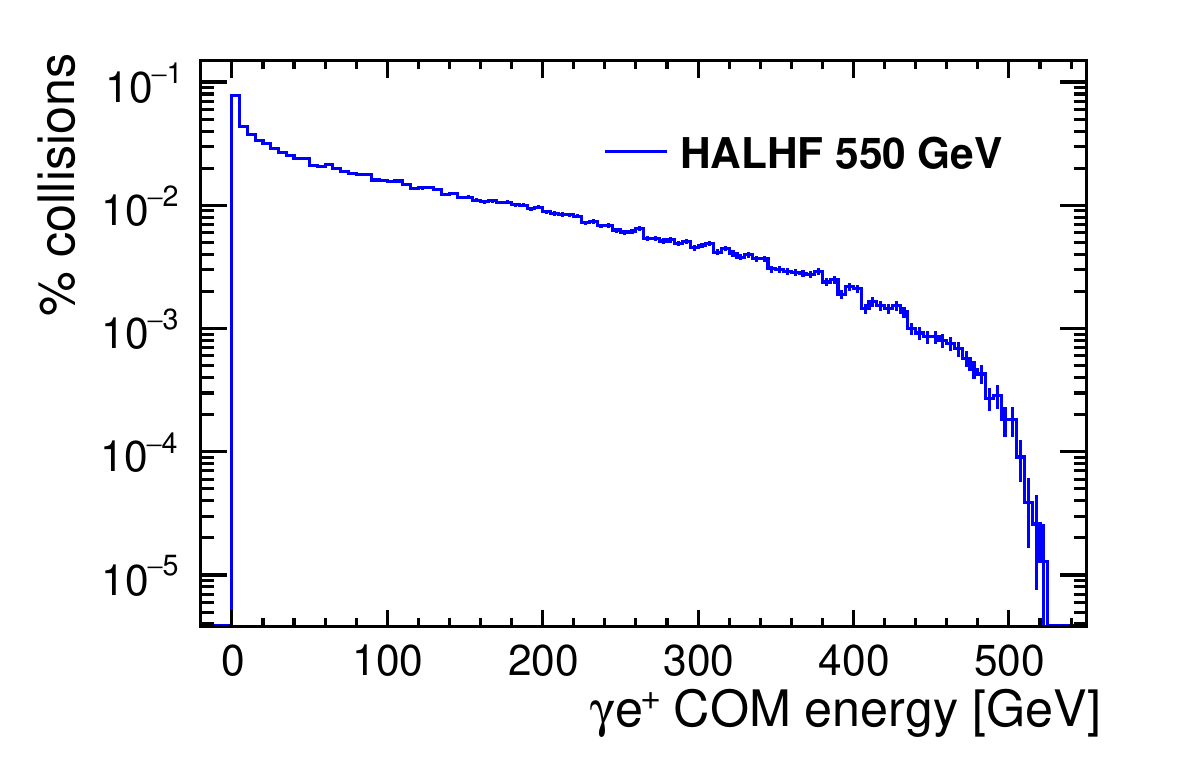}
    \caption{}
    \label{fig:lumi550:ge}
\end{subfigure}\hfill
\begin{subfigure}{.5\textwidth}
    \centering
    \includegraphics[width=0.95\textwidth]{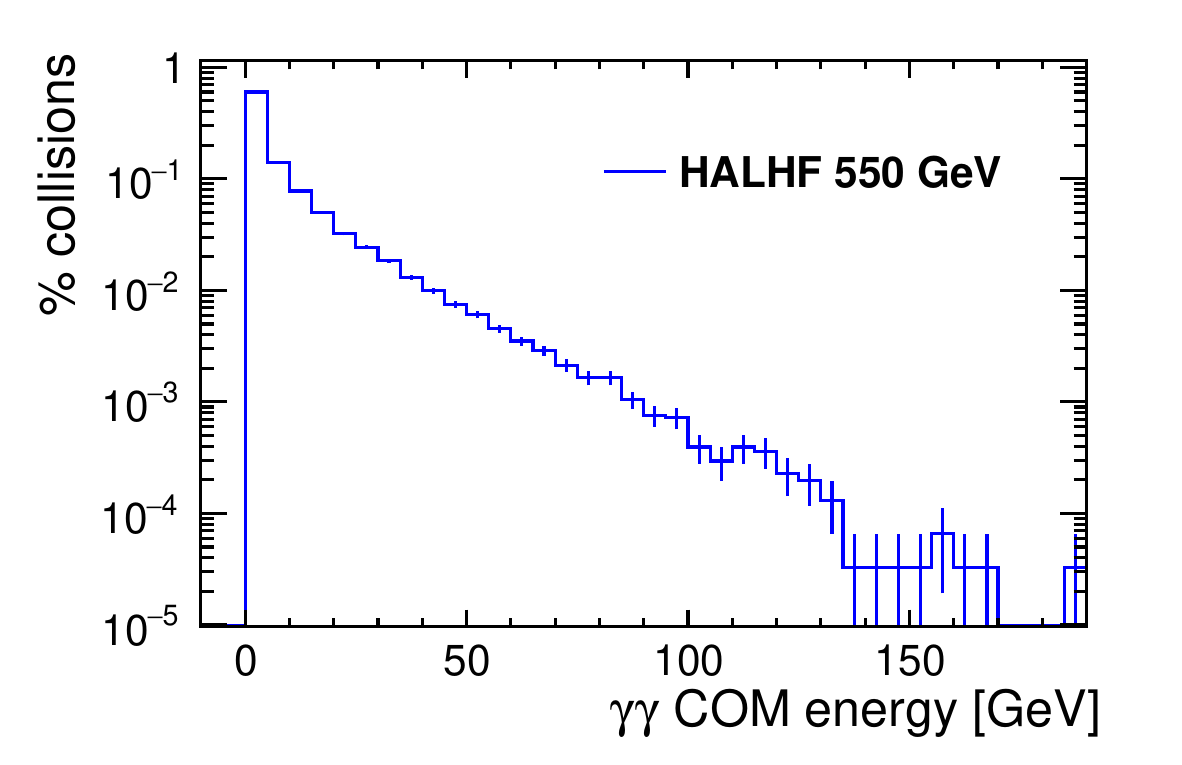}
    \caption{}
    \label{fig:lumi550:gg}
\end{subfigure}
\caption{Luminosity spectra for (a) $e^+e^-$, (b) $e^-\gamma$, (c) $\gamma e^+$, and (d) $\gamma\gamma$ collisions at HALHF550 obtained from {\sc GuineaPig}.} 
\label{fig:lumi550}
\end{figure}

The corresponding effect can be seen in the $e^-\gamma$ (Figs.~\ref{fig:lumi250:eg1} and~\ref{fig:lumi250:eg2}) and $\gamma e^+$ (Figs.~\ref{fig:lumi250:ge1} and~\ref{fig:lumi250:ge2}) spectra: The $e^-\gamma$ (i.e.\ with the $\gamma$ emitted by the positron beam) luminosity spectrum is much softer than the $\gamma e^+$ (i.e.\ with the $\gamma$ emitted by the electron beam) spectrum, despite the fact that the electron beam has a much higher energy than the positron beam.
The $\gamma\gamma$ luminosity spectrum, however, has a very similar shape to that at ILC. This means that e.g.\ the production of low-$p_T$ hadrons from $\gamma\gamma$ collisions, a.k.a.\ ``overlay background'', will have similar properties in the centre-of-mass system than what is known from ILC, but of course the produced hadrons will be forward-boosted in the lab frame.
Figures~\ref{fig:lumi380} and~\ref{fig:lumi550} show the analogous spectra for HALHF380 and HALHF550, respectively. 

\begin{figure}[htb!]
\centering
\begin{subfigure}{.5\textwidth}
    \centering
    \includegraphics[width=0.95\textwidth]{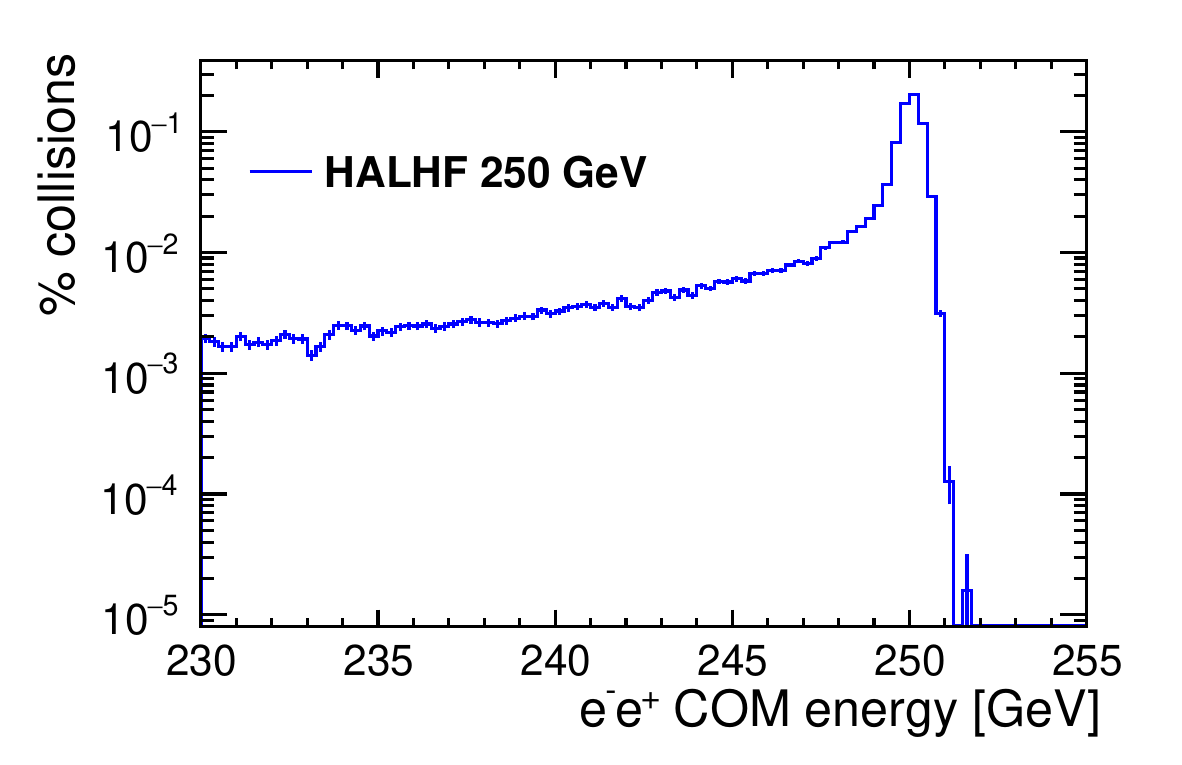}
    \caption{}
    \label{fig:lumi_zoom:250}
\end{subfigure}\hfill
\begin{subfigure}{.5\textwidth}
    \centering
    \includegraphics[width=0.95\textwidth]{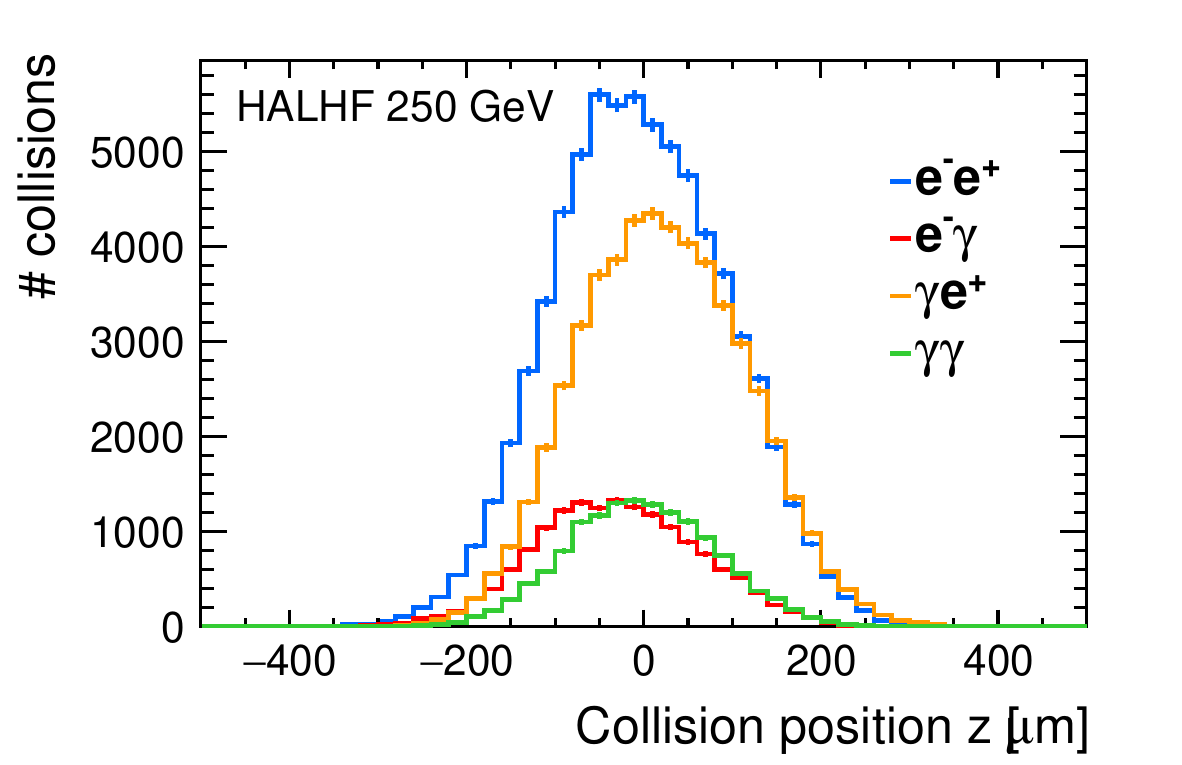}
    \caption{}
    \label{fig:zvtx:250}
\end{subfigure}
\vspace{0.1cm}
\begin{subfigure}{.5\textwidth}
    \centering
    \includegraphics[width=0.95\textwidth]{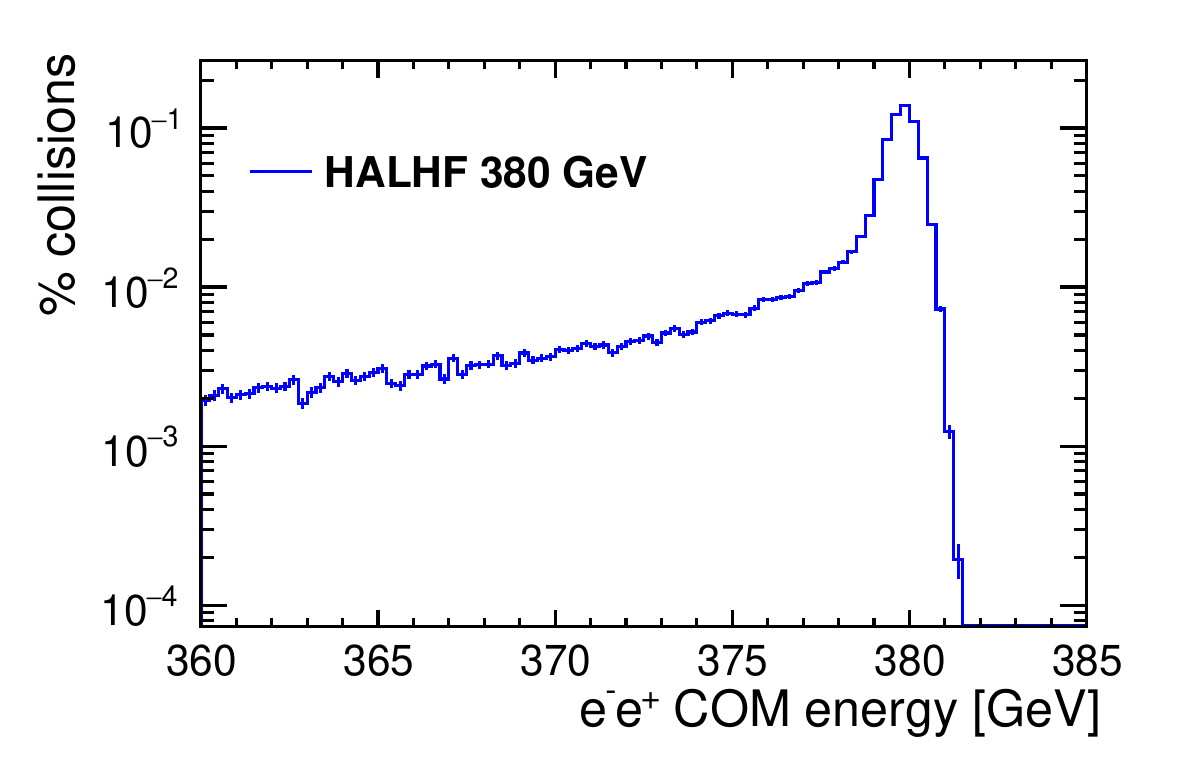}
    \caption{}
    \label{fig:lumi_zoom:380}
\end{subfigure}\hfill
\begin{subfigure}{.5\textwidth}
    \centering
    \includegraphics[width=0.95\textwidth]{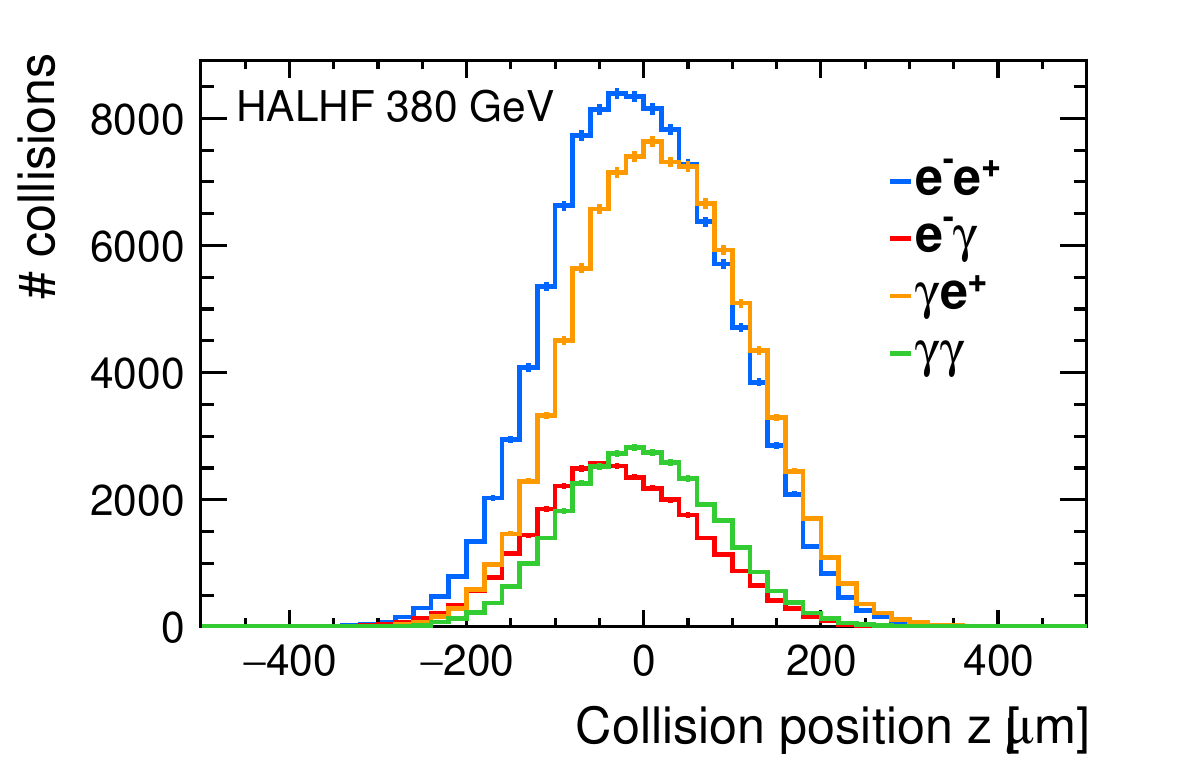}
    \caption{}
    \label{fig:zvtx:380}
\end{subfigure}
\vspace{0.1cm}
\begin{subfigure}{.5\textwidth}
    \centering
    \includegraphics[width=0.95\textwidth]{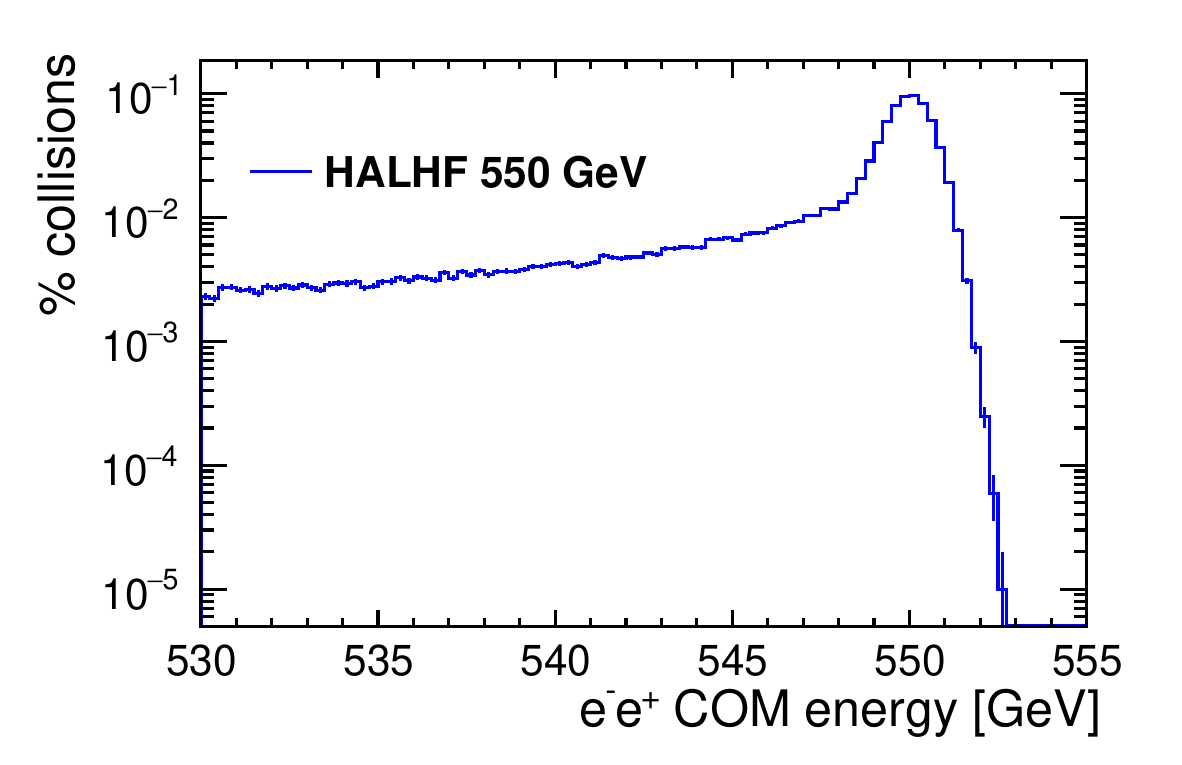}
    \caption{}
    \label{fig:lumi_zoom:550}
\end{subfigure}\hfill
\begin{subfigure}{.5\textwidth}
    \centering
    \includegraphics[width=0.95\textwidth]{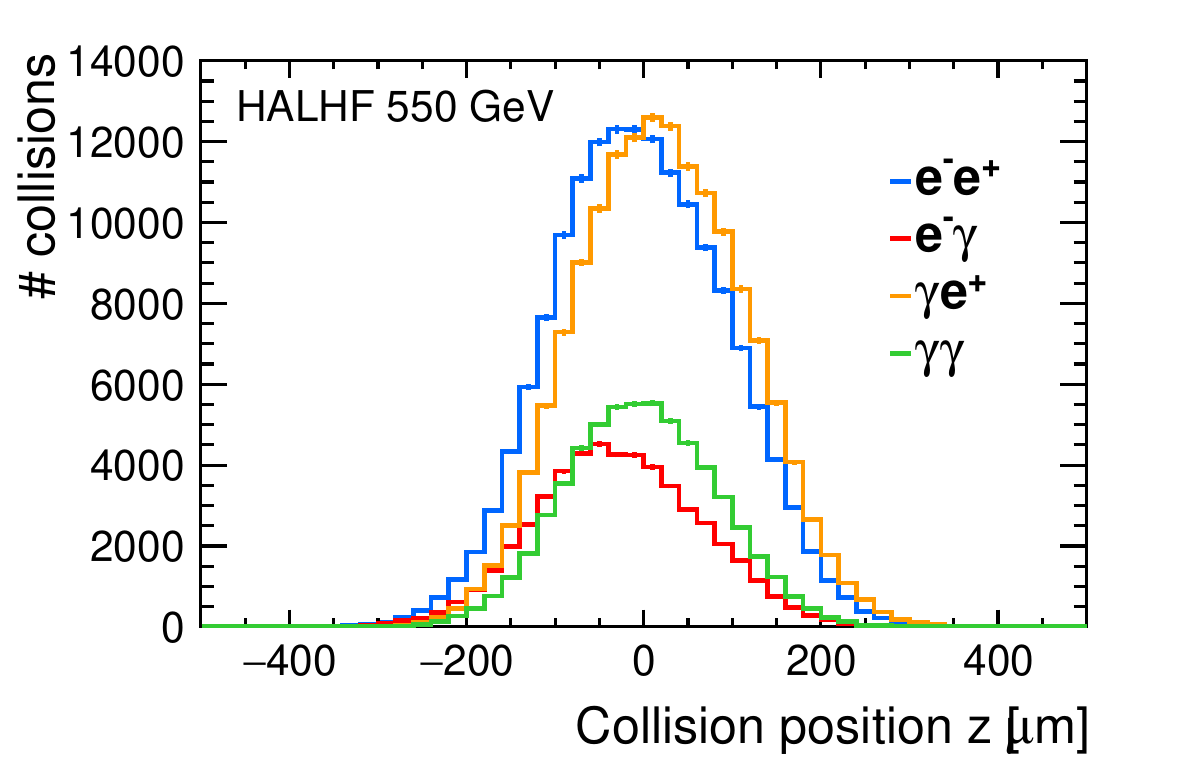}
    \caption{}
    \label{fig:zvtx:550}
\end{subfigure}
\label{fig:lumi_zoom_zvtx}
\caption{Zoomed luminosity spectra for $e^+e^-$ collisions at (a) 250\,GeV,  (c) 380\,GeV, and (e) 550\,GeV along with the distribution of the collisions along the $z$ axis for $e^+e^-$, $e^-\gamma$, $\gamma e^+$, $\gamma\gamma$ collisions at (b) 250\,GeV, (d) 380\,GeV, and (f) 550\,GeV obtained from {\sc GuineaPig}. The distributions are normalised to collisions per bunch crossing.}
\end{figure}

Zoomed views of peak regions of the $e^+e^-$ spectra, important for the physics performance, are shown in Fig.~\ref{fig:lumi_zoom:250},~\ref{fig:lumi_zoom:380} and~\ref{fig:lumi_zoom:550} for the three centre-of-mass energies, respectively, along with the corresponding distributions of the collisions along the $z$ axis in Fig.~\ref{fig:zvtx:250},~\ref{fig:zvtx:380} and~\ref{fig:zvtx:550}. The latter show $e^+e^-$, $e^-\gamma$, $\gamma e^+$ and  $\gamma\gamma$ collisions overlaid, now showing also the relative frequency of the four types of interactions. Here it becomes obvious that $\gamma e+$ collisions also occur at a much higher rate than $e^- \gamma$ collisions, and especially at the higher centre-of-mass energies become as frequent as $e^+e^-$ collisions.

\clearpage

\subsubsection{Beam-induced backgrounds}
\label{sec:physics:background}

For linear colliders, it has been known for a long time that $e^+e^-$ pairs from beamstrahlung are the dominant source of background hits in the detector~\cite{Vogel:2008zza}, while real photons and Compton-scattered electrons/positrons mostly leave the detector through the beampipe. Single-beam backgrounds (e.g.\ from beam-gas interactions) play a sub-leading role~\cite{Karl:2019hes}. 
Therefore, the $e^+e^-$ pairs from beamstrahlung are the focus here, while of course in the future, with a much more advanced design of the HALHF accelerator, a full evaluation of all background sources, including e.g.\ neutrons back-scattering from the beam dumps, should be done.

\begin{figure}[htb!]
\centering
\begin{subfigure}{.5\textwidth}
    \centering
    \includegraphics[width=0.95\textwidth]{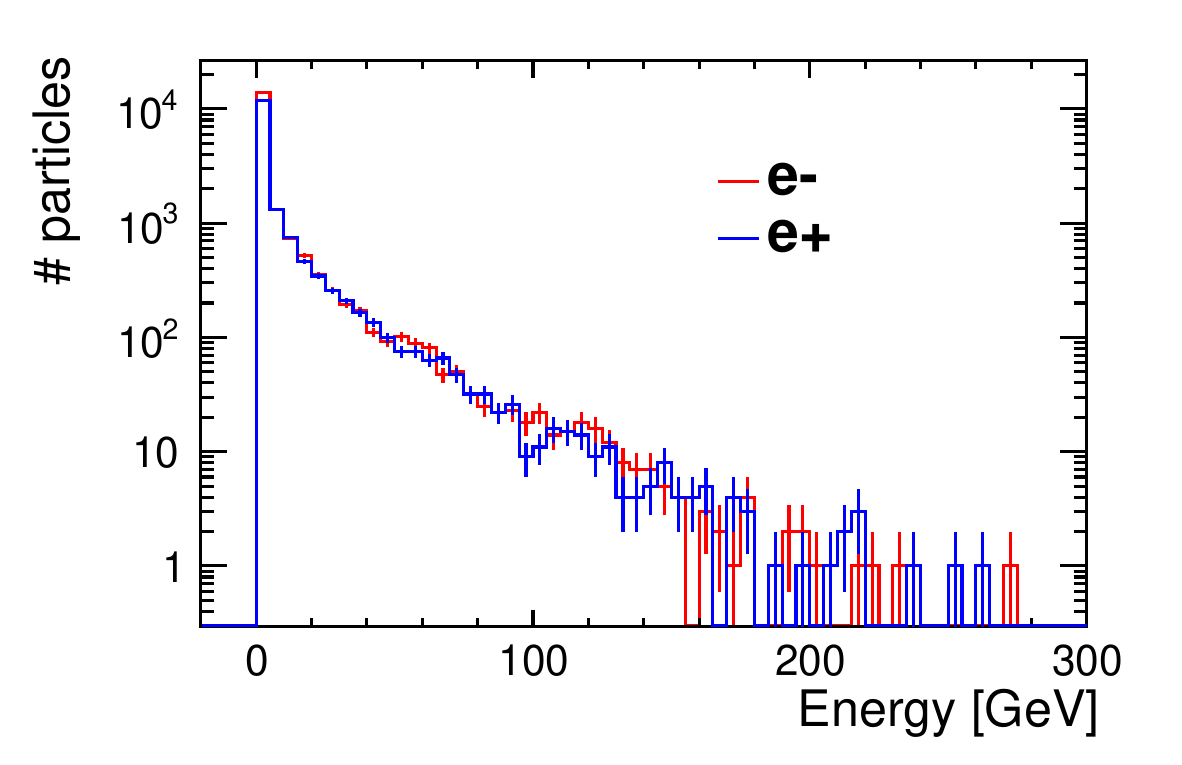}
    \caption{}
    \label{fig:pairs1D:250_fwd}
\end{subfigure}\hfill
\begin{subfigure}{.5\textwidth}
    \centering
    \includegraphics[width=0.95\textwidth]{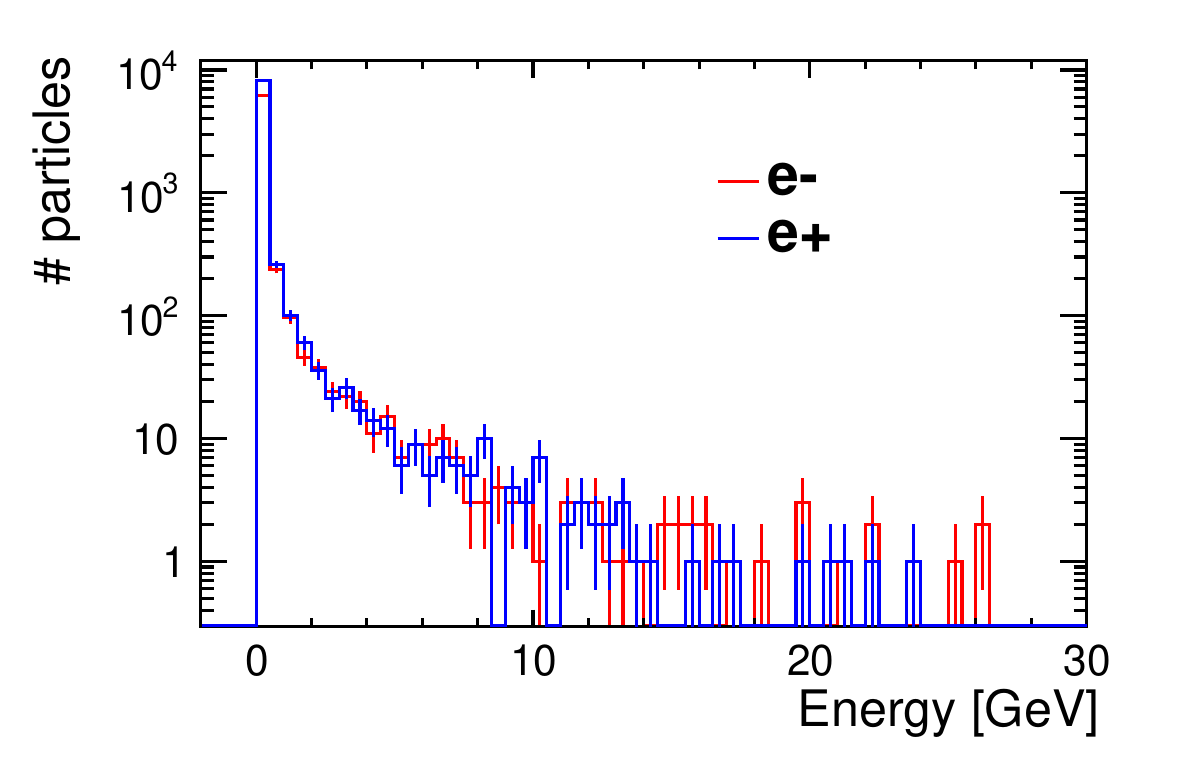}
    \caption{}
    \label{fig:pairs1D:250_bwd}
\end{subfigure}
\vspace{0.1cm}
\begin{subfigure}{.5\textwidth}
    \centering
    \includegraphics[width=0.95\textwidth]{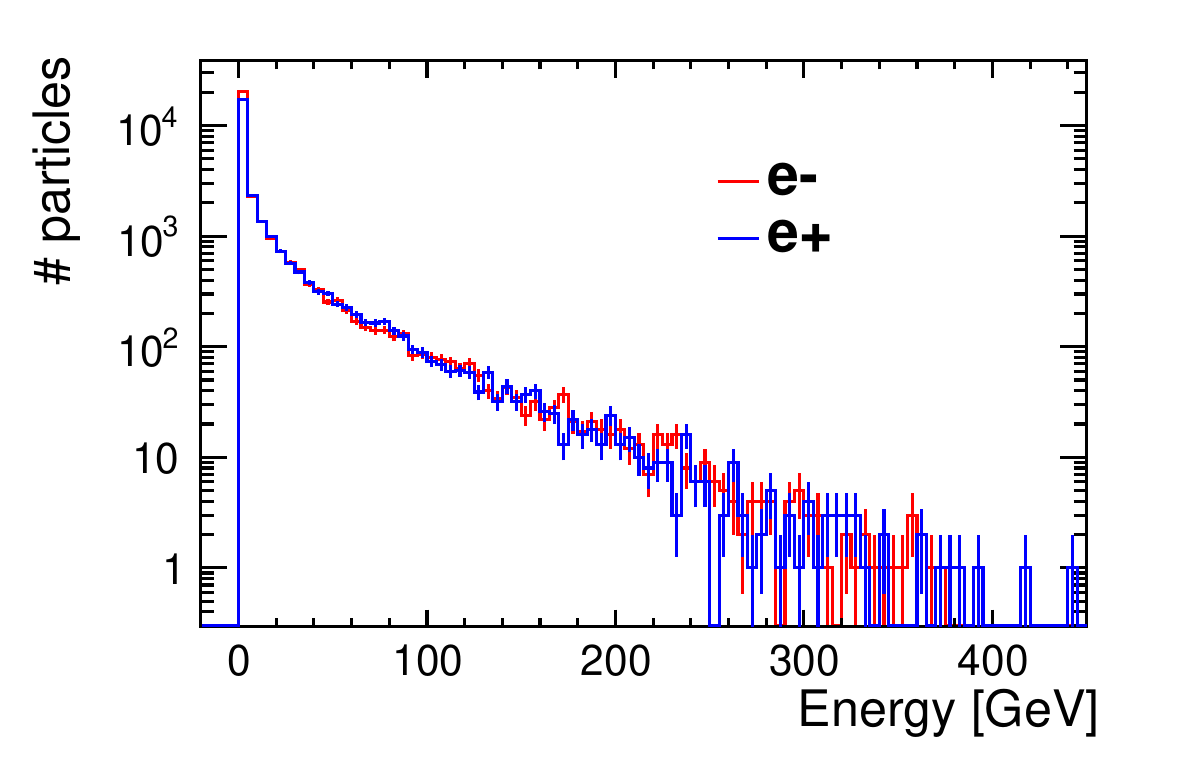}
    \caption{}
    \label{fig:pairs1D:380_fwd}
\end{subfigure}\hfill
\begin{subfigure}{.5\textwidth}
    \centering
    \includegraphics[width=0.95\textwidth]{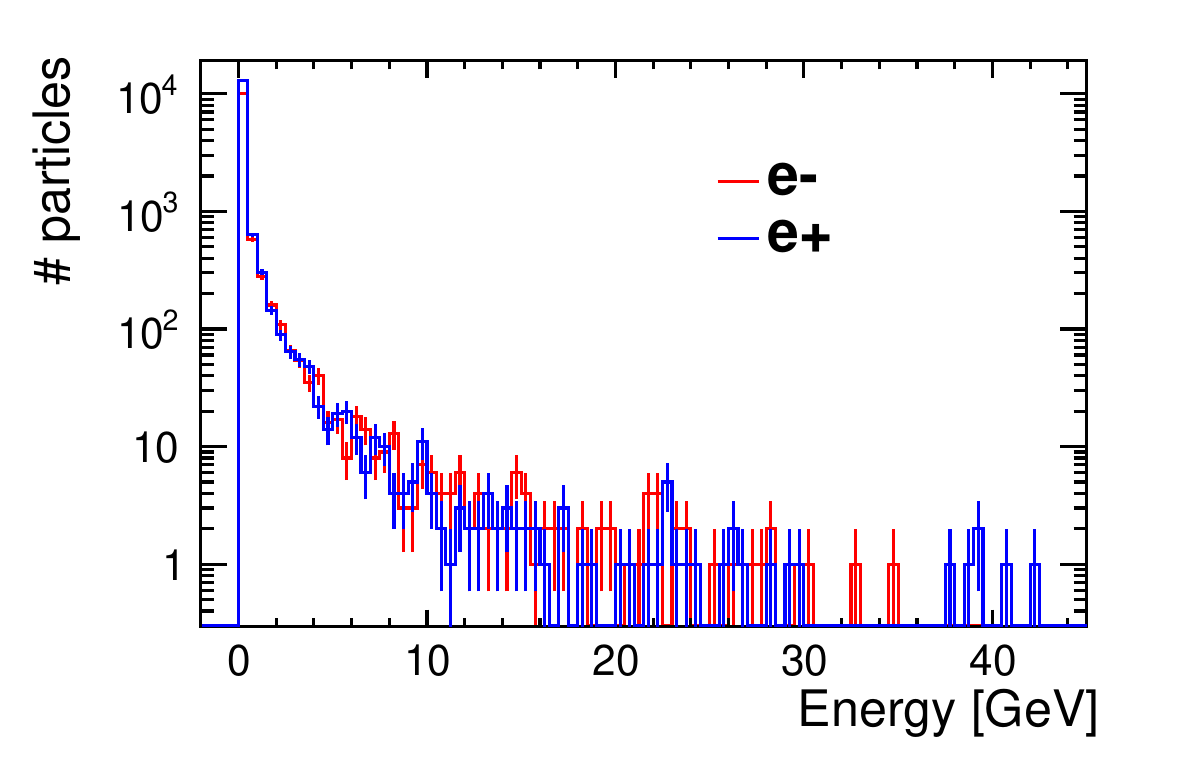}
    \caption{}
    \label{fig:pairs1D:380_bwd}
\end{subfigure}
\vspace{0.1cm}
\begin{subfigure}{.5\textwidth}
    \centering
    \includegraphics[width=0.95\textwidth]{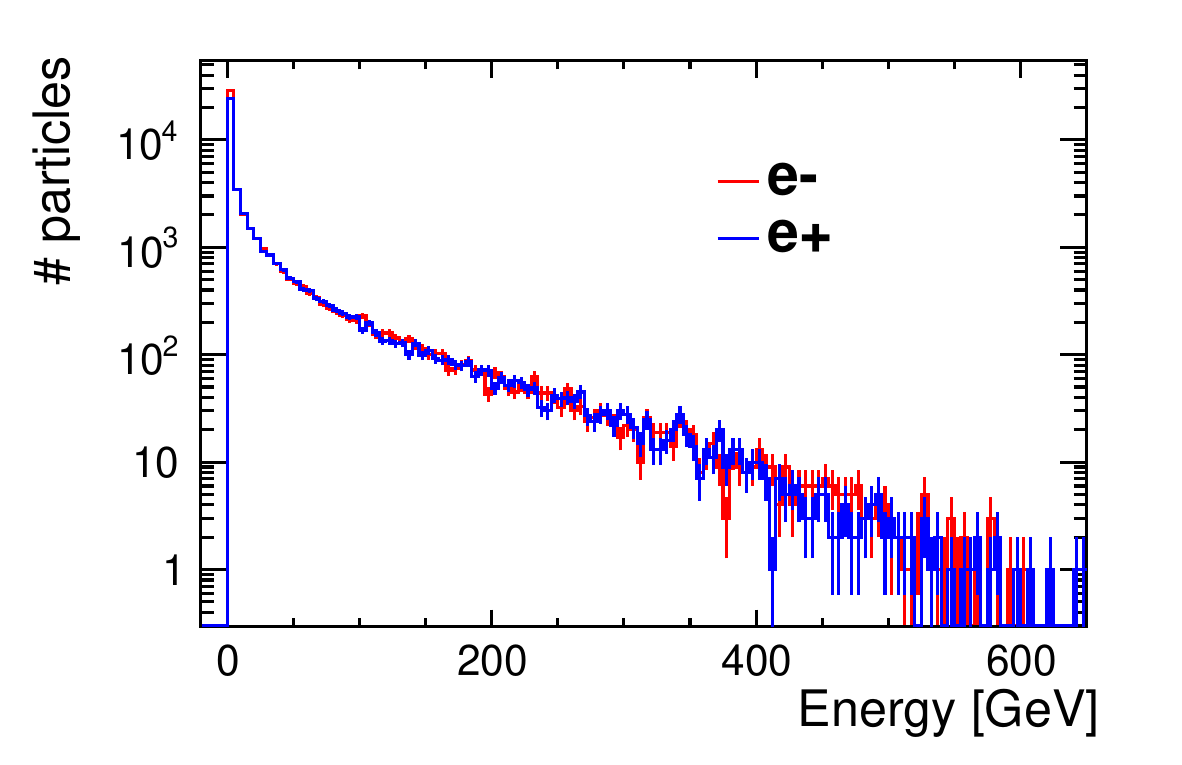}
    \caption{}
    \label{fig:pairs1D:550_fwd}
\end{subfigure}\hfill
\begin{subfigure}{.5\textwidth}
    \centering
    \includegraphics[width=0.95\textwidth]{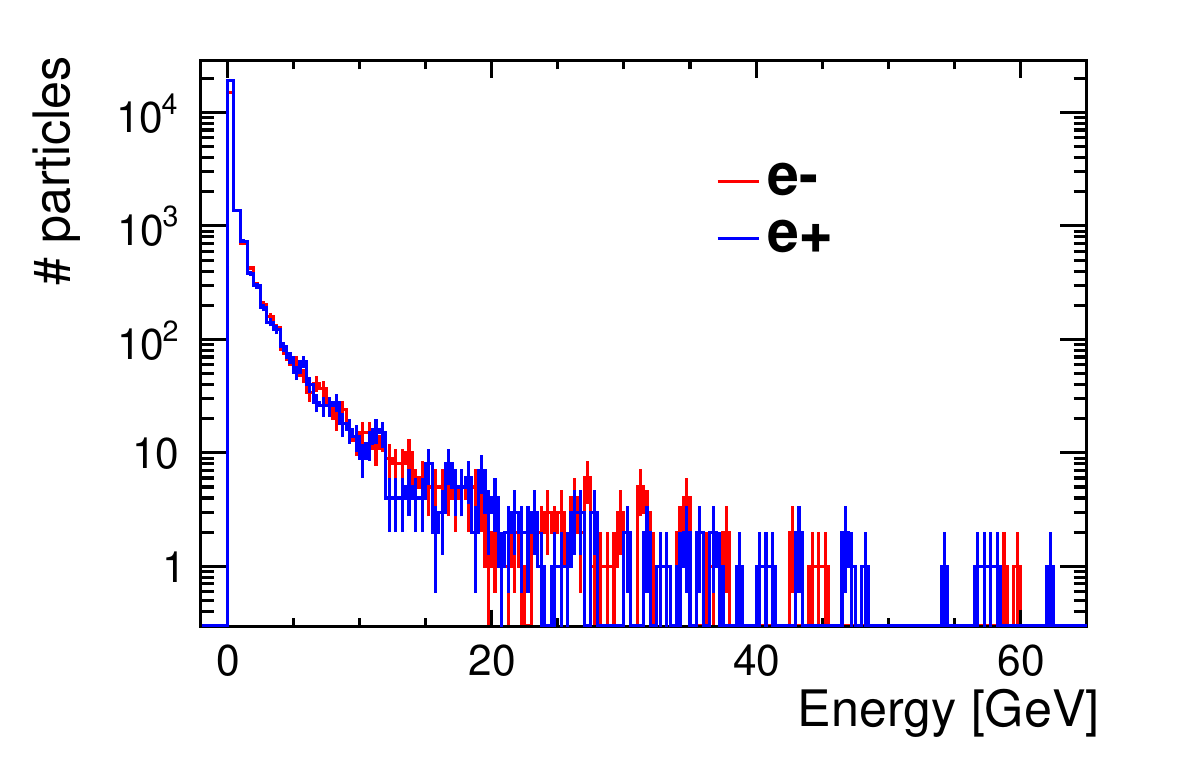}
    \caption{}
    \label{fig:pairs1D:550_bwd}
\end{subfigure}
\caption{Energy spectra of the $e^+e^-$ pair particles created from beamstrahlung in forward direction for (a) 250, (c) 380, and (e) 550 GeV centre-of-mass energies, and in backward direction for (b) 250, (d) 380, and (f) 550 GeV centre-of-mass energies. The distributions are normalised to collisions per bunch crossing.} 
\label{fig:pairs1D}
\end{figure}
\clearpage
Figure~\ref{fig:pairs1D} shows the energy of the electrons and positrons created from beamstrahlung for all three centre-of-mass energies, separately for the forward and backward direction. It can clearly be seen that in the direction of propagation of the electron beam many more energetic particles are expected.

\begin{figure}[H]
\centering
\begin{subfigure}{.5\textwidth}
    \centering  
    \includegraphics[width=0.95\textwidth]{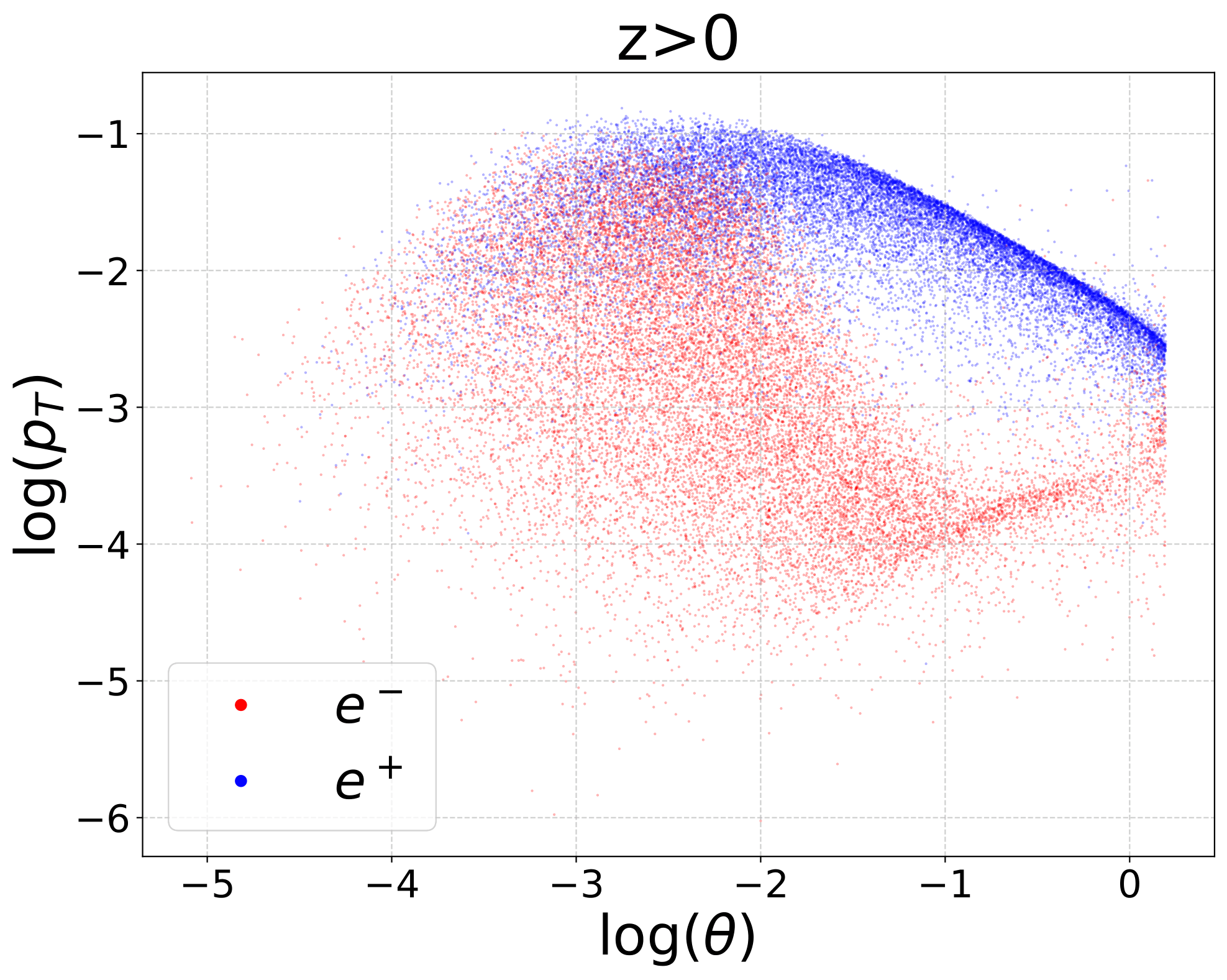}
    \caption{}
    \label{fig:pairs2D:fwd_250}
\end{subfigure}\hfill
\begin{subfigure}{.5\textwidth}
    \centering
    \includegraphics[width=0.95\textwidth]{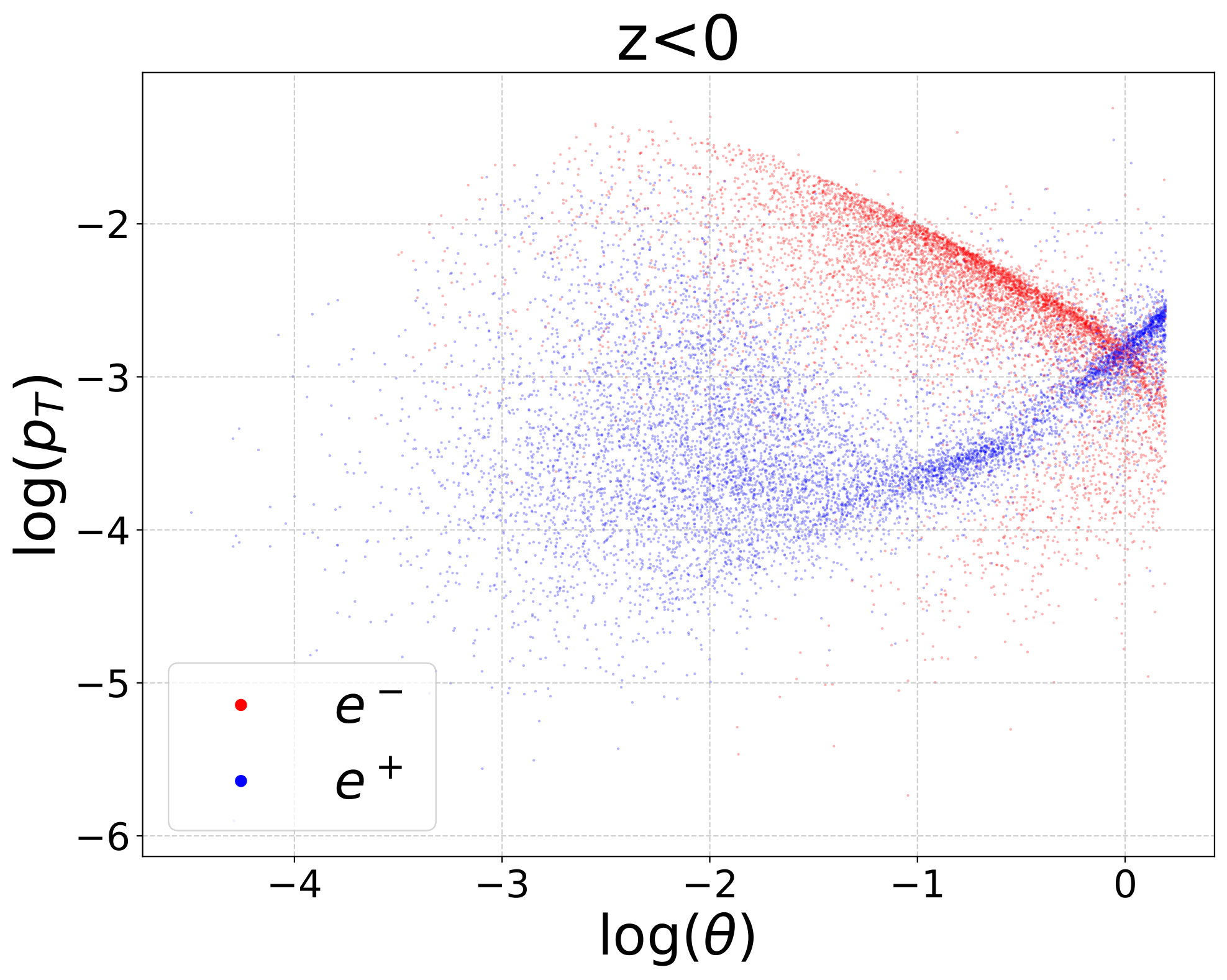}
    \caption{}
    \label{fig:pairs2D:bwd_250}
\end{subfigure}
\vspace{0.1cm}
\begin{subfigure}{.5\textwidth}
    \centering  
    \includegraphics[width=0.95\textwidth]{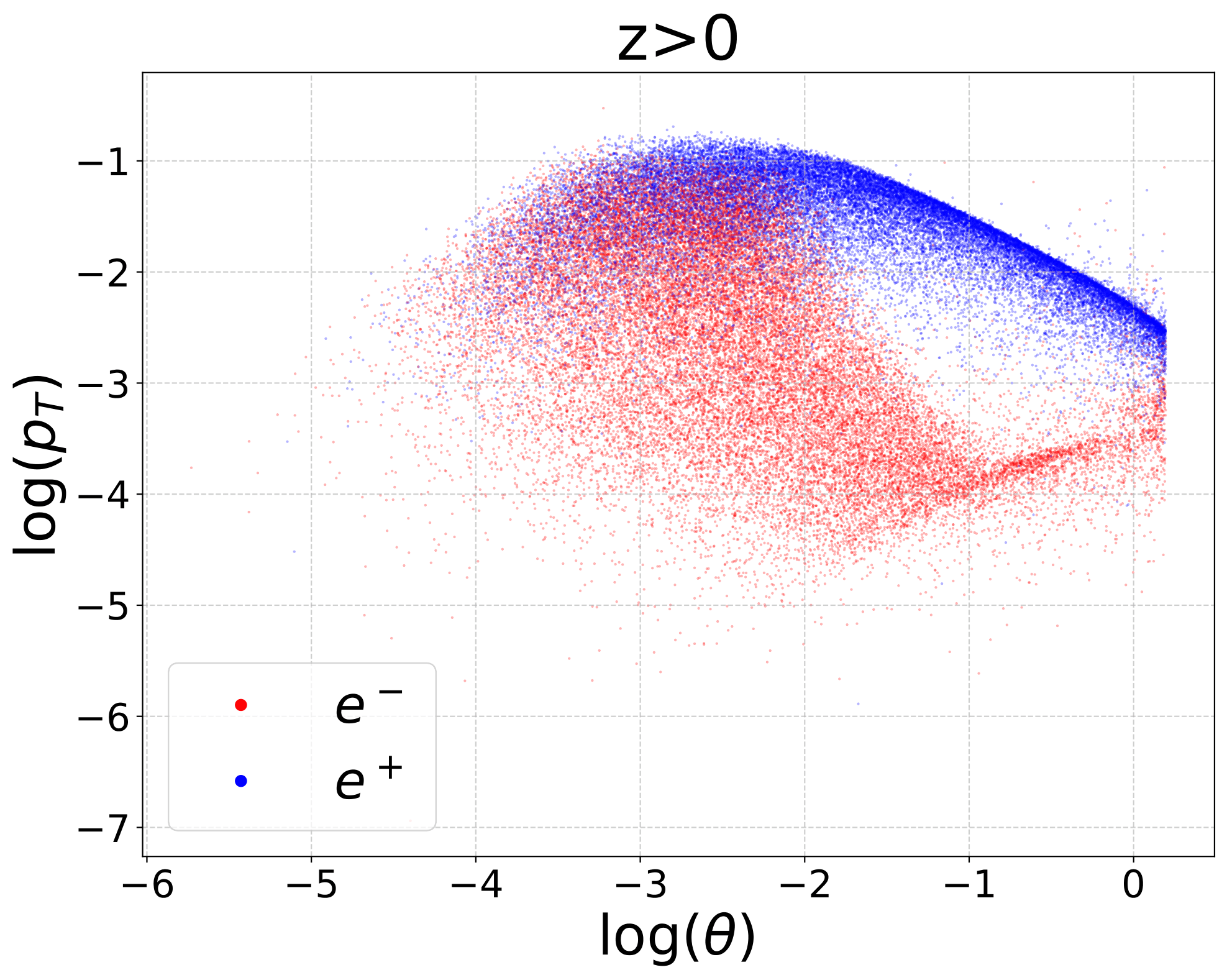}
    \caption{}
    \label{fig:pairs2D:fwd_380}
\end{subfigure}\hfill
\begin{subfigure}{.5\textwidth}
    \centering
    \includegraphics[width=0.95\textwidth]{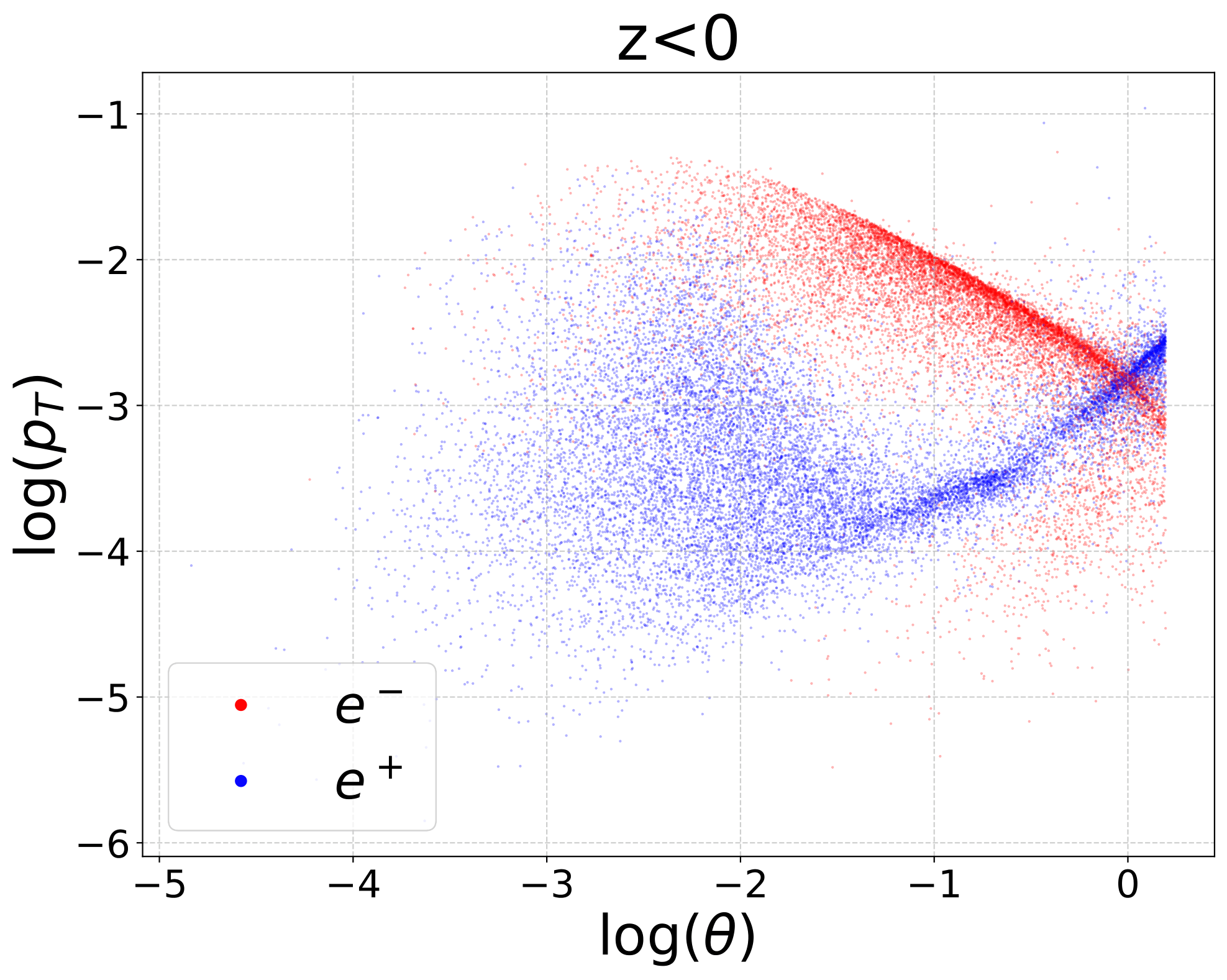}
    \caption{}
    \label{fig:pairs2D:bwd_380}
\end{subfigure}
\vspace{0.1cm}
\begin{subfigure}{.5\textwidth}
    \centering  
    \includegraphics[width=0.95\textwidth]{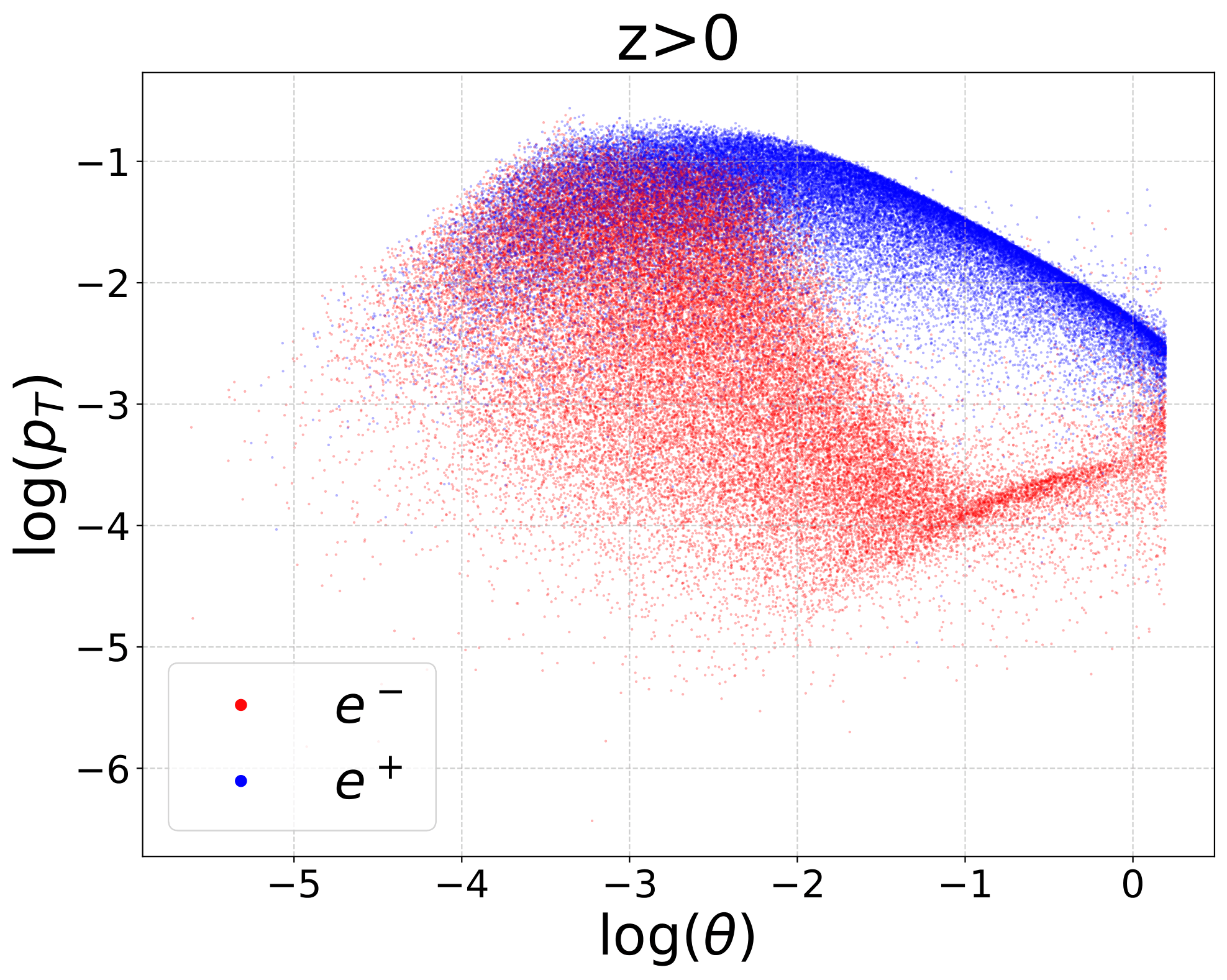}
    \caption{}
    \label{fig:pairs2D:fwd_550}
\end{subfigure}\hfill
\begin{subfigure}{.5\textwidth}
    \centering
    \includegraphics[width=0.95\textwidth]{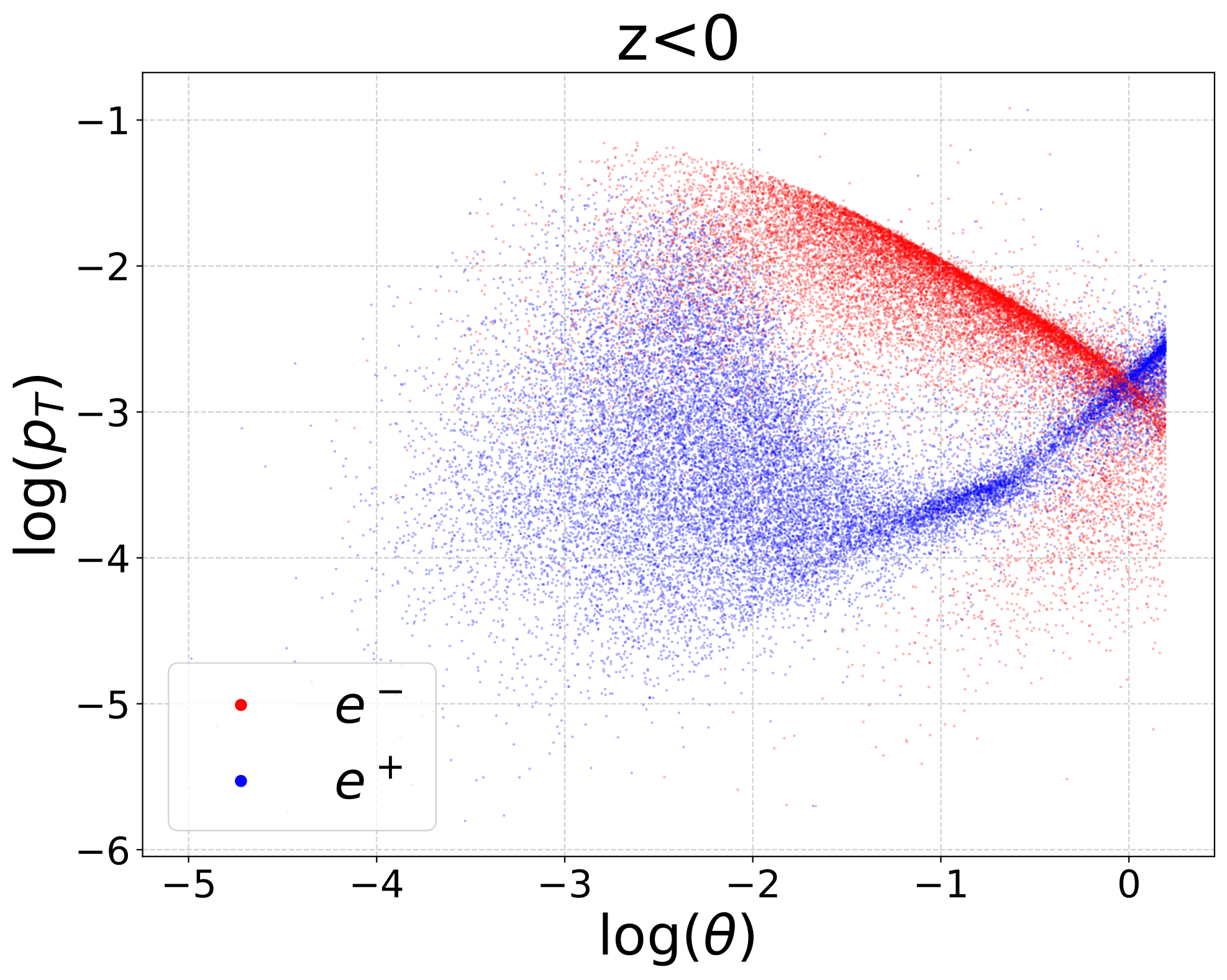}
    \caption{}
    \label{fig:pairs2D:bwd_550}
\end{subfigure}
\caption{Transverse momentum ($\log(p_T)$) vs polar angle ($\log(\theta$)) of the $e^+e^-$ pair particles created from beamstrahlung (a) at HALHF250 in forward direction, (b) at HALHF250 in backward direction, (c) at HALHF380 in forward direction, (d) at HALHF380 in backward direction, (e) at HALHF550 in forward direction, and (f) at HALHF550 in backward direction.} 
\label{fig:pairs2D}
\end{figure}

While the momentum of the $e^+e^-$ pairs extends up to the respective beam energy, the majority has components with very low momentum transversally to the beam axis. Thus, the pair particles curl within a rather small radius around the magnetic field lines of the detector solenoid while moving in the $\pm z$ direction. If the beampipe and the innermost sensitive components of the detector stay clear of the majority of these curling pairs, the residual backgrounds from single particles with higher momentum or secondaries can usually be handled by the detectors~\cite{Vogel:2008zza}.

Figure~\ref{fig:pairs2D} shows the distribution of the $e^+e^-$ pair particles in the plane of their transverse momentum vs their polar angle, both logarithmically. The detector acceptance typically starts at best at a few mrad, i.e.\ any particle below a few $10^{-3}$ in polar angle will exit through the beampipe. Likewise, particles below a few 10\,MeV in transverse momentum will curl up inside the central beampipe and follow a helical path towards the very forward (or backward) region of the detector. 
As a rule of thumb, the beampipe (and thus any  instrumentation) should be placed such that only particles with transverse momenta and polar angles in the upper-right corner, outside the sharp edge of the point cloud, can reach them.

\subsection{Detector design and physics program}
\label{sec:physics:detector}

Both the detector design for symmetric $e^+e^-$ linear colliders and the physics program of polarised high-energy $e^+e^-$ colliders have been studied for decades, resulting in a very robust understanding and advanced prototypes for most of the detector components~\cite{LinearColliderVision:2025hlt, ILCInternationalDevelopmentTeam:2022izu, Breidenbach:2021sdo, ILDConceptGroup:2020sfq, Dannheim:2019rcr, CLICdp:2018vnx, Fujii:2017vwa, Moortgat-Pick:2015lbx, ILC:2013jhg, Behnke:2013lya}.

For HALHF, all these considerations apply to first order, if a) the collider can produce similar top-level performance parameters, i.e.\ in terms of centre-of-mass energy, luminosity and beam polarization, and if b) a detector can be designed which -- considering the asymmetric background conditions and the forward boost of the physics events -- delivers the same effective performance. The canonical top-level detector performance goals for a linear  $e^+e^-$ collider have recently been summarised in~\cite{LinearColliderVision:2025hlt}:
\begin{description}
    \item[Impact parameter resolution:]  An impact parameter resolution of $ 5\,\mu \mathrm{m} \oplus 10\,\mu \mathrm{m} / [ p~[{\mathrm{GeV}/c}]\sin^{3/2}\theta$] has been defined as a goal in order to enable efficient identification of charm-quark decays. 
    \item[Momentum resolution:] An inverse momentum resolution of $\Delta (1 / p) = 2 \times 10^{-5}\,\mathrm{[(GeV/c)}^{-1}]$ asymptotically at high momenta should be reached in order not to limit the precision of the Higgs recoil measurement in the $Z\to\mu\mu$ channel. Maintaining excellent tracking efficiency and very good momentum resolution at lower momenta will be achieved by an aggressive design to minimise the detector's material budget.
    \item[Jet energy resolution:] A jet energy resolution $\Delta E/ E =~$\SI{3}{-}\SI{4}{\%} is required for the statistical separation of hadronically decaying $W$, $Z$, and Higgs bosons. Particle-flow-based detectors and corresponding reconstruction algorithms have been shown to meet this requirement. 
 
\end{description}

The canonical detector concepts for symmetric colliders typically perform best in the barrel detectors, i.e.\ for polar angles above $\simeq 30^{\circ}$, and worsening in the forward (and backward) regions of the detector. 

For HALHF, the high performance needs to extend significantly further forward than for the symmetrical ILD, while staying clear of the larger pair background in the forward region. One solution to this might be the design of dedicated forward detectors as was done for the HERA detectors ZEUS and H1~\cite{ZEUSH1}. Such a design is however outside the scope of this document and beyond the currently available effort on HALHF. The simpler solution of extending the length of the central TPC is therefore assumed in what follows. 

\subsubsection{The Higgs recoil measurement as guidance}
The measurement of the total $e^+e^- \to ZH$ cross-section, independently of the Higgs decay mode, solely via its recoil against the $Z$ boson, provides the normalisation for determining the Higgs boson's decay width and thus its absolute couplings to fermions and bosons. The cleanest channel, also providing the best determination of the Higgs-boson mass, is given by the $Z$ boson's decay into a pair of muons. The recoil mass distribution in this channel depends crucially on the momentum resolution of the detector, and thus provides a simple but stringent benchmark.

The recoil mass is given by
\begin{equation}
    m_{\mathrm{recoil}}^2 = (E_{\mathrm{CoM}} - (E_{\mu^+}  + E_{\mu^-}))^2 - ( \vec{p}_{\mu^+} + \vec{p}_{\mu^-} )^2,
\end{equation}
which shows that besides the centre-of-mass energy, only the measured momenta of the two muons enter the measurement of the recoil mass.  

Already a very first look (c.f.\ Fig.~10 in~\cite{Foster:2025gsj}) showed that placing one of the ILC detector concepts, in that case ILD, unmodified at the HALHF interaction point would wash out the recoil peak substantially -- but also that this can be recovered by choosing a detector design with better momentum resolution in the forward region. While providing an important starting point, this very first study had two shortcomings: the barrel of the ``e-ILD'' detector had been doubled in length, a ``brute force'' procedure that is not realistic in terms of beam backgrounds, nor very cost-effective. Neither had the beam-energy spectrum been taken into account at all, i.e.\ before initial-state radiation, each event had exactly a centre-of-mass energy of 250\,GeV, exaggerating the impact of the momentum resolution.

In this document, these earlier results are updated by taking into account for the first time not only the boost, but also a luminosity spectrum and the pair background. The ILD detector~\cite{ILDConceptGroup:2020sfq} is used as a starting point, and the length of the main barrel tracking device, the time projection chamber, as well as of the corresponding auxiliary silicon tracking layers, are varied.  

\subsubsection{Implications of the pair background}
The first requirement is to observe the stay-clear from the hottest region of the pair background, i.e.\ the sharp ``edges'' towards the upper-right corner in the panels in Fig.~\ref{fig:pairs2D}.

As a very efficient way of estimate, we propagate the electrons and positrons in the solenoidal magnetic field of the detector, by default 3.5\,T in ILD, and evaluate the $r$ and $z$ coordinates at which they (first) reach the maximum $r$ on their helical path. These can then be graphically compared to the corresponding coordinates of the beam pipe and the detector. 

\begin{figure}[htb]
\centering
\begin{subfigure}{.5\textwidth}
    \centering  
    \includegraphics[width=0.95\textwidth]{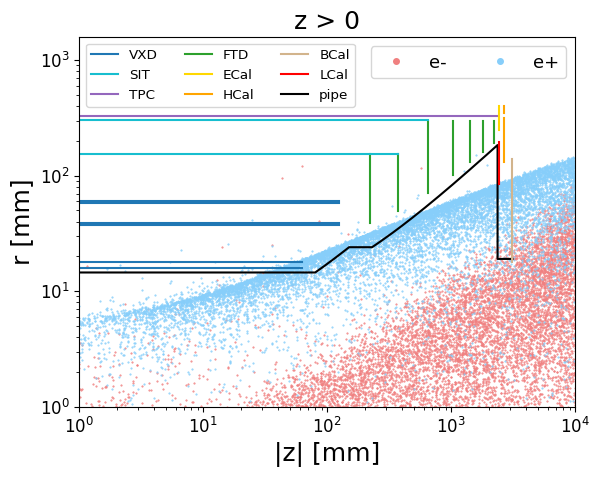}
    \caption{}
    \label{fig:pairs2D:fwd_250_ild}
\end{subfigure}\hfill
\begin{subfigure}{.5\textwidth}
    \centering
    \includegraphics[width=0.95\textwidth]{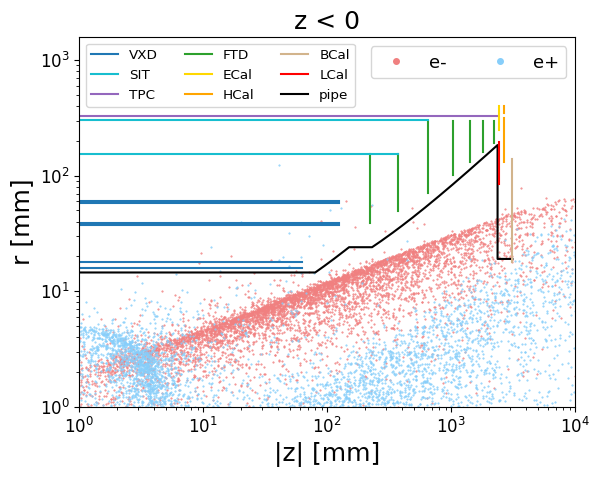}
    \caption{}
    \label{fig:pairs2D:bwd_250_ild}
\end{subfigure}
\vspace{0.1cm}
\begin{subfigure}{.5\textwidth}
    \centering  
    \includegraphics[width=0.95\textwidth]{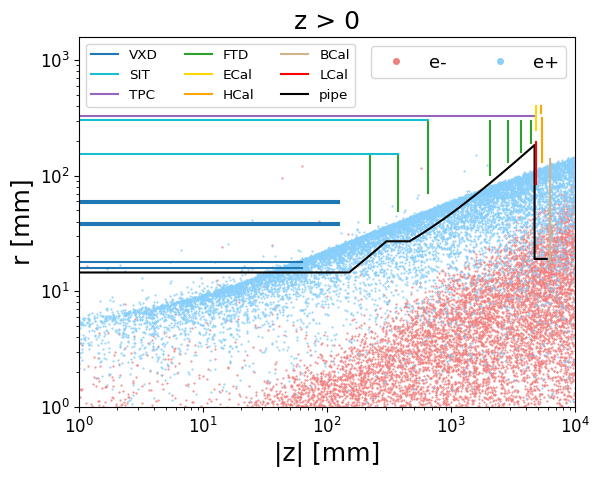}
    \caption{}
    \label{fig:pairs2D:fwd_250_eild1}
\end{subfigure}\hfill
\begin{subfigure}{.5\textwidth}
    \centering
    \includegraphics[width=0.95\textwidth]{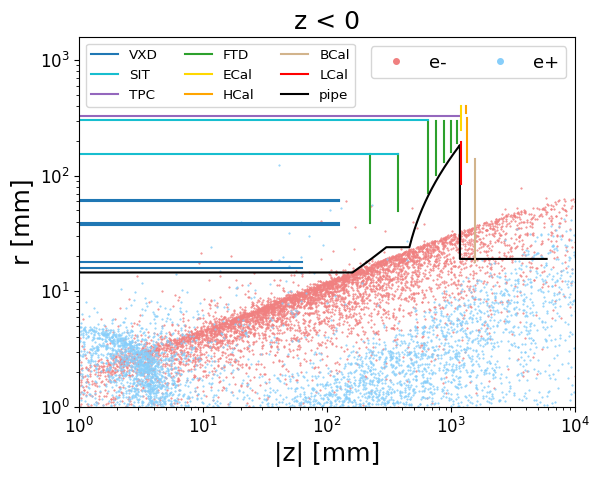}
    \caption{}
    \label{fig:pairs2D:bwd_250_eild1}
\end{subfigure}
\caption{Radial ($r$) vs longitudinal ($z$) coordinate of the maximum propagation of the $e^+e^-$ pair particles at HALHF250 with the ILD detector (a) in forward direction, and (b) in backward direction and with the e-ILD\_v1 detector (c) in forward direction, and (d) in backward direction.} 
\label{fig:pairs2D:250:olddet}
\end{figure}

The result is shown in Fig.~\ref{fig:pairs2D:250:olddet} for the symmetric standard ILD detector ((a) and (b)) as well as for the ``e-ILD'' detector from Fig.~10 in~\cite{Foster:2025gsj} ((c) and (d)),  which is referred to in the following as ``e-ILD\_v1'' for simplicity. In the forward direction, i.e.\  Fig.~\ref{fig:pairs2D:fwd_250_ild} and (c), the beam pipe as well as the barrel layers of the vertex detector sit within the high-$r$ edge of the positrons from beamstrahlung, which would destroy the detector in showers of secondaries. The only subdetector designed to sustain the direct fan of beamstrahlung pairs is the so-called BeamCal (or BCal, beige line in Fig.~\ref{fig:pairs2D:250:olddet}).
\clearpage

\begin{figure}[htb]
\centering
\begin{subfigure}{.5\textwidth}
    \centering  
    \includegraphics[width=0.95\textwidth]{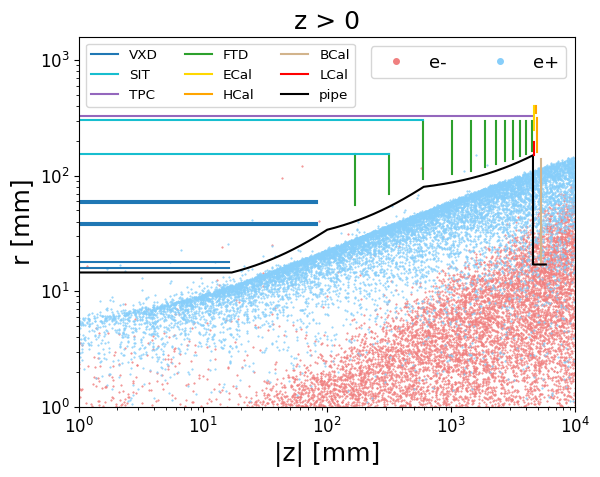}
    \caption{}
    \label{fig:pairs2D:fwd_250_eild2}
\end{subfigure}\hfill
\begin{subfigure}{.5\textwidth}
    \centering
    \includegraphics[width=0.95\textwidth]{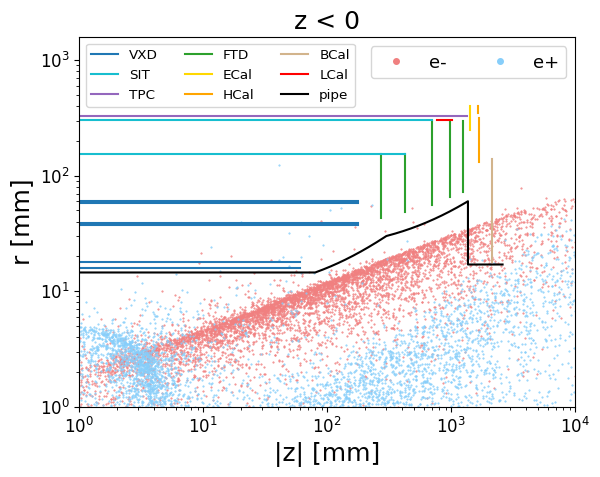}
    \caption{}
    \label{fig:pairs2D:bwd_250_eild2}
\end{subfigure}
\vspace{0.1cm}
\begin{subfigure}{.5\textwidth}
    \centering  
    \includegraphics[width=0.95\textwidth]{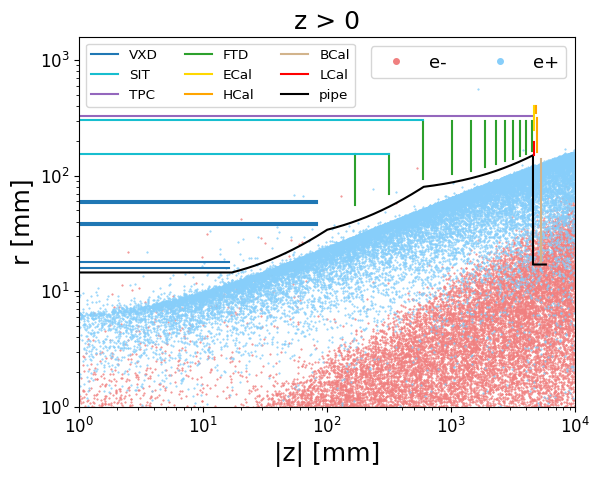}
    \caption{}
    \label{fig:pairs2D:fwd_550_eild2}
\end{subfigure}\hfill
\begin{subfigure}{.5\textwidth}
    \centering
    \includegraphics[width=0.95\textwidth]{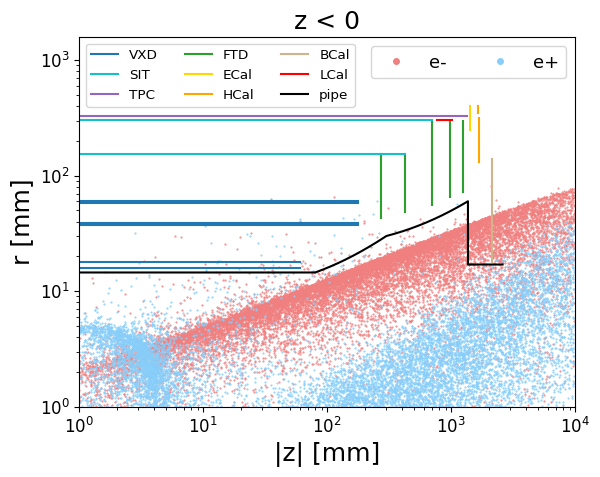}
    \caption{}
    \label{fig:pairs2D:bwd_550_eild2}
\end{subfigure}
\caption{Radial ($r$) vs longitudinal ($z$) coordinate of the maximum propagation of the $e^+e^-$ pair particles at HALHF250 with the e-ILD\_v2 detector overlaid (a) in forward direction, and (b) in backward direction and at HALHF550 with the e-ILD\_v2 detector overlaid (c) in forward direction, and (d) in backward direction.} 
\label{fig:pairs2D:newdet}
\end{figure}

Thus, a very first design action is to move the beam pipe and all tracking detectors beyond the high-$r$ edge of particles from beamstrahlung, in particular beyond the edge of the positrons in the forward direction. The result can be seen in Fig.~\ref{fig:pairs2D:newdet}; this detector is called e-ILD\_v2 in the following. Note that the number and position of the silicon pixel layers (VXD - blue, SIT - cyan, FTD - green) has not yet optimised; this will one of the first steps to take in future.

\subsubsection{Momentum and impact-parameter resolutions}
SGV~\cite{Berggren:2012ar} is used to assess the momentum and impact-parameter resolutions of e-ILD\_v2, based on a prediction of the track covariance matrices via an inverse Kalman filter, taking into account the position and point resolutions of the measurement layers as well as the material budget. 
\begin{figure}[H]
\centering
\begin{subfigure}{.9\textwidth}
    \centering  
\includegraphics[width=0.9\textwidth,clip=]{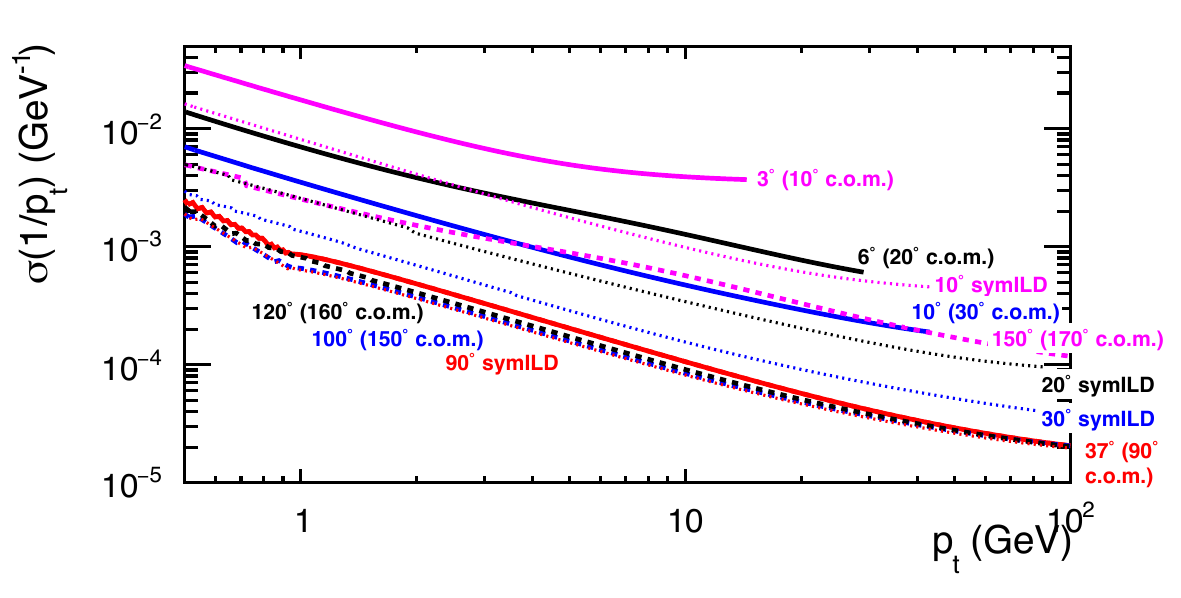}  
    \caption{}
    \label{fig:SGV_res_vs_pt:pt}
\end{subfigure}\hfill
\begin{subfigure}{.9\textwidth}
    \centering
    \includegraphics[width=0.9\textwidth]{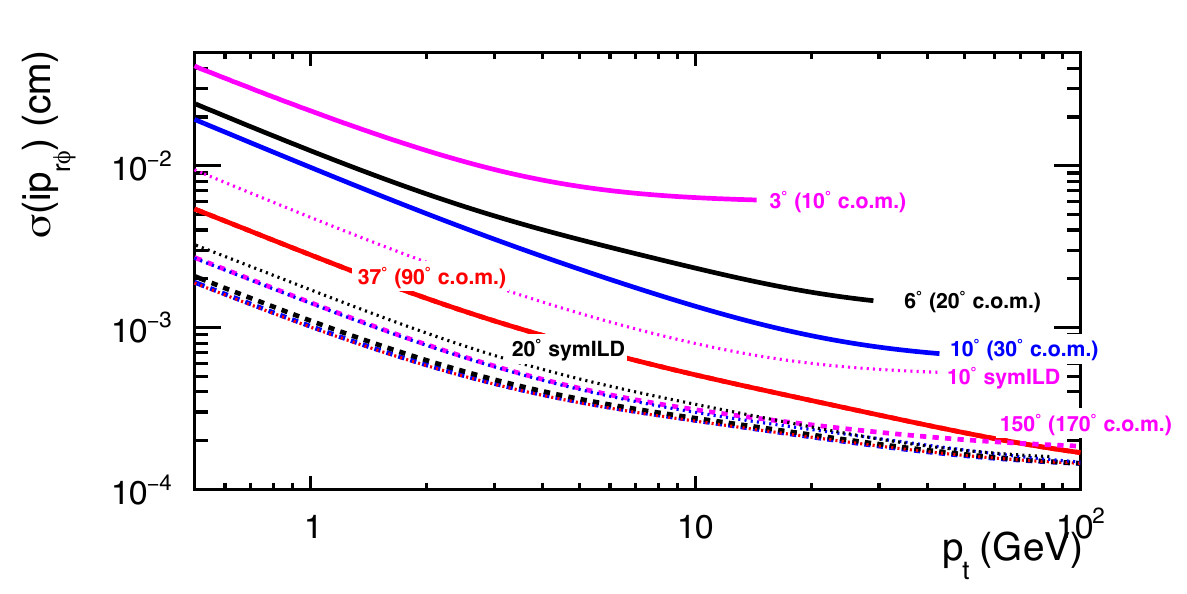}
    \caption{}
    \label{fig:SGV_res_vs_pt:IP}
\end{subfigure}
\caption{Resolution on (a) $1/p_t$ and (b) the impact parameter in the transverse plane as a function of $p_t$ for charged particles produced under fixed angles ($10^{\circ}$, $20^{\circ}$, $30^{\circ}$, $90^{\circ}$ and their corresponding mirrored angles in the backward direction, i.e.\ $150^{\circ}$, $160^{\circ}$, $170^{\circ}$) in the centre-of-mass system.
With the Lorentz boost of HALHF of $\gamma =1.67$ (c.f.\ Table~\ref{tab:1}), these translate to the following angles in the lab frame: $3^{\circ}$, $6^{\circ}$, $10^{\circ}$, $37^{\circ}$ (full lines) and $100^{\circ}$, $120^{\circ}$, $150^{\circ}$ (thick dashed lines). For comparison, the corresponding resolutions of the standard ILD detector are drawn for the symmetric case (i.e.\ lab frame equals centre-of-mass frame). All lines which correspond to the same (or mirrored) angle in the centre-of-mass frame have the same colour.}
\label{fig:SGV_res_vs_pt}
\end{figure}

Figure~\ref{fig:SGV_res_vs_pt} shows the transverse momentum and impact-parameter resolutions as a function of the transverse momentum at different angles in e-ILD\_v2. The angles are chosen such that they correspond to $10^{\circ}$, $20^{\circ}$, $30^{\circ}$, $90^{\circ}$ and their corresponding mirrored angles in the backward direction, i.e.\ $150^{\circ}$, $160^{\circ}$, $170^{\circ}$ \emph{in the centre-of-mass system}. For comparison, corresponding resolutions of the standard ILD detector for the symmetric case (i.e.\ lab frame equals centre-of-mass frame), are also indicated. Considering for example a pair of muons produced back-to-back in the centre-of-mass system, lines of the same colour show the corresponding resolutions at an asymmetric collider and for the forward- and backward-going muon at HALHF. For most (c.o.m.) angles, this means that in the boosted basis, one muon is measured better, as it is more central in the detector, while the other one is measured less well, as it is boosted to smaller angles in the detector.

Specifically, Fig.~\ref{fig:SGV_res_vs_pt:pt} shows that in the barrel, i.e.\ all lines for angles between $37^{\circ}$ and $120^{\circ}$ perform similarly to the $90^{\circ}$ case in standard (symmetric) ILD. It also can be seen that e-ILD\_v2 at $6^{\circ}$ performs still as well as the symmetric ILD at $10^{\circ}$, while e-ILD\_v2 at $10^{\circ}$ performs almost as well as the symmetric ILD at $20^{\circ}$. 

The impact-parameter resolution has not yet been optimised for e-ILD\_v2. This is clearly visible in Fig.~\ref{fig:SGV_res_vs_pt:IP}: while for the ``backward'' angles (thick dashed lines), the resolution is very comparable to the  performance of central tracks in standard ILD, already the resolution at $37^{\circ}$ (i.e.\ $90^{\circ}$ in the centre-of-mass) is significantly worse, and degrades further at even smaller angles (full lines). To optimise the impact-parameter resolution in the forward direction is one of the next steps, e.g.\ using $H\to c \bar{c}$ as a benchmark.

\subsubsection{Higgs recoil spectrum}
The Higgs-recoil measurement provides an excellent physics benchmark for the momentum resolution for isolated high-momentum tracks.
Based on the earlier result that a brute-force doubling of the barrel length of ILD would mitigate the effect of the boost on the Higgs recoil peak, the height of the recoil peak was studied as a function of the forward-side length of the main tracking device of ILD, the time-projection chamber (TPC). The positions of inner silicon detectors are stretched accordingly. However, in order not to overestimate the effect, the HALHF luminosity spectrum is folded in addition, meaning that MC events have been generated with the HALHF beam-energy spectra. Since it turned out that the standard LCC Generator Group setup~\cite{Berggren:2021sju} with Whizard\,3~\cite{Kilian:2007gr} generating the hard events according to tree-level matrix elements and handing them to Pythia\,6~\cite{Sjostrand:2006za} cannot handle an asymmetric-collider setup correctly, existing events with the ILC beam energy spectrum were used. A procedure to generate events with the HALHF beam-energy spectrum will be developed in the future (c.f.\ Sec.~\ref{sec:det:outlook}). 
The impact of using the ILC spectra can be estimated from Fig.~\ref{fig:lumi250:ee2}: It gives a slightly pessimistic result for  the recoil spectrum at HALHF, but is sufficiently close to the HALHF beam-energy spectrum that the impact of the detector resolution is estimated much more realistically than previously, when no energy spectrum was considered at all.

\begin{figure}[H]
\centering
\begin{subfigure}{.5\textwidth}
    \centering  
    \includegraphics[width=0.95\textwidth]{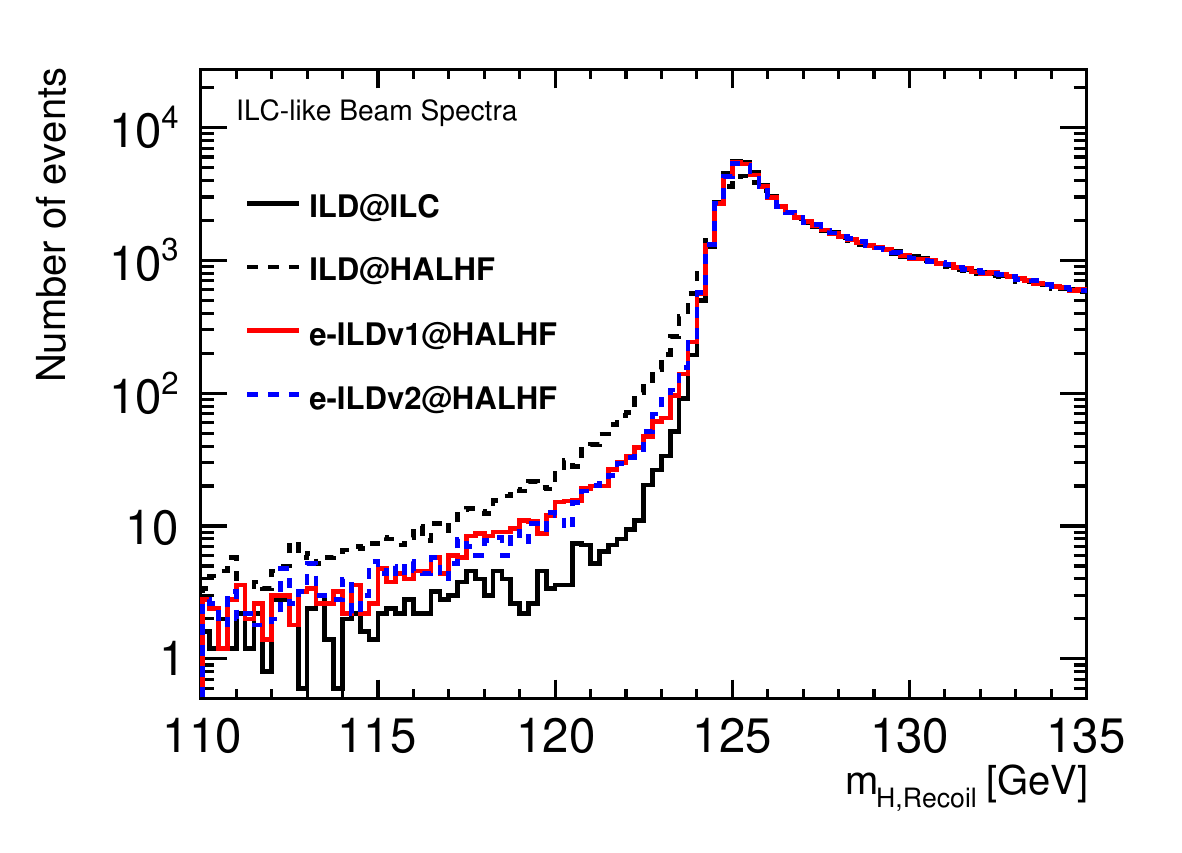}
    \caption{}
    \label{fig:m_rec:log}
\end{subfigure}\hfill

\begin{subfigure}{.5\textwidth}
    \centering  
\includegraphics[width=0.95\textwidth]{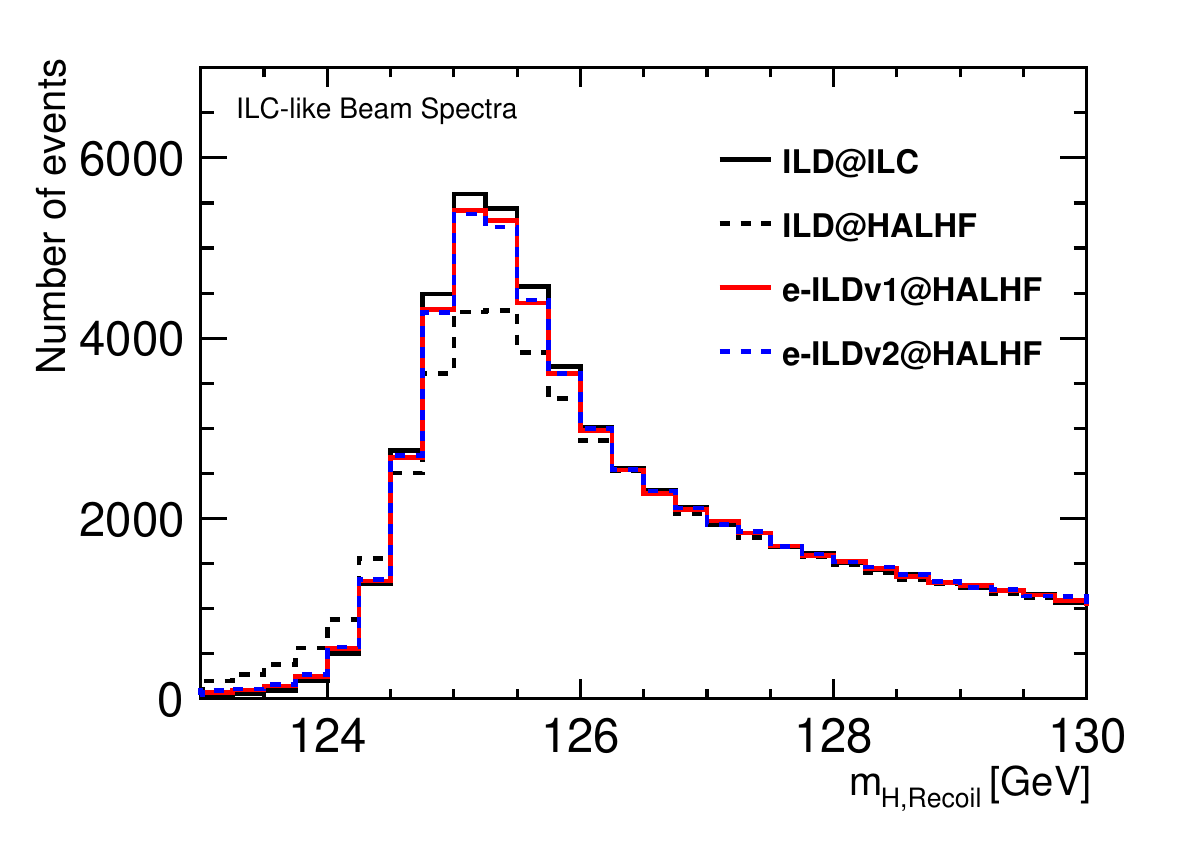}
    \caption{}
    \label{fig:m_rec:zoom:lin}
\end{subfigure}
\caption{(a) Wide-range and (b) Zoomed-in distribution of the Higgs recoil mass at $\sqrt{s}=250$\,GeV comparing the standard ILD detector at ILC and HALHF with the e-ILD\_v1 and e-ILD\_v2 detectors at HALHF. For each run, 100k $e^+e^-\rightarrow HZ$ events are generated and analysed.} 
\label{fig:m_rec}
\end{figure}

The simulation of different variants of the HALHF detector was performed with SGV, which reproduces the predictions of a Geant4-based simulation for tracking-dominated quantities with very high fidelity~\cite{Berggren:2012ar}. Figure~\ref{fig:m_rec} compares the resulting recoil-mass distributions as obtained from the standard ILD detector at ILC and HALHF as well as from e-ILD\_v1 and e-ILD\_v2. All cases use the ILC beam-energy spectrum. 

The consideration of the beam energy spectrum reduces the relative impact of the momentum resolution on the sharper of the recoil peak, but the symmetric ILD detector still performs significantly worse in the boosted situation at HALHF compared to the symmetric situation at ILC. While the e-ILD\_v1 design recovers the performance, it is not feasible due to the pair background. The e-ILD\_v2 detector is adapted to the background conditions and performs nearly as well as ILD at ILC. 

This is a clear example that with a detector optimised for the asymmetric experimental environment, HALHF can deliver a physics program comparable to a symmetric linear collider.

\begin{figure}[H]
\centering
    \includegraphics[width=0.75\textwidth]{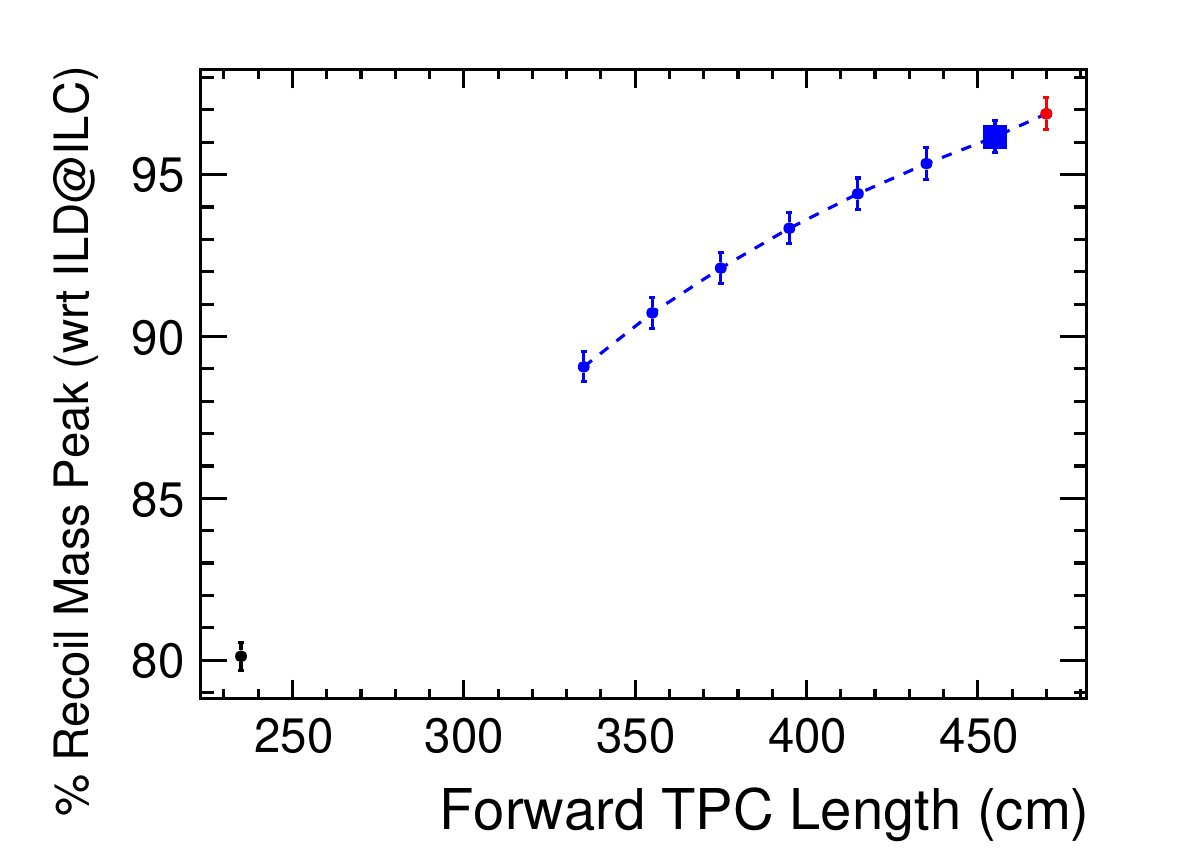}  
\caption{Ratio of the sum of the content of the three highest bins of the recoil mass distribution for a given detector setting over the same sum for the ILD@ILC case as a function of the forward length of the TPC. The backward length of the TPC has been fixed to 135\,cm. The black dot corresponds to the standard ILD detector, while the red dot and the blue square show  e-ILD\_v1 and e-ILD\_v2, respectively.}
\label{fig:mrec_vs_tpc}
\end{figure}

The performance differences are quantified in Fig.~\ref{fig:mrec_vs_tpc}, which shows the count of the highest three bins of the recoil peak, normalised to the ILC case, as a function of the length of the forward side of the TPC. The backward length of the TPC has been fixed to 135\,cm. The black dot corresponds to the standard ILD detector, which only reaches a peak-height-ratio of about 80\%. The red dot and the blue square show  e-ILD\_v1 and e-ILD\_v2, respectively. These models recover the peak-height-ratio to  97\% and 96\%, respectively. Although e-ILD\_v1 is unfeasible because of the pair background, e-ILD\_v2 gives very satisfactory performance.

\subsubsection{Luminosity measurement}
\label{sec:det:lumi}
Due to the asymmetric nature of the collisions at HALHF, the Bhabha-scattered electrons and positrons are no longer back-to-back but get a boost in the incoming $e^+$-direction. Therefore the luminosity measurement needs to be re-thought. A minimal change approach would still use Bhabha scattering, and see whether inserting special-performance sections with sufficient energy and angular resolution at the appropriate places in to the ECal can yield sufficient performance. For a more comprehensive discussion of possibilities for luminosity measurements at future $e^+e^-$ colliders, see e.g.~\cite{LinearColliderVision:2025hlt, Altmann:2025feg, Madison:2025myj}.

In the standard ILD detector~\cite{ILDConceptGroup:2020sfq}, the geometric acceptance of the LumiCal extends from 31 to 78\,mrad~\cite{BozovicJelisavcic:2013lni}, corresponding to pseudo-rapidities of $\pm 4.17$ and $\pm 3.24$. Given the Lorentz boost of HALHF of $\gamma = 1.67$ (c.f.\ Table~\ref{tab:1}), a rapidity of 1.10 should be added to these, yielding pseudo-rapidities between $5.27$ and $4.34$ in the forward direction and $-3.07$ and $-2.14$ in the backward direction, corresponding to $10-26$\,mrad as the forward and $166-175^{\circ}$ as the backward Bhabha polar angles in the lab frame. Polar angles of $10-26$\,mrad are instrumented by the BeamCal, designed to survive the harsh environment in the main fan of $e^+e^-$ pairs from beamstrahlung, but not at all suited for precision measurements. In order to instrument such small angles, in principle the detector could be moved much further from the IP, however details would need to be studied in an integrated final-focus system and machine-detector interface.    

Alternatively, larger scattering angles in the centre-of-mass frame could be used, so that the forward luminosity calorimeter is placed at a similar location in the detector as in standard ILD, while the backward LumiCal would become even more central. This configuration is sketched by the red ``LCal'' lines in Fig.~\ref{fig:pairs2D:newdet}. 
For such a configuration, the impact from the reduced statistics at larger polar angles and the resulting requirements on energy and angular resolution would need to be studied, including the effect of the tracker material (the ILD LumiCal has been positioned to minimize the material in front, and has in particular no tracking layers in front of it) and the deflection -- in the forward direction -- by the much stronger electromagnetic fields of the bunches. 

The real-time measurement of luminosity required to tune the accelerator using the singles rate of $e^-$ scattering, which peaks at very small angles, can however still be obtained via a ``standard'' luminosity calorimeter module as for a symmetric machine. Whether this is a dedicated, relatively crude detector or can use the luminosity monitor discussed above requires detailed design of the interaction region. Beyond low-angle Bhabha scattering, the full range of methods discussed e.g.\ in~\cite{Altmann:2025feg, Madison:2025myj} should be evaluated for HALHF.

\subsection{Outlook \& open questions}
\label{sec:det:outlook}

From  the discussions above, it is obvious that while significant progress has been made in understanding the beam-beam interaction and experimental environment at HALHF, a lot remains to be done towards a full physics- and cost-optimised detector design for HALHF. 

For future detector-optimization and physics-benchmarking studies,  a procedure to avoid the previously mentioned technical challenges of the boosted setup in the interface between Whizard and Pythia\,6 has been developed: the beam spectra obtained from {\sc GuineaPig} are boosted into the nominal centre-of-mass frame. Using these, symmetric $e^+e^-$ events can be generated with Whizard\,3 and Pythia\,6, following all the established procedures and tunings of the LCC Generator group~\cite{Berggren:2021sju}. Finally, the events can be boosted again into the HALHF lab frame upon reading them in to the detector simulation. Both the Geant4-based detector simulation and SGV provide this functionality.

The most important next steps include:
\begin{itemize}
    \item Demonstrating a concept for luminosity measurement at HALHF, addressing the challenges discussed in Sec.~\ref{sec:det:lumi};
    \item Optimizing the auxiliary silicon tracking for impact-parameter resolution and studying the flavour-tagging capabilities, which could be improved by the boost; $H\to c\bar{c}$ would be an obvious physics benchmark;
    \item Studying the impact of a higher magnetic field: this would improve the tracking resolution but more importantly make the pair background curl up at smaller radii, allowing the instrumentation to move somewhat closer to the beam;
    \item Introducing cost considerations: A longer TPC means also a longer solenoid, which is a significant cost-driver. As a rule of thumb, the magnetic volume should be kept constant for constant cost. 
\end{itemize}

In summary, the studies so far carried out imply that the physics goals of a symmetric linear collider can also be achieved with little or no degradation at HALHF.


\section{PWFA linac: beam-quality preservation}

Beam quality must be preserved in the PWFA electron linac to ensure sufficient luminosity, which constitutes one of the major challenges of the HALHF project. It involves preserving the charge, the transverse emittance, the energy spread, as well as the spin polarization. Much previous research has been performed on the preservation of beam quality within a single PWFA stage \cite{Benedetti2017,Zhao2024}, for which also a small-scale experimental demonstration was recently performed \cite{Lindstrom2024}. However, much less work has gone into demonstrating beam-quality preservation between stages. This is made difficult by the space needed for in- and out-coupling of drivers, a resulting chromatic aberration \cite{Lindstrom2021}, as well as the very long distances (over multiple stages) required.

This section describes progress toward beam-quality in the PWFA linac, including a detailed description of achromatic staging optics (Sec.~\ref{sec:achromatic-staging-optics}), the resulting multistage simulations made possible with this optics (Sec.~\ref{sec:multistage-simulations}), the transverse tolerances and the effect of transverse instabilities (Sec.~\ref{sec:transverse-tolerances}), and some first results regarding preservation of spin polarization (Sec.~\ref{sec:spin-polarization-preservation}).

All the simulation work reported below was performed using the newly developed start-to-end simulation framework ABEL \cite{Chen2025}, combining both reduced models and full-complexity codes for the plasma and interstages (in particular using HiPACE++ and ImpactX, respectively).

\subsection{Achromatic staging optics}
\label{sec:achromatic-staging-optics}

Several requirements apply to the optics between PWFA stages. Firstly, the depleted drive beam must be extracted from the previous stage and a new drive beam injected into the next; this is most easily done with a dipole magnet to separate/merge the main beam and the drive beam. The resulting gap allows the main beam to diverge to a significantly larger beta function than that inside the PWFA, leading to strong chromaticity when refocusing into the next stage.

\begin{figure}[t]
   \centering
   \includegraphics[width=0.7\columnwidth]{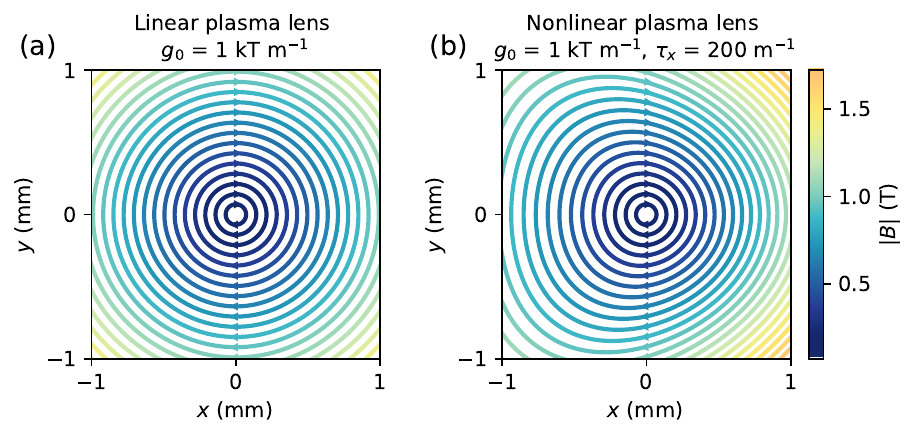}
   \caption{(a) A regular linear plasma-lens field versus (b) a nonlinear plasma-lens field (here with an exaggerated nonlinearity for clarity). The latter enables achromatic beam transport by placing high-energy particles on the right (where the focusing is stronger) and low-energy particles on the left (where the focusing is weaker). From Ref.~\cite{Lindstrom2026}.}
   \label{fig:interstage_nonlinearplasmalens}
\end{figure}

\begin{figure}[!htb]
   \centering
   \includegraphics[width=\columnwidth]{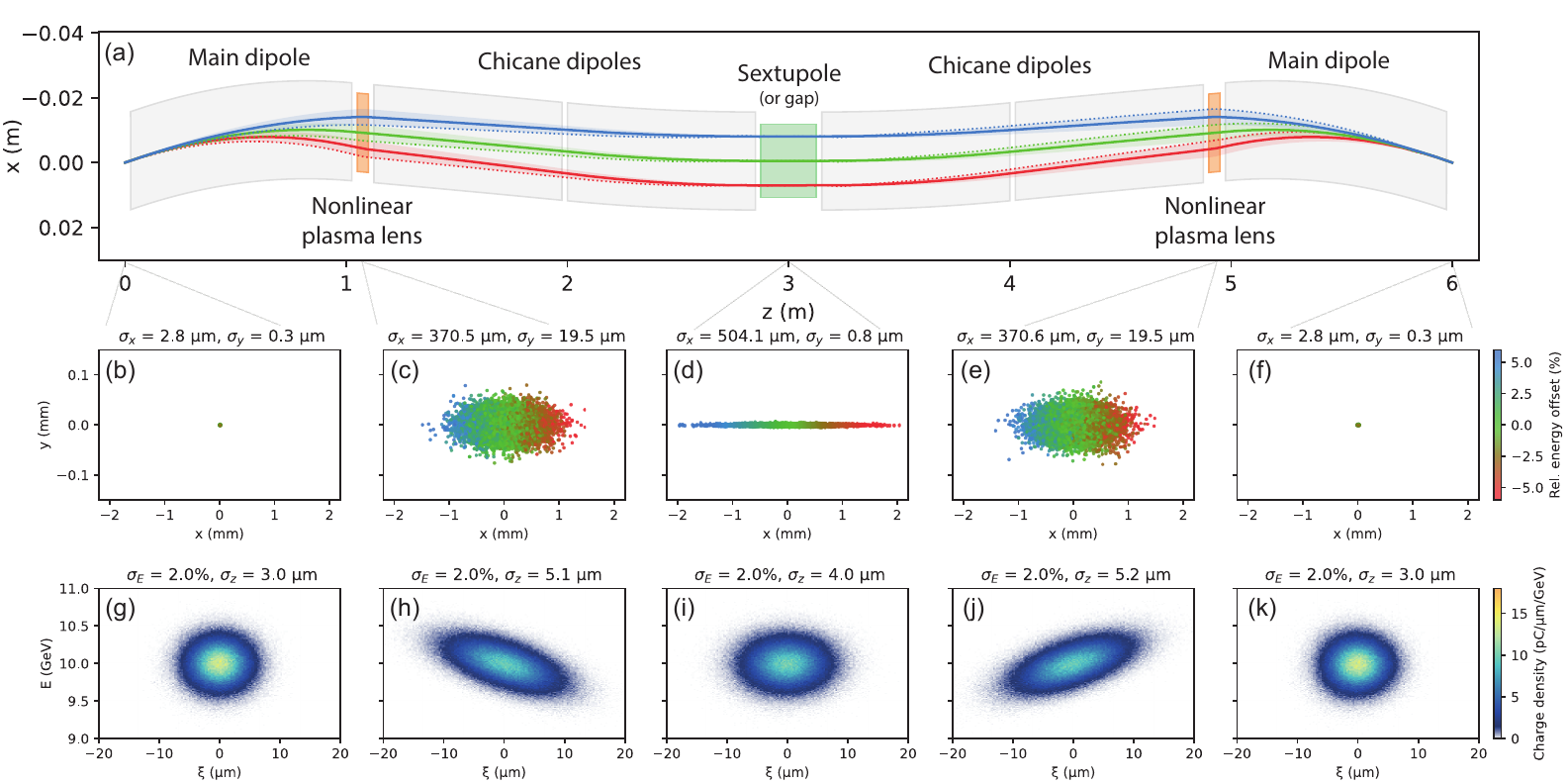}
   \caption{(a) Schematic overview of the interstage optics concept, as seen from above, showing how low- (red), medium- (green) and high-energy particles (blue) are transported and refocused. (b--f) The transverse profile in $x$--$y$ shows the evolution of the divergence of a dispersed beam at key points in the lattice. (g--k) Similarly, the evolution in longitudinal phase space shows that the $R_{56}$ is zero, and hence the bunch length is preserved. Here, $\xi$ is the co-moving longitudinal coordinate and $E$ is the energy. From Ref.~\cite{Lindstrom2026}.}
   \label{fig:interstage_schematic}
\end{figure}

A new staging optics solution is therefore required, to provide achromatic focusing in as compact a space as possible. This solution was recently reported in Ref.~\cite{Lindstrom2026}, and is summarised below. The key innovation is the use of nonlinear plasma lenses, where the magnetic field profile (see Fig.~\ref{fig:interstage_nonlinearplasmalens}) inside provides a focusing gradient that grows linearly with horizontal position. Combined with a dispersion (from the dipoles anyway used for in- and out-coupling), the local field experienced by each energy slice can be tuned such that the focal length is identical for all energies, at least within a 10\% energy bandwidth.

The full staging-optics lattice uses two such nonlinear lenses, as illustrated in Fig.~\ref{fig:interstage_schematic}. The example shown is not HALHF-specific, but represents a similar but more challenging setup with smaller initial beta functions (at \SI{10}{GeV}); HALHF-specific numbers are used in the following section (Sec.~\ref{sec:multistage-simulations}). In short, the lattice works by: (1) dispersing the diverging beam in energy using a strong dipole; (2) focusing it achromatically with a nonlinear plasma lens tuned exactly to the beam's dispersion; (3) using a central dipole chicane to control the longitudinal dispersion $R_{56}$ (here it is set to zero but it will generally be set to a negative value to allow a self-correction mechanism); (4) using a central sextupole to cancel the second-order dispersion; (5) mirroring the lattice to refocus achromatically and cancel the first- and second-order dispersion at the end of the lattice.

\begin{figure}[!th]
   \centering
   \includegraphics[width=0.60\columnwidth]{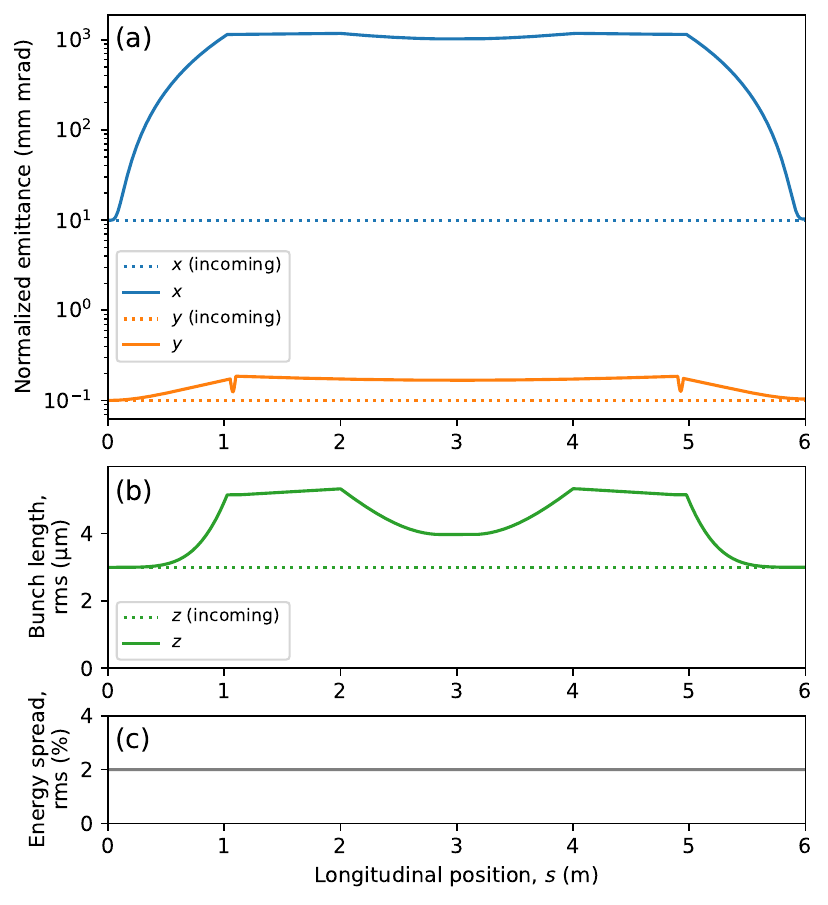}
   \caption{Full beam-quality preservation in the interstage optics, including (a) the transverse emittance, (b) bunch length and (c) energy spread. This simulation is performed with the following beam parameters: \SI{10}{GeV} energy, 2\% rms energy spread, \SI{15}{mm} initial beta function,  10$\times$\SI{0.1}{mm~mrad} emittance, \SI{3}{\micro\meter} rms bunch length, \SI{50}{pC} of charge (\SI{2}{kA} peak current). The strongest magnets are \SI{1}{m} long and have a field of \SI{1}{T}. Both coherent and incoherent synchrotron radiation is included in the simulation. From Ref.~\cite{Lindstrom2026}.}
   \label{fig:interstage_beamqualitypreservation}
\end{figure}

The beam quality is fully preserved in this lattice, including the transverse emittance, the bunch length and the energy spread, as demonstrated by a high-fidelity particle-tracking simulation using ImpactX (see Fig.~\ref{fig:interstage_beamqualitypreservation}).

An alternative optics setup using quadrupoles and sextupoles in place of the nonlinear plasma lens was also identified. While this provides a good backup solution, it is also approximately twice as long, has double the number of elements, and has half the energy bandwidth compared to the simpler plasma-lens-based solution described above.

\begin{figure}[!tb]
   \centering
   \includegraphics[width=\columnwidth]{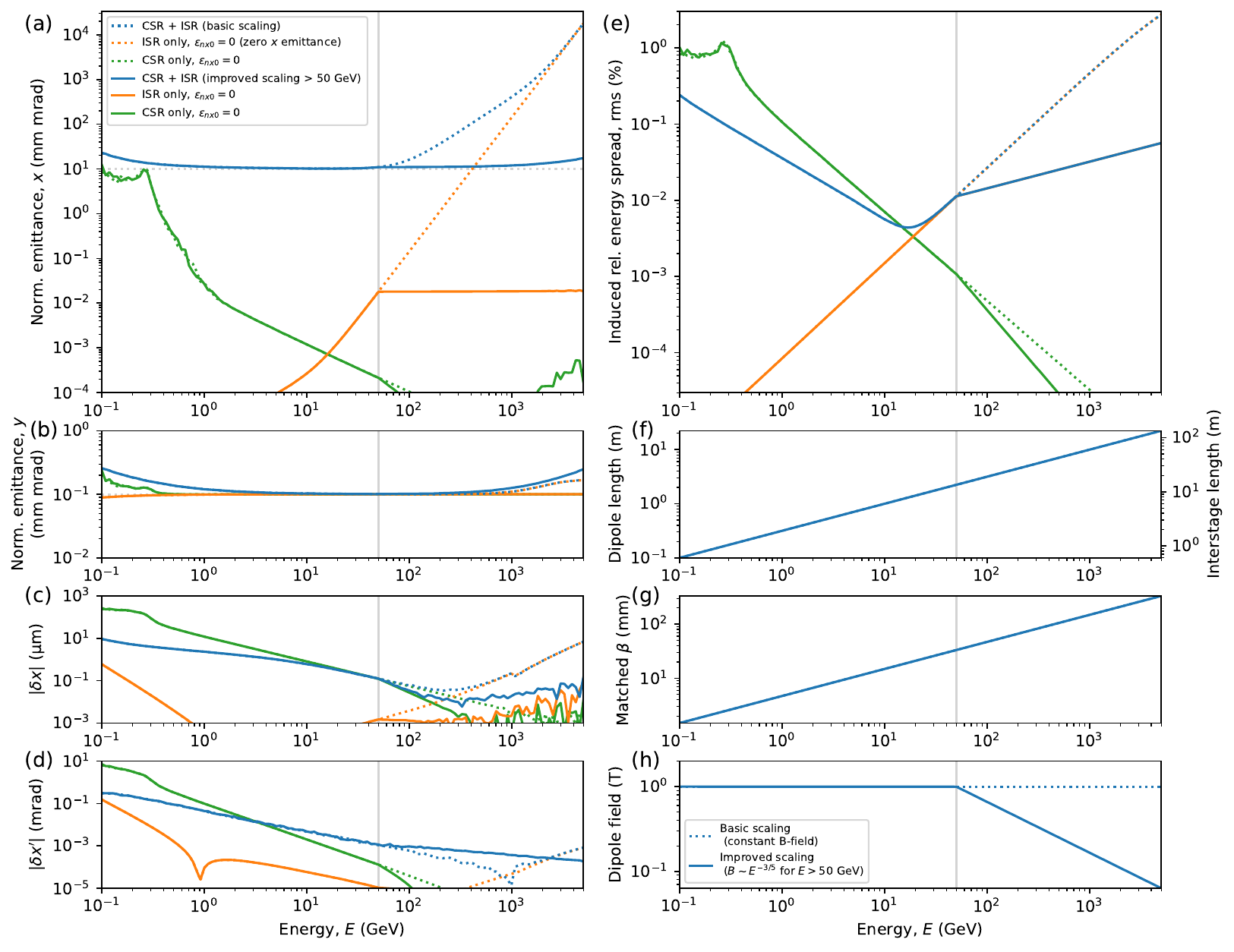}
   \caption{Energy scaling of the staging lattice, showing two different scaling solutions: basic scaling (dotted lines) and an improved scaling (solid lines). Various beam qualities are shown, including (a--b) emittances, (c) centroid offsets and (d) angles, (e) energy spread, as well as lattice parameters such as the (f) dipole length, (g) matched beta function at the start and end, and (h) the dipole field variation for the two solutions. From Ref.~\cite{Lindstrom2026}.}
   \label{fig:interstage_energyscaling}
\end{figure}

For a high-energy multistage PWFA linac, many of these staging lattices will be employed, each at a different energy. It is therefore important to understand how the lattice scales with energy. While many solutions exist for scaling to higher energy, the most natural solution maintains a constant field strength in all optical elements (including the dipole and sextupole fields, as well as the plasma-lens gradient and nonlinearity) but scales their length with the square root of the energy ($L\sim\sqrt{E}$, where $E$ denotes the main-beam energy in that interstage). The performance of this lattice is shown in Fig.~\ref{fig:interstage_energyscaling} for a range of energies from \SI{100}{MeV} to \SI{5}{TeV} (only 8 to \SI{800}{GeV} is relevant for HALHF). We observe that emittance is largely preserved in the vertical plane (which is most critical), whereas the emittance can start to grow in the horizontal (bending) plane at high energies \SI{>50}{GeV} due to incoherent synchrotron radiation effects. This can be avoided by slowly ramping down the dipole field as the  energy increases, specifically as $B\sim E^{-3/5}$ starting from \SI{50}{GeV} (denoted in the figure as an ``improved scaling"). Since this leads to a smaller dispersion, the plasma-lens nonlinearity must be made correspondingly stronger at higher energy, which eventually leads to emittance growth from nonlinear kicks -- here only seen beyond \SI{1}{TeV}, which is not relevant for HALHF.

The nonlinear plasma lens, as well as the full staging optics setup, is being developed experimentally at the University of Oslo as part of the SPARTA project \cite{Lindstrom2025_SPARTA}. A prototype of a nonlinear plasma lens is currently undergoing testing with beam at the CLEAR test facility at CERN \cite{Drobniak2025a,Drobniak2025b}, with plans to expand testing to the full lattice soon.

\subsection{Multistage simulations}
\label{sec:multistage-simulations}

The current HALHF \SI{250}{GeV} baseline design has 48 stages, each adding \SI{7.8}{GeV} of energy. Beam-quality preservation is required both in all the PWFA stages and in all the interstages, as well as in the interaction between them. The ultimate goal is to demonstrate all this in one full-complexity 48-stage simulation. Recent progress toward this goal is presented below, in the form of a full-complexity two-stage simulation (Sec.~\ref{sec:multistage-simulations-twostages}) and a 48-stage simulation with simplified PWFA stages but full-complexity interstages (Sec.~\ref{sec:multistage-simulations-manystages}).

\subsubsection{Two-stage full-complexity simulation}
\label{sec:multistage-simulations-twostages}

\begin{figure}[!htb]
   \centering
   \includegraphics[width=\columnwidth]{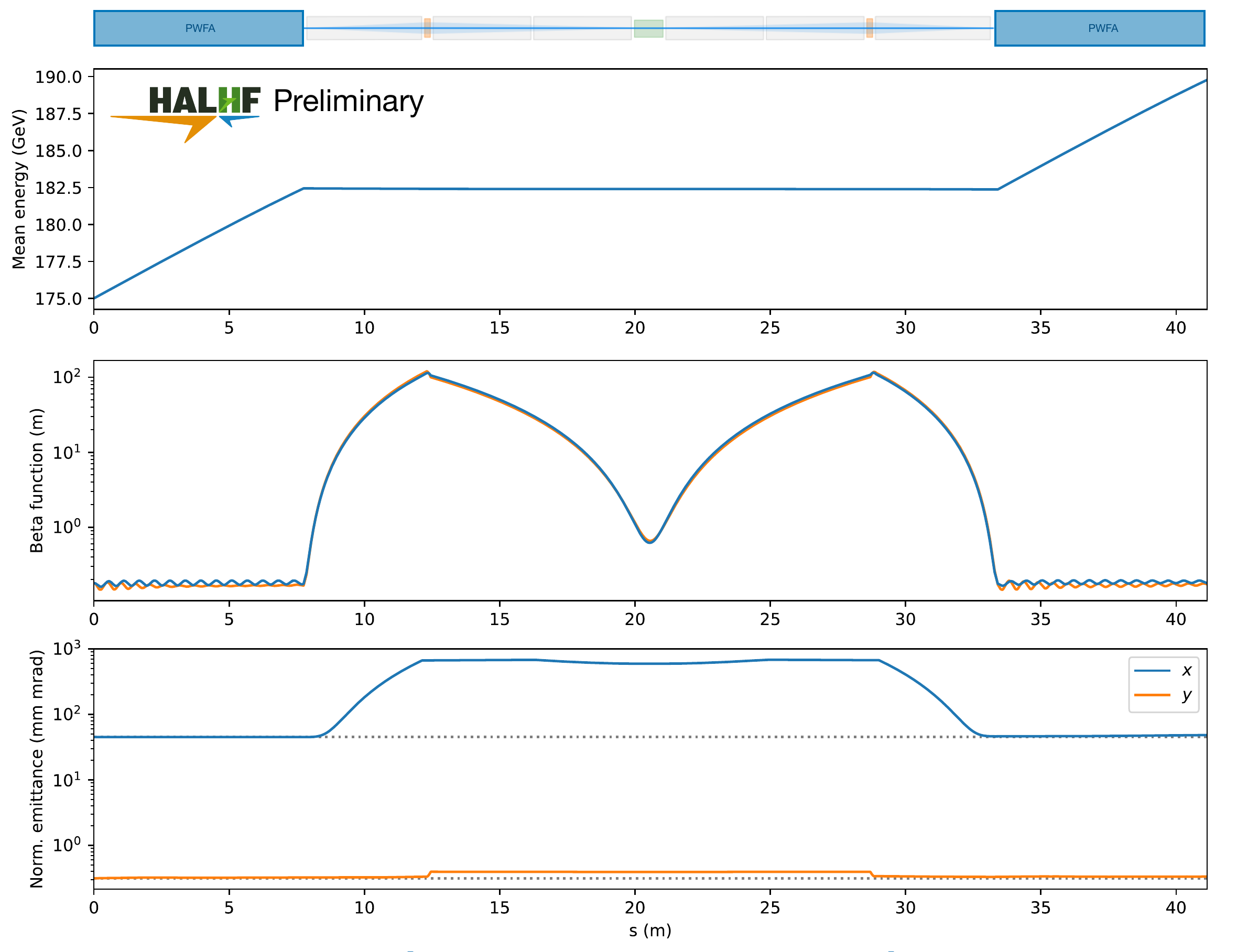}
   \caption{Full-complexity two-stage simulation using HiPACE++ and ImpactX, here performed in the middle of the linac (175 to 190 GeV). The emittance is preserved in both transverse planes across both stages and the interstage. From Ref.~\cite{ALEGRO2026}.}
   \label{fig:two-stage-simulation}
\end{figure}

A two-stage simulation was performed in ABEL, using HiPACE++ for the PWFA stages and ImpactX for the interstage particle tracking. Here, the initial beam energy was \SI{175}{GeV} (increasing to \SI{190}{GeV}), representing two stages close to the middle of the PWFA linac. A helium plasma is used, resulting in a moderate amount of ion motion (forming a density spike on axis), balancing emittance growth from nonlinear focusing and the suppression of the beam-breakup instability. As seen in Fig.~\ref{fig:two-stage-simulation}, the emittance is preserved to within a few percent in both transverse planes.

\subsubsection{48-stage interstage simulation with simplified PWFA stages}
\label{sec:multistage-simulations-manystages}

\begin{figure}[!htb]
   \centering
   \includegraphics[width=\columnwidth]{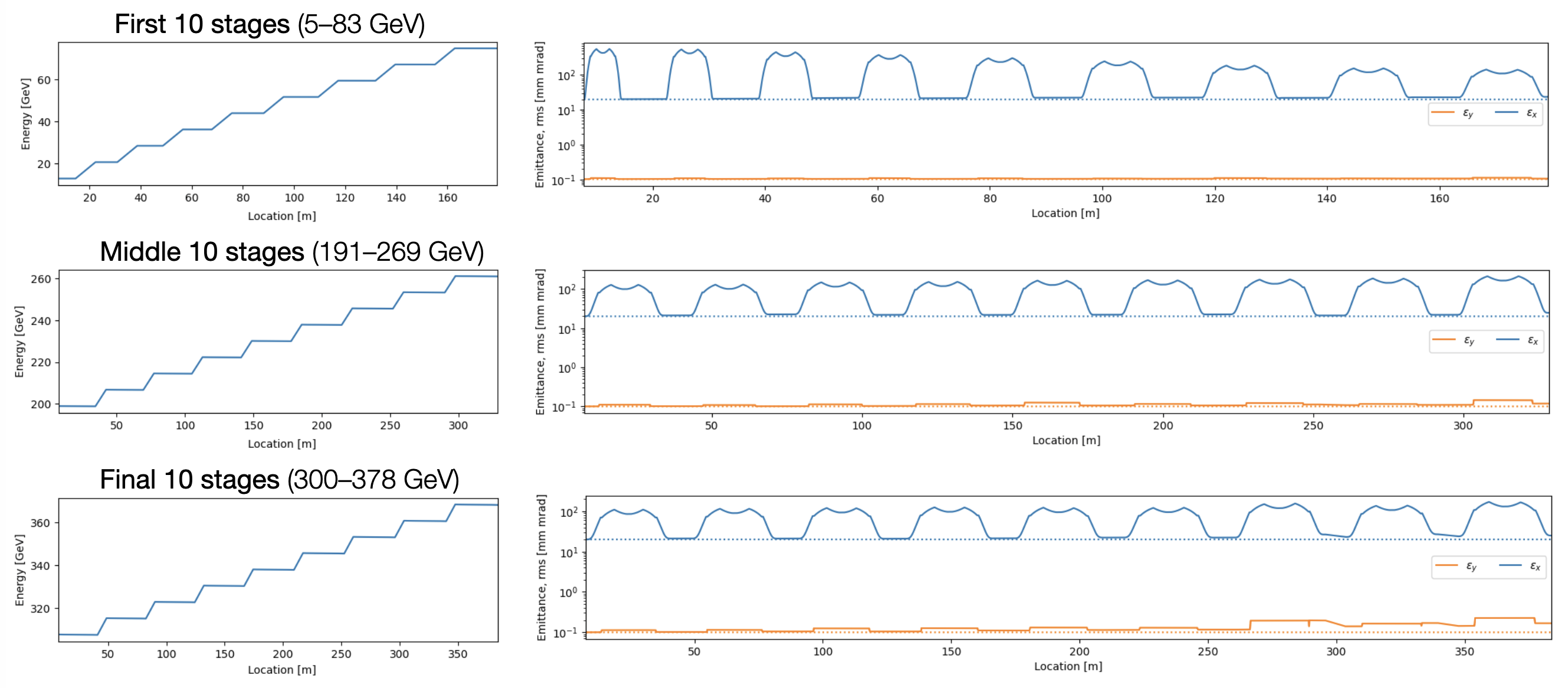}
   \caption{Emittance preservation over 10 stages, initialized at low, middle and high energy.}
   \label{fig:different-energies-10stages}
\end{figure}

In order to understand the more subtle effects of repeatedly employing the staging optics a large number of times, simulations were also performed using simplified PWFA stages (i.e., using a fast single-step calculation assuming a constant wakefield, as calculated using WakeT) but with full complexity in the interstages. Starting with three separate 10-stage simulations at low, medium and high energy -- in order to understand whether emittance can in principle be preserved at any energy -- it is found that emittance can indeed be preserved to a high degree (see Fig.~\ref{fig:different-energies-10stages}).

\begin{figure}[!htb]
   \centering
   \includegraphics[width=\columnwidth]{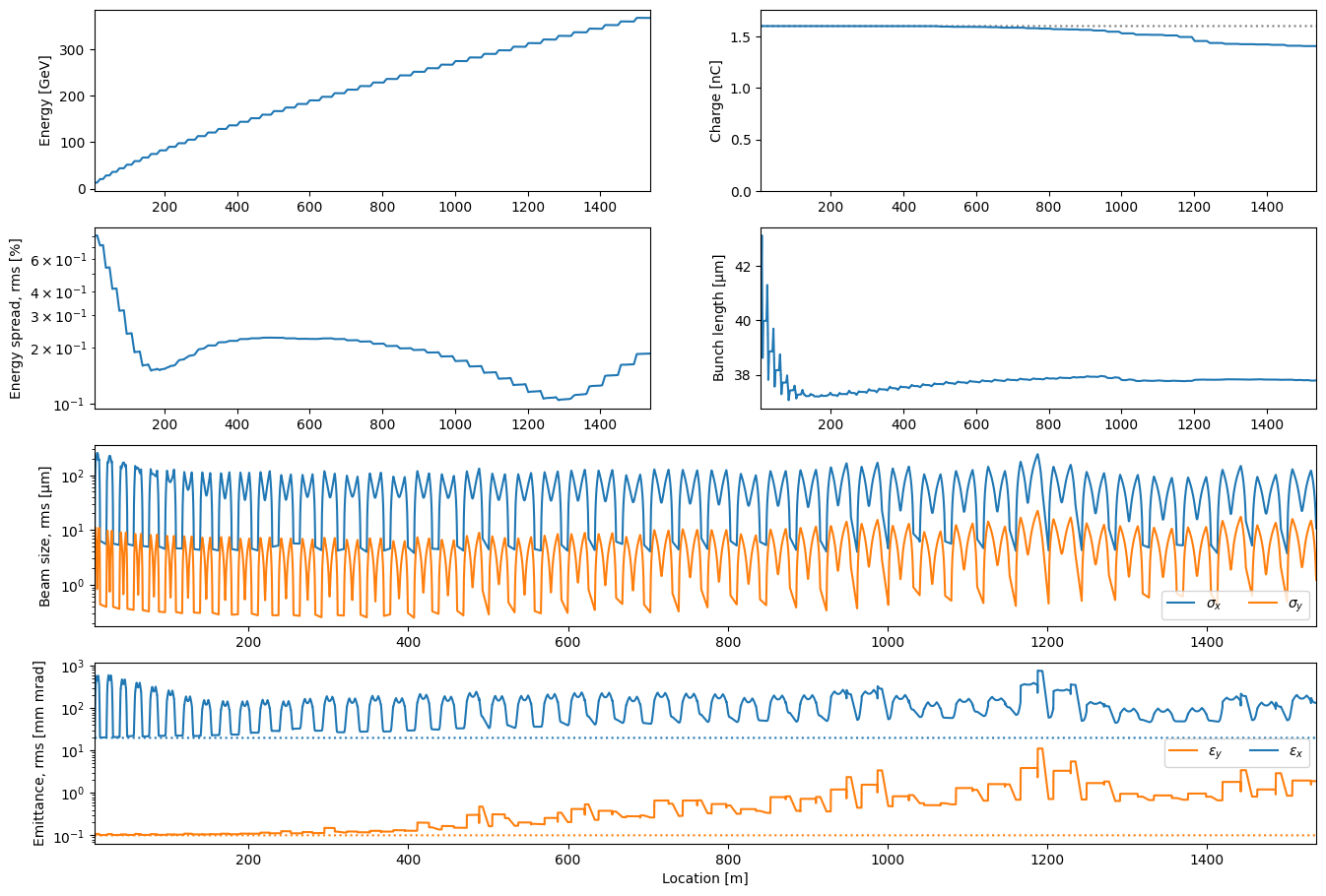}
   \caption{48-stage simulation with full complexity in the interstage simulation, but reduced complexity in the PWFA stages. The emittance can be seen to increase gradually, and eventually lead to some charge loss.}
   \label{fig:all-48-stages-simulation}
\end{figure}

However, when extending to the full 48-stage simulation, as shown in Fig.~\ref{fig:all-48-stages-simulation}, the simulation shows a slowly growing emittance. Starting (deliberately) from an emittance value lower than the design value, it ends up somewhat above the design values. This emittance growth needs to be more fully understood, and is currently being studied in detail, but appears to be related to a very gradual build up of particle amplitude via nonlinear kicks from the nonlinear plasma lenses. Note that emittance growth directly from ISR and CSR has been fully suppressed.

\subsection{Transverse instabilities and tolerances}
\label{sec:transverse-tolerances}

The transverse instability and subsequent emittance growth may lead to very tight tolerances for the drive-beam jitter. Independent of wakefield effects, the strong focusing fields themselves may also lead to tight jitter tolerances.  A first estimate of drive-beam jitter tolerances for HALHF, based on start-to-end simulations, using ABEL, is found in \cite{adli_trans_ipac_2025}, indicating nm-level tolerances.  The estimates reported assumed that the driver was guided by an external magnetic field, such that angular driver-injection jitter was mitigated. A further  strategy for mitigating transverse jitter is to match the length of the plasma acceleration stage to an integer or half-integer number of betatron oscillations of the main beam, as discussed in \cite{schulte_application_2016}, for example.

Recently, robust schemes for both driver-guiding and main-beam betatron-length matching have been studied~\cite{chen_guiding_temp}; 
they are summarised below.  

\subsubsection{Driver guiding}

One of the most critical tolerances is the extreme angular tolerance: the drive beam creates an ion column that defines the transverse axis for the trailing main beam. Any angular offset of the drive beam causes the main beam to perform betatron oscillations around a tilted trajectory, leading to a large transverse displacement at the end of a meter-scale plasma-acceleration stage, as pointed out in e.g.~\cite{schulte_application_2016} and~\cite{white_transverse_2019}.  A uniform current can be used to induce an azimuthal magnetic field. 
The gradient is tuned such that the radial focusing force will steer the drive beam to perform an integer number of half oscillations along a plasma-acceleration stage.

\begin{figure}[!htb]
   \centering
   \includegraphics[width=0.80\columnwidth]{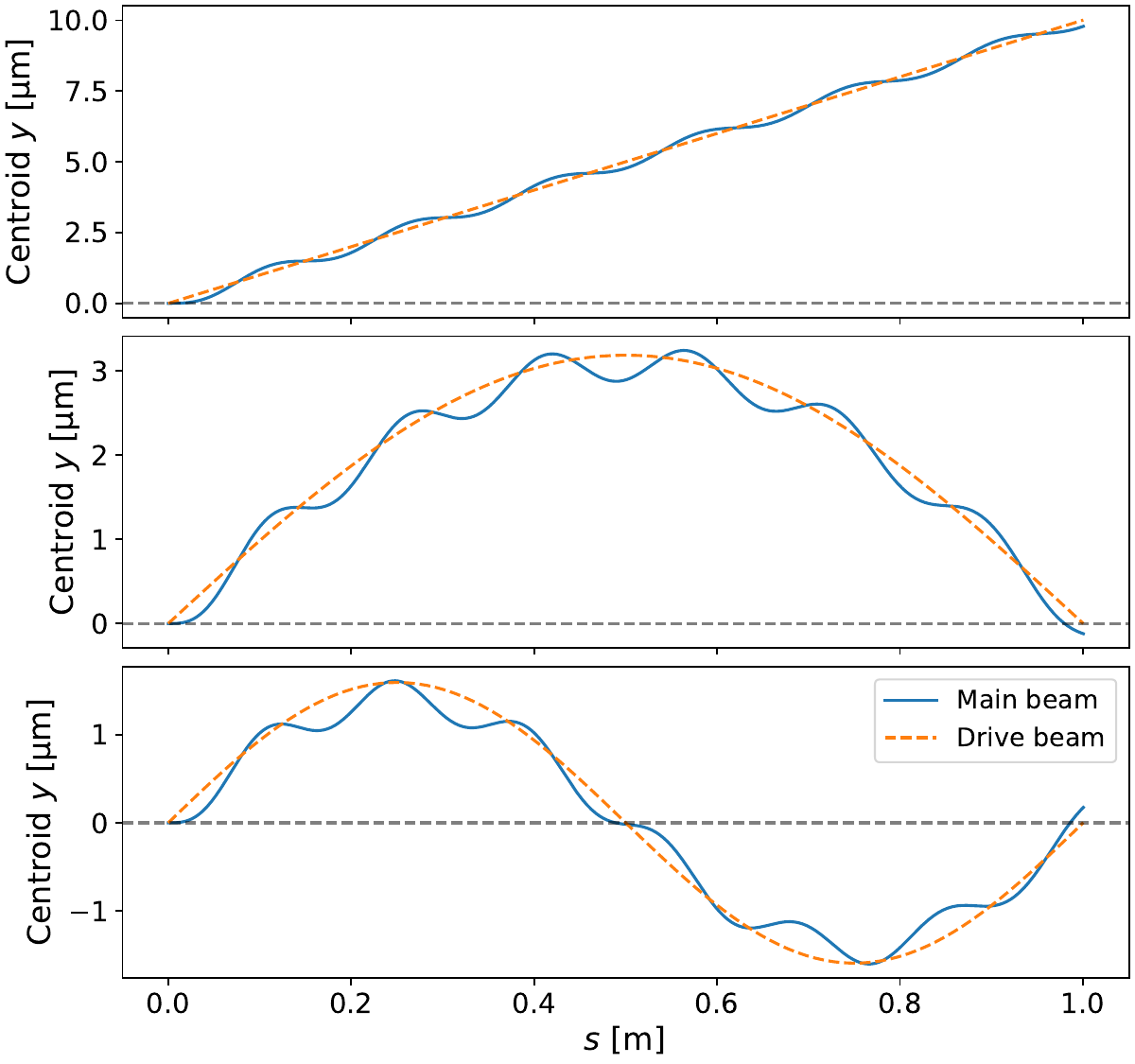}
   \caption{Illustration for guiding of a drive beam using an external magnetic field. The top panel shows the trajectories for the drive beam and main beam when no guiding is applied. The middle panel shows the case of driver guiding, with the driver undergoing half a betatron oscillation, while the bottom panel shows the case of one full betatron oscillation. From Ref.~\cite{chen_guiding_temp}.}
   \label{fig:driver_guiding_illustration}
\end{figure}

Fig. \ref{fig:driver_guiding_illustration} illustrates the effect; without driver guiding, the main beam performs betatron oscillation around a straight drive-beam trajectory and gains a large transverse offset after propagation. In contrast, when applying an external magnetic field, the driver propagates along a sinusoidal trajectory with the main beam oscillating around the driver trajectory. The driver is eventually guided back to axis, which drastically mitigates the transverse offset of the main beam at exit~\cite{chen_guiding_temp}.

\subsubsection{Betatron-length matching}

\begin{figure}[!htb]
   \centering
   \includegraphics[width=0.8\columnwidth]{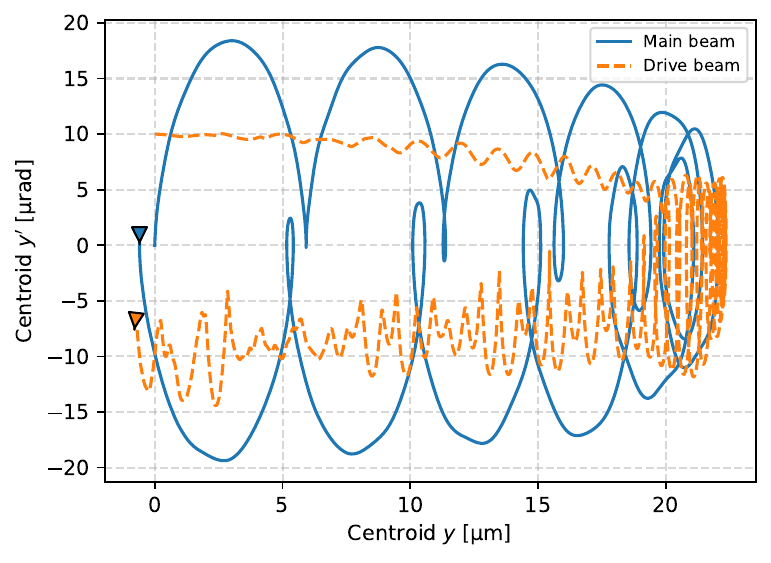}
   \caption{Phase-space beam-centroid trajectory of the drive beam (\SI{4.5}{\giga\electronvolt}) and main beam (\SI{55}{\giga\electronvolt}) with stage length matched to 11.5 main beam betatron oscillations. This is calculated using a PIC simulation. The triangles represent the end point of the propagation of each beam. Here, the origin defines the ideal axis of the entire multistage accelerator. The noise visible in the drive beam centroid comes from the onset (and subsequently suppression by energy spread) of the hosing instability. From Ref.~\cite{chen_guiding_temp}.}
   \label{fig:phase_trajectories_y_4.5GeV_55GeV-11.5beta-osc-beam}
\end{figure}

Under stable conditions, the main beam performs betatron oscillations in a predictable manner, so that it is possible to adjust the length of the plasma-acceleration stage to minimise the transverse position and angular offsets of the main beam at stage exit. A proposed strategy for mitigating transverse jitter is to match the length of the plasma-acceleration stage to an integer or half-integer number of betatron oscillations of the main beam, such that it exits the stage at a betatron phase with minimal angular and position offsets~\cite{chen_guiding_temp}. An example, based on full HiPACE++ simulations, including hosing and ion motion effects~\cite{chen_guiding_temp}, is displayed in Fig. \ref{fig:phase_trajectories_y_4.5GeV_55GeV-11.5beta-osc-beam}. The phase-space trajectory of the main beam is close to closing on itself after traversing the plasma stage,  which demonstrates a significant mitigation in the final offsets of the main beam. 

\subsection{Spin-polarization preservation}
\label{sec:spin-polarization-preservation}

The preservation of spin polarization is a key requirement for future high-energy electron–positron colliders like HALHF, where polarised beams enhance the sensitivity of precision electroweak measurements and improve the production rates. Preservation of spin polarization in plasma accelerators, in particular in the blow-out regime, has been studied in e.g.~Refs.~\cite{Polarisation, Thomas_2020,Reichwein_2025}. In Ref.~\cite{horlyk2026spin} this topic is investigated for HALHF.

\subsubsection{Tracking model}

The spin evolution of a relativistic particle in an electromagnetic field, in the laboratory frame, is described by the Thomas--BMT equation \cite{orginaltbmt}
\begin{align}\label{eq:spin} 
    \frac{d \mathbf{s}}{dt} = \mathbf{\Omega}_{\mathrm{T-BMT} } \times \mathbf{s},
\end{align}
with the precession vector
\begin{align}\label{eq:TBMT}
    \mathbf{\Omega}_{\mathrm{T-BMT} } =  -\frac{e}{m} \left[\left(a_e + \frac{1}{\gamma}\right) \mathbf{B} - \frac{a_e \gamma}{1 + \gamma}\mathbf{v} \cdot \mathbf{B} \frac{\mathbf{v}}{c^2} - \left(a_e + \frac{1}{1 + \gamma}\right) \frac{\mathbf{v} \times \mathbf{E}}{c^2}\right],
\end{align}
where $e$ is the particle charge, $m$ its mass, $a_e$ the anomalous magnetic moment $(g-2)/2$, $\gamma$ the Lorentz gamma factor, $\mathbf{v}$ the particle velocity, $\mathbf{B}$ and $\mathbf{E}$ magnetic and electric fields in the laboratory frame and $c$ the velocity of light {\it in vacuo}.

A dedicated spin-tracking module based on the Thomas--BMT equation was implemented to propagate spin vectors alongside phase-space coordinates. In addition, a simplified model of an interstage was implemented into the ABEL framework.  The spin evolution was computed using a Boris-type integration scheme \cite{Decyk2023AnalyticBoris}.

\subsubsection{First results}

\begin{figure}[h!]
    \centering
    \includegraphics[width=\linewidth]{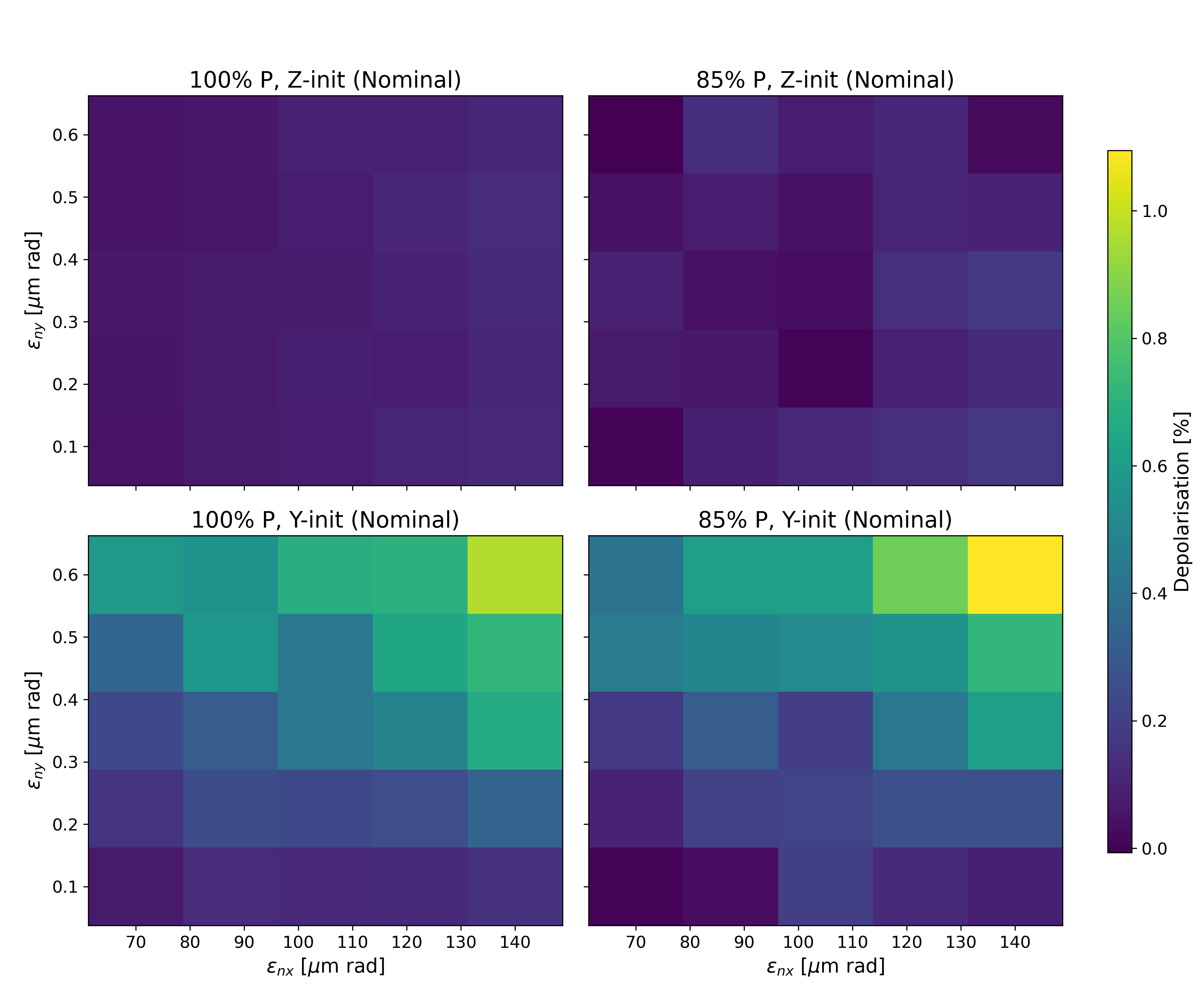}
    \caption[48-stage \texttt{StageQuasistatic2d} scan for emittance, $\epsilon_{n,x} \neq \epsilon_{n,y}$]{Final beam depolarization as a function of both horizontal emittance ($\epsilon_{n,x}$) and vertical emittance ($\epsilon_{n,y}$) after 48 plasma stages using the ABEL \texttt{StageQuasiStatic2d} model, where $\epsilon_{n,x} \neq \epsilon_{n,y}$ at each step. The nominal values are $\epsilon_{n,x} \in [70-110]\hspace{1mm} \mu \text{m}$ and $\epsilon_{n,y} \in [0.1-0.6] \hspace{1mm}\mu \text{m}$. The top plots start from longitudinal ($z$) polarization, whereas the bottom plots show vertical ($y$) polarization (100\% and 85\% on the left and right, respectively). From Ref.~\cite{horlyk2026spin}.}
\label{fig:fullemittancescanasymnom}
\end{figure}

Figure~\ref{fig:fullemittancescanasymnom}, from Ref.~\cite{horlyk2026spin}, shows the dependence of final beam polarization on transverse emittance after 48 plasma stages, scanned around the nominal HALHF 2.0 emittance values. These first results indicate a depolarization of less than one percent across the 48 plasma stages of the HALHF linac (when polarized in the vertical $y$ plane; much less if longitudinally polarized in $z$). Similarly, the beam energy spread also results in very limited, sub-percent level, depolarization inside the linear focusing fields in plasma stages ~\cite{horlyk2026spin}. Stern-Gerlach forces and the Sokolov-Ternov effect (i.e., spin flip from radiation reaction) were not included in these simulations, as they are expected to be negligible. However, at very high energy, the Sokolov-Ternov effect should be included and further studied. 

\begin{figure}[h!]
    \centering
    \includegraphics[width=0.8\linewidth]{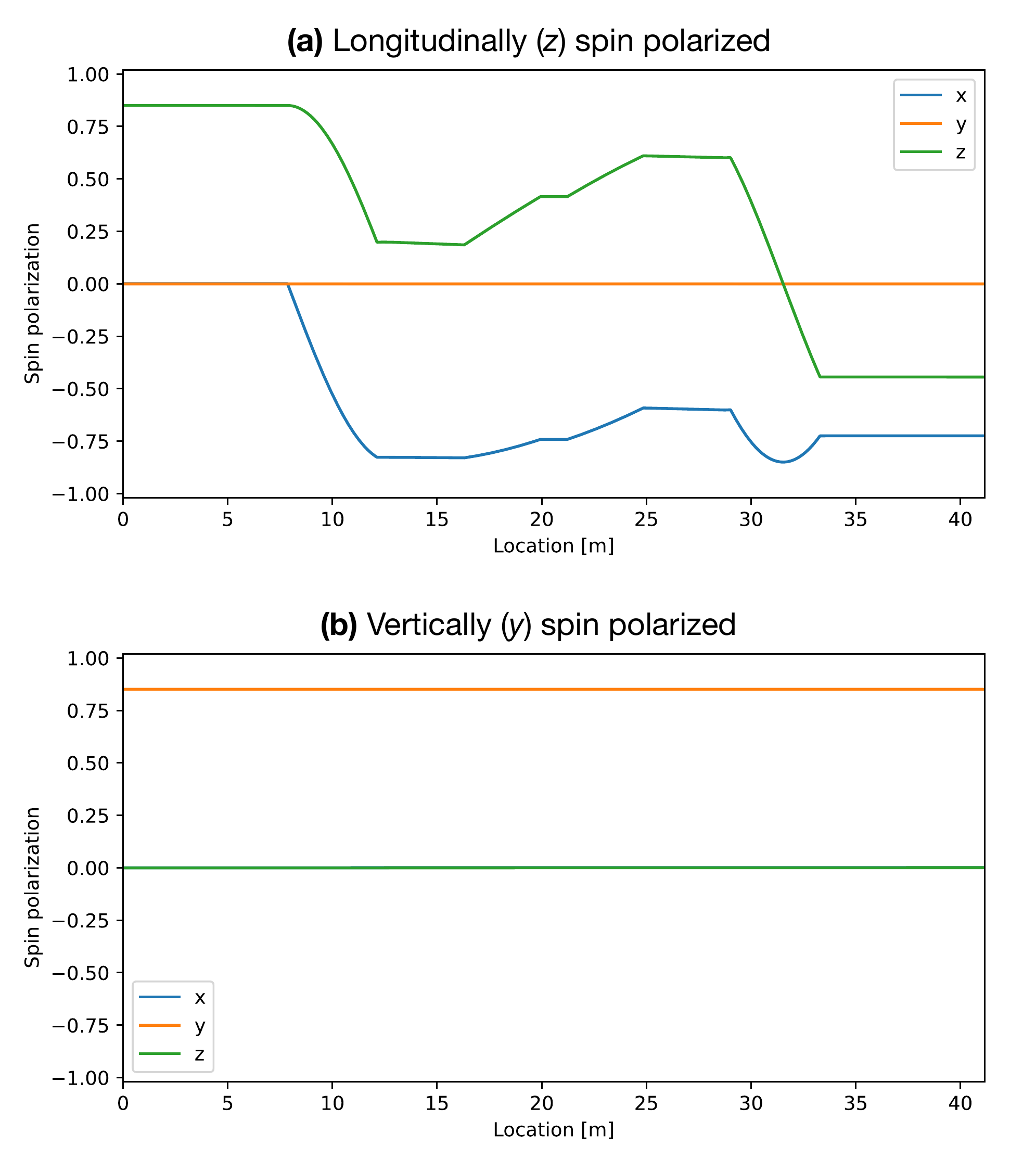}
    \caption{Two-stage simulation using HiPACE++ and ImpactX of a spin-polarized beam across both PWFA stages (0--\SI{7.8}{m} and 33.4--\SI{41.1}{m}) and the interstage optics (7.8--\SI{33.4}{m}). The beam is 85\% spin polarized in either (a) the $z$ direction, leading to significant spin precession, or in (b) the $y$ direction, where the spin polarization is fully preserved.}
\label{fig:spin_interstage}
\end{figure}

Finally, an integrated full-complexity two-stage simulation was performed with the same configuration as in Fig.~\ref{fig:two-stage-simulation}, in order to see the effect of the dipoles (which bend particles in the $x$ plane) in the interstage lattice. Figure \ref{fig:spin_interstage} shows the spin being preserved inside the PWFA stages, regardless of spin polarization direction ($z$ or $y$), but that the spin precesses significantly when initially polarized in the $z$ direction. In conclusion, the beam should be polarized in the $y$ direction in order to be preserved across the PWFA linac.

\section{PWFA linac: high beam power}

Moving from the current state of the art to HALHF will require an orders-of-magnitude increase in both operational repetition rate and average power. Achieving this increase requires a deeper understanding of the underlying physics governing the long-term evolution of the plasma, and in particular an answer to the fundamental question of what happens to the energy deposited in the plasma following each wakefield event. Understanding the subsequent transport and dissipation of this energy will be essential for optimising and managing the deposited power during high-repetition-rate operation. This includes determining how the resulting increase in plasma energy, or equivalently plasma temperature, modifies the wakefield experienced by subsequent bunches in the train, as well as establishing how the accumulated heat can be effectively managed within the surrounding plasma source and accelerator infrastructure.

This section describes the research status and plans to bridge this gulf, including: numerical developments towards a physical description of plasma evolution in HALHF and how that will partly define the bunch-train pattern of the machine (Sec.~5.1); the expected effects of extreme temperature (arising as a result of high-repetition-rate operation) on the wakefield and the plan to measure this experimentally (Sec.~5.2); a preliminary design of a HALHF discharge-generated plasma source taking in to account collisional physics effects, expected heating, and first attempts to measure and manage it (Sec.~5.3); radiation implications of high-average-power operation (Sec.~5.4); and beam-dump requirements for safe management of the various high-average-power beams at HALHF (Sec.~5.5).

\subsection{Long-term plasma evolution}

A major challenge in modelling the HALHF bunch train is the enormous range of temporal scales that must be captured. The initial wakefield interaction evolves on femtosecond-to-picosecond timescales, whereas modelling the complete bunch train requires following the plasma evolution over many microseconds. Conventional computational approaches become prohibitively expensive when applied across such a large dynamic range, motivating the development of an alternative numerical methodology.

This challenge will be addressed by combining the computational speed-up afforded by GPU-based operation with the quasi-static approximation (QSA) in particle-in-cell (PIC) codes. This combination will enable the initial plasma evolution to be followed into the nanosecond regime using a PIC model, after which the simulation will be handed over to a Hybrid-PIC code (affording larger time steps) for propagation until the arrival of the subsequent HALHF bunch, currently expected to occur 16 ns after the first. This numerical cycle can then be repeated for each successive bunch, allowing the evolution of the plasma to be followed throughout the complete bunch train without requiring a fully electromagnetic PIC simulation over the entire duration.

More specifically, HiPACE++ will initially be used to simulate the temporal evolution of a single transverse plasma slice. The simulation will include the necessary physics beyond the first plasma period and continue until the plasma reaches a sufficiently thermalised state. At this point, the evolved plasma distribution will be transferred to WarpX, which has a Hybrid-PIC model employing kinetic ions and fluid electrons. This will be used to evolve the system further, up to the arrival of the subsequent bunch at 16 ns. The resulting plasma state can then provide the initial conditions for the next wakefield interaction, allowing this HiPACE++–to-WarpX cycle to be repeated throughout the train.

An important advantage of this approach is that it provides opportunities to benchmark the framework across three distinct regimes. First, the initial HiPACE++ calculations can be benchmarked directly against fully three-dimensional electromagnetic PIC simulations. Secondly, consistency can be tested at the interface between HiPACE++ and WarpX, where the two numerical descriptions overlap. Finally, the complete framework can be compared with experimental measurements of plasma recovery from FLASHForward~\cite{Darcy2022}.

\begin{figure}[htp!]
\centering
\vspace{-1cm}
\includegraphics[width=\textwidth]{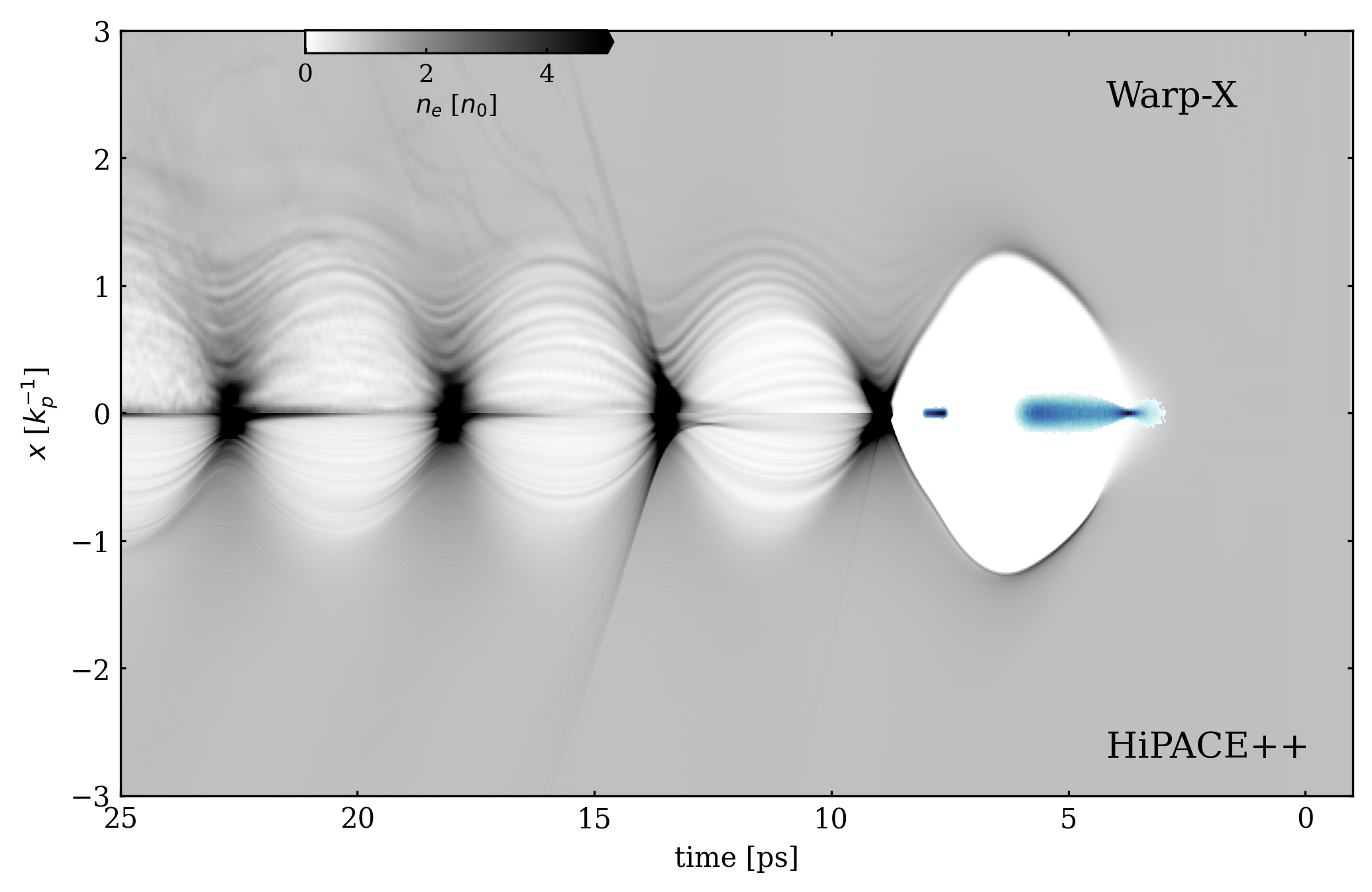}
\includegraphics[width=\textwidth]{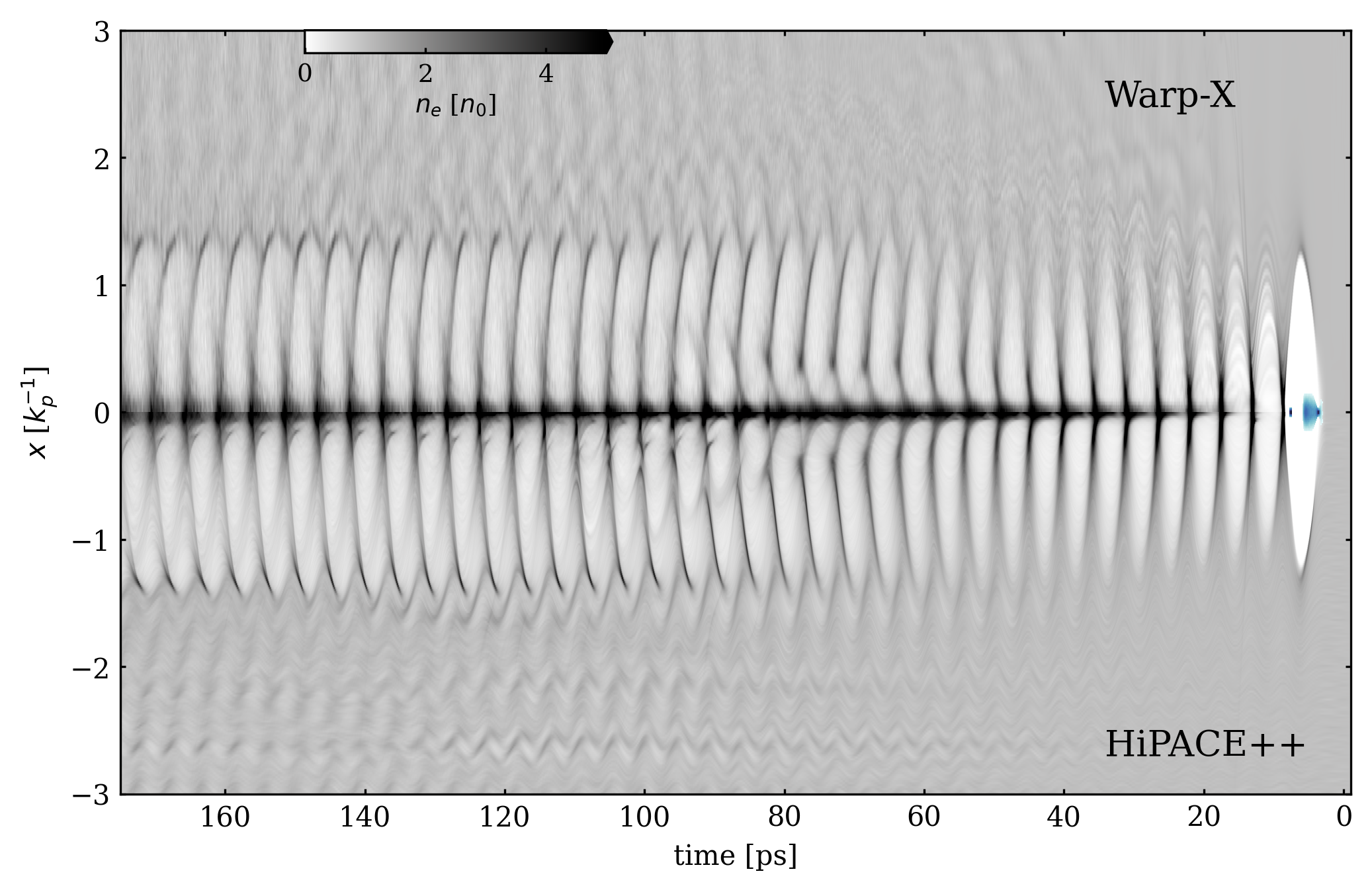}
\caption{(Top) Comparison of the first 4.5 plasma periods of a wake driven by the HALHF drive bunch simulated in the fully-3D EM PIC code WarpX and the QSA PIC code HiPACE++. The top half of each figure shows the simulation of the plasma blowout from WarpX, the lower half that from HiPACE++. (Bottom) The same as above but simulated up to 175 psec (approximately 25 plasma periods). Good agreement between the density and field evolution over both timescales can be seen.}
\label{fig:benchmarking}
\end{figure}

Benchmarking of the first of these regimes has already been performed using HALHF beam and plasma parameters. 'Out of the box' versions of the code (i.e. as-yet undeveloped for this specific application) of WarpX (in full 3D EM PIC mode) and HiPACE++, were compared over the first 175 ps following the excitation of the wakefield, with WarpX providing the fully three-dimensional electromagnetic PIC reference (Fig. \ref{fig:benchmarking}). Good agreement is observed between the two codes. Importantly, this agreement was achieved alongside a substantial reduction in computational cost, from 11.32 GPU-hours for WarpX to only 0.32 GPU-hours for HiPACE++, corresponding to a speed-up of approximately a factor of 35. This combination of agreement and computational efficiency provides encouraging initial evidence for the viability of the proposed methodology.

A key objective of this study will be to use the new simulation pipeline to inform the bunch-train pattern for HALHF. At present, the proposed 16 ns bunch spacing represents a compromise between the time required for the plasma to recover following a wakefield interaction, estimated by scaling from the FLASHForward experimental result, and the RF-motivated desire to minimise the separation between successive bunches for operational reasons. Establishing whether this spacing is physically viable therefore requires a more complete understanding of how the recovery time scales with the relevant beam and plasma parameters.

\begin{figure}[h!]
\centering
\includegraphics[width=\textwidth]{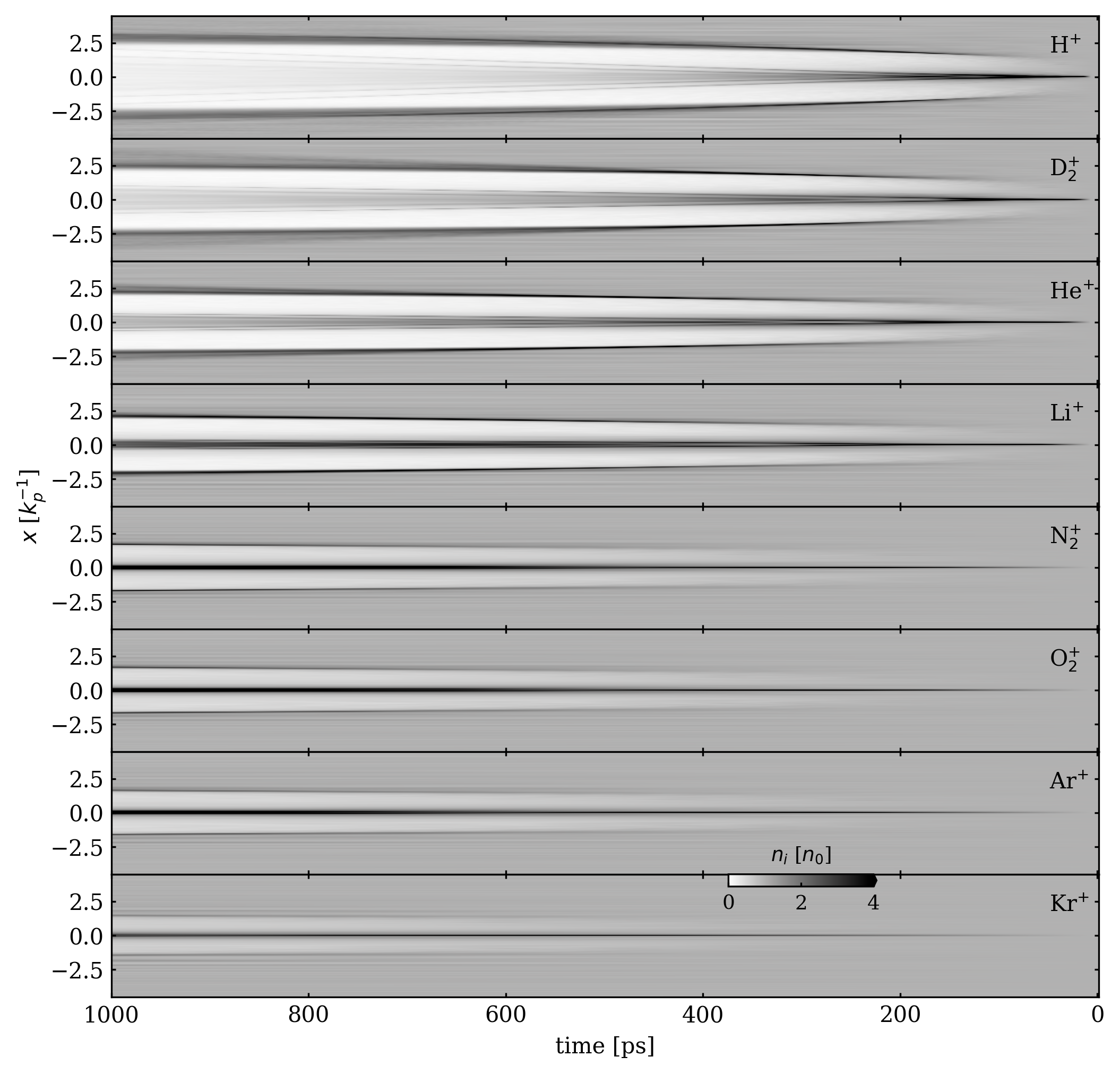}
\caption{Simulations of long-term plasma motion after a wake is driven by the HALHF drive bunch. The same simulation but for different plasma species indicates the strong dependence of the evolution timescale on the ion mass.}
\label{fig:ion-motion}
\end{figure}

The first application of the simulation pipeline will consequently be to investigate the scaling of the FLASHForward plasma-recovery result under conditions relevant to HALHF. A natural starting point is the dependence on plasma species and, in particular, ion mass. Figure \ref{fig:ion-motion} shows the same HiPACE++ simulation presented in Fig. \ref{fig:benchmarking}, but extended to approximately 1 ns after the wake is driven. Additional simulations performed around the helium baseline demonstrate that reducing the ion mass produces qualitatively similar ion motion but causes the evolution to occur on a substantially different timescale. Extending these calculations to the point at which the helium plasma has recovered, and thereby determining whether a 16 ns bunch separation is feasible, will provide a first important constraint on the achievable HALHF repetition rate.

\subsection{Plasma heating and temperature effects}\label{sec:heatingcalcs}

Simulating the long-term plasma dynamics following wake excitation will not only be important for defining the HALHF bunch-train pattern, but will also be essential for quantifying the challenges associated with the transition to high-average-power plasma-wakefield operation. These challenges arise from the fraction of the drive-beam energy that is transferred to the plasma but is not subsequently extracted by the accelerating beam.

The potential magnitude of this energy deposition illustrates the importance of the problem. Each HALHF drive bunch is expected to deposit approximately 13 J in the plasma, based on a 4 GeV, 8 nC drive beam and an assumed 60$\%$ energy transfer. If none of this energy leaves the plasma between successive bunches, the cumulative energy deposition could approach 2 kJ over a period of less than 3 $\mu$s. As an intentionally conservative upper-bound estimate, if this energy were distributed uniformly between every electron and ion within a 7.8 m-long, 1.5 mm-diameter plasma source with a density of \(6\times10^{14}\,\mathrm{cm^{-3}}\), the resulting temperature would approach 5 keV after a single bunch and approximately 800 keV over the complete bunch train. These values correspond to temperatures approximately three times greater than that at the centre of the Sun after a single bunch and around 50 times greater than characteristic temperatures in a fusion reactor after the full train.

\begin{figure}[h!]
\centering
\includegraphics[width=\textwidth]{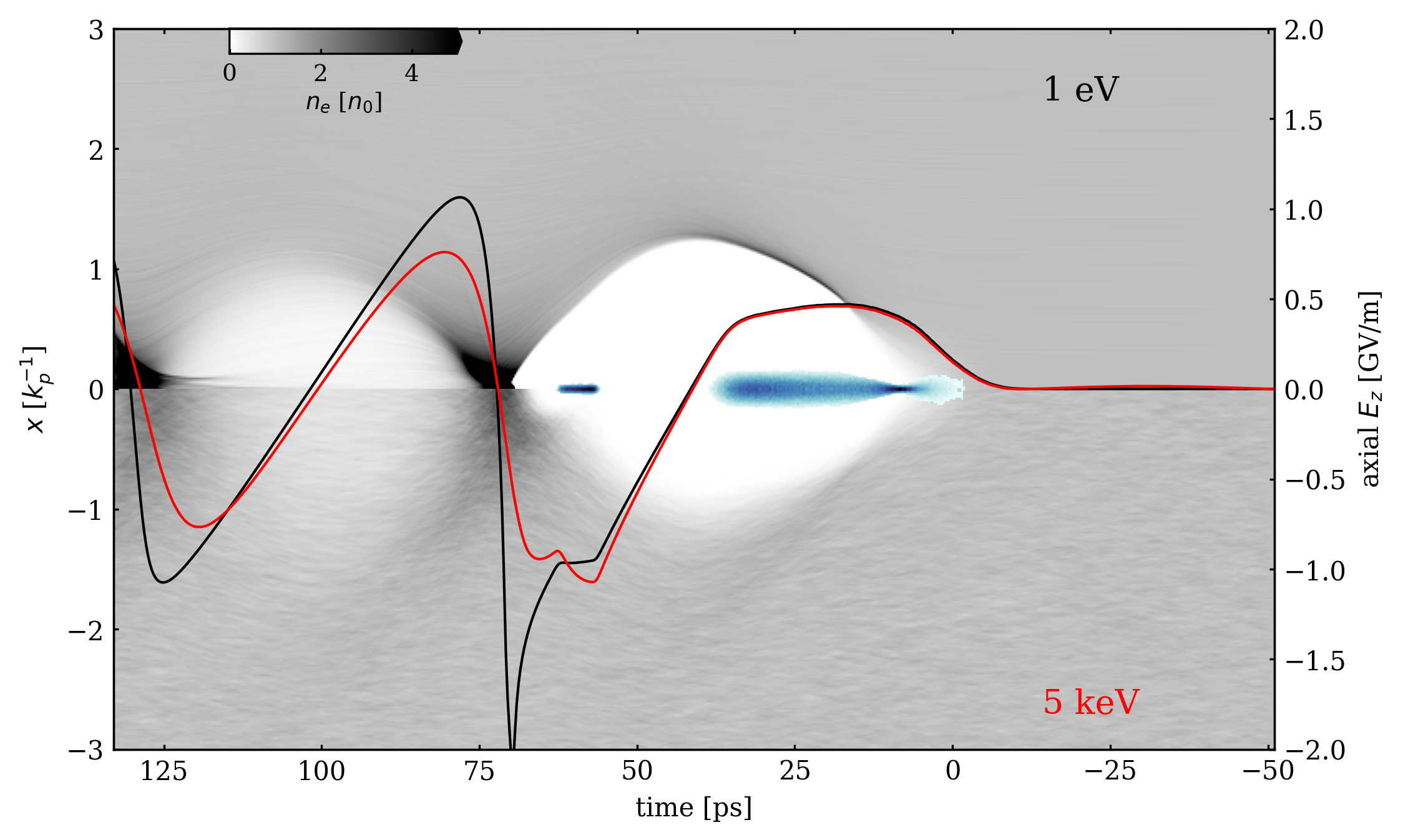}
\caption{Comparison of the wakefield structure, density map, and on-axis electric field for a wake driven in a 1~eV versus a 5~keV initial plasma-electron and plasma-ion temperature.}
\label{fig:temperature-effect}
\end{figure}

Such extreme plasma temperatures would have a significant effect on the properties of the subsequently driven wakefields and, in the absence of adequate heat-removal mechanisms, could also have destructive consequences for the plasma source itself (cf. Sec. \ref{sec:sourcedesign}). Figure \ref{fig:temperature-effect} illustrates the potential impact by comparing the wakefield generated by the first bunch, which encounters a 1 eV plasma, representative of that produced by a commonly used high-voltage discharge system, with that generated by a second bunch after imposing a 5 keV temperature increase associated with the preceding wakefield event. The wake structure is significantly altered in the latter case. In particular, the non-negligible transverse momentum of the heated plasma particles modifies the plasma response and consequently changes the electric fields experienced over the length of the witness bunch.

The magnitude of this estimated temperature increase is deliberately based on highly conservative and ultimately unphysical assumptions, most notably that no heat leaves the plasma source between bunches and that the deposited energy is distributed equally among all particles. At present, however, such assumptions are necessary to establish an upper bound on the potential heating because the relevant energy-transport channels have not yet been comprehensively investigated under these conditions. The simulation pipeline proposed and preliminarily investigated in Sec.~\ref{sec:sourcesim} will provide a means of determining how the deposited energy is redistributed and transported following each wakefield interaction.

Experimental validation of these simulations will nevertheless be essential. This will require direct measurements of the plasma temperature, rather than relying solely on the indirect diagnostics conventionally employed in plasma-wakefield experiments, such as electron spectrometers, which infer changes in plasma properties through their effects on the particle beams. Direct temperature measurements offer an additional advantage because they can help disentangle the otherwise similar effects of changes in plasma density and temperature on wakefield properties, including modifications to the plasma period and accelerating electric fields.

Ultimately, a complete experimental description of energy transport in a plasma accelerator will require temperature measurements of the plasma electrons, plasma ions, and the plasma source itself. Developing and implementing diagnostics capable of making these measurements will therefore form one of the primary strands of HALHF R\&D over the coming years.

\subsection{Plasma-source design \& cooling}
\label{sec:sourcedesign}
Plasma sources, used to generate the accelerating medium and house the accelerating plasma-cavities, are at the heart of the plasma-linac design. The beam-driven community has coalesced around discharge-capillary sources in recent years due to their reliability and scalability in terms of both length (AWAKE) and repetition rate (FLASHForward). Discharge capillaries are commonly constructed from optically transparent materials (e.g. sapphire) to enable access to plasma-emission light used for spectroscopic diagnostics. Sapphire is a robust material at low repetition rate but has a melting point far below the plasma temperatures expected at HALHF. Temperature stabilisation of future HALHF plasma sources must therefore be incorporated in any design, most likely with water cooling through manifolds attached to the capillary. An additional consideration is that the rate of energy deposition in the plasma is not dependent on the plasma-source diameter. Thus a very large diameter (much larger than the 1-10 mm diameters used in state-of-the-art machines) would decrease the required radial cooling rate.

Maintaining a 100$\%$ ionisation fraction of the plasma over the length of the bunch train will be key for controlling the plasma properties, for example by controlling collisional ionisation that may result from perturbations to the plasma caused by the wakefield. This will require a high-voltage discharge with a flat-top length that is longer than the bunch train e.g. 5 microseconds. Solid-state switches have recently been used in high-voltage systems to generate such pulse profiles~\cite{Lopes2026}. However, plasma recombination typically occurs over the microsecond timescale of the bunch train due to cooling of the plasma at the colder capillary walls. This can be mitigated by implementing magnetic radial confinement with a solenoid field such that the hot electrons remain within a central column of the plasma source. This would be aided by having a large inner capillary diameter such that the plasma column is surrounded by a large collar of neutral gas, which may act as a sink of energy for particularly hot plasma electrons. Finally, it will be necessary to control collisional-ionisation effects to maintain the plasma properties from bunch to bunch. A 100$\%$ ionisation fraction of the first level will aid in this. However, the perturbed plasma electrons and ions are expected to be energetic enough to ionise the 2nd (and higher) levels of gases typically used in plasma-acceleration experiments e.g. argon. It would therefore be desirable to completely remove the possibility of collisional ionisation of higher levels by using a gas species that only has one electron. A compromise between helium (the current baseline plasma species) and hydrogen would be deuterium. A gas recycling system would be required for such a choice.

\begin{figure}[h]
\centering
\includegraphics[width=\textwidth]{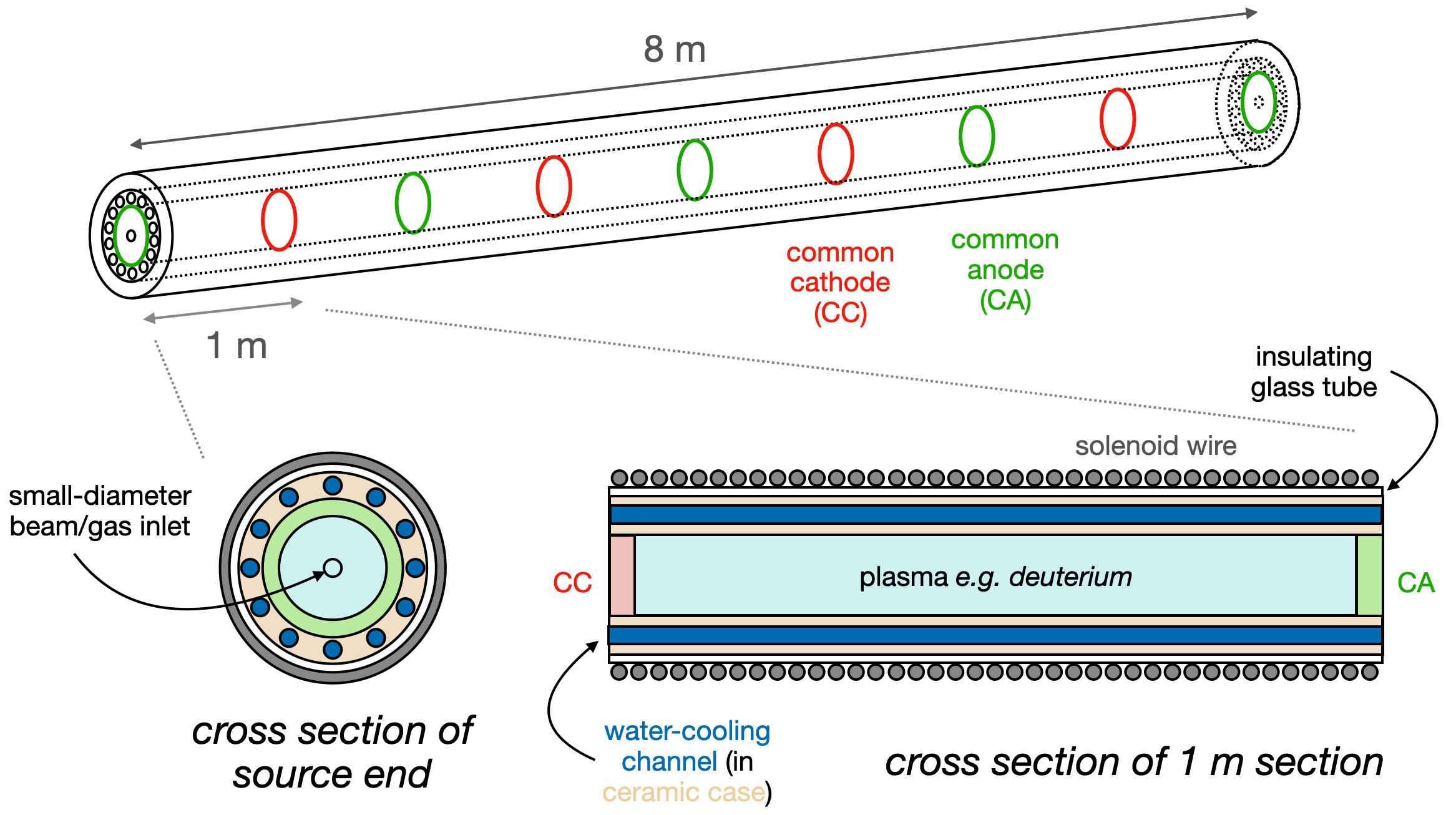}
\caption{Sketch of a possible 8-m-long segmented HALHF plasma-source design comprising: a large-diameter tube (with small diameter inlet to minimise expulsion) to contain the deuterium plasma; common anodes and cathodes to reduce discharge voltage and prevent ion drift; ceramic collar with water-cooling channels to thermally stabilise the source; a solenoid wire (on an electrically isolating glass collar) to mitigate cooling of the plasma, and thus recombination, at the capillary wall.}
\label{fig:plasma-source}
\end{figure}

All the aforementioned considerations have been incorporated in to a potential design for a plasma source, sketched out in Fig.~\ref{fig:plasma-source}. Much research and development is required to turn this sketch into a real-world prototype, for example the preliminary temperature studies described next. This R$\&$D is a key goal for HALHF in the coming years. It should also be noted that this preliminary design is not a catch-all for all the HALHF plasma source requirements. For example, the additional magnetic fields for driver guiding are not included and the possible detrimental impact of a strong solenoidal field on spin polarisation have not yet been fully studied.

\subsubsection{Source temperature measurements, simulations, and stabilisation concepts}
\label{sec:sourcesim}

As discussed in section \ref{sec:heatingcalcs}, a significant amount of heat will be left in the plasma by the plasma creation process and, especially for the HALHF baseline parameters, as the coherent wakefield energy dissipates into thermal energy.
From symmetry considerations, the majority of this heat will flow radially outwards towards the wall of the plasma source.
At repetition rates of $\sim \SI{100}{\Hz}$, this heat can be conducted away from the source by choosing materials with a high coefficient of thermal conduction \cite{warmcells}.
At the high repetition rates demanded by HALHF, to prevent expansion or even melting of the plasma source material, the source must be actively cooled.
Complementary research in this direction is being conducted at the ADVANCE Lab at DESY in order to support a high-average-power upgrade of the FLASHForward plasma accelerator experiment \cite{ADVANCELab, FFWD0, FFWD}.

\begin{figure}[h]
\centering
\includegraphics[width=\textwidth]{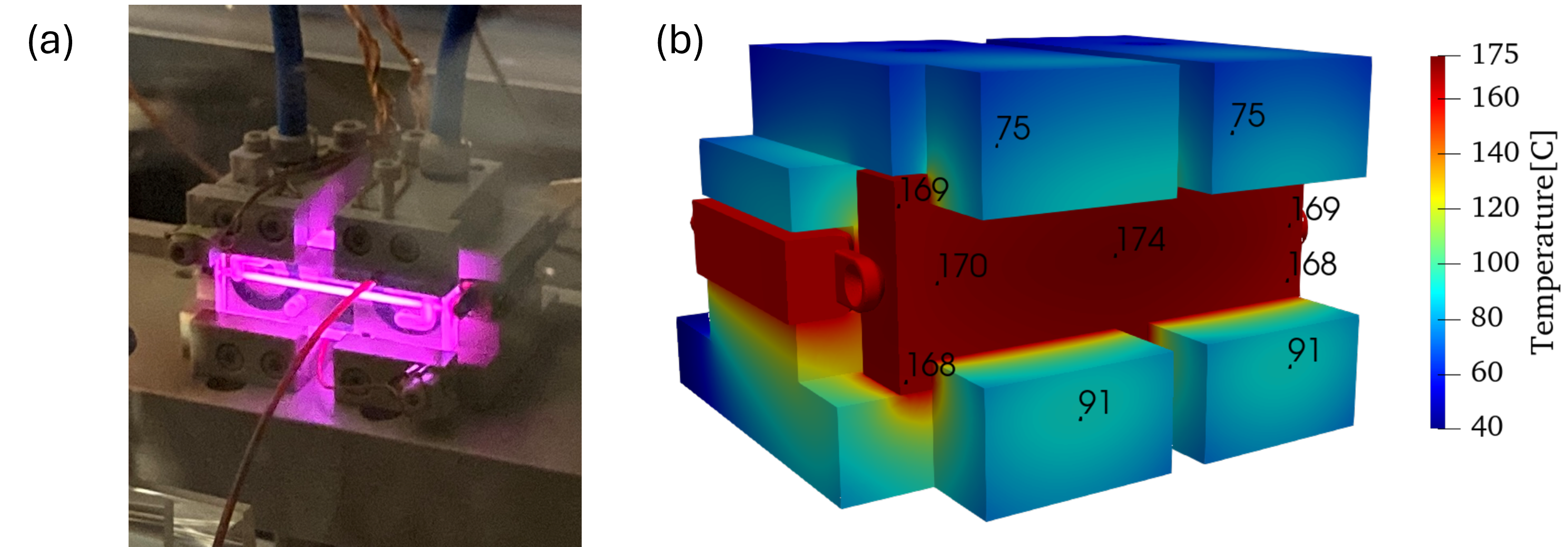}
\caption{Cell heating. (a) Photograph of a \SI{50}{\milli\metre} sapphire capillary containing hydrogen plasma (pink).
The wires above the capillary go to the PT-1000 sensors which are placed under the PEEK clamps (brown-grey).
The red fibre has a GaAs temperature sensor attached to it.
(b) Map of the asmyptotic capillary temperature for \SI{40}{\Hz} operation from an ANSYS simulation.}
\label{fig:heatedcells}
\end{figure}

There is currently no facility fully to test a (multi-)kHz plasma accelerator source, therefore simulations must be relied on. 
These simulations can be benchmarked at low repetition rate then extended to high repetition rate so that suitable cooling solutions can be designed.
Fig.~\ref{fig:heatedcells}~(a) shows a \SI{50}{\milli\metre}, open-ended, FLASHForward-style discharge capillary that was constructed by milling two slabs of sapphire then holding them together with PEEK clamps. 
The capillary was filled with Hydrogen to a pressure of approx.~\SI{2}{\milli\bar}.
Plasma was created by applying a \SI{20}{\kilo\volt} voltage pulse across the gas lasting $\sim \SI{0.5}{\us}$ at repetition rates of up to \SI{40}{\Hz}. 
This also acted as the heating source.
A plasma hydrodynamic model was used to estimate that \SI{180}{\milli\J} was deposited into the capillary walls by every discharge pulse, corresponding to \SI{3.6}{\J\per\metre} per pulse \cite{HYQUP}.
The evolution of the plasma temperature while discharging was measured by PT-1000 sensors placed between the sapphire slab and the PEEK clamp, and by a GaAs temperature sensor clamped directly to the surface of the sapphire.
The full results will be presented in a future publication.
The temperature measurements were used to benchmark an ANSYS model of the capillary, where a CW heating power, with the same average power delivery as the experiment, was applied uniformly over the inner surface of the capillary.
A map of the asymptotic ($t \rightarrow \infty$) temperature distribution from the simulation is shown in Fig.~\ref{fig:heatedcells}~(b).

\begin{figure}[h]
\centering
\includegraphics[width=\textwidth]{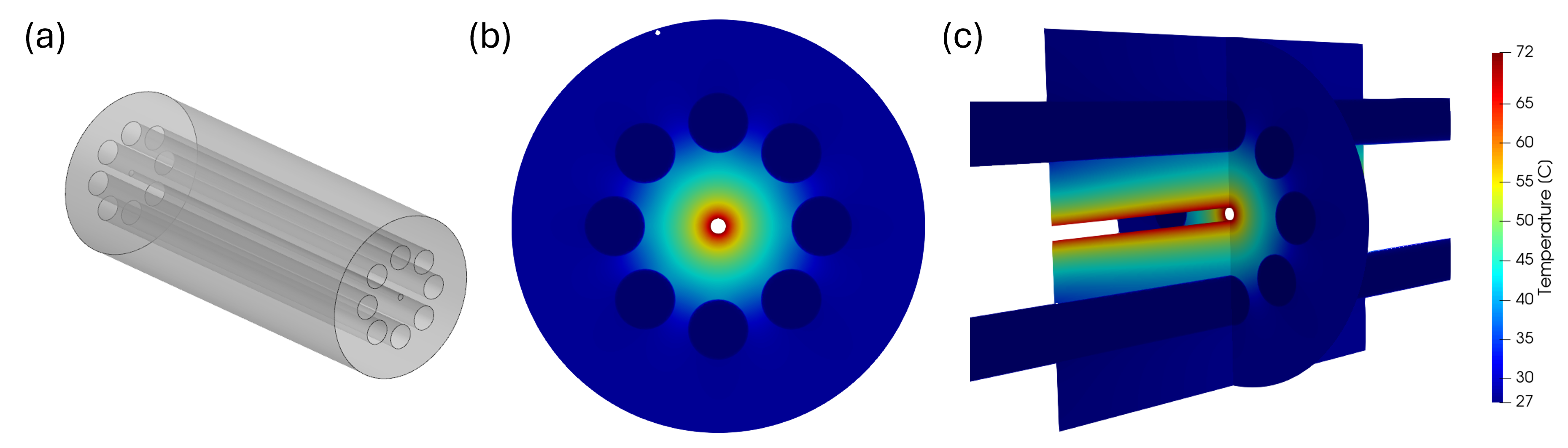}
\caption{Cell cooling. (a) The geometry of the simulated cooled source. 8 water pipes surround the central capillary tube in which the plasma would be formed. (b) and (c) show two views from the 3D simulated temperature distribution with \SI{3.6}{\kilo\W\per\metre} heating applied at the capillary tube surface. The capillary diameter is \SI{1.5}{\milli\metre} and the cooling channels are \SI{8}{\milli\metre} in diameter.}
\label{fig:cooledcells}
\end{figure}

Scaling the ANSYS simulation to discharge repetition rates of \SI{1}{\kilo\Hz}, corresponding to heating of \SI{3.6}{\kilo\W\per\metre}, results in final temperatures in excess of \SI{1300}{\kelvin}.
The impracticality of dealing with these high temperatures led to the consideration of actively cooled capillaries.
A simplified design was studied, consisting of a cylindrical sapphire block with an on-axis capillary channel where the plasma will be created (or in the simulated case, the heating source is applied), with 8 surrounding channels through which water can be pumped to cool the device.
DESY-standard cooling water (\SI{27}{\celsius}, \SI{2}{\metre\per\second}) was driven through these pipes in the simulation. 
The capillary geometry and two slices of the 3D source temperature map are shown in Fig.~\ref{fig:cooledcells}~(a), (b) and (c) respectively.
In this scheme, the limit to the applied heating power is vaporisation of water at the surface of the cooling pipes, which can lead to catastrophic source failure.
In Fig.~\ref{fig:cooledcells}~(b) \& (c) a safe operating temperature is shown for a constant heating power of \SI{3.6}{\kilo\W\per\metre}.
In a slightly modified design, \SI{7.2}{\kilo\W\per\metre} heating power could be safely applied.

Using the required cooling power per unit length from Table~\ref{tab:1}, the above design falls short by a factor of approximately 4. 
Considering that no optimisation has been performed, and easy improvements remain such as exploring other source materials and lowering the cooling water temperature, it seems likely that this deficit can be overcome.
However, this will only solve the problem of removing the average power deposited into the capillary. 
It must be understood how stresses caused by the rapid heating of the source by the bunch trains affects the source structure, and how the extremely hot plasma affects the capillary wall.

\subsection{Radiation environment}
\label{sec:halhf-radiation-environment}

The high average beam power of the particle beams in the HALHF project makes the control of beam-induced radiation an integral part of the accelerator design. In addition to radiation from the dedicated defined electron dumps, two aspects are of particular relevance: the energy deposited in, or emitted from, the plasma cells and the extraction of the subsequently strongly depleted driver bunches. The following discussion identifies the main challenges. It does not yet constitute a detailed shielding or component-lifetime assessment.

\subsubsection{Radiation Emission from Plasma Cells}
\label{sec:halhf-plasma-cell-emission}

During the driver--plasma interaction, plasma electrons can acquire substantial longitudinal and transverse momentum and leave the plasma over a broad range of energies and angles, as previously reported for laser--plasma accelerators~\cite{Kaganovich2008,Yang2017}. Although the energies of these electrons are generally low compared with those of the driver and witness bunches, their potentially large number and broad angular distribution may produce significant radiation when they interact with the plasma-cell structure, vacuum chamber or nearby beamline components. Such electron emission has also been observed in plasma accelerators and identified as a significant contributor to the overall radiation field~\cite{Bohlen2026Radiation}.

For HALHF, the magnitude and distribution of this source need to be investigated as part of the overall plasma-cell design. This should not be treated solely as a shielding problem, but as part of the energy balance, efficiency and power-management considerations for high-average-power operation. Dedicated PIC simulations should quantify the partition of the driver energy between the witness bunch, the depleted driver, plasma heating and particles escaping from the cell. These results should subsequently be coupled to thermal and radiation-transport calculations.

Plasma heating, recovery between bunches, cooling of the cell structure and radiation generation are closely interconnected. A suitable HALHF plasma-cell concept must therefore simultaneously provide stable plasma conditions, sufficient heat removal, acceptable material lifetime and shielding of emitted radiation.

The magnitude of this radiation source and its implications for shielding and component lifetime remain unquantified for HALHF. A viable concept addressing these requirements therefore needs to be demonstrated.

\subsubsection{Stray Radiation Fields}
\label{sec:halhf-stray-radiation}

After each plasma stage, the spent driver must be separated from the main beam and transported to a dedicated dump. The driver has transferred most of its energy to the wake and consequently has an energy spread approaching 100\,\%, with a spectrum strongly weighted towards energies of a few hundred MeV~\cite{Foster2025}. This broad distribution makes loss-free extraction challenging. If the driver is instead simply lost on the side of a magnet, the beam pipe or another component, its high average power produces an intense electromagnetic shower and an extensive secondary radiation field.

A preliminary Monte-Carlo study using the code FLUKA \cite{Ferrari2005FLUKA2005,Bohlen2014TheApplications} was performed to illustrate this problem. Particle distributions from a PIC simulation were used as the source in the Monte-Carlo simulations. The simplified model contains a magnet chicane including dipole magnets, a sextupole and active plasma lenses in an XFEL-like accelerator tunnel. It is intended to represent a generic extraction section rather than an optimised HALHF beam line. Both driver and witness bunches were included: the witness traverses the chicane with negligible chamber interactions, whereas the depleted driver is intercepted by the magnet chambers because its energy distribution is incompatible with the magnet settings.

\begin{figure}[tbp]
  \centering
  \includegraphics[width=\linewidth]{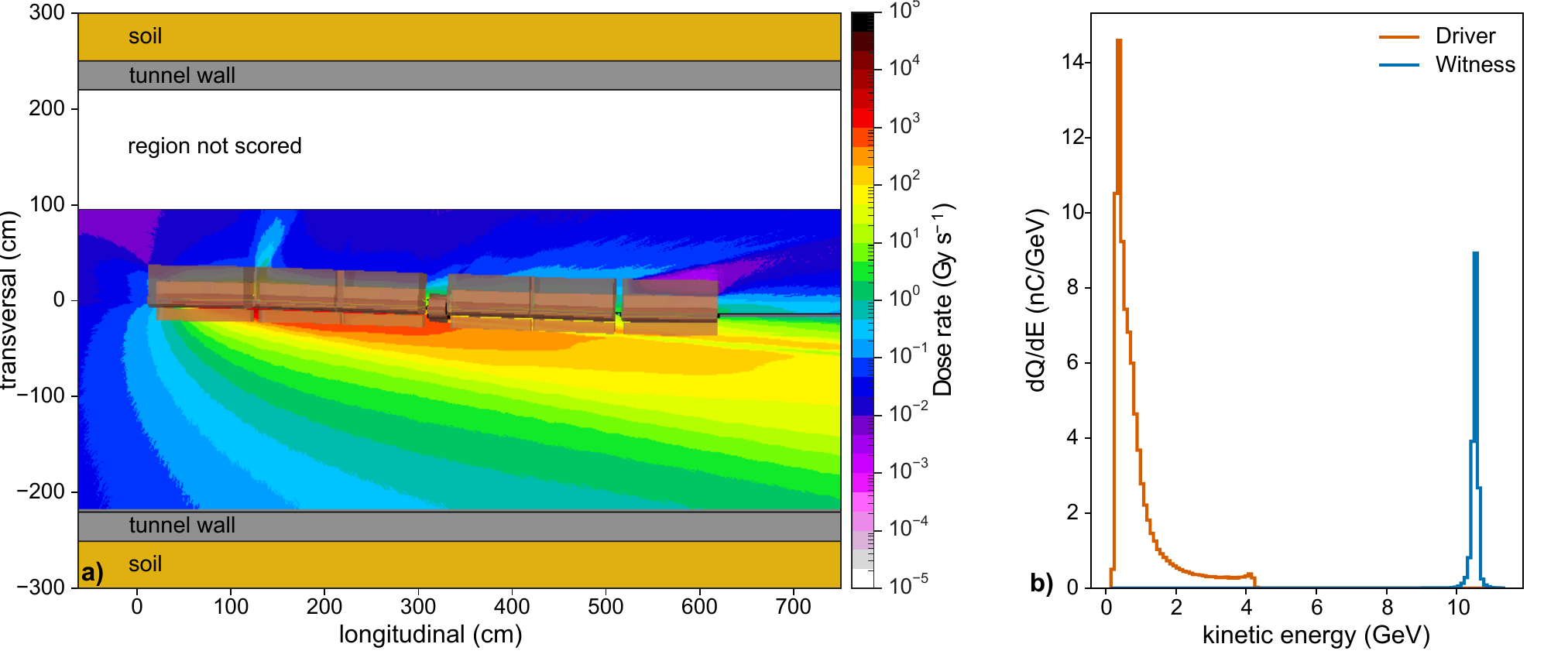}
  \caption{Preliminary assessment of the spent-driver extraction problem.
    \textbf{(a)} Top-view FLUKA simulation of the local absorbed-dose-rate distribution resulting from the loss of the spent driver in a simplified extraction section. The magnet section is optimised for the witness energy shown in panel (b). The source is scaled to 100 bunches per train at a train repetition rate of \SI{100}{\hertz}.
    \textbf{(b)} Kinetic-energy spectra of the spent driver and witness bunch obtained from PIC simulation which were used as input for the FLUKA simulation shown in panel (a).}
  \label{fig:halhf-driver-dose}
\end{figure}

Figure~\ref{fig:halhf-driver-dose} illustrates the radiation-safety challenge associated with the extraction of the spent driver. Panel~(a) shows the simulated absorbed-dose-rate distribution for a case in which the driver is not transported to a dedicated dump, but instead just lost in the magnet chambers. Panel~(b) shows the corresponding energy spectra of the spent driver and accelerated witness bunch. For the dose-rate calculation, the source was scaled to 100 bunches per train and a train repetition rate of \SI{100}{\hertz}. Local absorbed-dose rates reach the order of \SI{1000}{\gray/\second} within approximately \SIrange{10}{20}{cm} of the magnets, while rates of \SIrange{10}{100}{\gray/\second} persist throughout the modelled tunnel. The calculation includes only one acceleration stage and contributions from other stages would add to this field wherever they overlap. Radiation arising from plasma-electron emission, discussed above, is not included and would constitute an additional source. While the presented values describe local dose in material, not dose to personnel, they still demonstrate that uncontrolled driver losses would expose surrounding accelerator equipment to radiation levels incompatible with sustained operation of ordinary components.

For comparison, dose rates of the order of \SI{1}{\milli\gray/\hour} provide a useful preliminary engineering benchmark at locations hosting distributed accelerator electronics, subject to the particle spectrum, required lifetime and radiation qualification. This is not a universal acceptance limit, but the calculated tunnel-wide rates exceed it by approximately eight to nine orders of magnitude.

The extraction optics, dump layout and shielding must therefore be developed as one system. Although the simulated configuration is illustrative rather than a proposed design, it identifies an unresolved challenge: a viable extraction and disposal concept must demonstrate acceptable power deposition, component lifetime, electronics reliability and maintenance conditions. Further work is required to establish realistic loss scenarios, transport the depleted driver over its full phase space and assess the combined radiation fields and activation from multiple stages. Engineered loss points, local shielding, beam-loss monitoring and fast machine protection will be essential elements of the design.

\subsection{Beam-Dump Systems}
\label{sec:halhf-beam-dumps}

HALHF requires distributed dumps for the depleted drivers and main dumps for the electron and positron beams after the interaction region. In both cases, the average power, bunch structure, transverse beam size, energy distribution and credible worst-case-scenario conditions must be considered.

\subsubsection{Driver Beam Dumps}
\label{sec:halhf-driver-dumps}

In the published \SI{250}{\giga\electronvolt} centre-of-mass baseline, the driver system provides \SI{23.8}{\mega\watt} to 48 plasma stages. This corresponds to approximately \SI{0.50}{\mega\watt} of incident driver power per stage and, for the assumed 80\,\% driver depletion, approximately \SI{0.10}{\mega\watt} remaining in each spent driver~\cite{HALHF-EPPSU}. The corresponding nominal residual powers for the published \SI{380}{\giga\electronvolt} and \SI{550}{\giga\electronvolt} configurations are approximately \SI{0.15}{\mega\watt} and \SI{0.22}{\mega\watt} per stage.

These nominal powers are within the \SI{300}{\kilo\watt} rating of the solid-state electron dumps used at the European XFEL, providing a credible technical basis for the HALHF driver dumps~\cite{XFELDump2006}. The principal challenge is therefore expected to be the controlled transport of the broadly distributed spent driver to the dump, as discussed in Section~\ref{sec:halhf-stray-radiation}, rather than the development of the dump technology itself. The European XFEL design could serve as a starting point, with adaptations to the HALHF bunch structure, energy distribution and potentially larger beam size at the dump entrance. Furthermore an undepleted or incorrectly extracted driver could exceed the XFEL dump rating. Consequently, further research is required to investigate whether solid-state beam dumps could withstand such conditions or whether a suitable machine-protection strategy would be required.

\subsubsection{Main Beam Dumps}
\label{sec:halhf-main-dumps}

For the \SI{250}{\giga\electronvolt} baseline, the powers delivered to the interaction region are approximately \SI{9.6}{\mega\watt} for electrons and \SI{3.2}{\mega\watt} for positrons~\cite{HALHF-EPPSU}. The \SI{18}{\mega\watt} pressurised-water dump developed for the ILC therefore provides an appropriate starting point for the HALHF main dumps \cite{ILCDump2012}. The concept uses vortex water flow and beam rastering to distribute the deposited power and requires a cooled beam window, activated-water containment, gas management, shielding and remote-maintenance provisions.

The HALHF implementation must be verified for the asymmetric and disrupted post-collision beams. The ILC power rating also covers the published \SI{380}{\giga\electronvolt} configuration, but not the \SI{21.1}{\mega\watt} electron beam of the \SI{550}{\giga\electronvolt} configuration without modification, which would require an uprated or distributed dump solution.

\section{Driver-distribution system}
\subsection{Distribution concept and geometry}

The staging of plasma-accelerators intended for HALHF requires multiple drivers, each distributed to a specific plasma stage. With all drivers being generated at an RF linac with a bunch spacing of $\Delta t$, the distribution system must not only deliver the drivers to the stages, but also ensure proper timing of the bunches.

Due to the short bunch spacing of the drivers exiting the RF linac, and the achievable rise and fall times of kickers, the current solution is to separate the drivers into separate beamlines using RF deflectors. The driver bunch train will be separated into 3 beamlines using a single RF deflector, then each of those trains will again be separated into 2 separate trains using another RF deflector. This results in 6 different beamlines with 8 drivers per beamline per accelerated witness bunch. Each beamline will have a certain number of delay chicanes (described below) that ensure the injection of each driver is synchronous with the witness beam during the accelerating phase.

The preliminary lattice design consists of 9 quadrupoles and 8 dipoles, but a more robust version will likely include 2 sextupoles to combat the second order dispersion. The lattice can be scaled to achieve the same $\beta$/$\alpha$ functions, as well as the same periodic dispersion and bunch-length evolution. The scaling is $k=k_0\frac{S_E}{S_L^2}$ and $B=B_0\frac{S_E}{S_L^{3/2}}$ for a delay that is approximately constant, or $B=B_0\frac{S_E}{S_L}$ for a delay that scales approximately as $\Delta t_0 S_L/c$, where $c$ is the speed of light. The $0$-subscript indicates a reference lattice with a matched solution that closes dispersion and bunch-length evolution and keeps $\beta$/$\alpha$-functions periodic. $S_E$ and $S_L$ are the scaling of the energy and all element lengths, respectively. These scaling laws were derived from making the Hamiltonian of the elements invariant under the scaling and from a second-order approximation of the delay equation, Eq.~4, given in Ref.~\cite{kalvik_driver-delay_2026}. In the case where the delay is kept approximately constant, the lattice will need to be rematched to close the dispersion and bunch-length evolution, as the dipole-scaling to maintain a similar dispersion is incompatible with keeping a constant delay.

\subsection{Delay-chicane simulations}
\begin{figure}[h!]
    \centering
    \includegraphics[width=.9\linewidth]{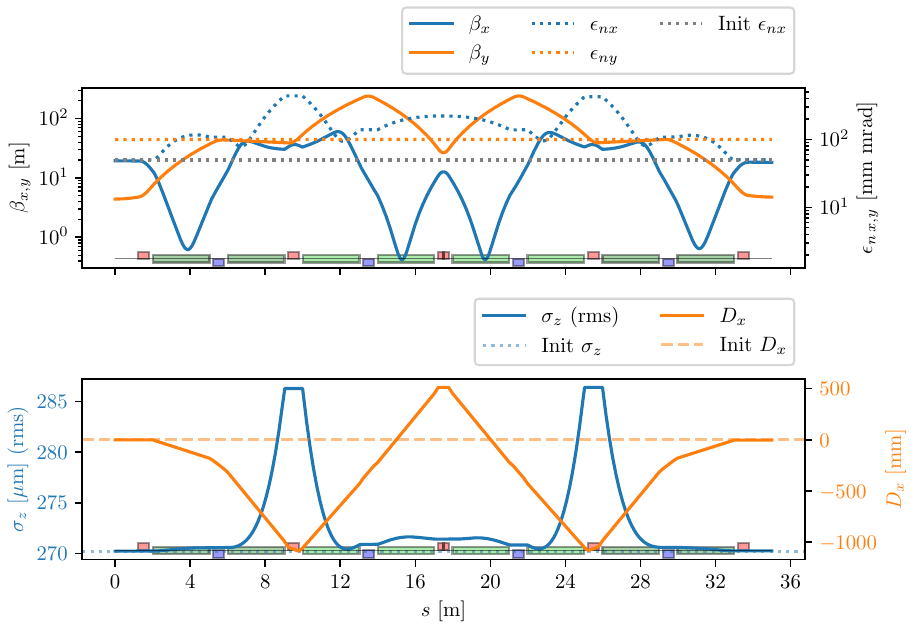}
    \caption{The evolution of the transverse phase-space parameters (top), and the evolution of the bunch length and horizontal first-order dispersion (bottom). The lattice elements are shown at the bottom of each plot, where the pink squares are (horizontally) focusing quadrupoles, the purple squares are defocusing, the dark-green rectangles are dipoles bending in the positive horizontal direction and the light-green rectangles are dipoles bending in the negative horizontal direction. The bunch length is given by the root-mean-squared (rms) size of the current profile, but the current profile is a trapezoid whose length is \SI{1050}{\mu m}. The figure is a scaled version of Fig.~\SI{5}{} from Ref.~\cite{kalvik_driver-delay_2026} (CC-BY 4.0).}
    \label{fig:delay_chicane}
\end{figure}

\begin{figure}[h!]
    \centering
    \includegraphics[width=0.8\linewidth]{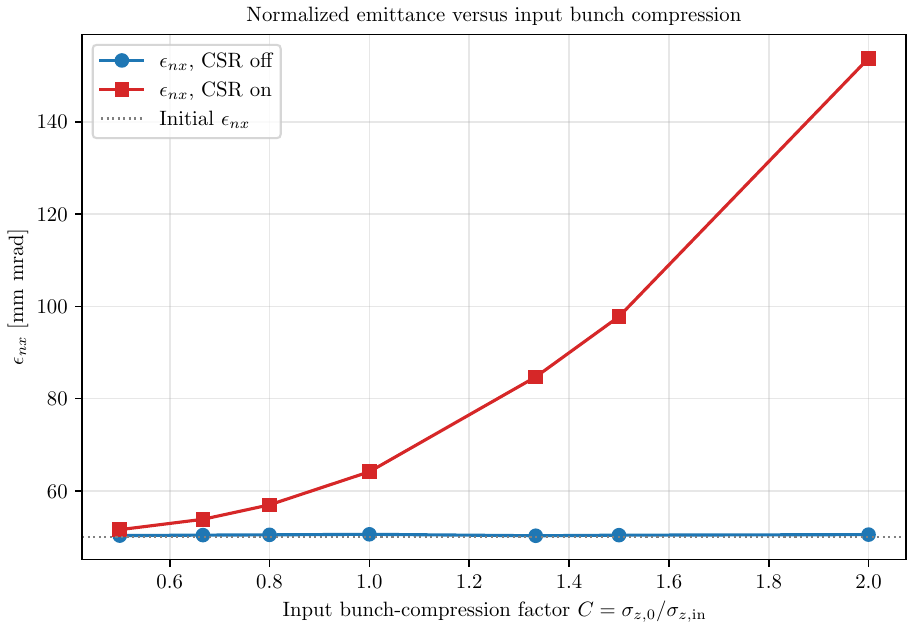}
    \caption{Horizontal emittance growth of an \SI{8}{nC} beam with and without coherent synchrotron radiation after a single delay chicane, as a function of bunch-length compression. The compression factor is defined as the reference length $\sigma_{z,0}$ over the incoming-bunch length $\sigma_{z,in}$. In the simulations, the reference length is \SI{1050}{\mu m}, or about \SI{270}{\mu m} rms. Incoherent synchrotron radiation is enabled in these scans, but has a negligible effect.}
    \label{fig:emittance_vs_bunch_length_delay_chicane}
\end{figure}

Figure \ref{fig:delay_chicane} shows a simulation of the delay-chicane lattice with the first-order beam parameters. The initial $\beta$-functions are \SI{19.38}{m} and \SI{4.397}{m} in the horizontal and vertical planes, respectively. The initial $\alpha$-functions are zero, as is the incoming dispersion. The initial current profile is a trapezoid spanning \SI{1050}{\micro\meter} or \SI{270}{\micro\meter} rms, with a total charge of \SI{8}{nC}. All parameters are periodic, which makes this a repeatable lattice, although sextupoles will be added to cancel second-order dispersion.
Figure \ref{fig:emittance_vs_bunch_length_delay_chicane} shows the effects of coherent synchrotron radiation due to the strong bends of the delay chicane. To maintain the emittance of the drivers, it is likely to be necessary to double the intended length of each driver, and compress them before injection.
\section{Particle sources}
\label{sec:sources}
\subsection{Physics with polarized beams}

The only operating linear collider
with polarized beams generated was the SLC at SLAC.
A high electron polarization of about $P(e^-) = 78\%$ at the Z- pole was achieved, leading to
the best single measurement of the electroweak mixing angle $\sin^2\theta_W = 0.23097\pm 0.00027$~\cite{SLD:2000leq},
although the luminosity was about one order of magnitude less than the corresponding LEP
data. This clearly demonstrates one crucial impact of polarized beams for precision physics. 

In general, polarized beams offer new observables like left-right asymmetries and are substantial for disentangling signal and background processes, for controlling systematic uncertainties, for achieving 
the best precision measurements, for analyzing the chiral and interaction structure of the underlying physics and is a real game changer in searches for Beyond Standard Model (BSM) physics.

Simultaneous polarization of the initial particles
enhances the effective polarization $P_{\rm eff}=\frac{P_{e^-}-P_{e^+}}{1-P_{e^-}P_{e^+}}$, as well as the effective luminosity by
${\cal L}_\text{eff}=\frac{1}{2}(1 - P_{e^-} P_{e^+}){\cal L}$  (, where the
ratio ${\cal L}_{\rm eff}/{\cal L}$ reflects the fraction of interacting particles). Please note, that polarized electrons only (even for $P_{e^-}=100\%$) do not enhance the effective luminosity
${\cal L}_\text{eff}$.

The polarization of the polarized beams is measured via Compton polarimetry with a precision of about
$\Delta P_{e^\pm}/P_{e^\pm}=0.25\%$-0.5\%. 
Since the measured left-right asymmetry $A_{\rm LR}$ significantly depends on the polarization, the experimental uncertainty  in the extracted $A_{\rm LR}$ is determined by the polarization accuracy. 
Therefore this quantity benefits greatly from the use of simultaneously polarized $e^{-}$ and 
$e^+$ beams (due to pure error propagation).

Furthermore, simultaneous polarization allows the exploitation of transversely polarized beams: transversely polarized beams generate interference terms between left- and right-helicity
amplitudes and provide therefore complementary information compared to longitudinally-polarized beams. In the limit $m_{\rm electron} \to 0$ and within the Standard Model (SM), the corresponding helicity amplitudes vanish~\cite{Moortgat-Pick:2005jsx}, so that if the
transverse polarization of the final state particles is not measured, the effects of transverse
polarizations are absent in the $\phi$-averaged cross section.

In the case of (axial)vector-like couplings between the electron and the gauge bosons (as in the Standard Model), transversely polarized contributions appear only bilinearly, i.e.\ are proportional to $P^T_{e^-} P^T_{e^+}$, and can only be exploited if both beams are simultaneously polarized.
This offers new observables to detect non-standard interactions, including possible new sources of CP-violation as well as access to specific triple-gauge couplings which cannot be extracted with only longitudinally polarized beams, for more details, see ~\cite{Moortgat-Pick:2005jsx,Moortgat-Pick:2015lbx,Fujii:2015jha,Fujii:2017vwa,Fujii:2018mli} and references therein.

\subsection{Polarized positrons}
\subsubsection{Introduction}

The undulator-based positron source is a mature design for a high-intensity collider source and allows even 
the generation of polarized positrons that would be of high physics impact, as briefly described in the previous section. Due to the long bunch trains, the ILC/LCF beam structure imposes the most challenging requirements for a positron source. 

In the following, a status report including recent progress and prototyping work for the actual ILC/LCF undulator-based positron source is given. The following positron areas 
still need further adjustments for the HALHF set-up compared to the ILC/LCF positron source:
\begin{itemize}
\item Drive-beam energy and undulator parameters,
\item Target material,
\item Optics-matching device.
\end{itemize}
but since the HALHF beam is not foreseen to have a long flat-top structure, some requirements are relaxed and the ILC/LCF positron prototyping results are also promising for a modified polarized HALHF positron-source design. These topics are discussed in Sec.~\ref{sec:positron-outlook}.
\subsubsection{Undulator-based positron source}

The undulator-based positron scheme provides a mature, reliable source with a rather large margin with respect to the damping-ring acceptance and fully matches the ILC/LCF requirements, namely a yield of 1.5 $e^+/e^-$ and about $1.3\times 10^{14}/$s at the IP. 

Detailed simulations have been performed including realistic electromagnetic fields, alignment studies and power deposition; 
a mask
system for wall protection for the ILC at $\sqrt{s}=250$, $500$, $1000$~GeV and GigaZ has been designed to protect the undulator walls from synchrotron-radiation absorption \cite{Alharbi:2022okv}. 

Simulations of an adapted undulator design for CLIC as well as for the HALHF design
have been performed. The baseline scheme consists of an undulator with 231 m active length and an electron driver beam of 126.5 GeV at 
5 Hz which is also capable of 10 Hz-running for high luminosity. Full-sized ILC helical-undulator prototypes have been built and field and energy measurements have been made at UK laboratories.\\
More than 91 undulators, each with 5m length, are in long-term operation at XFEL. The energy loss due to particle loss is negligible, the beam alignment is up to 10-20 microns for undulators up to 200 m length and is remeasured every 6 months. During beam operation, the beam trajectory is controlled to better than 3 microns with both slow and fast feedback systems~\cite{Moortgat-Pick:2024fcy}.
Therefore, stable operation and alignment experience has already been available for more than 10 years at XFEL. Here the  beam requirements are even more severe than for the corresponding ILC/LCF undulators for positron production, due to the FEL requirement for undulator photons .
\subsubsection{Target issues}

\begin{figure}
    \centering
    \includegraphics[width=1.0\textwidth]{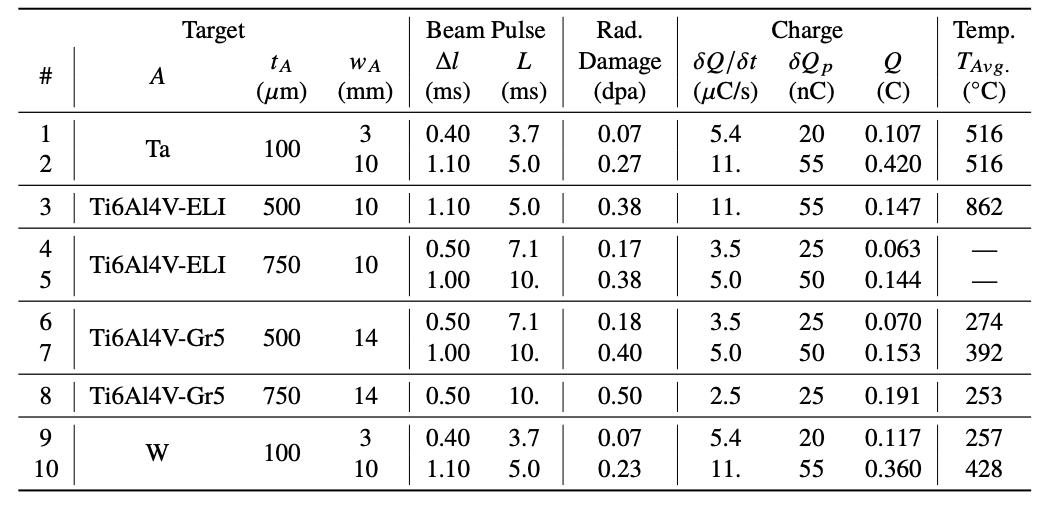}
    \caption{Material Irradiation Specifications: Positron-source candidate materials of different composition
($A$), thickness ($t_A$), and width ($w_A$) have been exposed to
the electron beam: two titanium alloys, pure tantalum, and pure
tungsten~\cite{Lengler:2024znl}.}
    \label{und-fig-target1}
\end{figure}

The incident photons have an energy of about 7.5 MeV (1st harmonic) and hit every 7-8 s on the same position at the target, i.e.\ in 5000 h roughly 5$\times 10^6$ load cycles.
For 250~GeV CMS operation, a target thickness of 7mm (0.2 radiation length) of Ti-alloy grade-5 is foreseen.
It is expected to result in a $\Delta T_{\rm max}=60$-120~K per pulse. 
Thus a maximum energy deposition  of about 60  J/g
is expected, 
with an effective pulse length on the material of about 25-55 $\mu$s and effective repetition rate of 0.17 Hz, causing displacement per atom (dpa) of about 0.3-0.5 per year~\cite{Lengler:2024znl}.
Comprehensive target-material tests have been performed at the Mainz Mikrotron (MAMI) accelerator, i.e. a target material has been exposed with   
a multiple annual peak energy-deposition density (PEDD) of the expected ILC target, 
generated via the electron beam. The results are shown in Fig.~\ref{und-fig-target1}. The irradiated target materials have been
precisely analysed afterwards both via scanning methods as well as via 
synchrotron-diffraction methods and have been compared
with corresponding target materials that have
been exposed with pure thermal loads only.
These tests have shown that the targets will stand the load,
since the different targets (different alloys, target thickness) did not show any material damage even under a multiple of the annual expected PEDD~\cite{Lengler:2024znl}.
A luminosity upgrade is feasible with the chosen Ti-alloy grade-5 target material.

The target wheel has a diameter of 1m and rotates at 2000 rpm (i.e. 100 m/s tangential speed) and is cooled via radiation into a stationary water-cooled cooler. The rotation system is in vacuum and therefore the turning system is set up on  magnetic bearings which allow long-term operations at high rotation speed without major maintenance. The whole system is sealed in an UHV-tank.
The deposited thermal power will be about 2~kW, accumulated close to the rim of the target. Therefore the heat deposition can be removed via thermal radiation from the spinning target to a stationary water-cooled Cu-cooler following 
$~T^4_{\rm radiator}-T^4_{\rm cooler}$, keeping $T_{\rm ave}<460^0$~C.
Mechanical drawings and engineering design for the rotating wheel are being performed in close contact with a potential manufacture of the corresponding magnetic bearings.

\begin{figure}
    \centering
    \includegraphics[width=1.0\textwidth]{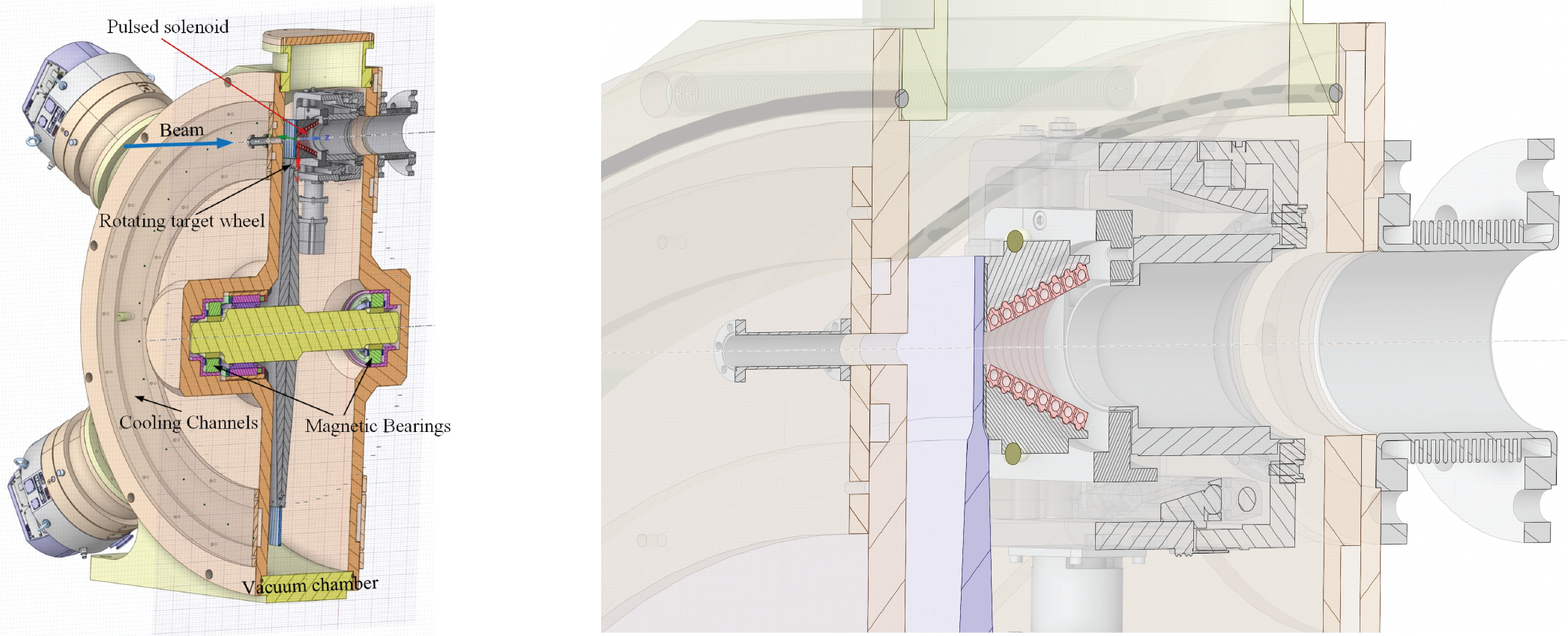}
    \hspace{3cm}  
    \caption{Design of the rotating wheel based on ANSYS simulations including magnetic bearings, pulsed solenoid and vacuum vessel (left panel); Picture detail of the pulsed solenoid device 
    (right panel)~\cite{Moortgat-Pick:2024gbw}.}
    \label{fig-rotwheel-engineering}
\end{figure}

\subsubsection{Optics-matching device}
\label{sec:omd}
Due to the long bunch-train length for ILC, an optics-matching device (OMD) matching  the flat-top pulse width (0.726 ms for 1312 bunches) is needed. 
Therefore a  flux concentrator cannot be used. Instead a pulsed solenoid is foreseen, 
generating a half-sine current pulse of about 4 ms and a peak field of 3-5~T at the target.
After the 4~ms, most of the eddy currents from the half-sine pulse have died out and a stable field over the duration of the beam pulse is achieved. In order to achieve multi-Tesla fields, peak currents
of about 50 kA or above are required. During the pulse, a heat load on the target is induced, generating
a force on the rotating target. Implementing a ferrite shield
inside the pulsed solenoid will only slightly reduce the peak magnetic flux but significantly reduce
the heat load (by more than a factor two) as well as the peak force on the rotating wheel. A yield of about $Y = ~1.7$ 
has been achieved in 2D- and 3D-simulations with COMSOL after implementing such a ferrite shield.

\begin{figure}[h]
\centering
   \includegraphics[width=0.45\textwidth,height=0.35\textwidth]{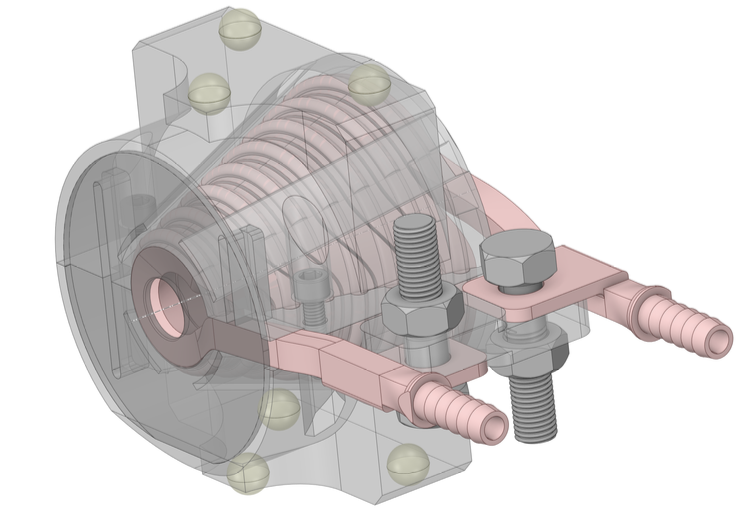} 
\hspace*{1cm}    
   \includegraphics[width=0.45\textwidth,height=0.35\textwidth]{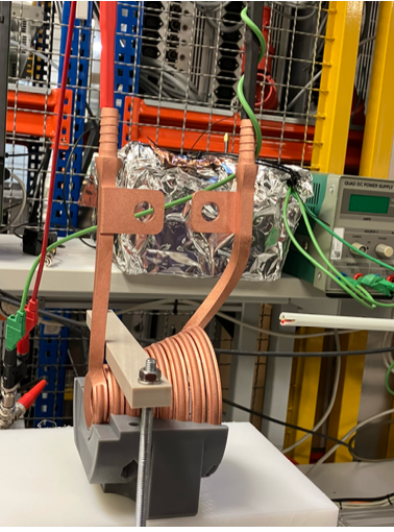}
    \caption{Design of the tapered solenoid coil with 8 windings, metal support bridges and
insulated support rods (left panel); 3d-printed copper prototype model (right panel) of the pulsed solenoid design~\cite{Hamann-thesis}.
    }
    \label{fig-pulsed-solenoid}
\end{figure}

Mechanical drawings for such a prototype have been produced,
(see the left panel of Fig.~\ref{fig-pulsed-solenoid}) .
It consists of a solenoid coil with tapered planar windings and interconnections.
Several prototypes for the copper pulsed solenoid have been produced via 3D printing and magnetic-field measurements are ongoing, see Fig.~\ref{fig-pulsed-solenoid} (right panel). 
An automated simulation tool for the pulsed solenoid including a chicane system has been set up in COMSOL in combination with ASTRA.

\subsubsection{Tracking and simulation tools and future work}
%
%
In order to establish a consistent simulation chain for the HALHF undulator-based positron source, several simulation tools developed or investigated in the context of the ILC/LCF positron source are currently being evaluated for their applicability to HALHF. The studies address both the generation of polarised photons and the subsequent positron production, capture and acceleration.

The optimisation of the positron-capture system is carried out with ASTRA (A Space Charge Tracking Algorithm)~\cite{Astra}. This simulates the beam dynamics from the conversion target through the initial pre-accelerator section. In combination with three-dimensional electromagnetic field maps calculated with COMSOL~\cite{COMSOL}, an automated optimisation framework has been developed in MATLAB. The framework employs a two-level Bayesian optimisation scheme, where the first level optimises the geometry and peak current of the optical-matching device (OMD) using COMSOL, while the second level optimises the RF phases of the accelerating structures within ASTRA. The objective is to maximise the positron yield within the damping-ring acceptance.
The particle tracking includes the three-dimensional electromagnetic field maps exported from COMSOL and the first section of the pre-accelerator, consisting of two standing-wave tubes and three travelling-wave structures. 

Simulations indicate that the longitudinal beam dynamics are largely determined 
downstream 
of these initial accelerating structures. Extending the simulated beamline further downstream would significantly increase the computational effort, since each additional accelerating structure introduces another RF phase to be considered in the optimisation.

The existing optimisation framework can be adapted to the HALHF positron source, although several modifications will be required. For the ILC/LCF, the long bunch trains and the corresponding requirement for a stable magnetic field over approximately 1 ms motivate the use of a pulsed solenoid instead of a flux concentrator, as discussed in Sec.~\ref{sec:omd}. Since the HALHF bunch trains are considerably shorter than for ILC, the use of a flux concentrator becomes a viable alternative and should be investigated as a potential OMD. Furthermore, the pre-accelerator layout and RF configuration will need to be adapted to the HALHF bunch structure and the requirements of the subsequent cool-copper RF linac.
A prerequisite for these optimisation studies is a realistic initial positron distribution obtained from simulations of the undulator radiation and photon conversion at the target. For the generation of polarised-photon spectra, the applicability of CAIN~\cite{Yokoya} is currently being investigated. CAIN allows the simulation of polarised-particle spectra, although the undulator radiation is approximated using Compton backscattering. Whether this approach provides a sufficiently accurate description of the HALHF undulator-based source remains to be established.

The subsequent conversion of photons into positrons requires a detailed simulation of the photon interactions with the target material. For this purpose, Geant4~\cite{Allison:2016lfl} has been selected as the simulation framework. Geant4-based simulations have already been employed in other positron-source studies, including the 
PPS-Sim code~\cite{Ushakov:2011zz,Ushakov:2010zz,Dollan:2005nj}, a further development of the simulation tool used in~\cite{Astra}, which has been updated to current Geant4, ROOT and Qt versions. PPS-Sim includes the spin transport of the positrons and provides yield and polarization at the damping-ring acceptance, and it allows systematic scans of the undulator, target and capture parameters.

The next steps will therefore focus on establishing a consistent simulation chain from the HALHF electron drive beam through the undulator and conversion target to the positron capture system. While Geant4 is foreseen for the photon-target interactions, the choice between CAIN~\cite{Yokoya}, PPS-Sim or a dedicated simulation code for the generation of the polarised photon spectra is still under discussion. The resulting positron distributions will subsequently be used as input to the existing COMSOL-ASTRA optimisation framework. This will allow a systematic investigation of the OMD design, the preaccelerator configuration and the achievable positron yield within the damping-ring acceptance~\footnote{It should be noted that, due to lack of effort, there is currently no design work on the HALHF damping rings.} for HALHF.

\subsubsection{HALHF set-up adjustments and Outlook}
\label{sec:positron-outlook}

At HALHF, the proposal is to use the fully accelerated $e^-$ beam of at least $375$~GeV as the undulator driver.
 As a starting point for further optimisation and overheads, a driver beam of 500~GeV~\cite{Ushakov:2013bm} is used to generate positrons, propagating through
an undulator with 174 m length, with a high magnetic
field of $K=2.5$ and with an undulator period of $\lambda=$43~mm. 

With such a parameter set, a polarization as high as $54\%$ might be achievable for the positron beam. 
However, due to the high K value, higher harmonics are involved, leading to a higher mean photon power and a higher energy spread, i.e.\ the $e^+$ capture might become more difficult. Furthermore, a substantially higher yield than for the ILC has to be achieved for the HALHF positron beam. In order to study these effects consistently, the full source, from photon generation in the undulator through the target and the optics-matching device to the capture section, will be simulated with the Geant4-based code PPS-Sim. As a next step, the 3D field maps of the pulsed solenoid from COMSOL will be implemented, so that the HALHF source parameters can be optimised for yield and polarization together with the OMD design.

\subsection{Polarized electrons}

\subsubsection{Electron source}
The electron source for the colliding beam should likely be a high-voltage direct-current (HVDC) gun, since these have been used to provide polarized beams at high bunch charges with sufficiently low emittances in a linear-collider setting \cite{Clendenin_2003_AIPconf}. Typically GaAs cathodes are used, achieving polarization degrees of $\geq$85\,\% and quantum efficiencies $\geq$1\,\% \cite{Wang_2024_APL}. The 16,000 1.6-nC bunches per second required by HALHF amount to an average current of 25.6\,$\mu$A, which is well within what HVDC guns have been designed to accommodate \cite{Wang_2022_PRAB}. The single-bunch charge of 1.6\,nC is also an order of magnitude below previous demonstrations \cite{Wang_2022_PRAB}, and more similar to the CLIC design \cite{Aicheler_2012_CLIC_CDR,Aicheler_2018_CLIC_PIP}. The design emittance in Ref. \cite{Wang_2022_PRAB} of 3.4\,mm$\cdot$mrad is also near the HALHF requirements of 90$\times$0.32\,(mm$\cdot$mrad)\textsuperscript{2}, provided that emittances can be exchanged between the transverse planes (discussed further below).

It is important to note that nano-Coulomb bunches were also recently obtained from a superconducting RF gun with GaAs cathodes \cite{Litvinenko_2026_NatPhys}. While this technology is still challenging, the potential advantages warrant monitoring of the developments in this direction.

\subsubsection{Injector linac}

\paragraph{High-level linac design}
The main beam energy into the PWFA linac is set to 3 GeV, and the chosen RF technology is warm copper at S-band (3\,GHz). These are, incidentally, the exact specifications of the MAX IV linac \cite{Thorin_2010_FEL,Thorin_2014_LINAC}, from which some design choices may be borrowed. Key components are the arc-like bunch compressors, which circumvent the need for harmonic RF cavities and allow the trivial generation of linearly ramped current profiles \cite{Bjorklund_Svensson_2026_EAAC}. The compressors have the opposite sign of longitudinal dispersion, $R_{56}$, to more common chicanes, which requires operation on the falling RF slope (the bunch head acquires higher energy than the tail). This causes short-range longitudinal wakefields to increase, rather than decrease, the correlated energy spread (chirp) of the bunch, allowing acceleration closer to the RF crest, even when compressing, which improves energy efficiency. This increase in usage of the available RF voltage, together with the omission of harmonic RF cavities in the X-band, is likely to reduce costs. The direction of the bunch chirp is also naturally advantageous for damping transverse instabilities via the BNS (transverse wakefield) mechanism \cite{Balakin_1983_HEACC}, which is important given the high bunch charge and current, which can drive strong short-range RF wakefields in the linac.

However, operation with HVDC guns requires additional components before the S-band linac. Because the bunch duration is normally on the 100-picosecond to nanosecond scale at particle energies in the few-hundred-keV range, additional pre-compression and acceleration are required after the source. This is typically achieved with buncher cavities and linacs operating at various sub-harmonics of the main linac RF frequency \cite{Aicheler_2012_CLIC_CDR,Willeke_2021_EIC_CDR}. The updated CLIC injector design \cite{Aicheler_2018_CLIC_PIP} foregoes the sub-harmonic systems and operates all linear accelerator RF systems at the same frequency. 

\begin{figure}[b!]
    \centering
    
    \begin{subfigure}[b]{\textwidth}
        \includegraphics[width=\textwidth,keepaspectratio]{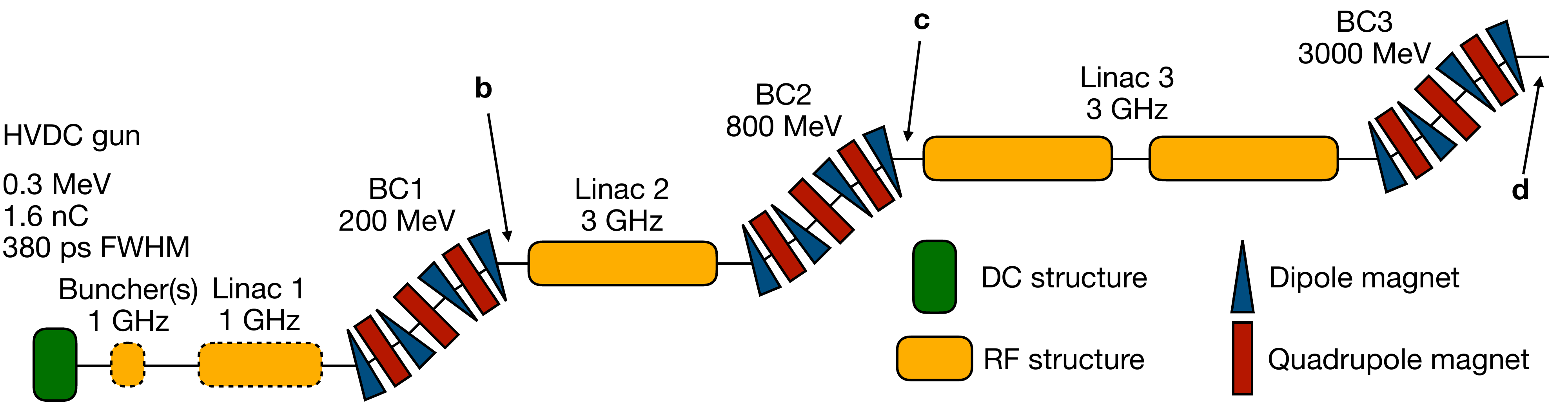}
        \caption{Schematic machine layout (not to scale).}
    \end{subfigure}
    
    \vspace{0.5em}
    
    \begin{subfigure}[b]{0.32\textwidth}
        \includegraphics[width=\linewidth,keepaspectratio]{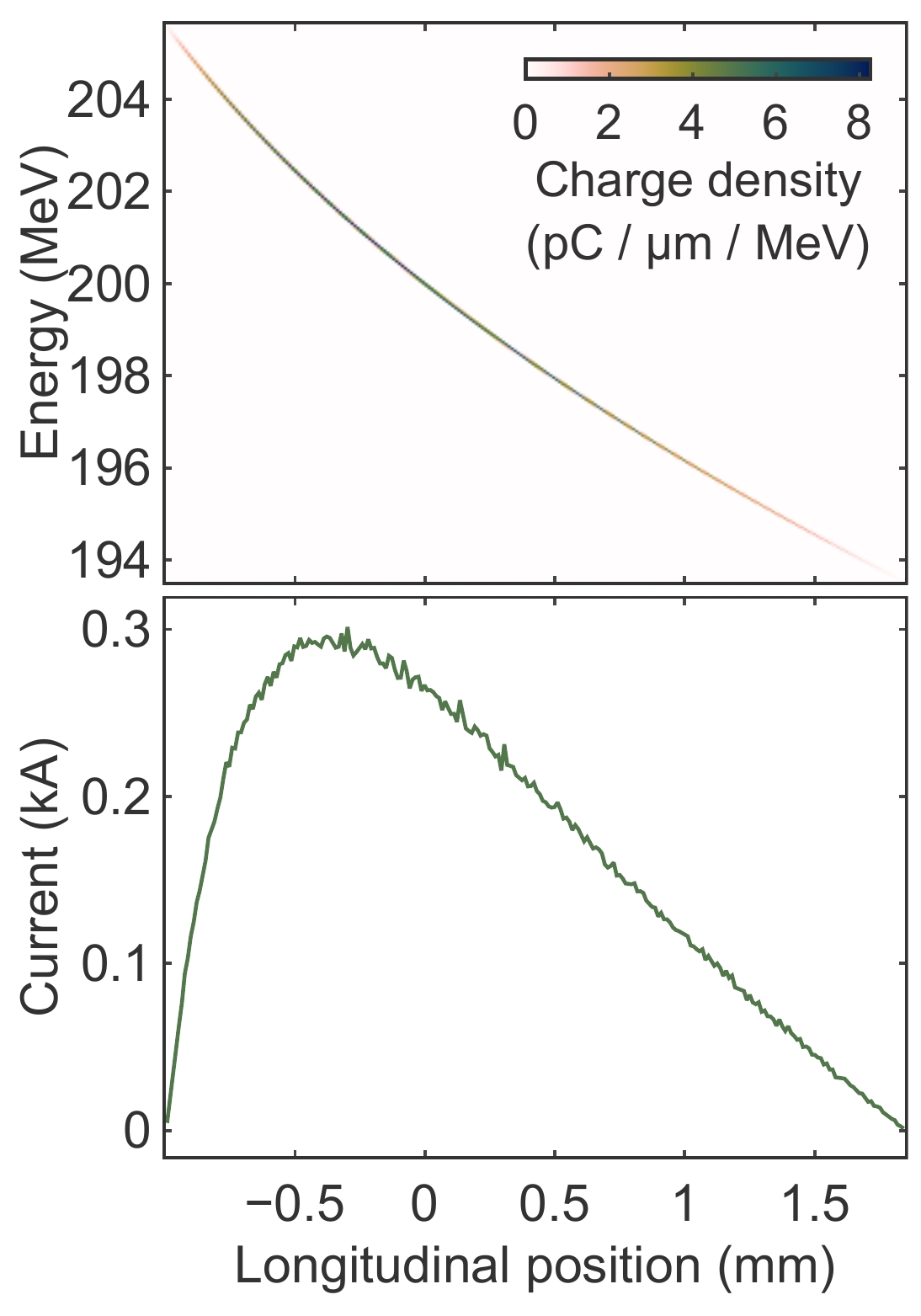}
        \caption{After BC1.}
    \end{subfigure}%
    \hfill
    \begin{subfigure}[b]{0.32\textwidth}
        \includegraphics[width=\linewidth,keepaspectratio]{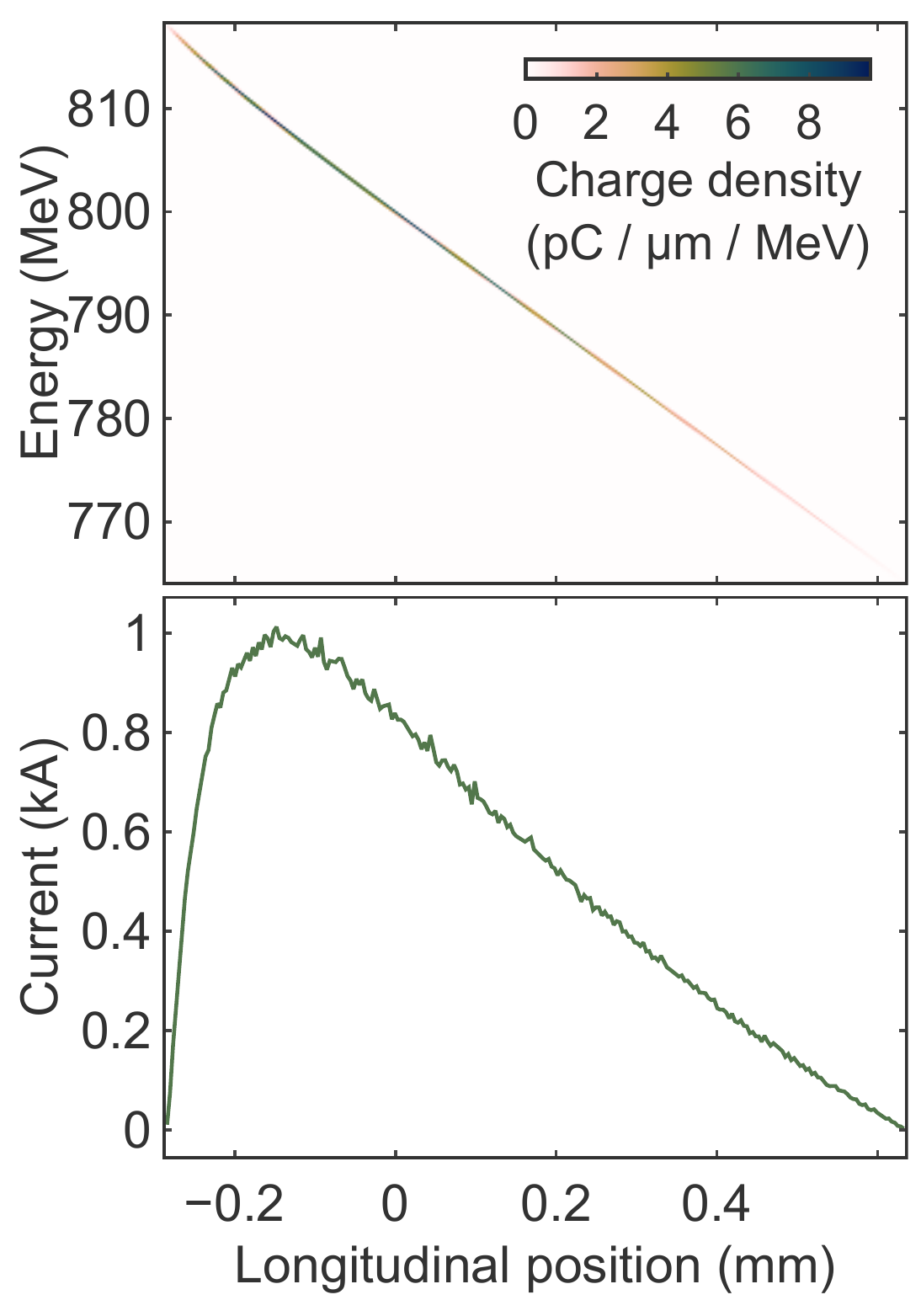}
        \caption{After BC2.}
    \end{subfigure}%
    \hfill
    \begin{subfigure}[b]{0.32\textwidth}
        \includegraphics[width=\linewidth,keepaspectratio]{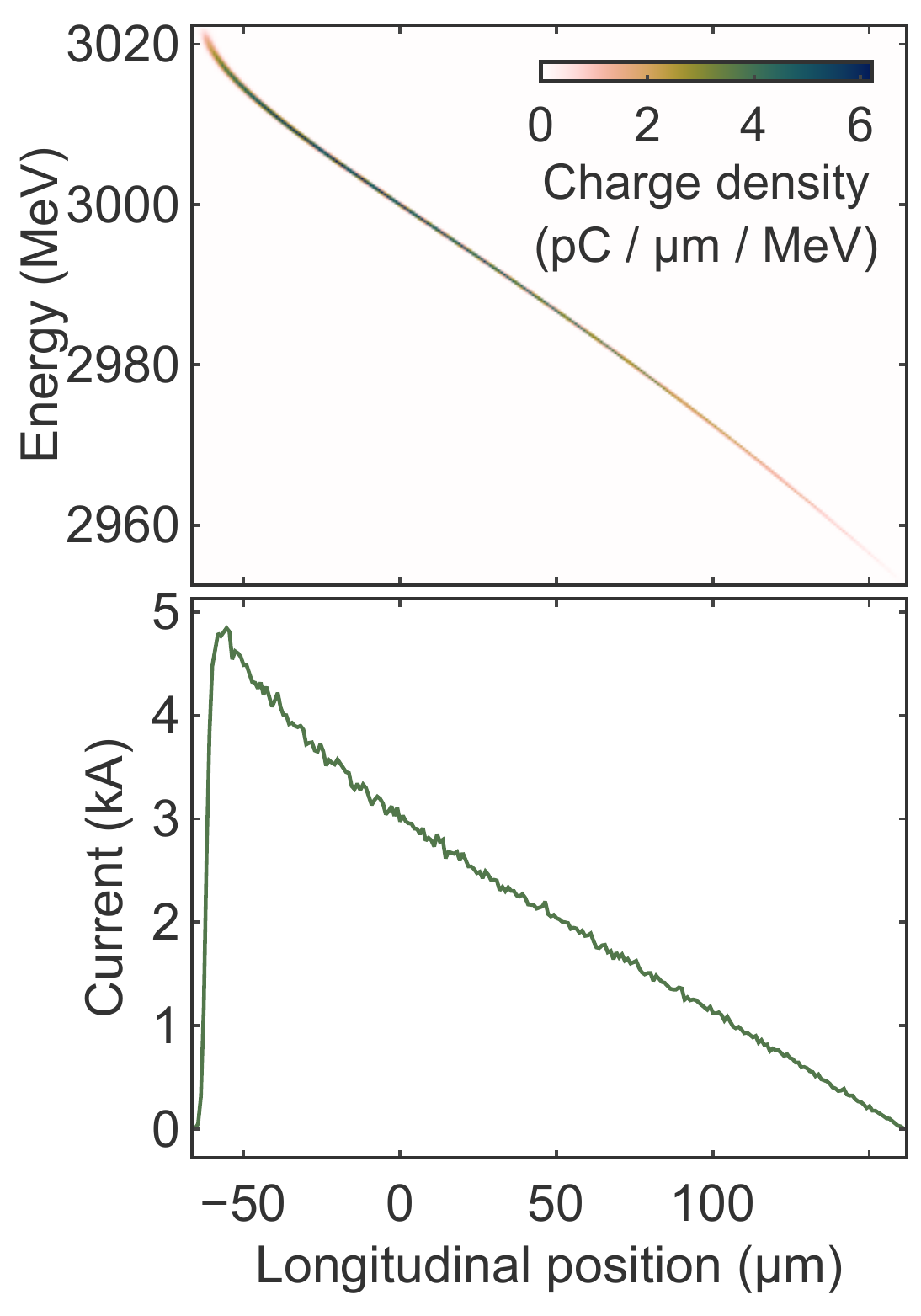}
        \caption{After BC3.}
    \end{subfigure}
    
    \caption{Main-beam linac design and simple beam dynamics.}
    \label{fig:main_beam_injector_dynamics}
\end{figure}

The baseline design for HALHF, shown in Fig. \ref{fig:main_beam_injector_dynamics}a, is therefore a 1-GHz buncher cavity (same RF frequency as for the drive-beam linac) and a 1-GHz pre-accelerator, perhaps separated by a second 1-GHz buncher, before the first bunch compressor. This all precedes the S-band linac. Velocity compression compresses the bunch longitudinally both between the buncher and the pre-accelerator and inside the pre-accelerator (or second buncher cavity). This compression must be modest, such that space-charge forces, which are significantly enhanced at non-relativistic and weakly relativistic particle energies where velocity bunching can be performed, do not degrade the beam.

Because of the short final bunch length of 40\,$\mu$m RMS and high final current of 4.8\,kA, the choice was made to use three bunch compressors. This allows for acceleration and compression in stages, reducing the maximum required energy spread at each stage and the impact of short-range wakes in the RF structures. It also offers additional stability in compression, mean energy, and arrival time. Compared to the CLIC and FCC-ee injector linacs, the bunch length at around 3\,GeV is significantly shorter for HALHF; 40\,$\mu$m RMS compared to 1.8\,mm--235\,$\mu$m RMS for CLIC and 1\,mm RMS for FCC-ee \cite{Bettoni_2025_PRAB}. The CLIC bunches are further compressed to 70\,$\mu$m RMS, but only following further acceleration to 9\,GeV, making the beam significantly stiffer. The additional compressor proposed for HALHF should therefore also allow greater mitigation of coherent synchrotron radiation (CSR).

\paragraph{Longitudinal beam dynamics}
Following the CLIC parameters of a 100 ps (assumed RMS) bunch duration at the cathode \cite{Aicheler_2018_CLIC_PIP}, which is also in line with extrapolation from, e.g., the simulations shown in Ref. \cite{Liu_2022_IJMPA}, the assumed Gaussian bunch duration at the cathode is here 160\,ps RMS (380\,ps FWHM) to keep the peak current similar to CLIC. The time-slice energy spread of the bunches out from the gun is assumed to be low (0.1\,\%).

The goal is to do the majority of the bunch current shaping in the first stage of the linac, where the energy spread is largest and higher-order optical effects are therefore strongest, and to compress more linearly in the following stages. This also allows a gradual tapering of the energy spread down approximately to the target of 0.5\,\% RMS at the final bunch compressor, which is also the end of the main-beam linac. This energy spread is manageable for focusing into the PWFA linac, when assuming that local chromaticity correction will be performed in the focusing system. As detailed further below, the arc-like compressor type can be designed to accommodate very large energy spreads without beam degradation.

A simple longitudinal beam-dynamics simulation was set up using exact trigonometric functions for the accelerating voltages, and transfer maps up to third order in energy offset for the bunch compressors. No collective effects were included at this stage (see discussion below). The results are shown in Fig. \ref{fig:main_beam_injector_dynamics}b-d. The particles were generated before linac 1, assuming a beam energy of 5\,MeV, with a parabolic distribution in time and a 20 ps FWHM duration (6-mm FWHM length) and a Gaussian distribution in energy with a spread of 0.1\,\% RMS. These post-buncher parameters were estimated based on the CLIC specifications \cite{Aicheler_2012_CLIC_CDR} and simulations presented in Ref. \cite{Liu_2022_IJMPA}.

The omission of RF buncher dynamics and collective effects is potentially significant, as these will clearly affect the overall beam dynamics, but given the flexibility of the compression system, accounting for them in a future iteration of the linac design is not foreseen to pose a major hurdle. Furthermore, optimising the bunch-current profile of the main-beam bunches is not strictly necessary, as the longitudinally self-stabilizing mechanism in the PWFA linac will also self-optimise the current profile. However, having an incoming current profile that is closer to nominal (ramped) than a Gaussian will lead to smaller energy offsets in the first stages of the PWFA linac, and small differences will self-correct more quickly than with a Gaussian starting point. This adds credibility to the assumption that future design iterations will be able to produce sufficiently optimised current profiles within existing constraints.

\paragraph{Transverse beam dynamics}
The arc-like bunch compressors chosen for the linac have significantly increased optical complexity compared to standard chicanes, but, in addition to delivering longitudinal properties advantageous for this implementation, this complexity also offers the opportunity to mitigate collective effects such as CSR. Analytical methods have been developed for the near-full suppression of CSR-induced emittance growth in double-bend achromats (the simplest substructure of arc-like bunch compressors) \cite{Zhang_2023_PRAB}; simulations have shown that these compressors can also be optically achromatic within a large range of energy offsets \cite{Bjorklund_Svensson_2019_PRAB}.

The target specifications also detail asymmetric transverse emittances of 90$\times$0.32\,(mm$\cdot$mrad)\textsuperscript{2}. Such ``flat" beams are usually achieved via damping rings, which incur a high additional cost and complexity. However, alternative methods based on linear lattices have been proposed \cite{Brinkmann_2001_PRSTAB}, and emittance ratios of approaching 200 (with absolute emittances in the range [2,1]\,mm$\cdot$mrad and [100,175]\,mm$\cdot$mrad in the two transverse planes) have been experimentally demonstrated using 1 nC bunches produced with 450 fs FWHM laser pulse durations, albeit in an RF photoinjector \cite{Xu_2022_PRAB}. These parameters are close to what is necessary for HALHF, and would only add a few meters of additional length to the injector, although it is not yet clear if the technique is compatible with HVDC guns and polarized bunches.


\section{Beam-delivery system}

\begin{figure}[!ht]
    \centering
    \includegraphics[width=\linewidth]
    {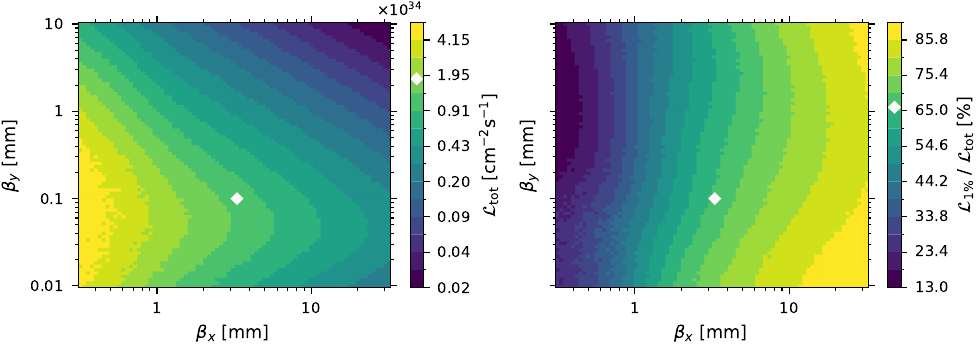}
    \caption{Total and peak luminosities versus horizontal and vertical beta functions at the IP with the nominal HALHF parameters at 250~GeV and assuming ideal Gaussian beams. Simulations are performed with GuineaPig.}
    \label{fig:lumis}
\end{figure}

The HALHF Beam-Delivery System (BDS) has to focus the beams at the IP to provide the desired luminosity. Figure~\ref{fig:lumis} shows total and peak luminosities versus horizontal and vertical beta functions at the IP with the nominal HALHF parameters assuming ideal Gaussian beams. Some reduction of the vertical IP beta function could still provide a gain for both total and peak luminosities.
The design horizontal beta function of 3.3~mm is about a factor 3 lower than in typical linear collider Final Focus Systems (FFS). This suggests to explore configurations with lower horizontal emittance and larger horizontal beta function. 

The BDS must transport and focus beams with normalized emittances substantially larger than those assumed in the CLIC or ILC designs. In particular, the large horizontal emittance increases the beam size throughout the FFS, while the large phase-space volume enhances the sensitivity to nonlinear geometric and chromatic aberrations. These effects can limit the achievable beam size at the interaction point (IP) and ultimately reduce the luminosity. Synchrotron-radiation-induced emittance growth in the bending and final-focus magnets provides an additional limitation, particularly as the electron-beam energy is increased.

The CLIC FFS was therefore adopted as a starting point because it provides an established high-energy linear-collider optics design. However, direct evaluation with the HALHF beam parameters showed substantial beam-size growth due to higher-order geometric aberrations. Three FFS configurations employing different chromaticity-correction schemes were consequently analysed: a Traditional FFS with dedicated chromatic-correction sections and two Local FFS based on local chromaticity correction; a FFS developed for the CERN Linear Collider Facility (LCF) study 
and a FFS originally developed for the Next Linear Collider (NLC) at SLAC.

\subsection{Traditional FFS}

Earlier studies~\cite{IPAC_Traditional_FFS} investigated the use of a CLIC-based FFS with the large emittances relevant to HALHF, showing that the original compact CLIC optics suffered substantial higher-order beam-size growth and that improved performance could be obtained by adopting a longer Traditional FFS, as used in~\cite{CLIC_Traditional}, with dedicated chromatic-correction sections. The study also included mitigation of synchrotron-radiation-induced growth in the final doublet, as done e.g. in~\cite{blanco}, and evaluated the nonlinear aberrations up to fifth order. These results established the reference Traditional configuration used as the starting point for the additional optimization presented here using MAPCLASS~\cite{mapclass_prab, mapclass2}.

The reference Traditional FFS has a total length of approximately
\(L_{\mathrm{FFS}}=1500\)~m and \(L^*=2.2\)~m. At the entrance, the beta functions are
\(\beta_x=28.0\)~m and \(\beta_y=10.0\)~m, with \(D_x=0\). For a 375~GeV electron beam and the normalized emittances used in~\cite{IPAC_Traditional_FFS},
\(\epsilon_{n,x}/\epsilon_{n,y}=30\,000/200\)~nm\,rad, this configuration gives a PLACET beam size of approximately
\(710.5\times17.4\)~nm with synchrotron radiation enabled.

\begin{figure}[!ht]
    \centering
    \includegraphics[width=\linewidth]
    {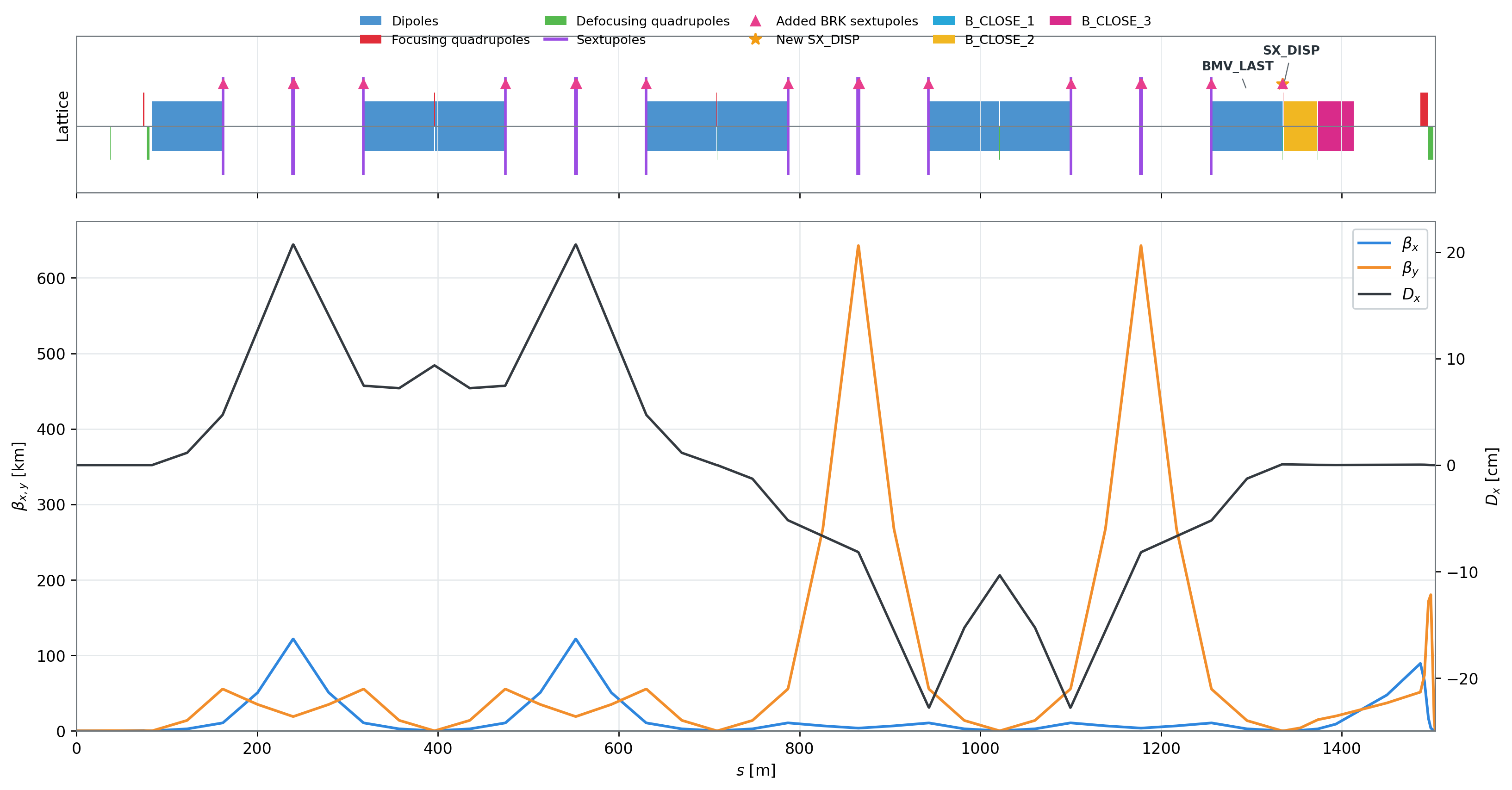}
    \caption{Linear optics of the optimised Traditional FFS for a 375~GeV electron beam. The lattice layout is shown together with the horizontal and vertical beta functions and the horizontal dispersion.}
    \label{fig:traditional_optics}
\end{figure}

The Traditional FFS was further optimized using MAPCLASS by retuning the sextupole system and introducing additional thin sextupole components at the existing sextupole locations. The corresponding optimized linear optics for a 375~GeV electron beam are shown in Fig.~\ref{fig:traditional_optics}, with
\(\beta_x^*/\beta_y^*=8.8/0.09\)~mm.

This optimised configuration reduces the PLACET beam size to
\(662.5\times7.93\)~nm. The largest improvement occurs in the vertical plane, where \(\sigma_y\) decreases by approximately 54\%, while \(\sigma_x\) decreases by approximately 7\%.

The optimised Traditional FFS was also evaluated with different normalized emittances. It remains viable with
\(\epsilon_{n,x}=45\,000\)~nm\,rad, corresponding to 50\% of the HALHF reference horizontal emittance of \(90\,000\)~nm\,rad, while retaining the nominal HALHF vertical emittance
\(\epsilon_{n,y}=320\)~nm\,rad. For this emittance configuration, the calculated beam size at 375~GeV is approximately
\(812.5\times11.9\)~nm.

For comparison, the HALHF reference emittances
\(\epsilon_{n,x}/\epsilon_{n,y}=90\,000/320\)~nm\,rad give a beam size of approximately \(1237.0\times18.3\)~nm in the same Traditional lattice.

\subsection{CLIC-style Compact Local FFS}

The Local FFS considered here was developed and optimised independently within the CERN Linear Collider Facility (LCF) study; it was not derived from the HALHF Traditional FFS. It was selected as an alternative starting point because its local chromaticity-correction scheme provides a substantially more compact final-focus system and therefore offers a useful comparison between achievable beam size and total system length. Its shorter layout may also reduce the accumulation of synchrotron-radiation effects in the bending system at high energy.

\begin{figure}[!ht]
    \centering
    \includegraphics[width=\linewidth]
    {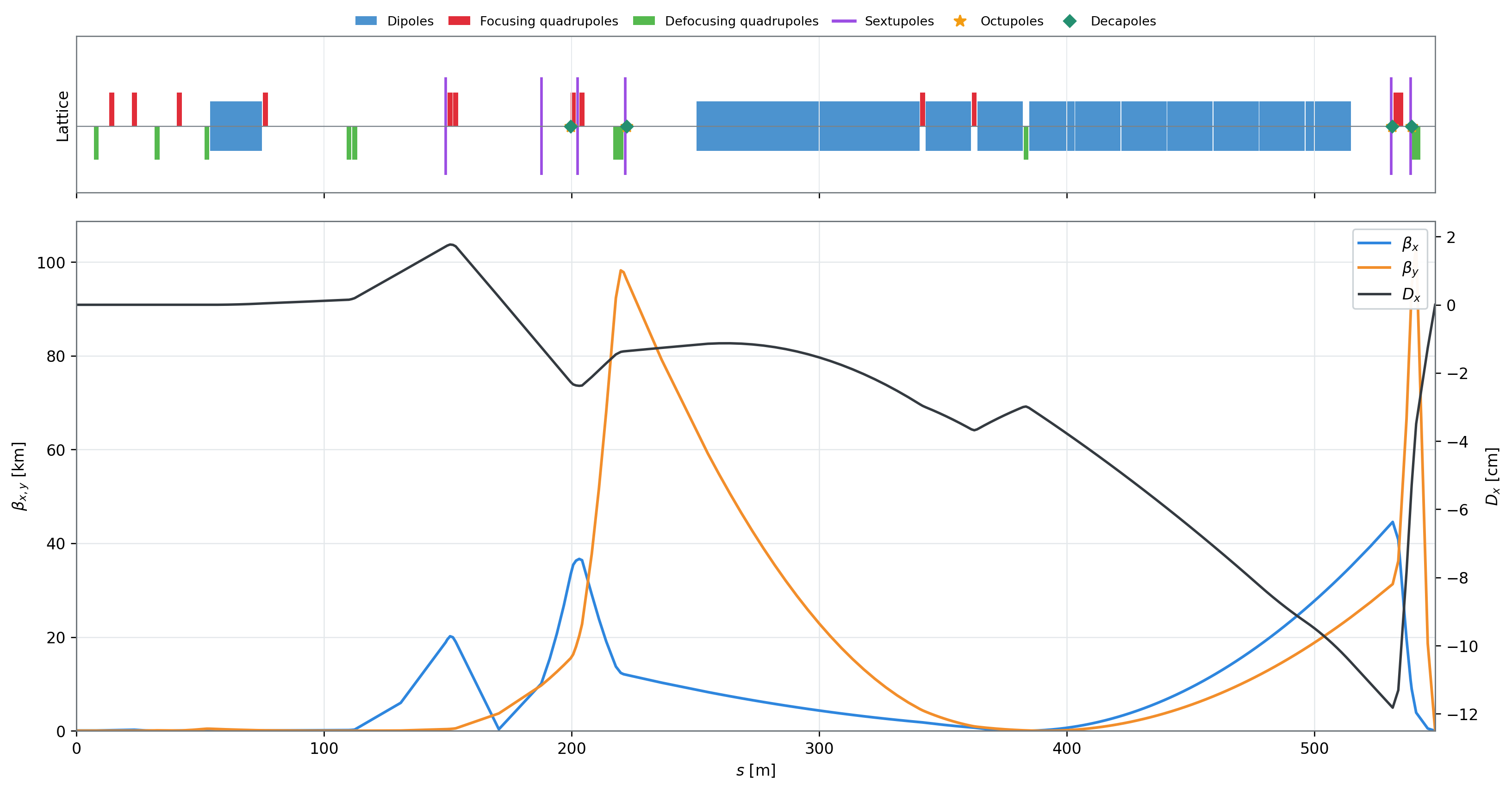}
    \caption{Linear optics of the optimised Local FFS for a 375~GeV electron beam. The lattice layout, horizontal and vertical beta functions, and horizontal dispersion are shown.}
    \label{fig:local_optics}
\end{figure}

The Local FFS has a total length of approximately 549~m, roughly one third of the length of the Traditional FFS, with \(L^*=6\)~m, as used in~\cite{andry,fabien,lewis,keegan}. At the entrance of the line, the beta functions are
\(\beta_x=66.15\)~m and \(\beta_y=17.92\)~m, with \(D_x=0\).
For the optimised lattice, the IP beta functions are
\(\beta_x^*=11\)~mm and \(\beta_y^*=0.48\)~mm. The corresponding linear optics for a 375~GeV electron beam are shown in Fig.~\ref{fig:local_optics}.

For a direct comparison with the Traditional FFS, the beam-size calculations use the same reference normalized emittances adopted in the IPAC study,
\(\epsilon_{n,x}/\epsilon_{n,y}=30\,000/200\)~nm\,rad. At 375~GeV, the original Local configuration gives a PLACET beam size of approximately
\(1174.3\times15.98\)~nm with synchrotron radiation enabled. After optimization, the beam size is reduced to approximately
\(1107.6\times15.91\)~nm. The optimization includes retuning of the sextupole, octupole and decapole components. It mainly improves the horizontal beam size, which decreases by approximately 5.7\%, while the vertical beam size remains nearly unchanged.

\subsection{NLC-style Compact Local FFS}

A third option considered here is a compact local FFS based on the design of the FFS for the Next Linear Collider (NLC) at SLAC for an electron beam energy of 500 GeV. This lattice provides the local chromaticity correction scheme, offering a 
compact FFS capable of delivering small beam sizes at high energies. It employs a similar layout to the Compact Local FFS described in the previous section, with bends occupying a larger 
fraction of the total FFS length and additional nonlinear optics to mitigate higher-order chromatic and geometric aberrations. To accommodate the different reference beam energy, all magnetic fields and gradients were scaled by the momentum ratio of the new energy to the 500 GeV reference energy. 
This procedure ensures preservation of the normalized magnet strengths and linear optics of the original design.

\begin{figure}[!ht]
    \centering
    \includegraphics[width=\linewidth]
    {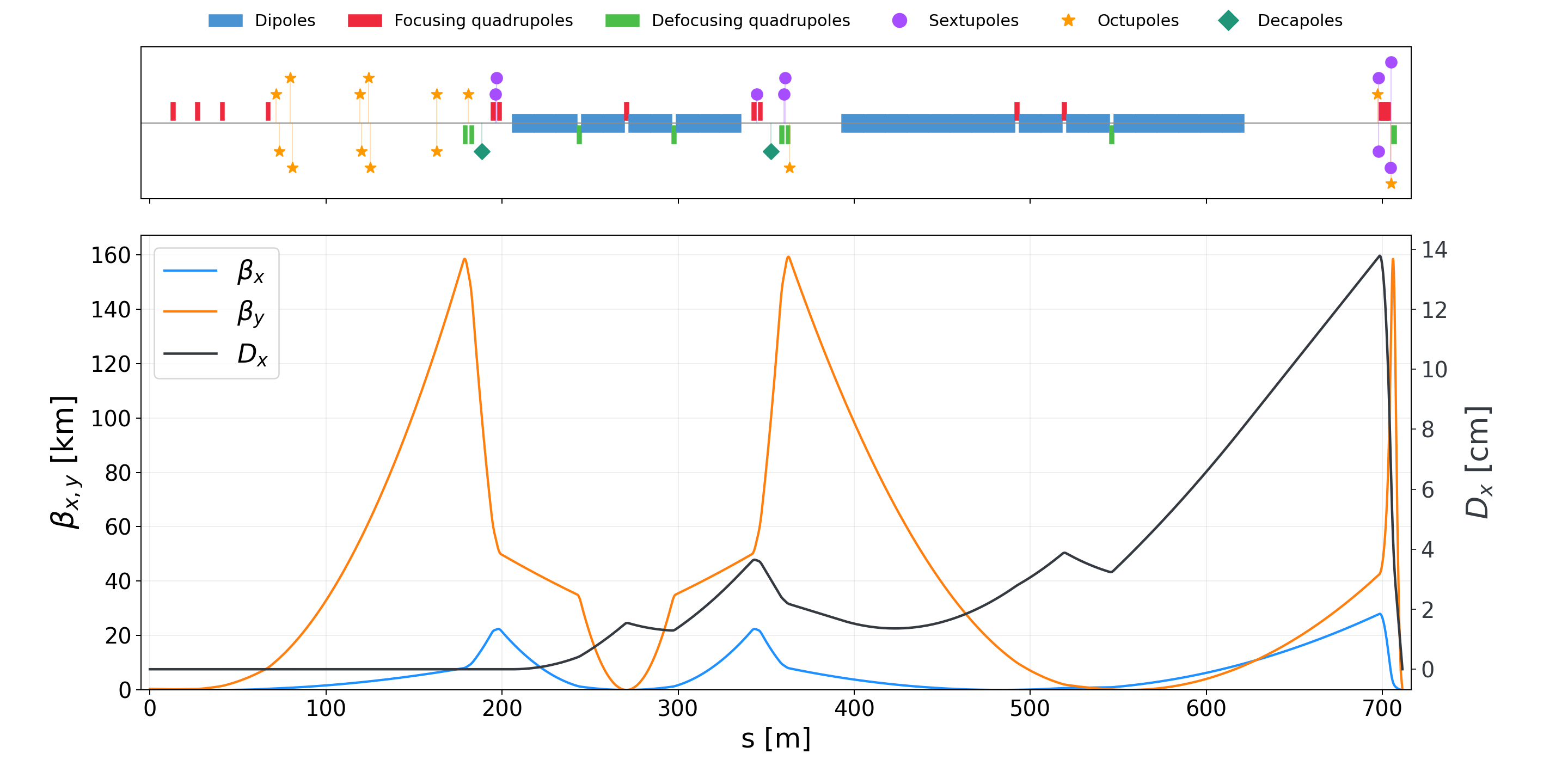}
    \caption{Linear optics of the NLC Local FFS for a 375~GeV electron beam. The lattice layout, horizontal and vertical beta functions, and horizontal dispersion are shown.}
    \label{fig:local_optics_nlc}
\end{figure}

The NLC local FFS has a total length of 711~m, approximately one 
half of the length of the Traditional FFS, with \(L^*=3.8\)~m. At the entrance of the FFS, the beta functions are \(\beta_x=60.0\)~m and \(\beta_y=391.9\)~m, with \(D_x=0\). At the IP, the NLC FFS gives \(\beta_x^*=13.33\)~mm and \(\beta_y^*=0.133\)~mm. The linear optics and survey for the local NLC FFS for a 375 GeV electron beam are presented in Figure~\ref{fig:local_optics_nlc}.

To compare with the Traditional and Local FFS, beam-size calculations were performed with the same normalized emittances, \(\epsilon_{n,x}/\epsilon_{n,y}=30\,000/200\)~nm\,rad. After scaling the magnet gradients down to accommodate the 375 GeV beam energy, the NLC local FFS produces a beam size of \(851.77\times11.17\)~nm with synchrotron radiation effects included. Apart from the scaling of the magnet gradients to 
account for the difference in beam rigidity, no further optimization was performed on the NLC FFS for this study. 

\subsection{Performance comparison at all energies}

Table~\ref{tab:ffs_beam_sizes} summarizes the main parameters for the different FFS configurations, including the PLACET or Xtrack beam sizes obtained with synchrotron radiation enabled. To isolate the dependence on the FFS configuration and beam energy, the normalized emittances are kept fixed as in~\cite{IPAC_Traditional_FFS},
\(\epsilon_{n,x}/\epsilon_{n,y}=30\,000/200\)~nm\,rad, throughout the energy comparison.

\begin{table}[htbp]
\centering
\small
\setlength{\tabcolsep}{4pt}
\scalebox{0.705}{
\begin{tabular}{lccccccc}
\toprule
FFS & $L^*$ [m] & FFS length [m]
& $\beta_x^*/\beta_y^*$ [mm]
& Energy [GeV] & $\sigma_{x,0}/\sigma_{y,0}$ [nm] & $\sigma_x^*/\sigma_y^*$ [nm]
& $(\sigma_{x}^{*}/\sigma_{x,0})/(\sigma_{y}^{*}/\sigma_{y,0})$ \\
\midrule
\multirow{3}{*}{Traditional}
& \multirow{3}{*}{2.2}
& \multirow{3}{*}{1500}
& \multirow{3}{*}{$8.8/0.09$}
& 375 & $599.79/4.953$ & $662.5/7.93$ & $1.105/1.601$ \\
& & & & 570 & $486.49/4.017$ & $856/13.13$ & $1.760/3.269$ \\
& & & & 825 & $404.38/3.339$ & $1770/26.28$ & $4.377/7.871$ \\
\midrule
\multirow{3}{*}{Local (NLC-style)}
& \multirow{3}{*}{3.8}
& \multirow{3}{*}{711}
& \multirow{3}{*}{$13.3/0.13$}
& 375 & $738.24/6.030$ & $851.77/11.17$ & $1.154/1.853$ \\
& & & & 570 & $598.79/4.891$ & $645.27/10.74$ & $1.078/2.196$ \\
& & & & 825 & $497.72/4.066$ & $553.09/21.88$ & $1.111/5.383$ \\
\midrule
\multirow{3}{*}{Local}
& \multirow{3}{*}{6.0}
& \multirow{3}{*}{549}
& \multirow{3}{*}{$11/0.48$}
& 375 & $670.58/11.437$ & $1107.6/15.91$ & $1.652/1.391$ \\
& & & & 570 & $543.91/9.277$ & $795/12.96$ & $1.462/1.397$ \\
& & & & 825 & $452.11/7.711$ & $637/15.07$ & $1.409/1.954$ \\
\bottomrule
\end{tabular}}
\caption{Parameter comparison for the optimised Traditional, Local, and NLC-style Local FFS at beam energies of 375 GeV, 570 GeV, and 825 GeV. The $\sigma_{x,0}/\sigma_{y,0}$ entries represent the ideal (lowest order) beam spot sizes based on the IP beta functions and emittances, while $\sigma_{x}^{*}/\sigma_{y}^{*}$ are the IP beam sizes from tracking with PLACET (Traditional and Local) or Xsuite's Xtrack package (NLC-style) with synchrotron radiation enabled. The normalized 
emittances are \(\epsilon_{n,x}/\epsilon_{n,y}=30\,000/200\)~nm\,rad.}
\label{tab:ffs_beam_sizes}
\end{table}

\begin{figure}[!ht]
    \centering
    \includegraphics[width=\linewidth]
    {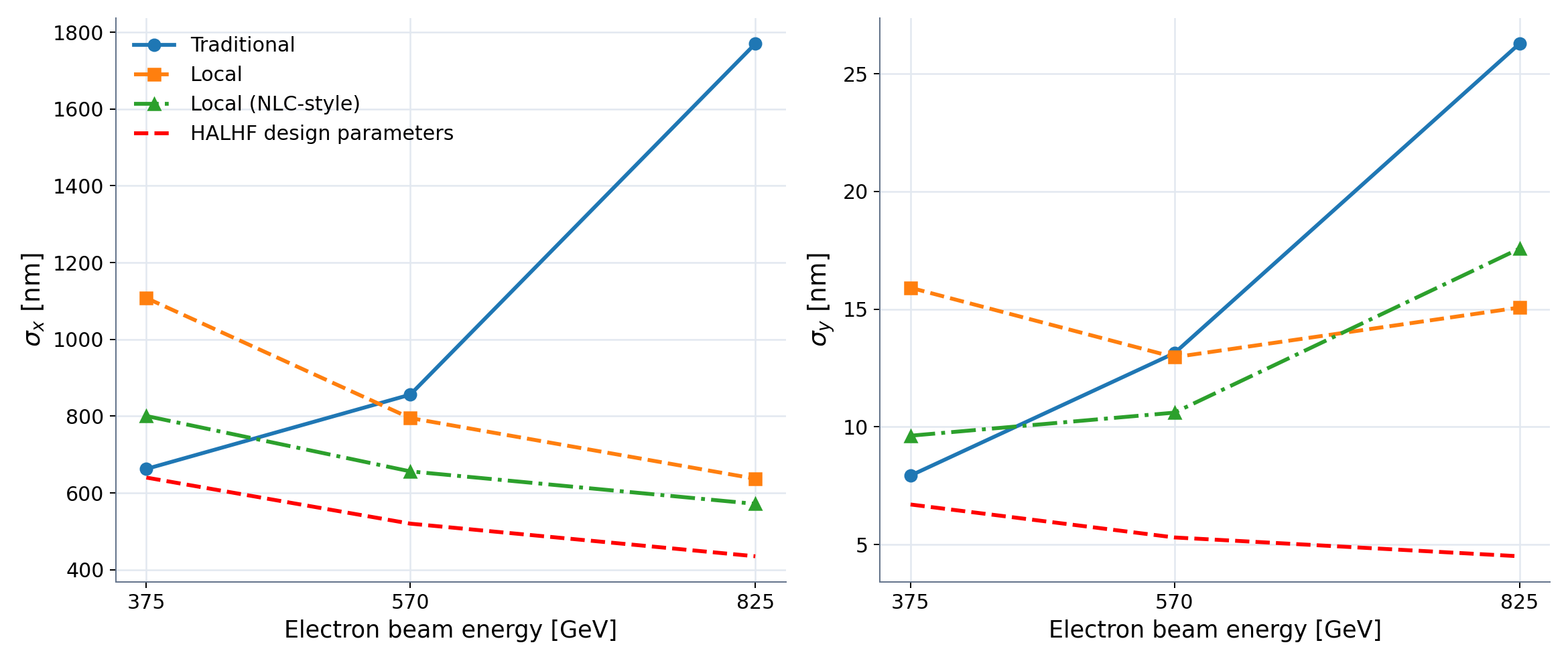}
    \caption{RMS beam sizes as a function of the electron-beam energy for the optimised Traditional and Local FFS schemes, and the NLC-style scheme, including synchrotron radiation. Beam sizes are calculated from PLACET for the Traditional and Local schemes, while for the NLC-style scheme the beam sizes are calculated from Xsuite's Xtrack package.
    All three FFS results use the reference normalized emittances adopted in the IPAC study, $\epsilon_{n,x}/\epsilon_{n,y}=30,000/200$~nm~rad. The dashed red curves show the HALHF design beam sizes, corresponding to the HALHF normalized emittances $\epsilon_{n,x}/\epsilon_{n,y}=90,000/320$~nm~rad.}
    \label{fig:ffs_energy_scaling}
\end{figure}

The energy dependence is shown in Fig.~\ref{fig:ffs_energy_scaling}. At 375~GeV, the Traditional FFS provides the smallest horizontal and vertical beam sizes. Its performance, however, deteriorates rapidly as the beam energy increases because of synchrotron-radiation-induced beam-size growth in the FFS.

\begin{figure}[!ht]
    \centering
    \includegraphics[width=\linewidth]
    {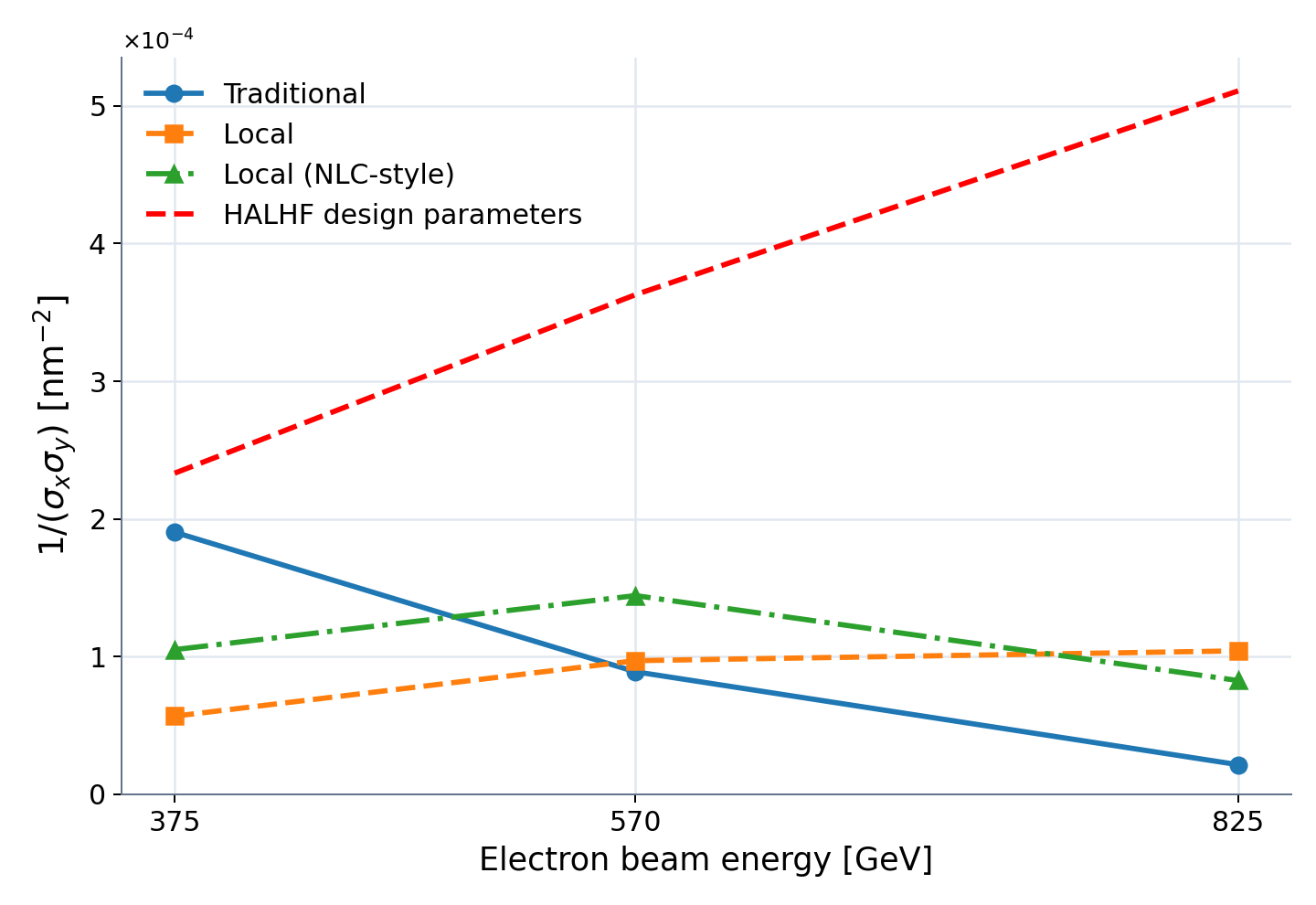}
    \caption{Figure of merit (\(1/\sigma_{x}^{*}\sigma_{y}^{*}\)) as a function of the electron-beam energy for the optimised Traditional, Local, and NLC-style FFS schemes. The beam sizes for the Traditional and Local FFS are calculated in PLACET, while the NLC-style lattice values are calculated using Xsuite's Xtrack package. Synchrotron radiation effects are included. The Traditional, Local, and NLC-style FFS results use the reference normalized emittances adopted in the IPAC study, $\epsilon_{n,x}/\epsilon_{n,y}=30,000/200$~nm~rad. The dashed red curves show the HALHF design beam sizes, corresponding to the HALHF normalized emittances $\epsilon_{n,x}/\epsilon_{n,y}=90,000/320$~nm~rad.}
    \label{fig:ffs_figure_of_merit}
\end{figure}

At 570~GeV, the NLC FFS has the best performance, while the Local FFS becomes competitive with the Traditional scheme. At 825~GeV, both the Local FFS and NLC FFS perform substantially better than the traditional, giving beam sizes of approximately \(637\times15.07\)~nm and \(553\times22.08\)~nm, respectively, compared with
\(1770\times26.28\)~nm for the Traditional FFS. An approximate figure of merit for the performance of the three FFS configurations is given by the inverse product of the tracked beam sizes, \(1/\sigma_{x}^{*}\sigma_{y}^{*}\), 
which is shown in figure~\ref{fig:ffs_figure_of_merit}.
The performance at 825~GeV remains at about factor 5 below than the design.

The comparison reveals a clear trade-off between low-energy performance, compactness, and robustness against synchrotron-radiation-induced beam-size growth. The optimised Traditional FFS provides the smallest beam sizes at 375~GeV, but its approximately 1.5~km length and strong degradation with increasing energy make it unsuitable as a general solution over the full HALHF energy range. The CLIC-style Local FFS is the most compact configuration considered, with a length of approximately 549~m, and performs substantially better than the Traditional scheme at high energy. Its relatively large \(L^*=6\)~m and its horizontal beam size at 375~GeV, however, indicate that further optimisation is required.

The NLC-style FFS provides the most promising overall starting point for a future HALHF baseline. With a length of approximately 711~m and \(L^*=3.8\)~m, it combines a substantially shorter lattice than the Traditional FFS with good performance over the investigated energy range. At 570~GeV, it gives the smallest tracked horizontal and vertical beam sizes among the three configurations. At 825~GeV, it provides a smaller horizontal beam size than the CLIC-style Local FFS, although its larger vertical beam size shows that the nonlinear correction remains to be improved. Importantly, these results were obtained after scaling the magnet strengths to the HALHF energy, without a dedicated re-optimisation of the NLC lattice for the HALHF beam parameters.

More generally, the nominal HALHF combination of
\(\epsilon_{n,x}=90\,000\)~nm\,rad and
\(\beta_x^*=3.3\)~mm places exceptionally demanding requirements on the horizontal focusing and chromatic correction. The results presented here indicate that a more robust design could be obtained through a joint optimisation of the accelerator emittance and the interaction-region optics. In particular, a reduced horizontal emittance would allow the use of a moderately larger \(\beta_x^*\), relaxing the nonlinear constraints on the FFS while retaining an acceptable horizontal spot size and luminosity. Similarly, a \(L^*\) shorter than the 6~m adopted for the CLIC-style Local scheme appears desirable, provided that it remains compatible with the detector and machine--detector-interface requirements.

The next design iteration could therefore concentrate on the NLC-style FFS, including a dedicated optimisation of \(\beta_x^*\) and \(\beta_y^*\), probably increasing the total FFS  and Final Doublet lengths to mitigate synchrotron radiation effects at the higher energy, followed by retuning  the sextupole and higher-order corrections.  A final baseline decision will also require luminosity calculations, detector-background studies, realistic magnet constraints, and tolerance  studies. 

The feasibility of achieving the 10~nm vertical beam sizes in the presence of realistic imperfections  should be studied in the future as experience in various high-energy machines indicates that this is a major experimental challenge~\cite{Vsize}.

\subsection{Ideas for IR and MDI further optimization}

The geometry of the interaction region has not been discussed in this paper, as indeed, the configuration largely depends on the beam parameters, such as emittances, which are still converging, coupled to optimisation of detector and the handling of beam-beam and pairs background. 

The first key parameter for determining the IR geometry is the divergence of the incoming beam, which for 250 GeV CM energy is given by Table~\ref{tab:1} as $\theta_{x}^{*}= \sigma_{x}^{*}/\beta_{x}^{*}$ = 636 nm/3.3 mm = 0.193 mrad; and $\theta_{y}^{*}= \sigma_{y}^{*}/\beta_{y}^{*}$ = 6.6 nm/0.1 mm = 0.066 mrad. These values are not dissimilar to divergences of earlier linear-collider designs, which indicate that similar IR configurations should be possible. For example, assuming that the collimation depth in the horizontal plane should be not less than ten sigmas, the minimum crossing angle should be around 8 mrad (where the final doublet coils and cryostat thickness are assumed to be about the same size as its clear aperture). 

It seems likely, however, that the asymmetrical design of HALHF opens up additional opportunities for optimisation of the Interaction-Region geometry and Machine-Detector Interface. In particular, at this stage both head-on and crossing-angle IR configurations can be considered.  

The detailed choice between selected configurations will be determined based on consideration of detector performance, background handling, collimation system, machine protection, and potential upgradability to TeV and multi-TeV energies. Notably, one of the multi-TeV gamma-gamma collider schemes \cite{gg-XFEL}, where an X-ray FEL is used for the effective conversion of particles to photons, requires a very small crossing angle of around a milliradian, which may be more compatible with HALHF tunnels optimised for head-on collisions. 

\begin{figure}[!ht]
    \centering
    \includegraphics[width=\linewidth]
    {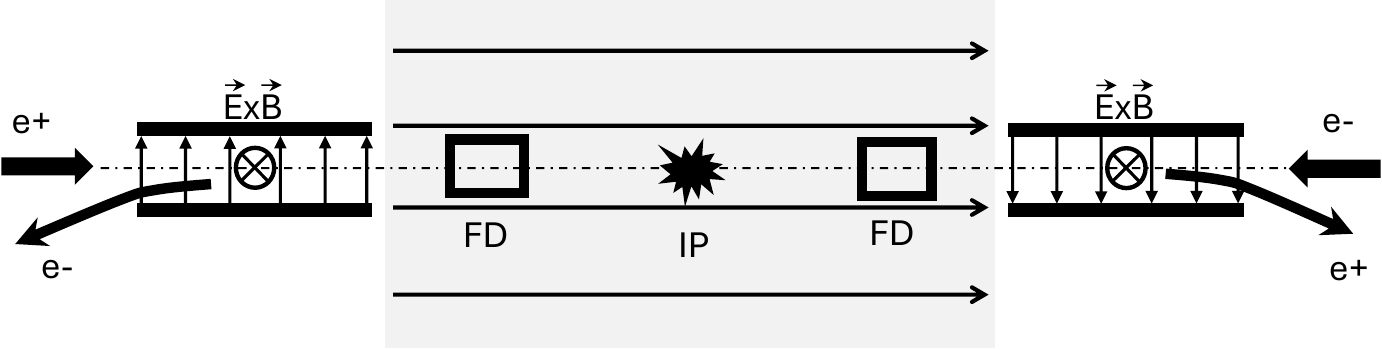}
    \caption{Head-on collisions can be considered as an option for HALHF MDI.}
    \label{fig:head_on}
\end{figure}

The head-on IR configuration was considered for the TESLA collider proposal \cite{TESLA-CDR}, where beam separation is achieved by electrostatic separators with crossed electric and magnetic field, using electrostatic-separator technology similar to that developed for LEP \cite{LEP-separator}. In this case, the effect of the electric and magnetic fields cancels for the incoming beams, while it doubles for the outgoing beams, deflecting them into an extraction beamline after IP collisions (see Fig.~\ref{fig:head_on}).

The head-on collision scheme was eventually abandoned for TESLA, due to beamstrahlung SR losses on a septum blade \cite{TESLA-septum}; however this shortcoming is not fundamental and can be reoptimised.

\begin{figure}[!ht]
    \centering
    \includegraphics[width=0.49\linewidth]{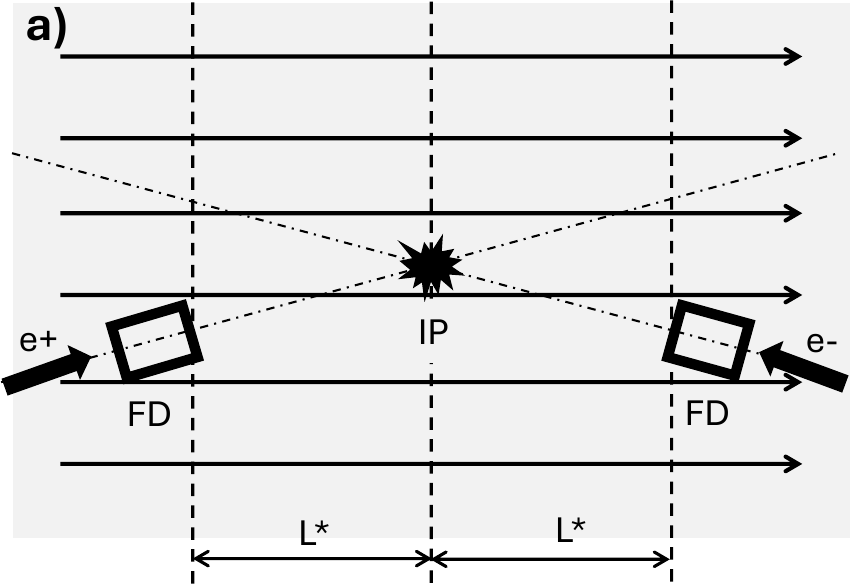}
    \includegraphics[width=0.49\linewidth]{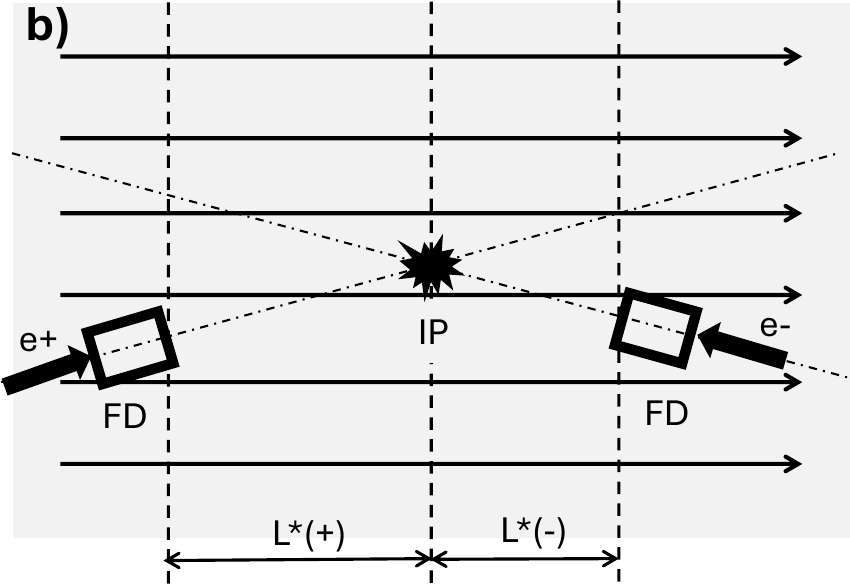}
    \includegraphics[width=0.49\linewidth]{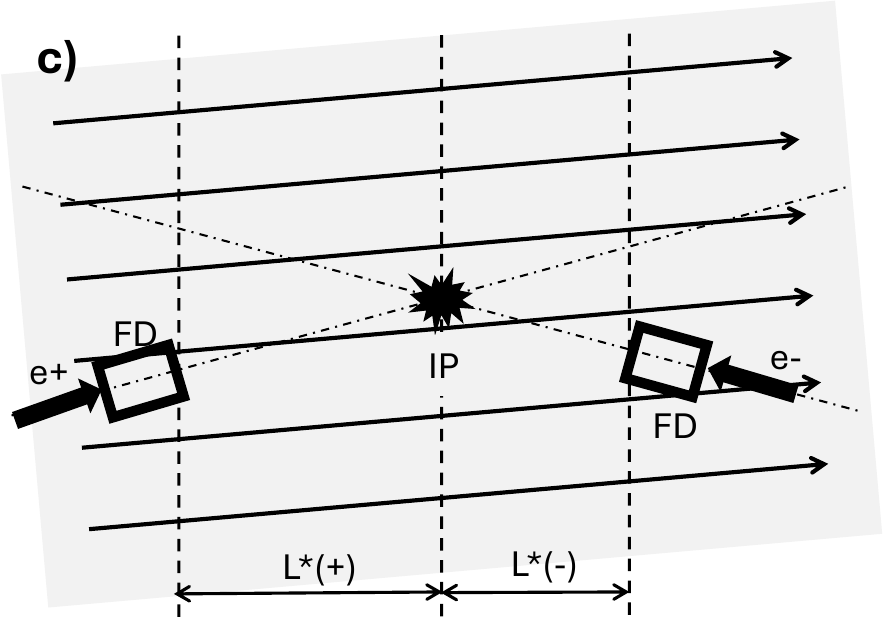}
    \includegraphics[width=0.49\linewidth]{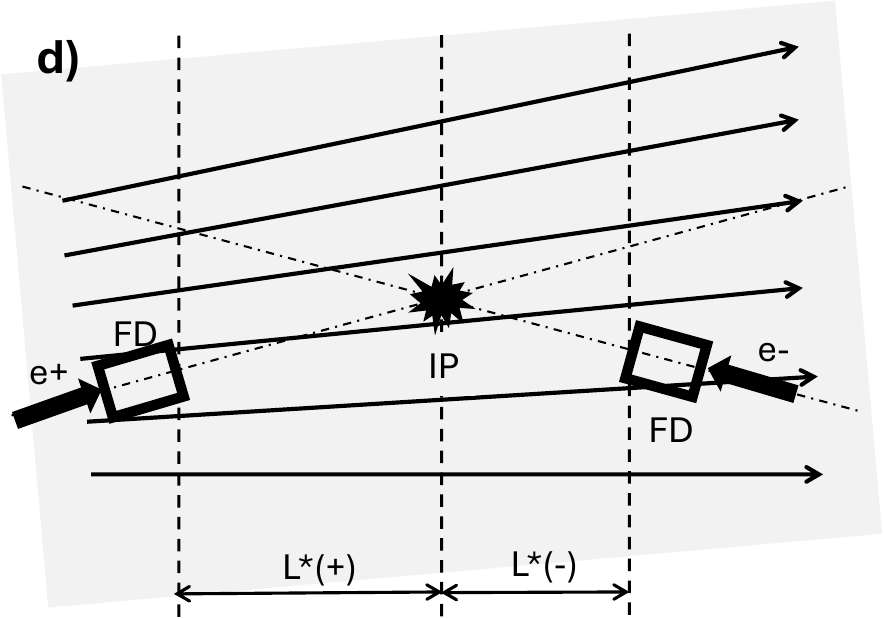}
    \caption{Optimization options for HALHF MDI with crossing-angle geometries: a) symmetrical case; b) asymmetry of $L^*$ added; c) asymmetry of crossing angle added; d) tapering of detector-solenoid field added.}
    \label{fig:crossing_options}
\end{figure}

The crossing angle IR geometry for the HALHF Machine-Detector Interface (MDI) will also be considered. 
In this case (as well as for the head-on case), there are additional optimisation opportunities due to the asymmetry of HALHF beam energies, as illustrated in Fig.~\ref{fig:crossing_options}.

In this optimisation, the $L^*$ parameter of the FFS may be allowed to be different for positrons and electrons (Fig.~\ref{fig:crossing_options}, a). In this case, one of the FFS will have lower chromaticity, and its performance may be relatively improved, in terms of handling a large-emittance beam or synchrotron radiation. 

The next step of optimisation may allow asymmetry of the crossing-angle configuration with respect to the axis of the detector-solenoid field. Two optimisation effects may be of relevance here: first, one of the beams will be less impacted by crossing the fringe field of the detector solenoid, easing compensation of the effect of this field on the beam, and secondly, background pairs produced by this beam will see lower transverse solenoidal field, possibly easing the handling of this background. 

The ultimate step of possible optimisation may involve adding to the above-mentioned asymmetries an asymmetry of the detector solenoid field, namely tapering of its field value along the axis, which may be achieved by gradual change of the winding density on the cylindrical body of the detector solenoid. In this case, for one of the beams, the incoming particles will see a lower solenoid field, while the background pairs produced at IP will see a larger solenoid field. 

Obviously, a more trivial asymmetry would be, while keeping $\sigma_x^*$ and $\sigma_y^*$ at IP the same, to allow beams to have different geometrical emittances and different beta functions at the IP. The enormously large parameter space opened for optimization by allowing these asymmetries, while insurmountable for traditional methods of design and optimisation, may be within reach due to emerging AI-assisted accelerator-design tools.

\section{Future perspectives}
\subsection{Strategy}
\label{sec:Strategy}
If FCC is approved, then HALHF will not proceed in the near-term. However, work on HALHF is necessary to underpin design work towards a 10 TeV collider. If approval for FCC is delayed or considered unachievable, the cause is likely to be the large cost inherent in such a project. This predicates a shift towards a much cheaper machine, or machines. There is no doubt that LEP3 would be much cheaper and affordable within the CERN budget. However, it suffers from the clear lack of perspective beyond a Higgs factory. Such a future perspective could be provided by HALHF, which has many advantages, including a smaller amount of civil construction than other machines. Revision of the HALHF design to produce a significantly cheaper machine would pay dividends. Such a revision of HALHF (v3.0) is under consideration for the future, some ideas for which are outlined in Sec.~\ref{sec:CRS}.

A further possibility is construction of a scaled-down radio-frequency-based linear collider, for which HALHF technology could act as an ``after-burner" to increase its energy. In any of these scenarios, work on HALHF should continue. 

\subsection{Medium-scale demonstrator}
\label{sec:demonstrator}
The SPARTA project~\cite{SPARTA}, initially funded as an ERC Starting Grant, can serve as an excellent starting point for a ``HALHF demonstrator", at least in part. Critical aspects that need to be demonstrated are multistage acceleration (the subject of the SPARTA project) and high-average-power operation. Such a demonstrator is not cheap, so it would be highly advantageous for it to have a ``stand-alone" physics motivation. Strong-field quantum electrodynamics (SFQED) provides such an interesting area of physics. Additional applications of the demonstrator could be to provide a high-energy electron test-beam facility, and directional muon beams such as recently demonstrated at LBNL~\cite{MuonPWFA2025}, which have enormous potential for applications including high-resolution imaging.

A cost-effective option to construct such a demonstrator is to reuse the great majority of elements of the CTF3 linac, which is currently scheduled to be disposed of at CERN. This would be well suited to provide the medium-energy electron beams required as the drive beams for the demonstrator.

This initial phase would also be suitable for testing important effects of the long HALHF-like bunch trains, as the first two stages are envisioned to be fed by drive beams separated by an RF deflector. It would also be vitally important to test prototypes of the plasma cell, which HALHF has already identified as a major area of R\&D for a PWFA-based collider.

\subsubsection{Phased approach}
A second stage of the HALHF demonstrator would lead to an upgrade in drive-beam energy to 750--1000 MeV by installing either more gradient-optimized accelerating structures or by adding additional structures and klystrons to the repurposed CTF3 hardware. This upgrade would allow final energy beams of potentially 20--30 GeV after 16 plasma-accelerator stages. This energy range is useful for both SFQED and the other applications indicated above. However, if the demonstrator uses fast kickers for drive-beam distribution into the plasma cells (which is much simpler than the all-RF-deflector-based concept envisioned for HALHF), this implies that using a HALHF-like bunch pattern (i.e., 16 ns between acceleration events) is precluded. This is because the recovery time of kickers is currently too slow. Nevertheless, useful tests of HALHF-like bunch patterns can be carried out in the first two stages only (in Phase I or II), which use only RF deflectors. The modest repetition rate in Phase II is however still of great interest to physics applications.

The cost estimate for Phase I of the demonstrator is in the range 5--15~MEUR, assuming that the CTF3 accelerator elements can be reused and reassembled at a suitable site, while Phase II would be in the range of 50--200 MEUR. The reuse of the CTF3 equipment saves up to around 10~MEUR.

\subsection{Cost-reduction strategy}
\label{sec:CRS}
An obvious area in which costs for the full HALHF Higgs factory can be reduced is in the BDS. A HALHF constructed in any of the scenarios discussed in Sec.~\ref{sec:Strategy} would not be a stand-alone machine but would co-exist with another Higgs factory of whatever technology, against whose results a single HALHF detector could be cross checked. Thus HALHF 3.0 would revert to the original single-detector/single-IP layout of the original design (as opposed to the dual-IP design of HALHF 2.0). 

The positron acceleration using $\rm{C}^3$ technology is relatively expensive. A system which utilizes an RF-based two-beam acceleration scheme also in the positron arm would certainly be considerably cheaper, even taking into account the long bypass necessary to bring the drive beams to the positron arm. 

Other possible cost savings will also be investigated.

\subsection{Roadmap to CDR}
\label{sec:RtoCDR}
Table 2 in the HALHF submission to the European Strategy Update 2026~\cite{HALHF-EPPSU} estimated the resources required to reach the capability of producing a HALHF TDR, concentrating on the critical R\&D items. Much of this component R\&D can and should be pursued in parallel to the overall design of the HALHF system. A CDR should demonstrate the plausibility, with as much experimental evidence as possible, of all components of HALHF together with an overall design that can be simulated ``cradle to grave" and shown to be in principle capable of meeting the design goals. This is a non-trivial amount of work, unfunded at the moment except for \textit{ad hominem} awards and general funds from the collaborating institutions. 

The statement in Sec.~6D of the Final Recommendations of the EPPSU process states that ``The longer-term development of advanced technologies, such as high-gradient wakefield
acceleration and those underpinning bright muon beams, should be supported at an
appropriate level"~\cite{EPPSU_final}. Such dedicated support, as opposed to the current piecemeal and incoherent approach, would be very welcome. A rough estimate of the resources required to produce a HALHF CDR over the next five years, is shown in Table~\ref{tab:3}, which shows only FTEs required for design; any necessary capital expenditure was included in the estimate given in the EPPSU document, although that is partially superseded by the discussion on a HALHF demonstrator in Sec.~\ref{sec:demonstrator}. Note that some fraction of the resources required in this table will be found from the sources that are currently funding HALHF R\&D. For example, in contrast to Table 2 in the CERN EPPSU document, there is no entry for positron-source R\&D, which has been successfully funded from specific grant awards and which is relevant to many other collider projects. Other developments since the CERN EPPSU submission have included realisation of the importance of cell design, cooling and beam-dump design and coping with radiation dose, which is therefore included explicitly in Table~\ref{tab:3}.
Nevertheless, the overall total is roughly consistent with the EPPSU estimate. 

\begin{table}[h!]
\small
    \begin{tabular}{p{0.55\linewidth}>
    {\centering}p{0.35\linewidth}}%
    \hline
    {\bf R\&D element}& FTE years \cr
    \hline
    PWFA linac & 8--10  \cr
    Cell design \& radiation protection & 7--9 \cr  
    Damping rings & 4--6 \cr
    Optics design & 4--6 \cr
    Beam delivery system  & 2--4 \cr
    Physics simulations at the IP & 2--4 \cr
    System integration & 2--4 \cr
    RF linacs  & 2--4 \cr
    \hline
    \textbf{Total} &  31--47 \cr
    \hline
    \end{tabular}
   \caption{Estimate of effort needed to produce a Conceptual Design Report (CDR).}
      \label{tab:3}
\end{table}

Assuming roughly five years to carry out the work implies 6--9 FTEs/year. Assuming a cost of 120 kCHF/FTEyr gives an estimate of approximately 1M CHF/yr and a total of 5M CHF.

\vspace{0.5cm}
\section*{Acknowledgments}

The HALHF Collaboration is grateful for the support of its funding authorities. In particular, this work was supported by the Research Council of Norway (NFR Grant No.~313770), the European Research Council ERC-2023-StG call under grant agreement No 101116161 SPARTA. Colleagues from the John Adams and Cockcroft Institutes and Rutherford Appleton Laboratory acknowledge support from the Science and Technology Facilities Council, UK. This research was funded in whole, or in part, by the UKRI. For the purpose of Open Access, the authors have applied a CC BY public copyright licence to any Author Accepted Manuscript version arising from this submission.The authors are also grateful for continuing support from DESY. Authors from institutions in the United States of America acknowledge the support of the US Department of Energy. 
\vspace{1cm}

\appendix

\hspace{-0.7cm} 
{\huge {\bf Appendices}} 

\section{{\sc GuineaPig} Steering Files}
\label{gp_steeringfile}

The following is the steering file used for the {\sc GuineaPig} simulations. While this steering file corresponds to the 250 GeV configuration, it can easily be converted to other configurations via changing the beam energies and the ecm\_min parameter accordingly.

\begin{minted}[frame=lines, linenos, fontsize=\small]{c++}
$ACCELERATOR:: asym
{
energy.1=375;energy.2=41.7;
espread.1=0.0015;espread.2=0.0015;
which_espread=3;
particles.1=1.0;particles.2=3.0;
charge_sign=-1;
beta_x.1=3.3;beta_y.1=0.1;
beta_x.2=3.3;beta_y.2=0.1;
emitt_x.1=90.0;emitt_y.1=0.32;
emitt_x.2=10.0;emitt_y.2=0.035;
sigma_z.1=150;sigma_z.2=150;
f_rep=100.0;n_b=160;
trav_focus=0;
scale_step=1.0;
offset_x=0.0;offset_y=0.0;
waist_x=0.0;waist_y=0.5*sigma_z.1;
angle_y=0.0;
angle_phi=0.0;
}

$PARAMETERS:: par
{
n_x=128;n_y=128;n_z=128;
n_t=10;n_m=640000;
cut_x=4*sigma_x.1;
cut_y=4*sigma_y.1;
cut_z=3*sigma_z.1;
integration_method=2 ;
force_symmetric=0;
rndm_seed=#seed#;rndm_load=0;rndm_save=0;
do_lumi=7;
num_lumi=500000;lumi_p=1e-29;
num_lumi_gg=500000;lumi_p_gg=1e-29;
num_lumi_eg=500000;lumi_p_eg=1e-29;
load_beam=3;store_beam=1;
ecm_min=0.99*250;
electron_ratio=1;
do_photons=1;store_photons=1;photon_ratio=1;
do_hadrons=1;store_hadrons=0;hadron_ratio=10000;
do_jets=0;store_jets=0;jet_ratio=10000;
do_pairs=1;store_pairs=1;pair_ratio=1.0;
pair_q2=2;track_pairs=1;pair_ecut = 0.0005;
grids=7;
beam_size=1;
do_eloss=1;do_espread=1;do_isr=1;do_coherent=1;
do_compt=1;compt_emax=1.1*375;compt_x_min=0.01;
do_dump=0;dump_step=1;dump_particle=1;
}

\end{minted}

\section{{\sc WarpX} Steering Files}
\subsection{{\sc WarpX} Input File}
\label{warpx_steeringfile}

Following is an example steering file used for the {\sc WarpX} simulations at 250 GeV.

\begin{minted}[frame=lines, linenos, fontsize=\small]{text}
#################################
########## MY CONSTANTS #########
#################################
my_constants.nano = 1.0e-9
my_constants.giga = 1.0e9

warpx.random_seed = 0

# DIAGS

my_constants.bin_num_1d = 520
my_constants.bin_center_min_1d = 0.
my_constants.bin_center_max_1d = 260*giga
my_constants.bin_size_1d = (bin_center_max_1d - bin_center_min_1d) / bin_num_1d
my_constants.bin_edge_min_1d = bin_center_min_1d - 0.5 * bin_size_1d
my_constants.bin_edge_max_1d = bin_center_max_1d + 0.5 * bin_size_1d
my_constants.bin_num_1d_eff = 521


# SIGMAS

my_constants.sigmaxinj = 652*nano #645*nano
my_constants.sigmayinj = 45*nano #35*nano
my_constants.sigmaz = 150e-6


# BOX
my_constants.Lx = 10*sigmaxinj
my_constants.Ly = 10*sigmayinj
my_constants.Lz = 20*sigmaz    

my_constants.nx = 512
my_constants.ny = 1024
my_constants.nz = 512

# TIME

my_constants.nt = 6400
max_step = nt               
my_constants.dt = sigmaz/clight/640.

#################################
####### GENERAL PARAMETERS ######
#################################
amr.n_cell = nx ny nz
amr.max_grid_size = 512
amr.blocking_factor = 2
amr.max_level = 0
geometry.dims = 3
geometry.prob_lo = -0.5*Lx -0.5*Ly -0.5*Lz
geometry.prob_hi =  0.5*Lx  0.5*Ly  0.5*Lz

#################################
######## BOUNDARY CONDITION #####
#################################
boundary.field_lo = open open open
boundary.field_hi = open open open
boundary.particle_lo = Absorbing Absorbing Absorbing
boundary.particle_hi = Absorbing Absorbing Absorbing

#################################
############ NUMERICS ###########
#################################
warpx.do_electrostatic = relativistic
warpx.const_dt = dt
warpx.grid_type = collocated
algo.particle_shape = 3
algo.load_balance_intervals=100
algo.particle_pusher = vay
warpx.poisson_solver = fft
warpx.use_2d_slices_fft_solver = 1

#################################
########### PARTICLES ###########
#################################
particles.species_names = beam1 beam2 pho1 pho2 ele1 ele2 pos1 pos2 
particles.photon_species = pho1 pho2 

beam1.species_type = electron
beam1.injection_style = external_file
beam1.injection_file = openpmd_generated_electrons.h5
beam1.z_shift = -5*150e-6

beam1.initialize_self_fields = 1
beam1.self_fields_required_precision = 5e-10
beam1.self_fields_max_iters = 10000
beam1.do_qed_quantum_sync = 1
beam1.qed_quantum_sync_phot_product_species = pho1
beam1.do_classical_radiation_reaction = 0

beam2.species_type = positron
beam2.injection_style = external_file
beam2.injection_file = openpmd_generated_positrons.h5
beam2.z_shift = 5*150e-6

beam2.initialize_self_fields = 1
beam2.self_fields_required_precision = 5e-10
beam2.self_fields_max_iters = 10000
beam2.do_qed_quantum_sync = 1
beam2.qed_quantum_sync_phot_product_species = pho2
beam2.do_classical_radiation_reaction = 0

pho1.species_type = photon
pho1.injection_style = none
pho1.do_qed_breit_wheeler = 1
pho1.qed_breit_wheeler_ele_product_species = ele1
pho1.qed_breit_wheeler_pos_product_species = pos1

pho2.species_type = photon
pho2.injection_style = none
pho2.do_qed_breit_wheeler = 1
pho2.qed_breit_wheeler_ele_product_species = ele2
pho2.qed_breit_wheeler_pos_product_species = pos2

ele1.species_type = electron
ele1.injection_style = none
ele1.self_fields_required_precision = 1e-11
ele1.self_fields_max_iters = 10000
ele1.do_qed_quantum_sync = 1
ele1.qed_quantum_sync_phot_product_species = pho1
ele1.do_classical_radiation_reaction = 0

pos1.species_type = positron
pos1.injection_style = none
pos1.self_fields_required_precision = 1e-11
pos1.self_fields_max_iters = 10000
pos1.do_qed_quantum_sync = 1
pos1.qed_quantum_sync_phot_product_species = pho1
pos1.do_classical_radiation_reaction = 0

ele2.species_type = electron
ele2.injection_style = none
ele2.self_fields_required_precision = 1e-11
ele2.self_fields_max_iters = 10000
ele2.do_qed_quantum_sync = 1
ele2.qed_quantum_sync_phot_product_species = pho2
ele2.do_classical_radiation_reaction = 0

pos2.species_type = positron
pos2.injection_style = none
pos2.self_fields_required_precision = 1e-11
pos2.self_fields_max_iters = 10000
pos2.do_qed_quantum_sync = 1
pos2.qed_quantum_sync_phot_product_species = pho2
pos2.do_classical_radiation_reaction = 0


#################################
############# QED ###############
#################################
qed_qs.photon_creation_energy_threshold = 0.

qed_bw.chi_min = 1.e-2

qed_bw.lookup_table_mode = generate
qed_bw.tab_dndt_chi_min = 1.e-2
qed_bw.tab_dndt_chi_max = 5
qed_bw.tab_dndt_how_many = 512
qed_bw.tab_pair_chi_min = 1.e-2
ed_bw.tab_pair_chi_max = 5
qed_bw.tab_pair_chi_how_many = 512
qed_bw.tab_pair_frac_how_many = 512

qed_bw.save_table_in = my_bw_table.txt

qed_qs.chi_min = 1.e-3

qed_qs.lookup_table_mode = generate
qed_qs.tab_dndt_chi_min = 1.e-3
qed_qs.tab_dndt_chi_max = 5
qed_qs.tab_dndt_how_many = 512
qed_qs.tab_em_chi_min = 1.e-3
qed_qs.tab_em_chi_max = 5
qed_qs.tab_em_chi_how_many = 512
qed_qs.tab_em_frac_how_many = 512
qed_qs.tab_em_frac_min = 1.e-12
qed_qs.save_table_in = my_qs_table.txt


#################################
######### DIAGNOSTICS ###########
#################################

# REDUCED

warpx.reduced_diags_names = DiffLumi_beam1_beam2 

DiffLumi_beam1_beam2.type = DifferentialLuminosity
DiffLumi_beam1_beam2.intervals = 6400
DiffLumi_beam1_beam2.species = beam1 beam2
DiffLumi_beam1_beam2.bin_number = bin_num_1d_eff
DiffLumi_beam1_beam2.bin_max = bin_edge_max_1d
DiffLumi_beam1_beam2.bin_min = bin_edge_min_1d

\end{minted}

\subsection{Beam Generation Script for {\sc WarpX} Simulations}
\label{warpx_beamgenscript}

Following is the python script used to generate the electron beam for the {\sc WarpX} simulations at 250 GeV.
\begin{minted}[frame=lines, linenos, fontsize=\small]{python}
"""
Generates a Twiss-matched positron beam for WarpX HALHF sims.

Includes waist shift in y direction
"""

import numpy as np
from scipy.constants import c, e, m_e
r_e =   2.8179403205e-15

np.random.seed(0) 

# Physical constants 
mc2 = m_e * c * c

# Beam parameters
GeV           = e * 1e9
energy        = 375.0 * GeV
gamma         = energy / mc2          
npart         = 1e10
nmacropart    = 10_000_000
charge        = e * npart

# Fractional energy spread (rms)
delta_E = 0.15e-2   # 0.15%

# IP beam sizes (rms)
sigmax_star   = 636e-9               # m
sigmay_star   = 6.6e-9               # m
sigmaz        = 150e-6               # m

# Normalised emittances
emit_nx       = 90e-6                # m rad 
emit_ny       = 0.32e-6             # m rad

# Geometric emittances
eps_geo_x     = emit_nx / gamma
eps_geo_y     = emit_ny / gamma

# Beta functions at IP
beta_x_star   = sigmax_star**2 / eps_geo_x
beta_y_star   = sigmay_star**2 / eps_geo_y

# Twiss transport from IP to injection point
# After a drift of length d:
#   beta_init  = beta* + d^2 / beta*   =  beta* * (1 + alpha_init^2)
#   alpha_init = d / beta*
#   gamma_init = (1 + alpha_init^2) / beta*

focal_distance_x = 5 * sigmaz          #must match z_shift in WarpX
focal_distance_y = 5 * sigmaz - 75e-6
d              = focal_distance_x

alpha_x_init  = focal_distance_x / beta_x_star
alpha_y_init  = focal_distance_y / beta_y_star
beta_x_init   = beta_x_star * (1 + alpha_x_init**2)
beta_y_init   = beta_y_star * (1 + alpha_y_init**2)
gamma_x_init  = (1 + alpha_x_init**2) / beta_x_init
gamma_y_init  = (1 + alpha_y_init**2) / beta_y_init

# Beam sizes at injection point 
sigmax_init   = np.sqrt(eps_geo_x * beta_x_init)
sigmay_init   = np.sqrt(eps_geo_y * beta_y_init)

# Generate beams
#
#   x ~ N(0, sqrt(eps_geo * beta)) using sigma= sqrt(eps_geo * beta)

#   p ~ N(0, 1) then p = -alpha/beta * x + sqrt(eps_geo/beta) * p_uncorr 
#
# The momentum here is the GEOMETRIC divergence x' = px/(gamma*m*c)
# multiply by gamma to get the unitless gamma*beta_x stored by WarpX

# Independent Gaussians
rx = np.random.normal(0, 1, nmacropart)
ry = np.random.normal(0, 1, nmacropart)
rpx = np.random.normal(0, 1, nmacropart)
rpy = np.random.normal(0, 1, nmacropart)

# Positions (m)
x = np.sqrt(eps_geo_x * beta_x_init) * rx
y = np.sqrt(eps_geo_y * beta_y_init) * ry
z = np.random.normal(0, sigmaz, nmacropart)

# Geometric divergences (rad), with Twiss correlation applied
xp = (-alpha_x_init / beta_x_init) * x + np.sqrt(eps_geo_x / beta_x_init) * rpx
yp = (-alpha_y_init / beta_y_init) * y + np.sqrt(eps_geo_y / beta_y_init) * rpy

# Convert to unitless gamma*beta momenta
ux = gamma * xp
uy = gamma * yp
sigma_uz = gamma * delta_E
uz = np.random.normal(gamma, sigma_uz, nmacropart) 



# Write to openPMD 
from openpmd_api import (
    Access_Type,
    Dataset,
    Mesh_Record_Component,
    Series,
)

SCALAR = Mesh_Record_Component.SCALAR

f = Series("openpmd_generated_electrons.h5", Access_Type.create)
f.set_meshes_path("fields")
f.set_particles_path("particles")
cur_it = f.iterations[0]
electrons = cur_it.particles["electrons"]

d_set = Dataset(np.dtype("float64"), extent=(nmacropart,))

electrons["weighting"].reset_dataset(d_set)
electrons["weighting"][SCALAR].make_constant(npart / nmacropart)

electrons["mass"].reset_dataset(d_set)
electrons["mass"][SCALAR].make_constant(m_e)
electrons["charge"].reset_dataset(d_set)
electrons["charge"][SCALAR].make_constant(-e)

for comp, arr in [("x", x), ("y", y), ("z", z)]:
    electrons["position"][comp].reset_dataset(d_set)
    electrons["position"][comp].store_chunk(arr)

for comp, arr in [("x", ux), ("y", uy), ("z", uz)]:
    electrons["momentum"][comp].reset_dataset(d_set)
    electrons["momentum"][comp].store_chunk(arr)
    electrons["momentum"][comp].unit_SI = m_e * c  # gamma*beta -> SI

for comp in ["x", "y", "z"]:
    electrons["positionOffset"][comp].reset_dataset(d_set)
    electrons["positionOffset"][comp].make_constant(0.0)

f.flush()
del f
print("\nParticles written to openpmd_generated_electrons.h5")]

\end{minted}

\bibliographystyle{elsarticle-harv} 

\end{document}